# ELECTROMAGNETIC QUANTUM GRAVITY

## The Quantum Principle of Equivalence, Quantum Inertia, and the Meaning of Mass.

**Mike Trushyk & Tom Ostoma**

48 O'HARA PLACE, Brampton, Ontario, L6Y 3R8
emqg@rogerswave.ca

Friday, February 12, 1999

### ACKNOWLEGMENTS

We would like to also thank G. Gomes for the many lunch time discussions on the nature of quantum theory and space-time. We wish to thank R. Mongrain (P. Eng.) for his encouragement, constructive criticisms, and for the lengthy conversations on the nature of space, time, light, matter, and CA theory, and special thanks to L. Walker for his work in proof reading this document.

# ABSTRACT

*This paper provides a new approach to quantum gravity called Electro-Magnetic Quantum Gravity (EMQG) is described. It is manifestly compatible with Cellular Automata (CA) theory (ref. 1), and is based on a new theory of inertia (ref. 5) proposed by R. Haisch, A. Rueda, and H. Puthoff (which we modified and called Quantum Inertia). They show that Newtonian Inertia is due to the strictly local electromagnetic force interactions of matter (quantum particles) with the surrounding charged virtual particles of the quantum vacuum. The sum of all the tiny electromagnetic forces originating from each charged particle in the mass with respect to the vacuum, is the source of the <u>total inertial force</u> of a mass which opposes accelerated motion in Newton's law 'F = MA'. Their theory also resolves the problems and paradoxes of accelerated motion introduced in Mach's principle, by suggesting that the state of acceleration of the charged virtual particles of the quantum vacuum (with respect to a mass) serves as Newton's universal reference frame for the mass, which Newton called 'absolute' space. The (net statistical) acceleration of the charged virtual particles of the quantum vacuum (with respect to some test mass) can be used as an absolute reference frame to gauge inertial mass. Therefore, this frame can be used to define both absolute acceleration and <u>absolute mass</u>, which is equivalent to relativistic rest mass. However, this frame <u>cannot</u> be used to gauge absolute state of motion of an inertial reference frame. Thus, Einstein's principle of relativity is still applicable for inertial frames (frames of constant velocity motion, or where Newton's law of inertia applies). The special relativistic treatment of inertial force, acceleration, and inertial mass is revised here to acknowledge the existence of absolute mass. We found that the special relativistic variation of mass with relative velocity $m = m_0 (1 - v^2/c^2)^{-1/2}$ is actually caused by <u>the decrease in the effectiveness of the applied force,</u> where the applied force and destination mass have a large relative velocity 'v'. We show that this decrease of the applied force is caused by a relativistic timing effect of the received force exchange particles, which alters the received flux.*

*Einstein's principle of equivalence of inertial and gravitational mass was invoked to understand the origin of gravitational mass from the perspective of quantum inertia. We found that gravity also involves the same 'inertial' electromagnetic force component that exists for inertial mass. We propose that the general relativistic Weak Equivalence Principle (WEP) is a physical phenomenon, originating from common 'lower level' quantum processes occurring in both gravitational mass and inertial mass. Gravitational mass results from the interactions of <u>both</u> the electromagnetic force (photon exchanges) and the <u>pure</u> gravitational force (graviton exchanges) on matter, acting <u>simultaneously</u>. However, inertial mass is strictly the result of the electromagnetic force process given in quantum inertia (with negligible graviton processes present). For a gravitational test mass near the earth, the graviton exchange process occurring between the earth, the test mass, and the surrounding vacuum particles upsets perfect equivalence of inertial and gravitational mass, with the gravitational mass being slightly larger than inertial mass. Similarly, if a large and a tiny test mass are dropped simultaneously on the earth, the larger mass falls faster by a minute amount. The tiny deviation from perfect equivalence, known as the 'Ostoma-Trushyk' effect, might be detected experimentally. Of course any violation of the WEP also necessarily implies that the strong equivalence principle is no longer applicable.*

*All elementary (fermion) particles get their quantum mass numbers from combinations of just <u>one</u> fundamental matter (and anti-matter) particle called the 'masseon' particle. The masseon has one, fixed (smallest) quanta of mass (similar to the idea of a quanta of electric charge), which we call low level 'mass charge'. The masseon also carries either a positive or negative (smallest) quanta of electric charge. Furthermore, we propose a new universal constant "i", defined as the inertial force existing between the quantum vacuum and a single charged masseon particle accelerating at 1 g. At 1g, this force is the smallest possible quanta of inertial <u>and</u> gravitational force. The masseon particle generates a fixed flux of gravitons, with a flux rate being <u>completely unaffected</u> by relativistic motion (called 'low level mass'). The graviton is the vector boson exchange particle of the <u>pure</u> gravitational force interaction. In EMQG, the graviton exchanges are physically similar to the photon exchanges in QED, with the same concept of positive and negative gravitational 'mass charge' carried by masseons and*

*anti-masseons. The ratio of the graviton to photon exchange force coupling is about $10^{-40}$. Near a large mass like the earth, we find that photon exchanges occur between the charged masseon particles of a test mass and the surrounding (falling) quantum vacuum virtual particles. We also find graviton exchanges occurring between the earth and the surrounding virtual particles of the quantum vacuum, and also directly between the earth and the test mass. The photon exchange process is responsible for the equivalence of inertial and gravitational mass. The graviton exchange process is responsible for the distortion of the (statistical average) acceleration vectors of the virtual particles of the quantum vacuum near the earth. There are equal numbers of virtual masseon and anti-masseon particles existing in the quantum vacuum everywhere, and at any given instant of time. This is why the cosmological constant is almost zero, in the universe; there is an equal proportion of attractive and repulsive gravitational forces in the quantum vacuum..*

*Now, 4D curved Minkowski space-time is a consequence of the behavior of matter (particles) and energy (photons) under the influence of this (statistical average) downward accelerated 'flow' of charged virtual particles of the quantum vacuum. This coordinated 'accelerated flow' of the virtual particles can be thought of as a special 'Fizeau-like fluid' that 'flows' through all matter near a gravitational field (and also in matter undergoing accelerated motion). Like in the Fizeau experiment (which was performed with a constant velocity water flow) the behavior of photons, clocks, and rulers are now affected by the accelerated flow of the virtual particles of the quantum vacuum. Einstein interpreted this phenomenon as being the result of 4D Minkowski space-time curvature. In EMQG, we take an alternative interpretation. Space-time measurements, which are based on measuring instruments made of matter, are now affected by the action of this accelerated 'Fizeau-like fluid'. At tiny quantum distance scales, there exists a kind of secondary (quantized) absolute 3D space and separate absolute (quantized) time as required by CA theory. This is represented by a rectangular array of numbers or cells; C(x,y,z). These cells change state after every new CA clock operation Dt. The array of numbers C(x,y,z) is called CA space, which acts like the Newtonian version of Cartesian absolute space. There also exists a separate absolute time needed to evolve the numerical state of the CA. CA space and time are not effected by any physical interactions, and are also not accessible through direct measurement. In CA space, the Plank distance ($1.6 \times 10^{-35}$ meters) roughly corresponds to the minimum event distance (or cell 'size'), and Plank time ($5.4 \times 10^{-44}$ sec) to the minimum time interval possible. We show that any CA model automatically leads to a maximum speed limit for the transfer of information from place to place in CA space, and hence leads to strict physical locality of all interactions. Furthermore, we show that the Lorentz transformation follows mathematically from CA theory. Hence, Minkowski 4D flat space-time of special relativity can be seen as the direct consequence of universal Cellular Automata, as seen by inertial observers, who are not aware, and cannot measure the absolute units of CA space and time.*

*Gravitational Waves (GW) are not directly quantized by the gravitons. A periodic accelerating mass causes a corresponding periodic variation in the graviton flux at any point surrounding the mass. This is responsible for the initial periodic disturbance in the (net statistical) average acceleration vectors of the virtual particles of the quantum vacuum (masseons and anti-masseons in the immediate vicinity of the mass). This periodic disturbance of the average acceleration vectors of the nearby virtual particles with respect to the original mass is the origin of the gravitational wave, which can carry energy to a very distant detector. The GW carries a large energy density (due to the huge numbers of virtual particles involved), which is quite capable of explaining the large stiffness of the GW predicted by general relativity (ref. 13). Once the periodic virtual particle disturbance is started, it is self-propagating by the mutual quantum vacuum virtual particles electromagnetic force interactions. The undisturbed acceleration vectors of the virtual particles ahead of the GW, now becomes disturbed by the electromagnetic force interaction when the GW arrives. Thus, in EMQG, the outward propagating GW is really a time-varying periodic increase and decreases in the net acceleration of the virtual particle acceleration vectors (with respect to the original mass), that was started by the time varying graviton flux. Once the GW is initiated, however, it is self-sustaining as is propagates throughout space electromagnetically. A very simple physical interpretation of the Lense-Thirring effect is given based on the propagation of gravitons and virtual particles. The finite speed of the graviton allows a massive rotating object to carry an observer a small amount during rotation. The virtual particles take the same*

*path as the gravitons. Since 4D space-time curvature is in the direction of the virtual particle flow, inertial frames appear to be dragged by the earth by 42.5 milli-arcseconds.*

# **TABLE OF CONTENTS**

# CONCISE SUMMARY OF ELECTROMAGNETIC QUANTUM GRAVITY

*"Experiment has provided numerous facts justifying the following generalization: absolute motion of matter, or, to be more precise, the relative motion of weighable matter and ether, cannot be disclosed. All that can be done is to reveal the motion of weighable matter with respect to weighable matter…"*

<div align="right">H. Poincare (1895)</div>

*"The interpretation of geometry advocated here cannot be directly applied to submolecular spaces … it might turn out that such an extrapolation is just as incorrect as an extension of the concept of temperature to particles of a solid of molecular dimensions"*

<div align="right">A. Einstein (1921)</div>

## Introduction to EMQG Theory

A new approach to the unification of quantum theory with general relativity referred to as Electro-Magnetic Quantum Gravity (EMQG)) is described. It has its origins in Cellular Automata (CA) theory (ref. 1,2,3, and 4), and is also based on the new theory of inertia that has been proposed by R. Haisch, A. Rueda, and H. Puthoff (ref. 5) known here as the HRP Inertia theory. These authors suggested that inertia is due to the strictly local force interactions of charged matter particles with their immediate background virtual particles of the quantum vacuum. They found that inertia is caused by the magnetic component of the Lorentz force, which arises between what the author's call the charged 'parton' particles in an accelerated reference frame interacting with the background quantum vacuum virtual particles. The sum of all these tiny forces in this process is the source of the resistance force opposing accelerated motion in Newton's F=MA. We have found it necessary to make a small modification of HRP Inertia theory as a result of our investigation of the principle of equivalence. The modified version of HRP inertia is called "Quantum Inertia" (or QI). In EMQG, a new elementary particle is required to fully understand inertia, gravitation, and the principle of equivalence (described in the next section). This theory also resolves the long outstanding problems and paradoxes of accelerated motion introduced by Mach's principle, by suggesting that the vacuum particles themselves serve as Mach's universal reference frame (for <u>acceleration</u> only), without violating the principle of relativity of constant velocity motion. In other words, our universe offers no observable reference frame to gauge inertial frames, because the quantum vacuum offers no means to determine absolute constant velocity motion. However for accelerated motion, the quantum vacuum plays a very important role by offering a resistance to acceleration, which results in an inertial force opposing the acceleration of the mass. Thus, the very existence of inertial force reveals the absolute value of the acceleration with respect to the net statistical average acceleration of the virtual particles of the quantum vacuum. Reference 14 offers an excellent introduction to the motion of matter in the vacuum, and on the history of the virtual particles of the quantum vacuum.

There have been various clues to the importance of the state of the virtual particles of the quantum vacuum, with respect to the accelerated motion of real charged particles. One example is the so-called Davies-Unruh effect (ref. 15), where uniform and linearly accelerated charged particles in the vacuum are immersed in a heat bath, with a temperature proportional to acceleration (with the scale of the quantum heat effects being very low). However, the work of reference 5 is the first place we have clearly seen the identification of inertial forces as the direct consequence of the interactions of real matter particles with the virtual particles of the quantum vacuum.

There has also been a suggestion that the virtual particles of the quantum vacuum are involved in gravitational interactions. The prominent Russian physicist A. Sakharov proposed in 1968 (ref. 16) that Newtonian gravity could be interpreted as a van der Waals type of force induced by the electromagnetic fluctuations of the virtual particles of the quantum vacuum. Sakharov visualized ordinary neutral matter as a collection of electromagnetically, interacting polarizable particles made of charged point-mass subparticles (partons). He associated the Newtonian gravitational field with the Van Der Waals force present in neutral matter, where the long-range radiation fields are generated by the parton 'Zitterbewegung'. Sakharov went on to develop what he called the 'metric elasticity' concept, where space-time is somehow identified with the 'hydrodynamic elasticity' of the vacuum. However, he did not understand the important clues offered by the equivalence principle, nor the role that the quantum vacuum played in inertia and Mach's principle.

In 1974, Hawkings (ref. 17) announced that black holes are not completely black. Black holes emit an outgoing thermal flux of radiation due to gravitational interactions of the black hole with the virtual particle pairs created in the quantum vacuum near the event horizon. At first sight, the emission of thermal radiation from a black hole seems paradoxical (since nothing can escape from the event horizon). However, the spontaneous creation of virtual particle and anti-particle pairs in the quantum vacuum near the event horizon can be used to explain this effect (ref. 18). Heuristically, one can imagine that the virtual particle pairs (created with wavelength $\lambda$, approximately equal to the size of the black hole) 'tunnel' out of the event horizon. For virtual particle with wavelength comparable to the size of the hole, strong tidal forces operate to prevent re-annihilation. One virtual particle escapes to infinity with positive energy to contribute to the Hawking radiation, while the corresponding antiparticle enters the black hole to be trapped forever by the deep gravitational potential. Thus, the quantum vacuum is important in order to understand Hawking Radiation. In EMQG, the quantum vacuum plays an *extremely* important role in both inertia and gravitation. Anybody who believes in the existence of the virtual particles of the quantum vacuum, and accepts the fact that many of them carry mass (virtual fermions in particular) will have no trouble in believing that they are falling in the presence of a large gravitational mass. The existence of the downward accelerating virtual particles (during their brief lifetime) under gravitational fields turns out to be a central theme of EMQG. This idea turns out to be the missing link between inertia and gravity, and leads us directly to a full understanding of the principle of equivalence.

### EMQG and the Quantum Theory of Inertia

EMQG theory presents a unified approach to Inertia, Gravity, the Principle of Equivalence, Space-Time Curvature, Gravitational Waves, and Mach's Principle. These apparently different phenomena are the common results of the quantum interactions of the real (charged) matter particles (of a mass) with the surrounding virtual particles of the quantum vacuum through the exchange of two force particles: the photon and the graviton. Furthermore, the problem of the cosmological constant is solved automatically in the framework of EMQG. This new approach to quantum gravity is definitely *non-geometric* on the tiniest of distance scales (Plank Scales of distance and time). This is because the large scale relativistic 4D space-time curvature is caused purely by the accelerated state of virtual particles of the quantum vacuum with respect to a mass, and their discrete interactions with real matter particles of a mass through the particle force exchange process. Because of this departure from a universe with fundamentally curved space-time, EMQG is a complete change in paradigm over conventional gravitational physics. This paper should be considered as a framework, or outline of a new approach to gravitational physics that will hopefully lead to a full theory of quantum gravity.

We modified the HRP theory of Inertia (ref. 5) based on our detailed examination of the principle of equivalence. In EMQG, the modified version of inertia is known as "Quantum Inertia", or QI. In EMQG, a new elementary particle is required to fully understand inertia, gravitation, and the principle of equivalence. All matter, including electrons and quarks, must be made of nature's most fundamental mass unit or particle which we call the 'masseon' particle. These particles contain one fixed, fundamental 'quanta' of both inertial and gravitational mass. The masseons also carry one basic, smallest unit or quanta of electrical charge as well, of which they can be either positive or negative. Masseons exist in the particle or in the anti-particle form (called anti-masseon), that can appear at random in the vacuum as virtual masseon/anti-masseon particle pairs of opposite electric charge and opposite 'mass charge'. The earth consists of ordinary masseons (with no anti-masseons), of which there are equal numbers of positive and negative electric charge varieties. In HRP Inertia theory, the electrically charged 'parton' particles (that make up an inertial mass in an accelerated reference frame) interact with the background vacuum electromagnetic zero-point-field (or virtual photons) creating a resistance to acceleration called inertia. We have modified this slightly by postulating that the real masseons (that make up an accelerating mass) interacts with the surrounding, virtual masseons of the quantum vacuum, electromagnetically (although the details of this process are still not fully understood). The properties of the masseon particle and gravitons are developed later.

Thanks to Quantum Inertia, we now have a new and clearer understanding of Newtonian momentum. We maintain that under QI theory, it is force that is truly a fundamental concept, not momentum (or conservation of momentum). The Newtonian momentum, which is defined by 'mv', is simply a bookkeeping value used to keep track of the inertial mass 'm' (defined as F/A) in the state of constant velocity motion 'v' with respect to another mass that it might collide with at a future time. In this way, momentum is a relative quantity. Momentum simply represents information (with respect to some other frame) about what will happen in later (possible) force reactions. This fits in with the fact that inertial mass cannot be measured for constant velocity mass in motion (in an inertial frame for example, away from all other masses) without introducing test accelerations. If a mass is moving at a constant velocity, there are no forces present from the vacuum. Furthermore, momentum requires some other reference frame to gauge the velocity 'v'. The higher the velocity of the mass 'm', the greater will be the subsequent deceleration (and the greater the inertial force present during a later collision when it meets with another object). If the velocity doubles with respect to a wall ahead, for example, then the deceleration doubles in a later impact. Before doubling the velocity, the acceleration $a_0 = (v_0 - 0)/t$; and after doubling, $a = (2v_0 - 0)/t = 2a_0$. Therefore we find that $f = 2f_0$, the force required from the wall (assuming the time of collision is the same). Similarly, if the mass is doubled, the force required from the wall doubles, or $f=2f_0$. Recall that inertial force comes from the opposition of the quantum vacuum to the acceleration of mass (or deceleration as in this case). Similarly, the kinetic energy '$1/2mv^2$' of a mass moving at a constant relative velocity 'v', is also a bookkeeping parameter (defined as the product of force and the time a force is applied). This quantity keeps track of the subsequent energy reactions that a mass will have when later accelerations (or decelerations) occur with respect to some other mass. It is important to remember that it is the electromagnetic quantum vacuum force that acts against an inertial mass to oppose any change in its velocity. To summarize, momentum simply keeps track of any possible future inertial force reactions with the quantum vacuum, with respect to a mass and a pair of observers.

### EMQG and the Quantum Origin of Newton's Laws of Motion

We are now in a position to understand the quantum nature of Newton's classical laws of motion. According to the standard textbooks of physics (ref. 19) Newton's three laws of laws of motion are:

(1) **An object at rest will remain at rest and an object in motion will continue in motion with a constant velocity unless it experiences a net external force.**
(2) **The acceleration of an object is directly proportional to the resultant force acting on it and inversely proportional to its mass. Mathematically: $\Sigma F = ma$, where F and a are vectors.**
(3) **If two bodies interact, the force exerted on body 1 by body 2 is equal to and opposite the force exerted on body 2 by body 1. Mathematically: $F_{12} = -F_{21}$.**

Newton's first law explains what happens to a mass when the resultant of all external forces on it is zero. Newton's second law explains what happens to a mass when there is a nonzero resultant force acting on it. Newton's third law tells us that forces always come in pairs. In other words, a single isolated force cannot exist. The force that body 1 exerts on body 2 is called the action force, and the force of body 2 on body 1 is called the reaction force.

In the framework of EMQG theory, Newton's first two laws are the direct consequence of the (electromagnetic) force interaction of the (charged) elementary particles of the mass interacting with the (charged) virtual particles of the quantum vacuum. Newton's third law of motion is the direct consequence of the fact that all forces are the end result of a boson particle exchange process.

**Newton's First Law of Motion:**

In EMQG, the first law is a trivial result, which follows directly from the quantum principle of inertia (postulate #3). First a mass is at relative rest with respect to an observer in deep space. If no external forces act on the mass, the (charged) elementary particles that make up the mass maintain a *net acceleration* of zero with respect to the (charged) virtual particles of the quantum vacuum through the electromagnetic force exchange process. This means that no change in velocity is possible (zero acceleration) and the mass remains at rest. Secondly, a mass has some given constant velocity with respect to an observer in deep space. If no external forces act on the mass, the (charged) elementary particles that make up the mass also maintain a *net acceleration* of zero with respect to the (charged) virtual particles of the quantum vacuum through the electromagnetic force exchange process. Again, no change in velocity is possible (zero acceleration) and the mass remains at the same constant velocity.

**Newton's Second Law of Motion:**

In EMQG, the second law <u>is</u> the quantum theory of inertia discussed above. Basically the state of *relative* acceleration of the charged virtual particles of the quantum vacuum with respect to the charged particles of the mass is what is responsible for the inertial force. By this we mean that it is the tiny (electromagnetic) force contributed by each mass particle undergoing an acceleration 'A', with respect to the net statistical average of the virtual particles of the quantum vacuum, that results in the property of inertia possessed by all masses. The sum of all these tiny (electromagnetic) forces contributed from each charged particle of the mass (from the vacuum) <u>is</u> the source of the total inertial resistance force opposing accelerated motion in Newton's F=MA. Therefore, inertial mass 'M' of a mass simply represents the total resistance to acceleration of all the mass particles.

**Newton's Third Law of Motion:**

According to the boson force particle exchange paradigm (originated from QED) all forces (including gravity, as we shall see) result from particle exchanges. Therefore, the force that body 1 exerts on body 2 (called the action force), is the result of the emission of force exchange particles from (the charged particles that make up) body 1, which are readily absorbed by (the charged particles that make up) body 2, resulting in a force acting on body 2. Similarly, the force of body 2 on body 1 (called the reaction force), is the result of the absorption of force exchange particles that are originating from (the charged particles that make up) body 2, and received by (the charged particles that make up) body 1, resulting in a force acting on body 1. An important property of charge is the ability to readily emit <u>and</u> absorb boson force exchange particles. Therefore, body 1 is both an emitter and also an absorber of the force exchange particles. Similarly, body 2 is also both an emitter and an absorber of the force exchange particles. This is the reason that there is both an action and reaction force. For example, the contact forces (the mechanical forces that Newton was thinking of when he formulated this law) that results from a person pushing on a mass (and the reaction force from the mass pushing on the person) is really the exchange of photon particles from the charged electrons bound to the atoms of the person's hand and the charged electrons bound to the atoms of the mass on the quantum level. Therefore, on the quantum level there is really is no contact here. The hand gets very close to the mass, but does not actually touch. The electrons exchange photons among each other. The force exchange process works both directions in equal numbers, because all the electrons in the hand and in the mass are electrically charged and therefore the exchange process gives forces that are equal and opposite in both directions.

**Quantum Inertia and Special Relativity**

Special and General Relativity were developed before (new version) quantum theory, and before the realization that the universe consists *mostly* of virtual particles of the quantum vacuum (with real matter making up the tiniest fraction of the total). Relativity also predates the knowledge that force results from the exchange of quantum particles called bosons. Furthermore, no one has thoroughly examined the consequences of the particle exchange process when viewed from the perspective of relativity theory. We show that when the force exchange process occurs between inertial frames with a very high velocity is examined carefully, then the particle exchange process itself is responsible for the relativistic mass-velocity relationship. We shall see that the flux rate of the force exchange particles from the source is unaffected by any kind of motion, and is an absolute constant of the CA. However, the received flux rate of exchange particles is affected by the state of motion of the mass receiving the force, with a relative velocity 'v' with respect to the source of the force. This can be viewed as a direct consequence of the special relativistic Lorentz time dilation formula for the moving force particle receiver, where the relativistic motion of the receiver upsets the timing of the received exchange particles. Therefore, the reduction of the force between two inertial frames with an extremely high relative velocity is what is perceived as an <u>increase</u> in inertial mass.

According to the principles of Quantum inertia theory, inertial mass is <u>absolute</u>. We have also seen that acceleration can be considered <u>absolute</u> as well. The state of *relative* acceleration of the charged virtual particles of the quantum vacuum with respect to the charged particles of the mass is what is responsible for the inertial force. By this we mean that it is only the acceleration 'a' of a mass 'm' with respect to the net statistical average acceleration of the virtual particles of the quantum vacuum, that accounts for the concept of inertia. Therefore, we conclude that inertial mass is <u>absolute</u> and equivalent to the special relativistic rest mass. Since inertial mass can be considered to be absolute in this framework, we must closely reexamine the principles of special relativity. Relativity is based on the premise that all constant velocity motion is relative, and also on the postulate of the constancy of light velocity. According to special relativity (which restricts itself to frames of constant velocity called inertial frames, or frames where Newton's law of inertia holds), the inertial mass 'm' is relative, and varies with the relative velocity 'v' with respect to an inertial observer, in accordance with the following formula: $m = m_0 / (1-v^2/c^2)^{1/2}$. Here, $m_0$ is defined as the rest mass (inertial mass measured in the same frame as the mass) and 'c' is the velocity of light. It appears that QI and special relativity are not compatible in regards to the meaning of inertial mass. From the point of view of quantum inertia, Einstein's definition of inertial mass cannot be fundamentally correct, because it is not related to the inertial quantum vacuum process described above. This is because we cannot associate the relative velocity 'v' directly to any quantum vacuum process. Recall that it is only the acceleration 'a' of a mass 'm', with respect to the net statistical average acceleration vectors of the virtual particles of the quantum vacuum that is the cause of inertial mass.

Most text book accounts of inertial force and mass in special relativity are based on the 'conservation of momentum approach' (ref. 20). Here, momentum conservation (and energy conservation) is assumed to be the fundamental aspect of nature. In order for momentum to be conserved with respect to all constant velocity reference frames, the mass must vary with the relative velocity of the two inertial frames according to special relativity. To see this, recall that momentum is defined as mass times velocity 'mv', and that the momentum is important in a collision only because it provides bookkeeping of the mass and relative velocity. The relative velocity between the two colliding masses will determine the amount of deceleration in the impact as follows: $a=(v_f -$

$v_i)/\Delta t$, were $v_f$ is the final velocity, and $v_i$ the initial velocity. Also, the mass is important because the subsequent force (and therefore energy $E = F \Delta t$) is determined by $F = m a$ through the quantum vacuum process described above. The more mass particles contained in a mass, the greater the resistance to the acceleration of the mass. Therefore, the product of mass and velocity is an indicator of the amount of future energy to be expected in a collision (or interaction) of the two masses. The total **incoming** momentum is defined as the momentum of the in-going masses $(m_1v_1 + m_2v_2)$, the total **out-going** momentum is $(m_1v_1' + m_2v_2')$. Here the two masses $m_1$ and $m_2$ are moving at velocities $v_1$ and $v_2$ before the collision, with respect to an observer, and velocity $v_1'$ and $v_2'$ after the collision. In Newtonian mechanics, the total momentum is conserved for any observer in a constant velocity reference frame. Therefore, $(m_1v_1 + m_2v_2) = (m_1v_1' + m_2v_2')$, even though different observers in general will disagree with each of the relative velocities of a pair of masses that are colliding. This is what we mean by conservation of momentum. If the inertial mass definition were left unmodified in special relativity, we would find that different observers in different constant velocity frames disagree on the conservation of momentum for colliding masses. However, it can be shown (ref. 20) that if the mass of an object 'm' (from the point of view of an observer in constant velocity motion 'v' with respect to a mass $m_0$, measured by an observer at relative rest) is redefined as follows:

$m = m_0 / (1-v^2/c^2)^{1/2}$, then the total momentum of the collision remains conserved as in Newtonian mechanics.

How does special relativity treat the definition of inertial mass from the point of view of inertial forces only, as given by Newton's law of inertia: $F = M A$? Since Einstein was aware that acceleration is not invariant, he knew that Newton's law had to be modified. An excellent description of this process is given in the text book titled 'Special Relativity' by A.P. French (ref. 21). To quote French:

"… the discovery and specification of laws of force is a central concern of physics. It is certainly important, therefore, to know how to transform forces and equations of motion so as to give a description of them from the point of view of different inertial frames. Since in special relativity the acceleration is not invariant, we know that we cannot enjoy the simplicity of Newtonian mechanics, but we can certainly arrive at some useful and meaningful statements.

The starting point, which we indeed made use of in the initial stages of our approach to relativity is Newton's law in the form

$F = dp/dt = d(mv)/dt$    where $m = m_0 (1-v^2/c^2)^{-1/2}$

We take this as a definition of F. It is a natural extension (and simplest extension) of the non-relativistic result. It is not a statement that can be independently proved."

Einstein had to modify Newton's inertial law during his attempt to revise all physics to be relativistic. However, he was not aware of the existence of the virtual particles of the quantum vacuum, or that force is a consequence of the exchange of particles at that time he formulated his theory. We now know that there is no direct contact when forces are transmitted, and forces do not act instantaneously. In order to understand what happens with inertial mass and force we must recall that the quantity of force transmitted between two objects very much depends on the **flux rate** of the exchange particles. In other words, the number of particles exchanged per unit of **time** represents the magnitude of the force transmitted between the particles. For example, imagine that there are two charged particles at relative rest in an inertial reference frame. There are a fixed number of particles exchanged per second at a fixed distance 'd'. Now imagine that particle B is moving away at a constant relative velocity 'v' with respect to particle A. If the relative velocity $v \ll c$ the exchange process appears almost the same as when the two particles are at rest. This is because the velocity of light (the speed of the exchange particles) is very high when compared to 'v', and the flux rate is unaffected. Now imagine that the relative recession velocity v -> c, which is comparable to the velocity of the force exchange particle. Does the received flux rate of particle B get altered from the perspective of particle B? The answer is yes!

We have reason to believe that the received flux rate varies with the relative velocity 'v'. It is clear from Lorentz time dilation that the timing of the exchange particle will be altered when there is a receding relativistic velocity. Recall the Lorentz time dilation formula of special relativity: $t = t_0 / (1 - v^2/c^2)^{1/2}$, which states that the timing of events varies with relative velocity 'v'. If the timing of the exchange particles is altered, then the flux rate is altered as well, since flux has units of numbers of particles per unit time. <u>It is important to realize that the flux rate of the emission of force exchange particles by particle A is not actually affected by it's motion.</u>

Now assume that particle A emits a flux of $\Phi_a$ particles per second, as seen by an observer in particle A's rest frame. When the force exchange particles are transmitted to particle B, particle B sees the flux rate decrease because of time dilation. Therefore, we find that particle B receives a smaller quantity of exchange particles per second $\Phi_b$ then when the particles are at relative rest. Thus, particle A acts like it transmits a smaller flux rate $\Phi_b$, such that $\Phi_b = \Phi_a (1 - v^2/c^2)^{1/2}$. Since the force due to the particle exchange is directly proportional to the flux of particles exchanged, we can therefore write:

$F = F_0 (1 - v^2/c^2)^{1/2}$; where is $F_0$ is the magnitude of the force when particle A and B are at relative rest, and F is the smaller force acting between particle A and B when the receding relative velocity is 'v'. Thus, we can conclude that when a force acts to cause an object to recede away from the source of the force, the force <u>reduces</u> in strength, and is less effective. With a similar line of reasoning, we find that the force increases in strength when a force acts to cause an object to move towards the source of the force.

We are now in a position to see the apparent relationship between the inertial mass and velocity. Since all forces are due to particle exchanges, we can use the method developed above to study inertial forces between to inertial frames. First, at relative rest where v=0, we have $F = F_0$. The rest mass '$m_0$' is defined by Newton's law: $F_0 = m_0 a$, where 'a' is a test acceleration that is introduced to measure the inertial rest mass. Now, if there is a relative velocity between the applied force and the mass 'm' given by 'v', which causes the mass to recede, we can write:

$F = F_0 (1 - v^2/c^2)^{1/2} = m a$, where the force is reduced in magnitude for the reasons discussed above, and the mass 'm' is considered absolute (or m= $m_0$, as in Newtonian Mechanics). In EMQG, we believe that this represents the actual physics of the force interaction. However, if one takes the position that the force does not vary with velocity, but that the mass is what actually varies, then the above equation can be re-interpreted as:

$F = m a = m_0 (1 - v^2/c^2)^{-1/2} a$, from which: $m = m_0 (1 - v^2/c^2)^{-1/2}$ as given by Einstein.

So we see that we are in a situation where it is experimentally impossible to distinguish between inertial mass variation with velocity, versus force variation with velocity. What velocities can a mass achieve through the application of an accelerating force? According to our analysis above, the answer is that the limiting speed is the speed of the exchange particles, or light velocity. At this limit, the accelerating force effectively becomes negligible in magnitude!

It is convenient to associate the variation of force with an increase in relativistic mass as Einstein proposed, for two important reasons. First, this restores the conservation of total momentum in collisions for all inertial observers (in fact, this is how Einstein derived his famous mass-velocity relationship). Secondly, if a mass is accelerated to relativistic velocities with respect to some observer by a force, and this force is removed, there will be no way to determine the subsequent energy release when a collision occurs later. In other words, when this mass collides with an object, a rapid deceleration occurs with a large release of energy (which is force multiplied by time). This energy release is greater then what can be expected from Newton's laws, as we have seen. In fact, the large energy release is due to the effective increase in the force during the collision due to increased numbers of force exchange particles (a relativistic timing effect opposite to time dilation) between the mass and the colliding wall (say). This effect acts to increase the apparent flux rate of the force exchange particles, and thus to a greater colliding force.

### Quantum Principle of Equivalence

Are virtual particle force exchange processes originating from the quantum vacuum also present for gravitational mass? The answer turns out to be a resounding yes! As we suggested, there is some evidence of the interplay between the virtual particles of the quantum vacuum and gravitational phenomena. In order to see how this impacts our understanding of the nature of gravitational mass, we found it necessary to perform a thorough investigation of Einstein's Principle of Equivalence of inertial and gravitational mass in general relativity under the guidance of the new theory of quantum inertia.

There are two main formulations of the principle of equivalence: the Weak Equivalence Principle (WEP) and the Strong Equivalence Principle (SEP). The Strong Equivalence Principle states that the results of any given physical experiment will be precisely identical for a uniformly accelerated observer in free space as it is for a non-accelerated observer in a perfectly uniform gravitational field. The Weak Equivalence Principle restricts itself to the laws of motion of masses only. It states that objects of different mass and of any type of material composition will fall at the same rate in a uniformly accelerated reference frame or under the influence of a uniform gravity field (ref. 22).

We have uncovered some theoretical evidence that the SEP may not hold for certain experiments. There are two basic theoretical problems with the SEP in regards to quantum gravity. First, if gravitons (the proposed force exchange particle) can be detected with some new form of a sensitive graviton detector, we would be able to distinguish between an accelerated reference frame and a gravitational frame with this detector. This is because accelerated frames would have virtually no graviton particles present, whereas gravitational fields like the earth have enormous numbers of graviton particles. Secondly, theoretical considerations from several authors (ref. 23) regarding the emission of electromagnetic waves from a uniformly accelerated charge, and the lack of radiation from the same charge subjected to a static gravitational field leads us to question the validity of the SEP for charged particles radiating electromagnetically. These ideas will be examined in detail in section 15.8.

How does the WEP hold out in EMQG? The WEP has been tested to a phenomenal accuracy (ref 24.) in recent times. Yet, in our current understanding of the WEP, we can only specify the accuracy as to which the two different mass values (or types) have been shown experimentally to be equal inside an inertial or gravitational field. There exists no physical or mathematical proof that the WEP is precisely true. It is still only a postulate of general relativity. We have applied the recent work on quantum inertia (ref. 5) to the investigation of the weak principle of equivalence, and have found theoretical reasons to believe that the WEP is not precisely correct when measured in extremely accurate experiments. Imagine an experiment with two masses; one mass $M_1$ being very large in value, and the other mass $M_2$ is very small ($M_1 \gg M_2$). These two masses are dropped simultaneously in a uniform gravitational field of 1g from a height 'h', and the same pair of masses are also dropped inside a rocket accelerating at 1g, at the same height 'h'. We predict that there should be a minute deviation in arrival times on the surface of the earth (only) for the two masses, known as the 'Ostoma-Trushyk effect', with the heavier mass arriving just slightly ahead of the smaller mass. This is due to a small deviation in the magnitude of the force of gravity on the mass pair (in favor of $M_1$) on the order of $(N_1-N_2)i * \delta$, where $(N_1-N_2)$ is the difference in the low level mass specified in terms of the difference in the number of masseon particles in the two masses (defined latter) times the single masseon mass 'i', and δ is the ratio of the gravitational to electromagnetic forces for a single (charged) masseon. This experiment is very difficult to perform on the earth, because δ is extremely small ($\approx 10^{-40}$), and $\Delta N = (N_1-N_2)$ cannot in practice be made sufficiently large to produce a measurable effect. However, inside the accelerated rocket, the arrival times are <u>exactly</u> identical for the same pair of masses. This, of course, violates the principle of equivalence, since the motion of the masses in the inertial frame is slightly different then in the gravitational frame. We will see that this imbalance is minute because of the dominance of the strong electromagnetic force which is also acting on the masseons of the two masses from the virtual particles of the quantum vacuum. This acts to stabilize the fall rate, giving us nearly perfect equivalence.

This conclusion is based on the discovery that the weak principle of equivalence results from lower level physical processes. Mass equivalence arises from the equivalence of the force generated between the net statistical average acceleration vectors of the charged matter particles inside a mass with respect to the immediate surrounding quantum vacuum virtual particles inside an accelerating rocket. This is almost exactly the same physical force occurring between the stationary (charged) matter particles and the immediate surrounding accelerating virtual particles of the same mass near the earth. It turns out that equivalence is not perfect in the presence of a large gravitational field like the earth. Equivalence breaks down due to an extremely minute force imbalance in favor of a larger mass dropped simultaneously with respect to a smaller mass. This force imbalance can be traced to the pure graviton exchange force component occurring in the gravitational field that is not present in the case of the identically dropped masses in an accelerated rocket. This imbalance contributes a minute amount of extra force for the larger mass compared to the smaller mass (due to many more gravitons exchanged between the larger mass as compared to the smaller mass), which might be detected in highly accurate measurements. In the case of the rocket, the equivalence of two different falling masses is perfect, since it is the floor of the rocket that accelerates up to meet the two masses simultaneously. Of course, the breakdown of the WEP also means the downfall of the SEP.

In EMQG, the gravitational interactions involve the same electromagnetic force interaction as found in inertia based on our QI theory. We also found that the weak principle of equivalence itself is a physical phenomenon originating from the hidden lower level quantum processes involving the quantum vacuum particles, graviton exchange particles, and photon exchange particles. In other words, gravitation is purely a quantum force particle exchange process, and is not based on low level fundamental 4D curved space-time geometry of the universe as believed in general relativity. The perceived 4D curvature is a manifestation of the dynamic state of the falling virtual particles of the quantum vacuum in accelerated frames, and gravitational frames. The only difference between the inertial and gravitation force is that gravity also involves graviton exchanges (between the earth and the quantum vacuum virtual particles, which become accelerated downwards), whereas inertia does not. Gravitons have been proposed in the past as the exchange particle for gravitational interactions in a quantum field theory of gravity without much success. The reason for the lack of success is that graviton exchange is not the only exchange process occurring in large-scale gravitational interactions; photon exchanges are also involved! It turns out that not only are there both graviton and photon exchange processes occurring simultaneously in large scale gravitational interactions such as on the earth, but that both exchange particles are almost identical in their fundamental nature (Of course, the strength of the two forces differs greatly). There is the other reason we named our theory of quantum gravity 'ElectroMagnetic Quantum Gravity' or EMQG; the great similarity between gravitons and photons exchange particles.

The equivalence of inertial and gravitational mass is ultimately traced down to the reversal of all the relative acceleration vectors of the charged particles of the accelerated mass <u>with respect</u> to the (net statistical) average acceleration of the quantum vacuum particles, that occurs when changing from inertial to gravitational frames. The inertial mass 'M' of an object with acceleration 'a' (in a rocket traveling in deep space, away from gravitational fields) results from the sum of all the tiny forces of the charged elementary particles that make up that mass with respect to the immediate quantum vacuum particles. This inertial force is in the opposite direction to the motion of the rocket. The (charged masseon) particles building up the mass in the rocket will have a net statistical average acceleration 'a' with respect to the local (charged masseon) virtual particles of the immediate quantum vacuum. A stationary gravitational mass resting on the earth's surface has this same quantum process occurring as for the accelerated mass, but with the acceleration vectors reversed. What we mean by this is that under gravity, it is now the virtual particles of the quantum vacuum that carries the net statistical average acceleration 'A' downward. This downward virtual particle acceleration is caused by the graviton exchanges between the earth and the mass, where the mass is not accelerated with respect to the center of mass of the earth. (Note: On an individual basis, the velocity vectors of these quantum vacuum particles actually point in all directions, and also have random amplitudes. Furthermore, random accelerations occur due to force interactions between the virtual particles themselves. This is why we refer to the statistical nature of the acceleration.) We now see that the gravitational force of a stationary mass is also the <u>same</u> sum of the tiny forces that originate for a mass undergoing accelerated motion in a gravitational field from the virtual particles of the quantum vacuum according to Newton's law 'F = MA'. In other words, the same inertial force F=MA is also found hidden inside gravitational interactions of masses! Mathematically, this fact can be seen in Newton's laws of inertia and in Newton's gravitational force law by slightly rearranging the formulas as follows:

$F_i = M_i (A_i)$ ... the inertial force $F_i$ opposes the acceleration $A_i$ of mass $M_i$ in the rocket, caused by the sum of the tiny forces from the virtual particles of the quantum vacuum.

$F_g = M_g (A_g) = M_g (GM_e/r^2)$ ... the gravitational force $F_g$ is the result of a kind of an inertial force given by '$M_g A_g$' where $A_g = GM_e/r^2$ is now due to the sum of the tiny forces from the virtual particles of the quantum vacuum (now accelerating downwards).

Since $F_i = F_g$, and since the acceleration of gravity is chosen to be the same as the inertial acceleration, where the virtual particles now have: $A_g = A_i = GM_e/r^2$, therefore $M_i = M_g$, or the inertia mass is equal to the gravitational mass ($M_e$ is the mass of the earth). Here, Newton's law of gravity is rearranged slightly to emphasis it's form as a kind of an 'inertial force' of the form F=MA, where the acceleration ($GM_e/r^2$) is now the net statistical average downward acceleration of the quantum vacuum virtual particles near the vicinity of the earth.

This derivation is not complete, unless we can provide a clear explanation as to why $F_i = F_g$, which we know to be true from experimental observation. In EMQG, both of these forces are understood to arise from an almost identical quantum vacuum process! For accelerated masses, inertia is the force $F_i$ caused by the sum of all the tiny electromagnetic forces from each of the accelerated charged particles inside the mass; with respect to the non-accelerating surrounding virtual particles of the quantum vacuum. Under the influence of a gravitational field, the <u>same force</u> $F_g$ exists as it does in inertia, but now the quantum vacuum particles are the ones undergoing the same acceleration $A_i$ (through graviton exchanges with the earth); the charged particles of the mass are stationary with respect to earth's center. The same force arises, but the arrows of the acceleration vectors are reversed. To elaborate on this, imagine that you are in the reference frame of a stationary mass resting on the surface with respect to earth's center. An average charged particle of this mass 'sees' the virtual particles of the quantum vacuum in the same state of acceleration, as does an average charged particle of an identical mass sitting on the floor of an accelerated rocket (1 g). In other words, the background quantum vacuum 'looks' exactly the same from both points of view (neglecting the very small imbalance caused by a very large number of gravitons interacting with the mass directly under gravity, this imbalance is swamped by the strength of the electromagnetic forces existing).

These equations and methodology illustrates equivalence in a special case: i.e., between an accelerated mass $M_i$ and the same <u>stationary</u> gravitational mass $M_g$. In EMQG, the weak equivalence principle of gravitational and inertial frames holds for many other scenarios such as for free falling masses, for masses that have considerable self gravity and energy (like the earth), for elementary particles, and for the propagation of light. However, equivalence is *not* perfect, and in some situations (for example, antimatter discussed in section 15.4) it simply does not hold at all!

An astute observer may question why all the virtual particles (electrons, quarks, etc, all having different masses) are accelerating downwards on the earth with the same acceleration. This definitely would be the case from the perspective of a mass being accelerated by a rocket (where the observer is accelerating). Since the masses of the different types of virtual particles are all different according to the standard model of particle physics, why are they all falling at the same rate? Since we are trying to derive the equivalence principle, we cannot invoke this principle to state that all virtual particles must be accelerating downward at the same rate. It turns out that the all quantum vacuum virtual particles are accelerating at the same rate because all particles with mass (virtual or not) are composed of combinations of a new fundamental "masseon" particle which carries just one fixed quanta of mass. Therefore, all the elementary virtual masseon particles of the quantum vacuum are accelerated by the same amount. These masseons can bind together to form the familiar particles of the standard model, like virtual electrons, virtual positrons, virtual quarks, etc. Recalling that the masseon also carries electrical charge, we see that all the constituent masseons of the quantum vacuum particles fall to earth at same rate through the electromagnetic interaction (or photon exchange) process, no matter how the virtual masseons combine to give the familiar virtual particles. This process works like a <u>microscopic principle of equivalence</u> for falling virtual particles, with the same action occurring for virtual particles as for large falling masses.

The properties of the masseon particle will be elaborated later (the masseon may be the unification particle sought out by physicist, in which case it will have other properties to do with the other forces of nature). For now, note that the masseon also carries the fundamental unit of electric charge as well. This fundamental unit of electric charge turns out to be the source of inertia for all matter according to Quantum Inertia. By postulating the existence of the masseon particle (which is the fundamental unit of 'mass charge' as well as 'electrical charge') all the quantum vacuum virtual particles accelerate at the same rate with respect to an observer on the surface of the earth. We have postulated the existence of a fundamental "low level gravitational mass charge" of a particle, which results from the graviton particle exchange process similar to the process found for electrical charges. This 'mass charge' is not affected when a particle achieves relativistic velocities, so we can state that 'low level mass charge' is an absolute constant. For particles accelerated to relativistic speeds, a high relative velocity between the source of the force and the receiving mass affects the ordinary measurable inertial mass, as we have seen (in accordance to Einstein's mass-velocity formula).

<u>Summary of the Basic Mass Definitions in EMQG</u>

In summary, EMQG proposes three different mass definitions for an object:

**INERTIAL MASS** is the measurable mass defined in Newton's force law F=MA. This is considered as the absolute mass in EMQG, because it results from force produced by the relative (statistical average) acceleration of the charged virtual particles of the quantum vacuum with respect to the charged particles that make up the inertial mass. The virtual particles of the quantum vacuum form Newton's absolute reference frame. In special relativity this mass is equivalent to the rest mass.

**GRAVITATIONAL MASS** is the measurable mass involved in the gravitational force as defined in Newton's law $F=GM_1M_2/R^2$. This is what is measured on a weighing scale. This is also considered as absolute mass, and is almost exactly the same as inertial mass!

**LOW LEVEL GRAVITATIONAL 'MASS CHARGE'** which is the origin of the pure gravitational force, is defined as the force that results through the exchange of graviton particles between two (or more) quantum particles. This type of mass analogous to 'electrical charge', where photon particles are exchanged between electrically charged particles. Note: this force is very hard to measure because it is masked by the background quantum vacuum electromagnetic force interactions, which dominates over the graviton force processes.

These three forms of mass are <u>not</u> necessarily equal! We have seen that the inertial mass is almost exactly the same as gravitational mass, but not perfectly equal. All quantum mass particles (fermions) have all three mass types defined above. But bosons (only photons and gravitons are considered here) have only the first two mass types. This means that photons and gravitons transfer momentum, and <u>do</u> react to the presence of inertial frames and to gravitational fields, but they do not emit or absorb gravitons. Gravitational fields affect photons, and this is linked to the concept of space-time curvature, described in detail later (Section 16). It is important to realize that gravitational fields deflect photons (and gravitons), but not by force particle exchanges directly. Instead, it is due to a scattering process (described later).

You might think that if a particle has energy, it automatically has mass; and if a particle has mass, then it must emit or absorb gravitons. This reasoning is based on Einstein's famous equation $E=mc^2$, which is derived purely from considerations of inertial mass (and Einstein's principle of equivalence extended to gravitational fields). In his famous thought experiment, a photon is emitted from a box, causing a recoil to the box in the form of a momentum change, and from this he derives his famous $E=mc^2$. In quantum field theory this momentum change is traceable to a fundamental QED vertex, where a electron (in an atom in the box) emits a photon, and recoils with a momentum equivalent to the photon's momentum '$m_p c$'. We have analyzed Einstein's thought experiment from the perspective of EMQG and concluded that the photon behaves as if it has an effective inertial mass '$m_p$' given by: $m_p = E/c^2$ in Einstein's light box. For simplicity, lets consider a photon that is absorbed by a charged particle like an electron at rest. The photon carries energy and is thus able to do work. When the photon is absorbed by the electron with mass '$m_e$', the electron recoils, because there is a definite momentum transfer to the electron given by $m_e v$, where v is the recoil velocity. The electron momentum gained is equivalent to the effective photon momentum lost by the photon $m_p c$. In other words, the electron momentum '$m_e v_e$' received from the photon when the photon is absorbed is equivalent to the momentum of the photon '$m_p c$', where $m_p$ is the effective photon mass. If this electron later collides with another particle, the same momentum is transferred. The rest mass of the photon is defined as zero. Thus, the effective photon mass is a measurable inertial mass. Note: the recoil of the light box is a backward acceleration of the box, which works against the virtual particles of the quantum vacuum. Thus, when one claims that a photon has a real mass, we are really referring to the photon's ability to impart momentum. This momentum can later do work in a quantum vacuum inertial process.

Does the photon have an effective gravitational mass? By this we mean; does it behave as if it carries a measurable gravitational mass in a gravitational field like the earth (as given by $E/c^2$)? The answer is yes! For example, when the photon moves parallel to the earth's surface, it follows a parabolic curve and deflects downwards. You might guess that this deflection is caused by the graviton exchanges originating from the earth acting on this photon, and that this deflection is the same as that inside an equivalent rocket accelerating at 1g. The amount of deflection is equivalent, but according to Einstein this is a direct result of the space-time curvature near the earth and in the rocket. Our work on the equivalence principle has shown however, that this is not true. The photon deflection is caused by a different reason, but ends up giving the <u>same</u> result. In the rocket, the deflection is simply caused by the accelerated motion of the rocket floor, which carries the observer with it. This causes the observer to perceive a curved path (described as curved space-time). In a gravitational field, however, the deflection of light is real, and caused by the <u>scattering</u> of photons with the downward accelerating virtual particles. The photon scatters with the <u>charged</u> virtual particles of the quantum vacuum, which are accelerating downwards (statistically). The photon moving parallel to the surface of the earth undergoes continual absorption and re-emission by the falling virtual (electrically charged) particles of the quantum vacuum. The vacuum particles induce a kind of '<u>Fizeau-like</u>' scattering of the photons (Note: this scattering is present in the rocket, but does not lead to photon deflection because only the rocket and observer are accelerated). The photons are scattered because of the electromagnetic interaction of the photons with the falling charged virtual particles of the vacuum. Since the downward acceleration of the quantum vacuum particles is the same as the up-wards acceleration of the floor of the rocket, the amount of photon deflection is equivalent. Under the influence of a gravitational field, photons take on the same downward component as the accelerating (charged) virtual particles of the vacuum. This, of course, violates the constancy of the speed of light; which we will explore further in section 16. For now, one should note that downward acceleration component is picked up by the photons only during the time the photons are absorbed by the quantum vacuum particles (and thus exist as charged virtual particles). In between virtual particle scattering, the photon motion is still strictly Lorentz invariant, and light velocity is still an absolute constant.

A similar line of reasoning as above applies to the motion of the graviton particle. The graviton has inertial mass because like the photon, it can transmit a momentum to another particle when absorbed in the graviton exchange process during a gravitational force interaction (although considerably weaker then photon exchanges). Like the photon, the graviton deflects when moving parallel to the floor of the rocket (from the perspective of an observer on the floor) and therefore has inertial mass. The graviton also has a gravitational mass (like the photon) when it moves parallel to the earth's surface, where it deflects under the influence of a gravitational field. Again, the graviton deflection is caused by the scattering of the graviton particle with the downward acceleration of the virtual 'mass-charged' particles of the quantum vacuum through an identical 'Fizeau-like' scattering process described above. Unlike the photon however, the scattering is caused by the 'mass charge' interaction (or pure graviton exchanges) of the quantum vacuum virtual particles, and not the electric charge as before. The end result is that the graviton has an effective gravitational mass like the photon. Again a graviton does not exchange gravitons with another nearby graviton, just as a photon does not exchange photons with other photons.

To summarize, both the photon and the graviton do not carry low level 'mass charge', even though they both carry inertial and gravitational mass. The graviton exchange particle, although responsible for a major part of the gravitational mass process, does not itself carry the property of 'mass charge'. Contrast this to conventional physics, where the photon and the graviton both carry a non-zero mass given by $M=E/C^2$. According to this reasoning, the photon and the graviton both carry mass (since they carry energy), and therefore both must have 'mass charge' and exchange gravitons. In other words, the graviton particle not only participates in the exchange process, it also undergoes further exchanges while it is being exchanged! This is the source of great difficulty for canonical quantum gravity theories, and causes all sorts of mathematical renormalization problems in the corresponding quantum field theory. Furthermore, in gravitational force interactions with photons, the strength of the force (which depends on the number of gravitons exchanged with photon) varies with the energy that the photon carries! In modern physics, we do not distinguish between inertial, gravitational, or low level 'mass charge'. They are assumed to be equivalent, and given a generic name 'mass'. In EMQG, the photon and graviton carry measurable inertial and gravitational mass, but neither particle carries the 'low level mass charge', and therefore do not participate in graviton exchanges.

We must emphasize that gravitons do not interact with each other through force exchanges in EMQG, just as photons do not interact with each other with force exchanges in QED. Imagine if gravitons did interact with other gravitons. One might ask how it is possible for a graviton particle (that always moves at the speed of light) to emit graviton particles that are also moving at the speed of light. For one thing, this violates the principles of special relativity theory. Imagine two gravitons moving in the same direction at the speed of light that are separated by a distance d, with the leading graviton called 'A' and the lagging graviton called 'B'. How can graviton 'B' emit another graviton (also moving at the speed of light) that gets absorbed by graviton 'A' moving at the speed of light? As we have seen, these difficulties are resolved by realizing that there are actually three different types of mass. There is measurable inertial mass and measurable gravitational mass, and low level 'mass charge' that cannot be directly measured. Inertial and gravitational mass have already been discussed and arise from different physical circumstances, but in most cases give identical results. However, the 'low level mass charge' of a particle is defined simply as the force existing between two identical particles due to the exchange of graviton particles only, which are the vector bosons of the gravitational force. Low level mass charge is not directly measurable, because of the complications due to the electromagnetic forces simultaneously present from the quantum vacuum virtual particles.

It would be interesting to speculate what the universe might be like if there were no quantum vacuum virtual particles present. Bearing in mind that the graviton exchange process is almost identical to the photon exchange process, and bearing in mind the complete absence of the electromagnetic component in gravitational interactions, the universe would be a very strange place. We would find that large masses would fall faster than smaller masses, just as a large positive electric charge would 'fall' faster then a small positive charge towards a very large negative charge. There would be no inertia as we know it, and basically no force would be required to accelerate or stop a large mass.

## The Masseon and Graviton Particles and EMQG Quantum Field Theory

EMQG addresses gravitational force, inertia, and electromagnetic forces only, and the weak and strong nuclear forces are excluded from consideration. EMQG is based on the idea that all elementary matter particles must get their quantum mass numbers from combinations of just one fundamental matter (and corresponding anti-matter particle), which has just one fixed unit or quanta of mass which we call the 'masseon' particle. This fundamental particle generates a fixed flux of gravitons that are exchanged during gravitational interactions. The exchange process is not affected by the state of motion of the masseon (as you might expect from the special relativistic variation of mass with velocity). We also purpose that nature does **not** have two completely different long-range forces, e.g. gravity and electromagnetism. Instead, we believe that there exists an almost perfect symmetry between the two forces, which is hidden from view because of the mixing of these two forces in all measurable gravitational interactions. In EMQG, the graviton and photon exchange process is found to be essentially the same, except for the strength of the force coupling (and a minor difference in the treatment of positive and negative masses discussed later). EMQG treats graviton exchanges by the same successful methods developed for the behavior of photons in QED. The dimensionless coupling constant that governs the graviton exchange process is what we call '$\beta$' in close analogy with the dimensionless coupling constant '$\alpha$' in QED, where $\beta \approx 10^{-40} \alpha$.

As we stated, EMQG requires the existence of a new fundamental matter particle called the 'masseon' (and a corresponding 'anti-masseon' particle), which are held together by a new unidentified strong force. Furthermore, EMQG requires that masseons and anti-masseons emit gravitons analogous with the electrons and anti-electrons (positrons) which emit photons in QED. Virtual masseons and anti-masseons are created in equal amounts in the quantum vacuum as virtual particle pairs. A masseon generates a fixed flux of graviton particles with wave functions that induce attraction when absorbed by another masseon or anti-masseon; and an anti-masseon generates a fixed flux of graviton particles with an opposite wave function (anti-gravitons) that induces repulsion when absorbed by another masseon or anti-masseon. A graviton is its own anti-particle, just as a photon is its own antiparticle. This process is similar to, but not identical to the photon exchange processes in QED for electrons of opposite charge. In QED, an electron produces a fixed flux of photon particles with wave functions that induce repulsion when absorbed by another electron, and induces attraction when absorbed by a positron. A positron produces a fixed flux of photon particles with wave functions that induce attraction when absorbed by another electron, and induces **repulsion when absorbed by a positron. From this it can be seen that if two sufficiently large pieces of anti-matter can be fabricated which are both electrically neutral, they will be found to repel each other gravitationally! Thus, anti-matter can be thought of as literally 'negative' mass (-M), and therefore negative energy. This violates the equivalence principle, and we will say more about this later.**

**These subtle differences in the exchange process in QED and EMQG produce some interesting effects for gravitation that are not found in electromagnetism. For example, a large gravitational mass like the earth does not produce vacuum polarization of virtual particles from the point of view of 'mass-charge' (unlike electromagnetism). In gravitational fields, all the virtual masseon and anti-masseon particles of the vacuum have a net average statistical acceleration directed downwards towards a large mass. This produces a net downward accelerated flux of vacuum particles (acceleration vectors only) that affects other masses immersed in this flux.**

**In contrast to this, an electrically charged object <u>does</u> produce vacuum polarization. For example, a negatively charged object will cause the positive and negative (electrically charged) virtual particles to accelerate towards and away, respectively from the negatively charged object. Therefore, there is no energy contribution to other real electrically charged test particles placed near the charged object from the vacuum particles, because the electrically charged vacuum particles contributes equal amounts of force contributions from both the upward and downward directions. The electrical forces from the vacuum cancels out to zero.**

**In gravitational fields, the vacuum particles are responsible for the principle of equivalence, precisely because of the lack of vacuum polarization due to gravitational fields. Recall that 'masseon' particles of EMQG are equivalent to the 'parton' particle concept that was introduced by the authors of reference 5 in regards to HRP Quantum Inertia. Recall that the masseons and anti-masseons also carry one quanta of electric charge of which there are two types; positive and negative charges. For example, masseons come in positive and negative electric charge, and anti-masseons also come in positive and negative charge. A single charged masseon particle accelerating at 1g sees a certain fixed amount of inertial force generated by the virtual particles of the quantum vacuum. In a gravitational field of 1g, a single charged masseon particle on the surface of the earth sees the same quantum vacuum electromagnetic force. In other words, from the vantage point of a masseon particle that makes up the total mass, the virtual particles of the quantum vacuum looks exactly the same from the point of view of motion and forces whether it is in an inertial reference frame or in a gravitational field. We propose a new universal constant "i" called the 'inertion', which is defined as the inertial force produced by the action of virtual particles on a single (real charged) masseon particle undergoing a relative acceleration of 1 g. This force is the lowest possible quanta of inertial mass. All other masses are fixed integer combinations of this number. This same constant 'i' is also the lowest possible quanta of gravitational force.**

**The electric charge that is carried by the electron, positron, quark and anti-quark originates from combinations of masseons, which is the fundamental source of the electrical charge. This explains why a fixed charge relationship exists between the quarks and the leptons, which belong to different families in the standard model. For example, according to the standard model, 1 proton charge precisely equals 1 electron charge (opposite polarity), where the proton is made of 3 quarks. This precise equality arises from the fact that charged masseon particles are present in the internal structure of both the quarks and the electrons (and every other mass particle).**

**The mathematical renormalization process is applied to particles to avoid infinities encountered in Quantum Field Theory (QFT) calculations. This is justified by postulating a high frequency cutoff of the vacuum processes in the summation of the Feynman diagrams. Recall that QED is formulated on the assumption that a perfect space-time continuum exists. In EMQG, a high frequency cutoff is essential because space is quantized as 'cells', specified by Cellular Automata (CA) theory. In CA theory there is quantization of space in the form of cells. If particles are sufficiently close enough,**

they completely lose their identity as particles in CA theory, and QFT does not apply at this scale. Since graviton exchanges are almost identical to photon exchanges, we suspect that EMQG is also renormalizable as is QED, with a high frequency cutoff as well. This has not been proven yet. The reason that some current quantum gravity theories are not renormalizable boils down to the fact that the graviton is assumed to be the only <u>boson</u> involved in gravitational interactions. The graviton must therefore exhibit all the characteristics of the gravitational field, including space-time curvature.

Should particles achieve very high densities by mutual attraction (as might exist in the center of a black hole, for example), they might completely lose their identity as separate matter and force particles. Here, the details of the universal cellular automata process would have to be known in order to understand what is going on. EMQG does not address these extreme conditions of mass concentration. EMQG works where interactions involve particle densities low enough to sustain the individual particle identity.

Past attempts at developing a quantum field theory of graviton exchanges have been unsuccessful. One of the reasons for this is that the gravitational force was assumed to by caused by the pure graviton exchange process. Yet, curved space-time must somehow emerge from this process. These theories were doomed from the start because the graviton particle exchange alone was responsible for general relativistic space-time curvature. In other words, the actual curvature of space-time that gravitons are presumed to be traveling in are caused by the gravitons themselves! When this is done, the results lead to infinities, to mathematical problems, and to a coupling constant that is not dimensionless as it is for photons. Furthermore, the nature of the long-range gravitational force becomes completely different to the (only) other long-range force we know, the electromagnetic force. Instead of this, EMQG treats gravitation interactions by both the photon exchange and graviton exchange processes, which are simultaneously occurring between the particles of ordinary matter and the surrounding virtual particles of the quantum vacuum.

<u>GravitoElectric and GravitoMagnetic Fields</u>

In EMQG, the photon exchange and graviton exchange process is virtually identical in its basic nature, which shows the great symmetry between these two forces. As a byproduct of this, the quantum vacuum becomes 'neutral' in terms of gravitational 'mass charge', as the quantum vacuum is known to be neutral with respect to electric charge. This is due to an equal number of positive and negative electrical charged virtual particles and 'gravitational charged' virtual particles created in the quantum vacuum at any given time. This in turn is due to the symmetrical masseon and anti-masseon pair creation process. (EMQG does not resolve the problem of why the universe was created with an apparent imbalance of real ordinary matter and anti-matter mass particles.)

It is known that in the context of general relativistic theory (which is somewhat similar to Maxwell's field theory of electromagnetism, ignoring curvature of space-time for now) that gravitational fields have what Kip Thorne calls the 'gravitoelectrical' and the 'gravitomagnetic' components (ref. 8) which arise out of Einstein's field equations. We take the existence of these fields as further evidence for the similarity between electric and magnetic fields and the gravitoelectric and gravitomagnetic fields, which originate from the similarities of the photon and graviton particle exchanges (except for coupling strength). This evidence supports the view that nature has two long range forces, electromagnetism and gravity that operate in virtually the same way. The gravitomagnetic force component is far weaker than the observable gravitoelectric force (which corresponds to the ordinary force of gravity), because this force is produced by pure graviton exchange process only. The gravitoelectric force is mixed up with the strong electrically charged component of all gravitational processes, and therefore not directly observable. If there were a complete absence of virtual particles, there would be no equivalence principle for masses. Since the electrical force from the virtual particles of the quantum vacuum exceeds the pure gravitational forces, the gravitoelectric force component (also caused by pure graviton exchanges only) is completely disguised and hard to observe. The ordinary measurable force of gravity we perceive everyday is dominated by the electromagnetic forces from the charged virtual particles of the quantum vacuum which equalizes the fall rate of all masses.

The Resolution of the Cosmological Constant Problem

EMQG solves the famous problem of the Einstein's cosmological constant, which was a term that he added to his gravitational field equations to make the universe remain in the steady state or non-expanding state as was believed in his time. Einstein later abandoned this constant when the astronomer Hubble discovered by astronomical observation that the universe is actually expanding. The cosmological constant can be thought of as the measure of the mass-energy density contained in empty space alone, which would of course be represented by the quantum vacuum virtual particles. In EMQG, the cosmological constant is essentially zero simply because of the existence of virtual masseon and anti-masseon particles in the quantum vacuum! The virtual masseon particles must be produced in the vacuum in symmetrical matter and anti-matter masseon pairs according to the principles of quantum field theory. The virtual masseons pairs are oppositely charged in both gravitational 'mass charge' and electrical charge. Approximately equal numbers of positive and negative electrical charge and gravitational mass charge masseons are present in the vacuum at any given time. The symmetrical combination of positive and negative mass-energy yields a net contribution of zero from the quantum vacuum, which is the solution to the cosmological constant problem. It is trivial to note that the <u>electric</u> charge of the vacuum is neutral as a whole.

To sum up, the gravitation force is now described at the quantum level as arising from ordinary matter (particles) moving primarily under the immediate electromagnetic influence of the distorted quantum vacuum particles (masseons). The distortion of the quantum vacuum means an unequal distribution of acceleration vectors of the surrounding virtual masseons and anti-masseons particles from place to place near the massive object. This distortion originates from graviton exchanges between masseons in the massive object and the surrounding virtual masseon and anti-masseon particles that constitute the quantum vacuum. On the earth, this causes a simple net accelerated (statistical average) 'stream' of virtual masseons and anti-masseons particles moving towards the earth's center along their radius vectors 'r' (where the net average acceleration vector 'a' of a virtual particle is given by: $a = GM/r^2$, where 'G' is the Newtonian Universal Gravitation constant, and 'M' is the mass of the earth). Of course, the flowing stream analogy is to help visualize this process, but in actual fact the virtual masseons are short lived, and are constantly replaced by newly created virtual masseons, endlessly.

NOTE: Strictly speaking, absolute CA space and time units must be used for distance and time measurements (of r and a).

This distortion of the acceleration vectors of the quantum vacuum 'stream' serves as an effective 'electromagnetic guide' for the motion of nearby test masses (themselves consisting of masseons) through space and time. This 'electromagnetic guide' concept replaces the 4D space-time geodesics (which is the path that light takes through curved 4D space-time) that guide light and matter in motion. Because the quantum vacuum virtual particle density is quite high, but not infinite (at least about $10^{90}$ particles/m$^3$), the quantum vacuum acts as a very effective and energetic guide for the motion of light and matter.

Introduction to EMQG Space-Time Curvature

The physicist A. Wheeler once said that: "space-time geometry 'tells' mass-energy how to move, and mass-energy 'tells' space-time geometry how to curve". In EMQG, this statement must be somewhat revised on the quantum particle level to read: large mass-energy concentrations (consisting of quantum particles) exchanges gravitons with the immediate surrounding virtual particles of the quantum vacuum, causing a downward acceleration (of the net statistical average acceleration vectors) of the quantum vacuum particles. This downward acceleration of the virtual particles of the quantum vacuum 'tells' a nearby test mass (also consisting of real quantum particles) how to move **electromagnetically,** by the exchange of photons between the electrically charged, and falling virtual particles of the quantum vacuum and the electrical charged, real particles inside the test mass. This new view of gravity is totally based on the ideas of quantum field theory, and thus acknowledging the true particle nature of both matter and forces. It is also shows how nature is non-geometric when examined on the smallest of distance scales, where Riemann geometry is now replaced solely by the interactions of quantum particles existing on a kind of quantized 3D space and separate time on the CA.

Since this downward accelerated stream of charged virtual particles also affects light or real photons and the motion of real matter (for example, matter making up a clock), the concept of space-time must be revised. For example, a light beam moving parallel to the surface of the earth is affected by the downward acceleration of charged virtual particles (electromagnetically), and moves in a curved path. Since light is at the foundation of the measurement process as Einstein showed in special relativity, the concept of space-time must also be affected near the earth by this accelerated 'stream' of virtual particles. Nothing escapes this 'flow', and one can imagine that not even a clock is expected to keep the same time as it would in far space. As a result, a radically new picture of Einstein's curved space-time concept arises from these considerations in EMQG.

The variation of the value of the net statistical average (directional) acceleration vector of the quantum vacuum particles from point to point in space (with respect to the center of a massive object) guides the motion of nearby test masses and the motion of light through electromagnetic means. This process leads to the 4D space-time metric curvature concept of general relativity. With this new viewpoint, it is now easy to understand how one can switch between accelerated and gravitational reference frames. Gravity can be made to cancel out inside a free falling frame (technically at a point) above the earth because we are simply taking on the same net acceleration as the virtual particles at that point. In this scenario, the falling reference frame creates the same quantum vacuum particle background environment as found in an non-accelerated frame, far from all gravitational fields. As a result, light travels in perfectly straight lines when viewed by a falling observer, as specified by special relativity. Thus, in the falling reference frame, a mass 'feels' no force or curvature as it would in empty space, and light travels in straight lines (defined as 'flat' space-time). Thus, the mystery as to why different reference frames produce different space-time curvature is solved in EMQG. It is interesting that in an accelerated rocket, space-time curvature is also present, but is now caused by another mechanism; the accelerated motion of the floor of the rocket itself. In other words, the space-time curvature, manifesting itself as the path of curved light, is really caused by the accelerated motion of the observer! The observer (now in a state of acceleration with respect to the vacuum), 'sees' the accelerated virtual particle motion in his frame. Furthermore, the motion appears to him to be almost exactly the same as if he were in an equivalent gravitational field. This is why the space-time curvature appears the same in both a gravitational field and an equivalent accelerated frame. These differences between accelerated and gravitational frames imply that equivalence is not a basic element of reality, but merely a result of different physical processes, which happen to give the same results. In fact, equivalence is **not** perfect!

We have seen that besides the electromagnetic force exchange, there is another force exchange component involved in gravitational interaction. This is the direct graviton exchange between the test mass itself and a massive object. Although negligible when compared to the electromagnetic component, its presence is still felt. This causes a tiny deviation of the principle of equivalence, which is not found in general relativity theory. If a large and small test mass is dropped simultaneously near a massive object, the larger test mass will arrive slightly earlier due to the greater number of gravitons exchanged there. This imbalance in the forces on the two masses upsets perfect equivalence. However, because of the ratio of the electromagnetic to gravitational force strength is on the order of $10^{40}$, the difference in arrival time is absolutely negligible for ordinary test mass combinations dropped on the earth. This imbalance of equivalence might be measured experimentally.

According to EMQG, all metric theories of gravity, including general relativity, have a limited range of application. These theories are useful only when a sufficient mass is available to significantly distort the virtual particle motion surrounding the mass; and only where the electromagnetic interactions dominates over the graviton processes (or where the graviton flux is not too large). For precise calculation of gravitational force interactions of small masses, EMQG requires that the gravitational interaction be calculated by adding the specific Feynman diagrams for both photon and graviton exchanges. Thus, the use of the general relativistic Schwarzchild Metric for spherical bodies (even if modified by including the uncertainty principle) is totally useless for understanding the gravitational interactions of elementary particles. The whole concept of space-time 'foam' is incorrect according to EMQG, along with all the causality problems associated with this complex mathematical concept.

## Space-Time Curvature as a Pure Quantum Vacuum Particle Process

4D Minkowski curved space-time takes on a radically new meaning in EMQG, and is no longer a basic physical element of our reality. Instead, it is merely the result of quantum particle interactions alone. The curved space-time of general relativity arises strictly out of the interactions between the falling virtual particles of the quantum vacuum near a massive object and a nearby test mass. The effect of the falling quantum vacuum acts somewhat like a special kind of "Fizeau-Fluid" or media, that affects the propagation of light; and also effects clocks, rulers, and measuring instruments. Fizeau demonstrated in the middle 1850's that moving water varies the velocity of light propagating through it. This effect was analyzed mathematically by Lorentz. He used his newly developed microscopic theory for the propagation of light in matter to study how photons move in a flowing stream of transparent fluid. He reasoned that photons would change velocity by frequent scattering with the molecules of the water, where the photons are absorbed and later remitted after a small time delay. This concept is discussed in detail in section 16. If Einstein himself had known about the existence of the quantum vacuum when he was developing general relativity theory, he may have deduced that space-time curvature was caused by the "accelerated quantum vacuum fluid". He was aware of the work by Fizeau, but was unaware of the existence of the quantum vacuum. After all, Einstein certainly realized that clocks were not expected to keep time correctly when immersed in an accelerated stream of water! We show mathematically in this paper that the quantity of space-time curvature near a spherical object predicted by the Schwarzchild metric is identical to the value given by the 'Fizeau-like' scattering process in EMQG.

In EMQG, anywhere we find accelerated vacuum disturbance, there follows a corresponding space-time distortion (including gravitational waves). We have seen that both accelerated and gravitational frames qualify for the status of curved 4D space-time (although caused by <u>different</u> physical circumstances). We have found that in EMQG there exists two, separate but related space-time coordinate systems. First, there is the familiar global four dimensional relativistic space-time of Minkowski, as defined by our measuring instruments, and is designated by the x,y,z,t in Cartesian coordinates. The amount of 4D space-time curvature is influenced by accelerated frames and by gravitational frames, which is the cause of the accelerated state of the quantum vacuum.

Secondly there is a kind of a quantized absolute space, and separate time as required by cellular automata theory. Absolute space consists of an array of numbers or cells $C(x,y,z)$ that changes state after every new clock operation $\Delta t$. $C(x,y,z)$ acts like the absolute three dimensional pre-relativistic space, with a separate absolute time that acts to evolve the numerical state of the cellular automata. The CA space (and separate time) is not effected by any physical interactions or directly accessible through any measuring instruments, and currently remains a postulate of EMQG. Note that EMQG absolute space does not correspond to Newton's idea of absolute space. Newton postulated the existence of absolute space in his work on inertia. He realized that absolute space was required in order to resolve the puzzle of what reference frame nature uses to gauge accelerated motion. In EMQG, this reference frame is **not** the absolute quantized cell space (which is unobservable), but instead consists of the net average state of acceleration of the virtual particles of the quantum vacuum with respect to matter (particles). A very important consequence of the existence of absolute quantized space and quantized time (required by cellular automata theory) is the fact that our universe must have a maximum speed limit!

### Mach's Principle and EMQG

The problem of accelerated motion introduced in Mach's principle has been resolved by the work of ref. 5 on the quantum origins of inertia. Mach's principle is a loose collection of ideas and paradoxes concerning the problem of what reference frame nature uses to judge accelerated (or rotating) motion. Their work replaces Newton's concept of absolute space with a kind of new universal reference frame, originated by the quantum vacuum particles, for <u>accelerated motions only</u>. The vacuum particles themselves serve as Mach's universal reference frame for acceleration, but without violating the principle of relativity. Matter actually 'knows' when their should be a force associated with acceleration because it's 'instructions' come from the state of the local virtual particles of the quantum vacuum itself. In other words, the forces experienced in matter (particles) actually originates from the state of <u>relative acceleration</u> with respect to the (net average) acceleration of the local virtual particles of the quantum vacuum, and this determines the quantity of the associated force. For uniform non-accelerated (constant velocity) motions, there is no preferred reference frame as given by Einstein's Special Relativity, because the quantum vacuum imparts no directed force on ordinary matter in uniform motion, in any direction, or at any speed. It is interesting to note that the cosmic distribution of all the other masses in the universe out there, ultimately contributes to the state of acceleration of the local vacuum here, by the long range gravitational force or graviton exchanges. Thus, Mach was absolutely right to suspect that cosmic distribution of all the matter in the universe affects the local accelerated or rotating motions in our own neighborhood!

### EMQG and the Physics of Gravitational Waves

The physics of gravitational waves (GW) in EMQG is definitely not the same as the physics of electromagnetic waves described by QED. For a periodically accelerating large mass (or masses), the fluctuating graviton flux is responsible for the periodic disturbance in the net acceleration of the quantum vacuum particles in the immediate vicinity, which will affect another test mass. For example, a close pair of relativistic neutron stars in orbit would both periodically disturb the state of the virtual particles of the quantum vacuum through periodically fluctuating graviton exchanges. Thus, the outward propagating GW is really a time varying periodic increase and decreases in the net acceleration of the virtual particle acceleration vectors with respect to the neutron stars. Once the GW is started, however, it is self sustaining throughout space (as is electromagnetic radiation). The reason that it is self-sustaining is due to the local electromagnetic vacuum process. The strong electromagnetic component in the GW is the gravitational waves primary means of acting upon other masses. The GW periodic excitation propagates electromagnetically because the quantum virtual particles are constantly redistributing their acceleration vector imbalance among each other via photon exchanges, as the GW travels outward. It is known in the context of general relativity theory that the GW's have a very large stiffness (in analogy with Hooke's law, ref. 16), and thus a large energy density. This fact follows from Einstein's Gravitational Field equations. This is easily explained in EMQG as arising from the very large energy density of the vacuum disturbance itself. The virtual particles of the quantum vacuum have a large particle density that is very roughly on the order of $10^{90}$ particles per cubic meter (Plank length cubed presents a rough upper limit to the particle density per cubic meter on a CA, but the exact value is not known). Because of this, the fluctuating quantum vacuum disturbance (GW) carries a large energy density, and is quite capable of explaining the stiffness of the GW wave. In fact, the GW is capable of vibrating a large aluminum cylinder after traveling hundreds of light years in distance.

**Emqg and the Lense-Thirring Effect**

We now apply the principles of EMQG to calculate the amount of inertial frame dragging (Lense-Thirring effect) on the earth. The Lense-Thirring effect is a tiny perturbation of the orbit of a particle caused by the spin of the attracting body, first calculated by the physicists J. Lense and H. Thirring in 1918 using general relativity (ref. 25). Einstein's general relativity predicts the perturbation in the vicinity of the spinning body, but the effect has not been accurately verified experimentally. However, recent work in ref. 26 using the LAGEOS and LAGEOS II earth orbiting satellites has rendered an unconfirmed experimental value that agrees with theory to an accuracy of about 20 %. The Lense-Thirring effect has also been interpreted as being due to gravitomagnetic fields (section 19.2), and also tied in with Mach's principle (ref. 27). It is hoped that with the launch of the Gravity Probe B (co-developed by Stanford University) by NASA the Lense-Thirring effect will be measured to an unprecedented accuracy of 1% or better. The Gravity Probe B (there was a different Gravity Probe A launched earlier by NASA) is a drag free satellite carry an four ultra-precise gyroscopes that will be put in a polar orbit around the earth at a height of about 400 miles (ref. 28).

An important consequence of the Lense-Thirring effect is that the orbital period of a test mass around the earth depends on the direction of the orbit! A test mass that has an orbit which revolves around the earth in the same direction of the spin rotation would be longer then the orbital period of the same test mass revolving opposite to the direction of the spin of the earth. The difference in the orbital period of the two test masses becomes smaller with increasing height until it disappears when the orbits are at infinity. The Lense-Thirring effect can also be be thought of as a kind of 'a dragging of inertial frames' first named by Einstein himself. The Lense-Thirring effect for a rotating mass is most pronounced as the angular velocity of the rotating mass increases. The basic reason for inertial frame dragging is the finite speed of propagation of the graviton particle (the speed of light). This allows time for a large rapidly spinning mass to rotate a small amount while the graviton is still in flight as it propagates outwards. The finite velocity of the graviton particle along with the downward $GM/R^2$ acceleration component of the charged virtual particles of the quantum vacuum is entirely responsible for inertial frame dragging.

We now examine inertial frame dragging for a weak gravitational field such as the earth (the strong field inertial frame dragging is a very formidable problem in EMQG). Gravitons are emitted in huge numbers from the earth. Since we know that gravitons are physically very similar to photons, we can predict the characteristics of this graviton flux. The graviton flux can be visualized as being the same as a rotating flashlight emitting light outwards as it rotates. Since the velocity of the source does not effect the velocity of light, we conclude that the velocity of the spinning earth does not affect the motion of the gravitons. (However, gravitons are affected by the state of accelerated motion of the virtual particles, because gravitons *do* scatter with the virtual masseon particles of the quantum vacuum. This scattering is small for a mass the size of the earth, but for an object with a very strong gravitational field that is rotating at relativistic speeds, the calculation of the Lense-Thirring effect is an extremely difficult problem in EMQG).

As the gravitons propagate outward, they encounter virtual masseon particles in the quantum vacuum. Since masseons posses 'mass-charge', the vacuum is accelerated down. The virtual masseons are therefore accelerated in the same direction of the motion of the graviton particle flux. The magnitude of the acceleration depends on the graviton flux at a point 'r' from the center, and is given by $GM/r^2$. If a test mass (composed of real electrically charged masseon particles) is dropped onto the surface of the earth, the electromagnetic interactions between the real masseons of the mass and the virtual masseons causes the test mass to accelerate downwards at 1g in the direction of the graviton flux. Thus, from the perspective of an external observer, the mass falls along the radius vectors. The observer on the surface of the equator 'sees' the graviton flux leaving the equator in curved paths he is carried along with the earth's rotation. From his frame of reference the gravitons appear to curve, and the average acceleration vector of the virtual particles follows this same curved path, with the acceleration vectors increasing in magnitude the closer to the center. These curved paths also represent the path of light that light will take if it propagates straight up (in other words, geodesic paths). In absolute CA units, the light velocity varies upwards (or downwards) along these curved paths due to the Fizeau-like scattering with the quantum vacuum, and thus these paths represent the direction of the 4D space-time curvature. These paths are deflected when compared to the non-rotating earth.

The equation of the outward propagating graviton curve (in the equatorial plain) for a clockwise rotating earth turns out to be Archimedes' Spiral, and takes the form $r = k\theta$ in polar coordinates, where k is a constant (r is the distance that the graviton travels). The constant k depends only on the velocity of the graviton 'c' and the velocity $v = 2\pi R/T_p$ of the earth's rotation, where $T_p$ is the rotation period of the earth and R is the earth's radius, and the time t of transit. If k is small then the spiral has a high curvature, and if k is large the curvature is small. The ratio 'c/v' determines the value of the spiral constant k. If c were to be very large, then $k \gg 1$ which causes the gravitons to 'unwind' slowly; and if v were to be very large, then $k \ll 1$ which causes the gravitons to 'unwind' rapidly.

Based on these considerations, the equation for the spiral is:

$r = c^2 \theta/2\pi v$ where $v = 2\pi R/T_p$, and $\theta = 2\pi t v/c$

Let us calculate the shape of the spiral for the observer A on the earth. The earth has a radius $R = 6.37 \times 10^6$ meters, and a rotation period $T_p = 24$ hours. Therefore, the $v = 463$ m/sec. The equation of the spiral for the earth is: $r = 103{,}176 \, c\theta$. We wish to solve for the angle $\theta$ at the earth's surface, where $r = 6.37 \times 10^6$ meters. Therefore: $\theta = 2.05 \times 10^{-7}$ radians = 42.5 milli-arcseconds. This agrees well with the standard prediction based on general relativity (ref. 28). However, the result is much easier to derive and visualize with EMQG than with general relativity theory. This angle represents the deflection of the downward accelerating virtual particles of the quantum vacuum with respect to the non-rotating earth. Furthermore, the deflection angle varies with height (along a spiral path) which causes inertial frame dragging.

<u>Experimental Verification of EMQG Theory</u>

EMQG proposes several new experimental tests that give results that are different from the conventional general relativistic physics, and can be used to verify the theory.

(1) EMQG opens up a new field of physics, which we call anti-matter gravitational physics. We propose that if two sufficiently large pieces of anti-matter are manufactured to allow measurement of the mutual gravitational interaction (with a torsion balance apparatus for example), then the gravitational force will be found to be repulsive! The force will be equal in magnitude to $-GM^2/r^2$ where M is the mass of each of the anti-matter masses, r is their mutual separation, and G is Newton's gravitational constant). This is in clear violation of the principle of equivalence, since in this case, $M_i = -M_g$, instead of being strictly equal. Antimatter that is accelerated in far space has the same inertial mass '$M_i$' as ordinary matter, but when interacting gravitationally with another antimatter mass it is repelled ($M_g$). Note: The earth will attract bulk anti-matter because of the large abundance of gravitons originating from the earth of the type that induce attraction. This means that no violation of equivalence is expected for anti-matter dropped on the earth, where anti-matter falls normally. However, an antimatter earth will repel a nearby antimatter mass. Recent attempts at measuring earth's gravitational force on anti-matter (e.g. anti-protons) will not reveal any deviation from equivalence, according to EMQG. However, if there were two large identical masses of matter and anti-matter close to each other, there would be no gravitational force existing between them because of the balance of "positive and negative" masses, e.g. equal numbers of gravitons that induce attraction and repulsion. This gravitational system is considered gravitationally 'neutral', as is the quantum vacuum, which is also gravitationally neutral.

**(2) For an extremely large test mass and a very small test mass that is dropped simultaneously on the earth (in a vacuum), there will be an extremely small difference in the arrival time of the masses on the surface of the earth in slight violation of the principle of equivalence. This tiny deviation from perfect equivalence is called the 'Ostoma-Trushyk effect'. This effect is on the order of $\approx \Delta N \times \delta$, where $\Delta N$ is the difference in the number of masseon particles in the two masses, and $\delta$ is the ratio of the gravitational to electric forces for one masseon. This experiment is very difficult to perform on the earth, because $\delta$ is extremely small ($\approx 10^{-40}$), and $\Delta N$ cannot be made sufficiently large. To achieve a difference of $\Delta N = 10^{30}$ masseons particles between the small and large mass requires dropping a molecular-sized cluster and a large military tank simultaneously in the vacuum in order to give a measurable deviation. Note: For ordinary objects that might seem to have a large enough difference in mass (like dropping a feather and a tank), the difference in arrival time may be obscured by background interference, or by quantum effects like the Heisenberg uncertainty principle which restrict the accuracy of time measurements.**

**(3) If gravitons can be detected by the invention of a graviton detector/counter in the far future, then there will be experimental proof for the violation of the strong principle of equivalence. The strong equivalence principle states that all the laws of physics are the same for an observer situated on the surface of the earth as it is for an accelerated observer at 1 g. The graviton detector will find a tremendous difference in the graviton count in these two cases. This is because gravitons are vastly more numerous here on the earth. Thus, this detector can distinguish between an inertial frame and a gravitational frame in violation of the strong equivalence principle.**

**(4) Since mass has a strong electromagnetic force component, mass measurements near the earth might be disrupted experimentally by manipulating some of the electrically charged virtual particles of the nearby quantum vacuum through electromagnetic means. If a rapidly fluctuating magnetic field (or rotating magnetic field) is produced under a mass it might effect the instantaneous charged virtual particle spectrum, and disrupt the tiny inertial forces for each masseon of the mass. This may reduce the measured gravitational (and inertial masses) of an object in the vicinity. In a sense, this device would act like a primitive and weak "anti-gravity" machine. The virtual particles are constantly being "turned-over" in the vacuum at different rates, with the high frequency virtual particles (and therefore, the high-energy virtual particles) being replaced the quickest. If a magnetic field is made to fluctuate fast enough so that it does not allow the new virtual particle pairs to replace the old and smooth out the disruption, the spectrum of the virtual particles will be altered. According to conventional physics, the energy density of virtual particles is infinite, which means that all frequencies of virtual particles are present. In EMQG there is an upper cut-off to the frequency, and therefore the highest energy according to the Plank's law: $E = h\upsilon$, where $\upsilon$ is the frequency that a virtual particle can have. We can state that the smallest wavelength that a virtual particle can have is about $10^{-35}$ meters, e.g. the plank wavelength (or a corresponding maximum Plank frequency of about $10^{43}$ hertz for very high velocity ($\approx c$) virtual particles). Unfortunately for our "anti-gravity" device, it is technologically impossible to disrupt the highest frequencies. According to the uncertainty principle, the relationship between energy and time is: $\Delta E \times \Delta t > h$. This means that the high frequency end of the spectrum consists of virtual particles that "turns-over" the fastest. To give maximum disruption to a significant percentage of the high frequency virtual particles requires magnetic fluctuations on the order of at least $10^{20}$ cycles per seconds. Therefore, only lower frequencies virtual particles of the vacuum can be practically affected, and only small changes in the measured mass can be expected with today's technology. As a result of this, a relationship should exist between the amount of gravitational (or inertial) mass loss and the frequency of electromagnetic fluctuation or disruption. The higher the frequency the greater the**

mass loss. Recent work on the Quantum Hall Effect by Laughlin (ref. 29) on fractional electron charge suggests that under the influence of a strong magnetic field, electrons might move in concert with swirling vortices created in the 2D electron gas. This leads to the possibility that this 'whirlpool' phenomena holds for the virtual particles of the quantum vacuum under the influence of a strongly fluctuating magnetic field. These high-speed whirlpools might disrupt the high frequency end of the distribution of electrically charged virtual particles in small pockets. Therefore, there might be a greater mass loss under these circumstances (idea this is very speculative at this time). Recent experiments on mass reduction with rapidly rotating magnetic fields are inconclusive. Reference 30 gives an excellent and detailed review of the various experiments on reducing the gravitational force with superconducting magnets.

# 1. INTRODUCTION

*" It always bothers me that, according to the laws as we understand them today, it takes a computing machine an infinite number of logical operations to figure out what goes on in no matter how tiny a region of space and no matter how tiny a region of time. .... why should it take an infinite amount of logic to figure out what one tiny piece of space-time is going to do?"*

*- Richard Feynman*

*"Truth is much to complicated to allow anything but approximations."*

*- John Von Neumann*

Quantum Gravity is defined as the unification of quantum field theory with Einstein's general relativity theory. It should describe the behavior of gravitational forces at quantum distance scales, in enormous gravitational fields, and for cosmological distance scales.

Various attempts at this unification have not been completely successful in the past, because these theories do not grasp the true nature of inertia, or the hidden physical processes behind Einstein's principle of equivalence. In developing a theory of quantum gravity, one might ask which of the existing approaches to quantum gravity is more relevant or fundamental, quantum field theory or Einstein's theory of general relativity? Currently, it seems that both of these theories are not compatible with each other.

Based on a postulate that the universe operates like a Cellular Automata (CA) computer, we assume that quantum field theory is in closer touch to the actual workings of the universe. General relativity is taken as a global classical description of space-time, gravity and matter. General relativity reveals the large-scale patterns and organizing principles that arise from the hidden quantum processes existing on the tiniest distance scales. Quantum field theory reveals to us that forces originate from a quantum particle exchange process, which transfers momentum from one quantum particle to another. The exchange process is universal, and applies to electromagnetic, weak and strong nuclear forces, and also for gravitational force, as we shall see. The generic name given to the force exchange particles is the 'vector boson'.

We have developed a quantum theory of gravity called ElectroMagnetic Quantum Gravity (or EMQG) that is manifestly compatible with a Cellular Automata computer model, which describes gravity as a pure particle exchange process. We will be dealing with the photon (the exchange particle for electromagnetic force) and the graviton (the exchange particle for the *pure* gravitational force). What is unique about EMQG is that gravitation involves <u>both</u> the photon and graviton exchange particles, where the photon plays a very important role! In fact, the photon exchange process is responsible for the principle of equivalence.

In order to formulate a theory of quantum gravity, a mechanism must be found that produces the gravitational force, while somehow linking to the principle of equivalence of inertial and gravitational mass. In addition, this mechanism should naturally lead to 4D space-time curvature and should be compatible with the principles of general relativity theory. Nature has another long-range force called electromagnetism, which has been described successfully by the principles of quantum field theory. This well-known theory is called Quantum ElectroDynamics (QED), and this theory has been tested for electromagnetic phenomena to an unprecedented accuracy. It is therefore reasonable to assume that gravitational force should be a similar process, since gravitation is also a long-range force like electromagnetism. However, a few obstacles lie in the way, which complicate this line of reasoning.

First, gravitational force is observed to be <u>always</u> attractive! In QED, electrical forces are attractive and repulsive. As a result of this, there are an equal number of positive and negative charged virtual particles in the quantum vacuum (section 15.4) at any given time because virtual particles are always created in equal and opposite charged particle pairs. Thus, there is a balance of attractive and repulsive forces in the quantum vacuum, and the quantum vacuum is electrically neutral, overall. If this were not the case, the virtual charged particles of one charge type in the vacuum would dominate over all other physical interactions.

Secondly, QED is formulated in a relativistic, flat 4D space-time with no curvature. In QED, electrical charge is defined as the fixed rate of emission of photons (strictly speaking, the fixed probability of emitting a photon) from a given charged particle. Electromagnetic forces are caused by the exchange of photons, which propagate between the charged particles. The photons transfer momentum from the source charge to the destination charge, and travel in flat 4D space-time (assuming no gravity). From these basic considerations, a successful theory of quantum gravity should have an exchange particle (graviton), which is emitted from a mass particle at a fixed rate as in QED. The 'mass charge' or just plain mass replaces the idea of electrical charge in QED, and the graviton momentum exchanges are now the root cause of gravitational force. Yet, the graviton exchanges must somehow produce disturbances of the normally flat space and time, when originating from a very large mass.

Since mass is known to vary with velocity (special relativity), one might expect that 'mass charge' must also vary with velocity. However, in QED the electromagnetic force exchange process is governed by a fixed, universal constant ($\alpha$) which is not affected by anything like motion (more will be said about this point later). Should this not be true for graviton exchange in quantum gravity as well? It is also strange that gravity, which is also a long-range force, is governed by same form of mathematical law as found in Coulomb's Electrical Force law. Coulomb's Electric Force law states: $F = KQ_1Q_2/R^2$ , and Newton's Gravitational Force law: $F=GM_1M_2/R^2$. This suggests that there is a deep connection between gravity and electromagnetism. Yet, gravity supposedly has no counterpart to negative electrical charge. Thus, there seems to be no such thing as negative 'mass charge' for gravity, as we find for electrical charge. Furthermore, QED also has no analogous phenomena as the principle of equivalence. Why should gravity be connected with the principle of equivalence, and thus inertia, and yet no analogy of this exists for electromagnetic phenomena?

To answer the question of negative 'mass charge', EMQG postulates the existence of negative 'mass charge' for gravity, in close analogy to electromagnetism. Furthermore, we claim that this property of matter is possessed by all anti-particles that carry mass. Therefore anti-particles, which are opposite in electrical charge to ordinary particles, are also opposite in 'mass charge'. In fact, negative 'mass charge' is not only abundant in nature, it comprises nearly half of all the mass particles (in the form of 'virtual' particles) in the universe! The other half exists as positive 'mass charge', also in the form of virtual particles. Furthermore, all familiar ordinary (real) matter comprises only a vanishing small fraction of the total 'mass charge' in the universe! Real anti-matter seems to be very scarce in nature, and no search for it in the cosmos has revealed anything to date.

Both positive and negative 'mass charge' appear in huge numbers in the form of virtual particles, which quickly pop in and out of existence in the quantum vacuum (section 15.4), everywhere in empty space. We will see that the existence of negative 'mass charge' is the key to the solution to the famous problem of the cosmological constant (section. 15.5), which is one of the great unresolved mysteries of modern physics. Finally, we propose that the negative energy or the antimatter solution of the famous Dirac equation of quantum field theory is also the *negative 'mass charge' solution.*

The question raised above regarding the principle of equivalence is much more difficult to answer. The principle of equivalence is one of the founding postulates of general relativity theory. Stated in a weaker form; objects of different mass fall at the same rate of acceleration in a uniform gravity field. Alternatively, it means that the inertial mass (mass defined by Newton's law of motion: $m_i = F_i/g$) is exactly equal to the gravitational mass (mass defined by Newton's universal gravitational law $m_g = (F_g r^2) / (GM)$). The equivalence principle requires that $m_i = m_g$. Why two such different physical definitions for these two mass types should give the same numerical result has remained a deep and unsolved mystery in modern physics, and deserves an explanation. This paper provides a solution to this difficult question, and is also an invitation to explore a new approach towards a full theory of quantum gravity. EQMG is based on a new understanding of both inertia and the principle of equivalence, which exists on the quantum particle level and is hidden from view.

Conventional wisdom in physics assumes that the principle of equivalence is **exact**, and somehow reflects a fundamental aspect of nature. It is assumed to be applicable under any physical circumstance. It is believed to hold true at the elementary particle level, and under enormously large gravitational fields such as neutron stars. As a consequence, Einstein's general relativity theory is also assumed to hold true under any physical condition. Unification attempts have focused on direct quantization of Einstein's gravitational field equations, which is supposed to apply at all distance scales and under all gravitational conditions. This assumption, along with the general principles of quantum mechanics, leads directly to the existence of the so called "space-time quantum foam", that results from the application of Schwarzschild Metric for matter particles on the quantum scale. As the distance r ---> 0 for a quantum particle, the space-time curvature approaches infinity and becomes dynamic or 'foamy' due to the uncertainty principle. The space-time foam concept leads to difficulties, such as problems with causality. It is also well known that Einstein's field equations themselves lead to bizarre objects such as black holes, or singularities in space-time, and to problems with the conservation of energy and momentum.

There have been past attempts at quantizing the gravitational field. They have focused on using the graviton force exchange particle directly as the quanta of the gravitational field, which is in direct analogy with the quantization of electromagnetic fields with photons. The graviton particle is chosen with the right mathematical characteristic to quantize gravity in accordance with quantum field theory and general relativity. These attempts, however, fail to account for the origin of space-time curvature. Specifically, how does a graviton 'produce' curvature when propagating from one mass to another? Does the graviton move in an already existing 4D space-time curvature? If it does, how is the space-time produced by the graviton? If not, how is 4D space-time curvature produced? In other words, if the 4D space-time curvature is not caused by the graviton exchanges, then what is the cause? There are more unanswered questions. For example, why do the virtual particles of the quantum vacuum **not** contribute a nearly infinite amount of curvature to the whole universe? After all, the force of gravity is universally attractive. According to quantum field theory, virtual gravitons should exist in huge numbers in the quantum vacuum, and should therefore contribute huge amounts of attractive forces and a large amount of space-time curvature.

Does graviton exchange processes get affected by high velocity motion (with respect to some other reference frame)? In other words, does the number of gravitons exchanged increase as the velocity of the mass increases, as seems to be required by the special relativistic mass increase with velocity formula? How does a graviton, which moves at the speed of light make it out of a black hole to attract other nearby masses? After all, even light does not move fast enough to escape from the gravitational pull of a black hole. If light cannot escape, how can a graviton escape? Why does the state of motion of an observer near a gravitational field affect his local 4D space-time curvature? For example, why does an observer in free fall near the earth affect his local space-time conditions in such a way as to match an observer in far space (who lives in flat space-time)? Many of these questions have remained unresolved in existing quantum gravity theories.

The principle of equivalence has become a major pillar of modern physics, and has been tested under a wide variety of gravitational field strengths and distance scales. It has been tested with different material types (ref. 6). It has been tested to high degrees of precision (up to 3 parts in $10^{-12}$ for laboratory bodies, and 1 part in $10^{-12}$ for the acceleration of the moon and earth towards the sun). Yet, the principle of equivalence has remained only as a postulate of general relativity. It cannot be proven from fundamental principles. Some of the better literature on general relativity have drawn attention to this fact, and admit that no explanation can be found as to; "why our universe has a deep and mysterious connection between acceleration and gravity" (ref. 8). After all, while standing on the surface of the earth, gravity appears like a static force holding your mass to the surface.

Yet, when your standing in an accelerated rocket moving with an acceleration of 1 g, the principle of equivalence states that there is an equivalent force exerted against the rocket floor by your inertial mass which is caused by the dynamic accelerated motion of the rocket. Why should there be such a deep connection between what appears to be two completely different physical phenomena?

The principle of equivalence means different things to different people; and to some it means nothing at all. The physicist Synge writes in his textbook on general relativity (ref. 31): "I have never been able to understand this principle (principle of equivalence) ... I suggest that it be now buried with appropriate honors". We will see what he meant by this statement in section 12. The equality of inertial and gravitational mass has been deduced strictly by observation, and by actual experience. But is equivalence exact? Since the principle of equivalence cannot be currently traced to deeper physics, we can never say that these two mass types are *exactly* equal. EMQG reveals the hidden phenomena that account for the principle of equivalence.

How is the principle of equivalence defined? Actually, there are two main formulations of the principle of equivalence. The Strong Equivalence Principle (SEP) states that the results of *any* given physical experiment will be precisely identical for a uniformly accelerated observer in free space, as it is for a non-accelerated observer in a perfectly uniform gravitational field. A weaker form of this postulate restricts itself to the laws of motion of masses only, or only to experiments that involve the motion of matter. This is called the Weak Equivalence Principle (WEP). The WEP implies that objects of different mass fall at the same rate of acceleration in a uniform gravity field.

Recently, there exists some theoretical basis to believe that the strong equivalence principle does ***not*** hold in general. First, we conclude that by the very definition of the existence of the graviton particle, the strong equivalence principle cannot hold! Let us assume that gravitons can be detected with some new form of a sensitive graviton detector. We would then be able to easily distinguish between an inertial frame and a gravitational frame with this detector. This is because inertial frames would have virtually no graviton particles present, whereas gravitational fields like the earth have enormous numbers of graviton particles (recall that the graviton exchange process ***is*** the source of the gravitational field). Secondly, recent theoretical considerations of the emission of electromagnetic wave radiation from a uniformly accelerated charge, and the *lack* of radiation from the same charge subjected to a static gravitational field has led one author (ref. 23) to also reconsider the validity of the strong equivalence principle. These issues will be examined later in section 15.8.

As for the weak equivalence principle, we can only specify the accuracy as to which the two different mass types have been shown *experimentally* to be equal inside an inertial frame and in a static gravitational field (technically at a point). Although we are not aware of any experiment that contradicts the WEP, we have theoretical reasons to believe that the WEP is not perfect. We will find that the WEP follows from lower level physical processes that gives near perfect equivalence. **Mass equivalence** arises from the equivalence of the electromagnetic forces generated between the net statistical average acceleration vectors of the (charged) matter particles inside a mass interacting with the surrounding virtual particles of the quantum vacuum inside an accelerating rocket. This is the *same* force occurring between the (charged) matter particles of a mass near the earth and the surrounding virtual particles of the quantum vacuum. Equivalence is ***not*** perfect, however, and breaks down when the accuracy of the measurement is extremely high (section 15.8)! This will be explained in some detail later.

Thus, we maintain that gravity also involves accelerated motion of a sort, like in a rocket undergoing uniform acceleration. However, under a gravitational field like the earth, this accelerated motion is hidden from our direct view. The motion that is hidden from us turns out to be the relative *accelerated motion of the particles of the quantum vacuum, which are falling*. The quantum vacuum consists of short-lived particles called virtual particles, which fills the surrounding space, everywhere (section 15.8). It is the hidden motion of virtual particles that turns out to be responsible for the equivalence between inertial and gravitational mass. We offer, for the first time, a detailed derivation of the principle of equivalence on the quantum scale, based on a few simple postulates. Because the most important physical process involved in gravity (and the principle of equivalence) is the ordinary electromagnetic force, we therefore call this new quantum gravity theory **'Electro-Magnetic Quantum Gravity'** or EMQG. Another reason for choosing this name is that the graviton particle turns out to have almost the identical characteristics as the photon particle. Figure 14 shows the relationship between EMQG and the rest of physics.

Existing quantum gravity theories that include gravitons as the exchange particle do not address the problem of variation of mass with velocity predicted by special relativity, which is very obvious when masses achieve velocities that are approaching the speed of light. But how can mass vary with velocity, when according to the principles of quantum field theory, mass (or 'mass charge') is the property of a particle to emit and absorb gravitons. Does this mean that the graviton emission rate varies with speed? However, if you are an observer in the same reference frame as a high velocity mass, then the mass appears to you to be the same as the rest mass, which would leave the graviton emission rate unchanged. What is going on here? The graviton emission cannot be relative. We cannot have different observers disagree on the graviton emission process! In fact, we believe that the graviton emission rate **must** be constant, and independent of velocity as for the precedent set by QED for the photon emission process of electrical charge.

EMQG provides a new understanding of gravitation, and is also testable, because it predicts new experimental results that cannot be explained by conventional theory. Before EMQG can be fully presented, however, a review of some very important background ideas is necessary.

## 2. SPACE, TIME, MATTER AND CELLULAR AUTOMATA THEORY

*"By getting to smaller and smaller units, we do not come to fundamental units, or indivisible units, but we <u>do</u> come to a point where <u>division has no meaning</u>."*

*- Werner Heisenberg*

Cellular Automata (CA) theory (references 1,2,3 and 4) forms the basis of EMQG, and has guided us towards our theory of quantum gravity. Cellular Automata theory will be discussed fully in section 3. We believe that the Cellular Automata is currently the *best* model we have for understanding the 'machinery' of our universe. It is CA theory that has led us along the torturous path towards EMQG theory, an *all* particle theory of gravity. CA theory suggests that **all** the laws of physics should be the result of interactions that are strictly local, which therefore forbids any action at a distance. If CA theory is correct, then the global laws of physics (like the Newton's law of inertia 'F=MA') should be the result of the local actions of matter particles, which exist directly as information patterns on the cells of the CA. In addition, CA theory suggests that space, time, matter, energy and motion are all the **same** thing, namely the result of information changing state according to some set of specific mathematical rules! We will see that special relativity is already manifestly compatible with the cellular automata model of the universe.

If CA theory is true, this implies that familiar space, time, and matter are not the basic elements of reality. In CA theory, all physical phenomena turn out to be the end result of a vast amount of numerical information being processed on a 'universal computer', and that the information *inside* this computer is in fact our *whole* universe, including ourselves! All elements of reality are due to numerical information being processed on the 'cells' at incredibly 'high speeds' (with respect to us) on this universal computer. The computer hardware is forever inaccessible to us, because we ourselves are also information patterns residing on the cells. The laws of physics that govern the 'hardware' functioning of the universal computer hardware do not even have to be the same as our own physical laws. In fact, the computer 'hardware' that governs our reality can be considered by definition of the word 'universe' to be outside our own universe, because it is **<u>inaccessible.</u>**

The computer model that best fits the workings of our universe is quite different from that of an ordinary personal desktop computer. In fact, our universe is implemented on the most massively parallel computer model currently known to computer science. This parallel computer model is called a 'CELLULAR AUTOMATA'. A CA consists of a huge array of 'cells' (or memory locations) that are capable of storing numerical data, which change state on every clock period, everywhere, according to the rules of the cellular automata (figure 18 shows a 3D CA, the model proposed for our universe). This type of computer was discovered theoretically by Konrad Zuse and Stanislav Ulam in the late 1940's, and later put to use by John von Neumann to model the real world behavior of complex spatially extended structures. The best known example of a CA is the game of life originated by John Horton Conway. In fact, the CA is so powerful that it is capable of updating the entire memory (no matter how big) in a single clock pulse! Contrast this to the desktop computer, which takes millions of clock cycles to update the entire memory.

CA's are inherently symmetrical, because one set of rules is programmed and repeated for each and every memory cell. We believe that this accounts for the high degree of spatial symmetry found in our universe. In other words, the laws of physics are the same no matter where you are, or how you are oriented in space. This is accountable by the perfect symmetry of the cell space. Not only is the CA a parallel computer, but it is also the fastest known parallel computer processor.

On remotely small scales of distance and time, called the Plank scale (about $10^{-35}$ meters distance and $10^{-43}$ seconds) the 'view' of the universe is completely unlike what we know from our senses. Figure 18 shows what space 'looks' like at the lowest possible scales. Space, time, and matter no longer exist. Elementary particles of matter reveal themselves as oscillating numeric information patterns. Motion turns out to be an illusion, which results from the 'shifting' of particle-like information patterns from cell to cell. However, there is no real movement! Forces, which are understood to result from vector boson particle exchanges, can be viewed as the exchange of oscillating information patterns, which are readily emitted and absorbed by matter particles (fermions). The absorption of a vector boson information pattern changes the internal oscillation of a particle, and causes an 'acceleration' to occur along a particular direction (usually towards or away from the source). The quantization of space reveals itself as cells, or storage locations for the numbers of the computer. What causes the numbers to change state as time progresses? It's the logical (and local) rules that are preprogrammed in all the storage cells of the computer. All the cells change state at the same time (not our time, but during a CA clock transition) and at regular CA 'clock' intervals (not to be confused with clocks in our universe). So, in CA theory at the lowest level, space isn't 'nothing', it is something; it's memory cells. Particle information patterns (numbers) residing in the cells are dynamic, and shifting (but **not actual** moving)! They simply change state as the computer simulation evolves. In order to understand CA theory and its relationship to modern physics we need to take a brief review of CA computer theory.

3.     CELLULAR AUTOMATA THEORY

*"Digital Mechanics is a discrete and deterministic modeling system which we propose to use (instead of differential equations) for modeling phenomena in physics....  We hypothesize that there will be found a single cellular automaton rule that models all of microscopic physics; and models it exactly. We call this field Digital Mechanics."*

*- Edward Fredkin*

In this section, we give a brief summary of Cellular Automata theory and it's connection with modern physics (ref. 1,2,3 and 4). We also include some **new** material we originally developed in regards to the CA space dimensionality. Edward Fredkin (ref. 2) has been first credited with introducing the idea that our universe is a huge cellular automata computer simulation. The cellular automata (CA) model is a special computer that consists of a large number of cells, which are storage locations for numbers. Each cell contains some initial number (say 0 or 1 for example), and the same set of rules are applied for *each* and every cell. The rules specify how these numbers are to be changed at the next computer 'clock' interval. Mathematically, a 'clock' is required in order to synchronize the next state of all the cells. The logical rules of a cell specifies the new state of that cell on the next 'clock' period, based on the cells current state, and on that of all the cell's immediate neighbors (each cell in CA has a fixed number of neighboring cells). The number of neighbors that influence a given cell is what we call the connectivity of the cellular automata. In other words, the number of neighbors that connect (or influence) a given cell is called the CA connectivity. The connectivity can be any positive integer number.

We define a 'geometric' cellular automata as a CA configuration where each cell has only the correct number of neighbors such that the CA connectivity allows a simple cubic geometric arrangement of cells (a cube in 3D space for example, see figure 18). This can be visualized as the stacking of cells into squares, cubes, hypercubes, etc. This structure is well suited for constructing 3D space, as we know it. The mathematical spatial dimensions required to contain the geometric CA is defined as the 'dimensionality' of the geometric cellular automata. For example, in a 2D geometric CA each cell has 8 surrounding neighbors, which can be thought of as forming a simple 2D space. One set of rules exists for every given cell, based on the input from its immediate 8 neighbors (and possibly on the state of the cell itself). The result of this 'computation' is then stored back in the cell on the next 'clock' period. In this way, all the cells are updated simultaneously in every cell, and the process repeats on each and every clock cycle. In a 1D geometric CA, each cell has 2 neighbors, a 2D geometric CA has 8 neighbors, a 3D geometric CA has 26 neighbors (figure 18), a 4D has 80 neighbors, and a 5D has 242, and so on. In general, if $C_D$ is the number of neighbors of an Nth dimensional geometric Cellular Automata, and $N_{D-1}$ is the number of neighbors of the next lower N-1th dimensional geometric Cellular Automata, then: $C_D = C_{D-1} + C_{D-1} + C_{D-1} + 2$, or $C_D = 3 C_{D-1} + 2$, which is the number of neighbors of an Nth dimensional geometric CA expressed in terms of the next lowest space.

The geometric CA model that is explored here for our universe is a simple geometric 3D CA, where each cell has 26 neighboring cells (figure 18). Your first impression might be that the correct geometric CA model would be a 4D geometric CA, so that it is directly compatible with relativistic space-time. There are several problems with this approach, however. First, 3D space and time have to be united in this CA model, which is not easy to do. More difficult still, is that 4D space-time is relative, and in some cases it is curved. This means that on the earth a falling reference frame observers flat 4D space-time, while a stationary observer exhibits curved 4D space-time. Furthermore, the 4D space-time curvature is directional near the earth. Curvature varies along the radius vectors of the earth, but does not vary parallel to the earth's surface (over small distances).

We will see that there are actually two space-time structures in our universe. First, there is the familiar relativistic 4D space-time that is measured with our instruments, and influenced by motion or gravity. Secondly, there is an absolute 'low level' 3D space, and separate 'time' that is not directly accessible to us, which exists at the lowest scale of distance and time, which is Cellular Automata level. The 3D space comes in the form of cells, which are automatically quantized. Time is also automatically quantized, and is extremely simple. Our time is the unidirectional evolution of the numeric state of the CA based on the local rules on each 'clock cycle'. The numeric state of the CA changes state on each and every 'clock' pulse (an external synchronizer of the CA), which can be thought of as a primitive form of CA time ('clock cycles' should not to be confused with our clocks, or our measure of time). Thus, our low-level space and separate time is simply a 3D geometric CA. This model restores utter simplicity to the structure of our universe. Everything is deterministic, and the future evolution of the universe can in principle be determined, if the exact numeric state of the Cellular Automata is known at this point in time, along with the rules that govern each cell.

What is the scale of this 3D geometric CA computer with respect to our distance and time scales? Assuming that the quantization scale corresponds to the Plank Scale (ref. 10) the number of cells per cubic meter of space is astronomically large: roughly $10^{105}$ cells. Later, we will see that the quantization scale is much finer than the Plank Scale of distance and time, according to EMQG theory. Remember that all the cells in the universe are all updated in one single 'clock' cycle! This is a massive computation indeed! The number of CA 'clock' pulses that occur in one of our seconds is a phenomenal $10^{43}$ clock cycles per second (based on the assumption that the Plank distance of $1.6 \times 10^{-35}$ meters is the rough quantization scale of space, and the Plank time is $5.4 \times 10^{-44}$ seconds). Because of the remotely small distance and time scales of quantization, we as observers of the universe are very far removed from the low-level workings of our CA computer. Why do humans exist at such a large scale as to be remotely removed from the CA cell distance scale? The simple answer to this is that life is necessarily complex! Even an atom is remarkably complex. A lot of storage locations (cells) are required to support the structure of an atom, especially in light of the complex QED processes going on. It is not possible to assemble anything as complex as life forms without using tremendous numbers of atoms and molecules.

Chaos and complexity theory teaches us that simple rules can lead to enormous levels of complexity. We can see this in a simple 2D geometric CA called the game of life. Being a 2D geometric CA, there are 8 neighbors for each cell, which forms a primitive geometric 2D space, which can be viewed on a computer screen. Here the rules are very simple. The rules for the game of life are summarized below:

**Rules for the famous 2D Geometric CA - Conway's Game of Life**:

(1) If a given cell is in the one state, on the next clock pulse it will stay one when it is surrounded by precisely two or three ones among it's eight neighbors on a square lattice. If the cell is surrounded by more than three neighbors, it will change to zero; if fewer than two neighbors have a one, it changes to zero.

(2) If a given cell is in the zero state, on the next clock pulse: it will change to a one when surrounded by precisely three ones, otherwise for any other combination of neighbor states the cell will remain a zero.

The rules are simple enough for a child to understand, yet the game of life leads to an endless number of different patterns, and to significant complexity. We see gliders, puffers, guns, 'oscillating' particles with different rates of translation and spontaneous particle emission from some oscillating patterns. We have even seen a pattern that resembles a particle exchange process.

How many different 2D geometric cellular automata's can be constructed from all possible rules? This number is unimaginably large. For simple binary cells, with 8 neighboring cells there are 8+1 cells that influence a given cell (previous state of a cell can influence it's next state), which leads to $2^{512}$ possible binary combinations or approximately $10^{154}$ different CA's, of which the game of life is but one. In general, for an Nth Dimensional Geometric CA with (m) neighbors, there are $2^k$ possible rules available for the Cellular Automata, where $k = 2^{(m+1)}$. Assuming our universe is a simple 3D geometric CA, then there are $2^{134,217,728}$ possible rules to choose from! You can give up trying to find the rules that govern our universal CA by simple trial and error.

In the early 1900's Max Plank (ref. 10) defined a set of fundamental scales based on his then newly discovered quantum of energy h (E = hv). Three fundamental constants G (Gravitational Constant), h (Plank's Quantum of Action), and c (the speed of light) were assembled in a set of equations that define natural physical units independent of any man made units. The Plank length is $1.6 \times 10^{-35}$ meters and the Plank time is $5.4 \times 10^{-44}$ seconds. The Plank length has often been suggested as the fundamental quantization scale of our universe. This suggests that the 'size' of a cell in the CA computer is the Plank length, and that the Plank time is the period of the cellular automata 'clock' (the smallest possible time period for a change in the state of the CA). Note: The cells actually have no real size, since they represent storage locations for numbers. Instead, this represents the smallest distance that you can increment as you move along a ruler. We will see that the actual quantization scale of the CA is much finer than the Plank Scale (by some currently unknown amount).

Even in conventional physics, there is growing evidence accumulating that suggests that the plank units of distance and time somehow represent the quantization scale of space-time itself. In CA theory, quantization is automatic! We will see as EMQG is developed that the quantization scale of the universal CA is much finer than the Plank Scale (by some unknown amount). There are also some other physical units that can be derived from the three fundamental constants listed above which includes: Plank Energy, Plank Temperature, Plank Mass, Plank Speed, and Plank Wavelength, which also represent fundamental universal limits to these parameters.

All physical things are the result of CA processes; including space, time, forces and matter (particles). Although, the exact rules of the CA that is our universe is unknown at this time, some very general physical conclusions can be drawn from the CA model. For example, matter is constructed from elementary particles, which move in space during some period of time. Particles interact via forces (exchange particles) which bind particles together to form atoms and molecules. Quantum field theory tells us that forces are also particles called vector bosons, which are readily exchanged between matter particles (fermions), and that these exchanges cause momentum changes and accelerations that we interpret as forces. Elementary particles and forces on the CA consist of oscillating information patterns, which are numbers changing state dynamically in the cells of the CA. These numerical information patterns roam around from cell to cell in given directions.

The shifting rates of a particle or information pattern, relative to some other particle, is interpreted by us as the state of relative motion of the particle. As we have seen, particles can interact by exchanging particles, which are also information patterns. Exchange particles are readily emitted at a given fixed rate by the source particle, and absorbed by target particle. When these force particles are absorbed, the internal state changes, which we interpret as a change in the particle momentum. The result of this process is that the shifting rates or motion of the matter particle changes by undergoing a positive or negative acceleration with respect to the source. This is what we observe as a fundamental unit of force. Of course, when we observe forces on the classical scale, the astronomical number of particle exchanges occurring per second blurs the 'digital' impact nature of the force exchange, and we perceive a smooth force reaction. All 'motion' is relative in CA theory, since all cells are identical and indistinguishable. In other words, we cannot know the specific cell locations that a particle occupies.

The closest CA elements that correspond to our space and time are the empty cells and the clock cycles that elapse. But this correspondence is not exact, as we shall find. The cells, which are storage locations for numbers, really form a low-level basis of the physical concept of space. Because the information patterns can roam freely in various directions that are determined by the dimensionality of the CA, we interpret this freedom of motion as space. Similarly, while matter patterns are in motion, a definite time period elapses. We can only sense the elapse of time when matter is in motion, by the changing state of the CA. The ultimate cause of change is the CA clock, and the common rules that govern the cells. But, it is important to realize that the internal clock required for the CA to function is not the same as our measure of time in our universe. Our time is based on physical phenomena only. This fact is the origin of much confusion on the nature of time in physics. We will investigate this important concept in EMQG theory later.

From these and other considerations, CA theory restores a great unity to all of physics. Where there used to be different phenomena described by different physical theories, now there is only one theory. Furthermore, CA theory is not only able to describe the way the universe works, but it also allows us to understand *how* it works in detail. Is there any evidence in the current laws of physics to support the idea that our universe really is a Cellular Automata computer simulation? The following sections will provide some rather speculative and sometimes circumstantial evidence to support this position.

**4.     THE QUANTUM VACUUM AND IT'S RELATIONSHIP TO CA THEORY**

*Philosophers:* *"Nature abhors a vacuum."*

One might think that the vacuum is completely devoid of everything. In fact, the vacuum is far from empty. In order to make a complete vacuum, one must remove all matter from an enclosure. However, this is still not good enough. One must also lower the temperature down to absolute zero in order to remove all thermal electromagnetic radiation. However, Nernst correctly deduced in 1916 (ref. 32) that empty space is still not completely devoid of all radiation after this is done. He predicted that the vacuum is still permanently filled with an electromagnetic field propagating at the speed of light, called the zero-point fluctuations (sometimes called vacuum fluctuations). This was later confirmed by the full quantum field theory developed in the 1920's and 30's. Later, with the development of QED, it was realized that all quantum fields should contribute to the vacuum state, like virtual electrons and positron particles, for example.

According to modern quantum field theory, the perfect vacuum is teeming with activity, as all types of quantum virtual particles (and virtual bosons or force particles) from the various quantum fields, appear and disappear spontaneously. These particles are called 'virtual' particles because they result from quantum processes that have short lifetimes, and are undetectable. Figure 12 shows a simplified schematic of the instantaneous velocity vectors of the virtual particles of the vacuum, in an extremely small volume of space.

One way to look at the existence of the quantum vacuum is to consider that quantum theory forbids the absence of motion, as well as the absence of propagating fields (exchange particles). In QED, the quantum vacuum consists of the virtual particle pair creation/annihilation processes (for example, electron-positron pairs), and the zero-point-fluctuation (ZPF) of the electromagnetic field (virtual photons) just discussed. The existence of virtual particles of the quantum vacuum is essential to understanding the famous Casimir effect (ref. 11), an effect predicted theoretically by the Dutch scientist Hendrik Casimir in 1948. The Casimir effect refers to the tiny attractive force that occurs between two neutral metal plates suspended in a vacuum. He predicted theoretically that the force 'F' per unit area 'A' for plate separation D is given by:

$$F/A = - \pi^2 h c /(240 D^4 ) \quad \text{Newton's per square meter} \quad (\text{Casimir Force 'F'}) \quad (4.1)$$

The origin of this minute force can be traced to the disruption of the normal quantum vacuum virtual photon distribution between two nearby metallic plates. Certain photon wavelengths (and therefore energies) in the low wavelength range are not allowed between the plates, because these waves do not 'fit'. This creates a negative pressure due to the unequal energy distribution of virtual photons inside the plates as compared to outside the plate region. The pressure imbalance can be visualized as causing the two plates to be drawn together by radiation pressure. Note that even in the vacuum state, virtual photons carry energy and momentum.

Recently, Lamoreaux made (ref. 12) accurate measurements for the first time on the theoretical Casimir force existing between two gold-coated quartz surfaces that were spaced 0.75 micrometers apart. Lamoreaux found a force value of about 1 billionth of a Newton, agreeing with the Casimir theory to within an accuracy of about 5%.

EMQG theory depends heavily on the existence of the virtual particles of the quantum vacuum, and so we present other evidence for the existence of virtual particles (briefly) below:

(1) The extreme precision in the theoretical calculations of the hyper-fine structure of the energy levels of the hydrogen atom, and the anomalous magnetic moment of the electron and muon are both based on the existence of virtual particles. These effects have been calculated in QED to a very high precision (approximately 10 decimal places), and these values have also been verified experimentally. This indeed is a great achievement for QED, which is essentially a perturbation theory of the electromagnetic quantum vacuum.

(2) Recently, vacuum polarization (the polarization of electron-positron pairs near a real electron particle) has been observed experimentally by a team of physicists led by David Koltick (ref. 33). Vacuum polarization causes a cloud of virtual particles to form around the electron in such a way as to produce charge screening. This is because virtual positrons migrate towards the real electron and virtual electrons migrate away. A team of physicists fired high-energy particles at electrons, and found that the effect of this cloud of virtual particles was reduced, the closer a particle penetrated towards the electron. They reported that the effect of the higher charge for the penetration of the electron cloud with energetic 58 giga-electron volt particles was equivalent to a fine structure constant of 1/129.6. This agreed well with their theoretical prediction of 128.5. This can be taken as verification of the vacuum polarization effect predicted by QED.

(3) The quantum vacuum explains why cooling alone will never freeze liquid helium. Unless pressure is applied, vacuum energy fluctuations prevent its atoms from getting close enough to trigger solidification.

(4) For fluorescent strip lamps, the random energy fluctuations of the vacuum cause the atoms of mercury, which are in their exited state, to spontaneously emit photons by eventually knocking them out of their unstable energy orbital. In this way, spontaneous emission in an atom can be viewed as being caused by the surrounding quantum vacuum.

(5) In electronics, there is a limit as to how much a radio signal can be amplified. Random noise signals are somehow added to the original signal. This is due to the presence of the virtual particles of the quantum vacuum as the photons propagate in space, thus adding a random noise pattern to the signal.

(6) Recent theoretical and experimental work in the field of Cavity Quantum Electrodynamics suggests that orbital electron transition time for excited atoms can be affected by the state of the virtual particles of the quantum vacuum surrounding the excited atom in a cavity.

What relationships exist between CA theory and the quantum vacuum? Recall that the quantum vacuum implies that **all** of empty space is filled with virtual particle processes. In simple 2D geometric CA's (such as the Conway's Game of Life), most random initial states or 'seed' patterns on the cells (and often from small localized initial patterns with all the remaining cells in the zero state) often evolve into a complex soup of activity, everywhere. This activity is very reminiscent of our quantum vacuum. In the game of life you can even see events that even look suspiciously like random 'particle' collisions, particle annihilation, and particle creation after a sufficiently long period of simulation time. Of course this is not hard evidence of CA theory, but it is highly suggestive.

## 5. THE BIG BANG (START OF SIMULATION AT T=0) AND CA THEORY

*"I want to know how God created this world (Universe)"*             *- A. Einstein*

*"Nothing can be created out of nothing"*             *- Lucretius*

If the universe is a CA computer simulation, then there must have been a point where the simulation was first started. This occurred ≈15 billion years ago (our time), according to the standard big bang theory. It is important to realize that the creation of the numeric state of our universe (if it were to be done *now,* as a single act of creation without evolution) would be **very** difficult to accomplish now. All the galaxies, stars, and planets, and life forms must be specified for all the states of the cellular automata cells, which for our universe is something on the order of something like $10^{100}$ cells per cubic meter of space!

Our universe contains on the order of a few billion galaxies, and many galaxies have on the order of 100 billion stars in it. Currently, there is also evidence for the possibility that a certain percentage of these stars have one or more planets circling around them. Each star and planet has it's own unique orbit, chemical composition, temperature, rotation rate, size, atmosphere, landscape and possibly even life forms. In the process of creating our universe, it is far more economical to start with just the "right" rules of the cellular automata so that stars and planets are the natural byproducts of the evolution of the CA (and possibly life as well). In other words, let the natural evolution of the CA run its course. It is also more "interesting" to start this process, and than "see" what comes out of it after a lot of computer processing. In fact, that is what the purpose of our universal CA computer is, it is to compute our universe! CA theory absolutely requires that our universe be an evolutionary process, with a simple beginning.

## 6. WHY OUR UNIVERSE IS MATHEMATICAL IN NATURE

*"Why is it possible that mathematics, a product of human thought that is independent of experience, fits so excellently the objects of physical reality."*
            *- Albert Einstein*

It is clear that all the known laws of physics are mathematical in nature. Many physicists like Einstein, for example, have commented on this mysterious fact. No good explanation has been given as to why this should be so. This fact is made even more mysterious when one considers that mathematics is strictly an invention or byproduct of intellectual activity. In a sense, mathematics is like art and music. For example, the mathematical concepts of infinity, the imaginary numbers, and the Mandelbrot set in the complex plane are all mathematical objects that are invented by mathematicians. In mathematics, you start with virtually any set of self-consistent axioms, and formulate new mathematics as you please. Mathematics is strictly a *creative* process. Yet, our universe definitely operates in a mathematical way. Every successful physical theory has been formulated in the language of mathematics, and a good theory can even predict new phenomena that was not expected from the original premises.

The cellular automata model provides a clear explanation as to why the universe is mathematical. Quite simply, everything in our universe is numerical information, which is governed by mathematical rules that specify how the numbers change as the computation progresses. In short, "**the universe is numbers**", as was once proclaimed by the great Greek philosopher and mathematician Pythagoras. The design of the cellular automata must have required intelligence, which was applied to the cellular automata in the form of the mathematical rules for the cells. CA theory claims that all the laws of physics that we know today are mathematical descriptions of the underling, discrete mathematical nature of the numeric patterns that are present in our universal cellular automata. Fredkin (ref. 34) once proposed that the universe should be modeled with a single set of cellular automaton rules, which will model all of microscopic physics exactly. He called this CA 'Digital Mechanics'. The laws of physics in this form are discrete and deterministic, and would replace the existing differential equations (based on the space-time continuum) for modeling all phenomena in physics.

Therefore, the current thrust to discover the theory of everything (or simply The Theory as it is now known) should not be looking for a set of partial differential equations incorporating relativity and quantum field theory. Instead, we should be looking for the correct structure of the universal CA and the set of logical rules that govern it's operation.

## 7.     QUANTUM PARTICLES IN SAME STATE ARE INDISTINGUISHABLE

*" Common sense is the layer of prejudice laid down in the mind prior to the age of 18"*
                                                                                                                *A. Einstein*

A particle physicist once remarked that elementary particles behave more like mathematical objects than like familiar point-like objects. Particles are able to transform from one species type to another. Particles seem to be spread-out in some sort of oscillatory wave, and at other times they seem like point-like objects. Particles can be readily annihilated and created. None of these processes seem familiar from our everyday experience. One of the most unfamiliar of all particle attributes is indistinguishability.

Quantum Mechanics teaches us that electrons in the same quantum state (or having the same quantum numbers) are absolutely identical, and indistinguishable from each other. You cannot mark one electron so that it is different than another. An electron is currently described by quantum mechanics as a particle with quantum numbers like: mass, charge, spin, position, and momentum, which are represented as numbers in the wave function of the electron. It is these properties alone tell you all there is to know about the electron. The electron has no size or shape. Quantum Mechanics has definitely ruled out any classical or 'mechanical' models to help us 'visualize' what an elementary particle really is.

Equality is strictly a mathematical concept. In mathematics, the equality 1+1=2 is exact. In classical physics, no two marbles can be constructed to be exactly the same. When it comes to elementary particles, however, two quantum particles can be **exactly** the same. According to quantum mechanics, two electrons in the same state of motion (and spin) are absolutely identical and indistinguishable. The cellular automata model explains this remarkable fact simply by stating that the two electrons in the same state have *exactly* the same numeric information pattern, and thus described by the same quantum wave function. Therefore, they are mathematically identical. In constructing a universe, it is very desirable to have building blocks that are identical, and exactly repeatable, so that large complex structures can be easily formed.

## 8. EINSTEIN'S RELATIVITY IN THE CONTEXT OF CA

**Special relativity is manifestly compatible with the CA model**! We have proposed that our universe is a vast Cellular Automaton. For this to be true, all physical phenomena must come from the strictly local interactions inside the CA. The very nature of the cellular automata model is totally incapable of any instantaneous action at a distance, since information can only be sent from cell to adjacent cell in any direction, only at each and every 'clock' pulse. This means that there can be no action at a distance in *all* the laws of physics. Einstein abolished action at a distance in special relativity with his famous velocity of light postulate. He also removed gravitational action at a distance in general relativity by replacing Newton's instantaneous gravitational force law with his space-time curvature concept. In the following sections, we will see how and why special relativity is manifestly compatible with our CA model!

Special relativity is one of the most successful theories of physics, and along with quantum theory forms one of the two great pillars of modern physics. However, it has failed to account for *why* the universe has a maximum speed, which has still remained as one of the two postulates of special relativity. CA theory provides a simple explanation for this. In fact, the CA model *demands* that the universe have a maximum speed limit! In addition to this, the second postulate regarding the relativity of inertial frames (constant velocity motion) can also be seen as a simple consequence of the basic structure of the CA.

However, general relativity as it is currently formulated, is *not* compatible with CA theory. First, general relativity is formulated with the classical continuum concept for matter-energy, and is also formulated with a space-time continuum. Both of these fields are not generally compatible with the CA model, or with quantum theory. Secondly, there is no known local action that couples a large mass to the surrounding space-time curvature. What is it about a large mass that causes space-time curvature around it? In general relativity, there exists a global tensor field called the 4D space-time metric, which merely describes the amount of the curved 4D space-time. Because of the relative nature of space-time (observers in free fall near the earth live in flat space-time), it is very difficult to conceive how relativistic 4D space-time can work on a CA. How does the principle of equivalence work on a CA? Why does the inertial mass is equal to the gravitational mass, especially since they are defined differently?

General relativity has failed to make any progress towards the understanding of inertia. Inertia is introduced in general relativity exactly as was conceived by Newton in his famous inertia law: F=MA. Associated with Newton's formulation of inertia are the problems introduced by Mach's principle, which is a loose collection of ideas and paradoxes that have to do with accelerated or rotating motion. Mach argued that motion would appear to be devoid of any meaning in the absence of some surrounding matter, and that the local property of inertia must somehow be a function of the cosmic distribution of all the matter in the universe. Mach's principle has remained as an *untestable* philosophical argument, even within the scope of general relativity. We will find that general relativity must be revised in order to be compatible with CA theory. These modifications of general relativity came about by a new understanding of inertia and the principle of equivalence. Inertia will be described on the quantum level in a model that is compatible with CA theory, and which also automatically resolves Mach's paradox (this new theory of inertial is called Quantum Inertia or QI). This new theory of inertia also explains the origin of the Einstein principle of equivalence, which is not really a fundamental principle of nature, but due to similar quantum processes occurring in accelerated frames and gravitational fields. To summarize, general relativity is reformulated with a new approach to inertia called Quantum Inertia, which explains the origin of the principle of equivalence, and in a form that is now manifestly compatible with CA theory. This new theory is called 'ElectroMagnetic Quantum Gravity' or EMQG. It is a quantum theory of gravity, because matter is treated as quantum particles, and 4D space-time is quantized and results from pure quantum particle processes. Furthermore, the action between a large mass and the surrounding 4D space-time is clearly understood. Before we formulate EMQG, a review of relativity is necessary from the point of view of compatibility with CA theory.

## 9.    SPECIAL RELATIVITY AND CELLULAR AUTOMATA

**"... space by itself and time by itself, are doomed to fade away into mere shadows ...."**

**- H. Minkowski**

**POSTULATES OF SPECIAL RELATIVITY**

Special Relativity theory is founded on two basic postulates:

**(1) The velocity of light in a vacuum is constant and is equal for all observers in inertial frames (inertial frame is one in which Newton's law of inertia is obeyed).**

**(2) The laws of physics are equally valid in all inertial reference frames.**

These postulates are used by Einstein to derive the famous Lorentz transformations, a set of equations that relate space and time measurements between different inertial frames. The second postulate implies that there are no absolute reference frames in the universe that can be used to gauge constant velocity motion. All inertial frames are equally valid in describing velocities. In a general sense, all the laws of physics are also equally valid in all inertial frames. Some of the important consequences of special relativity and the Lorentz transformations are:

(1) The universe is four dimensional, where 3D space and time now have to be united.
(2) There is a maximum speed to which matter can obtain.
(3) Mass and energy are interchangeable.
(4) Momentum (and mass) is relative. Mass varies with the relative velocity between two inertial frames.
(5) Spatially separated events that are simultaneous in one inertial frame are not generally simultaneous in another inertial frame.

The special theory of relativity also implies that the speed of light is the limiting speed for any from of motion in the universe. Furthermore, light speed appears constant no matter what inertial frame an observer chooses. However, nowhere in special relativity theory, (or any other theory we are aware of) is there an explanation as to why this might be so. It is simply a postulate, based on physical observations such as the Michelson-Morley experiment. The second postulate also implies that there are no experiments that can be performed that will reveal which observer is in a state of 'absolute rest'.

The second postulate of special relativity states that the laws of physics are equally valid in all inertial reference frames. Stated in a weaker form, there are no preferred reference frames to judge absolute constant velocity motion (or inertial frames). This latter form is easily explained in CA theory, by remembering that all cells and their corresponding rules in the cellular automata are absolutely identical everywhere. Motion itself is an illusion, and really represents information transfers from cell to cell. To assign meaning to motion in a CA, one must relate information pattern flows from one numeric pattern group with respect to another group (the actual cell locations are inaccessible to experiment). Therefore, motion requires reference frames. Unless you have access to the absolute location of the cells, all motion remains relative in CA theory. In other words, there is *no* reference frame accessible by *experiment* that can be considered as the absolute reference frame for constant velocity motion. (Later, we will see that virtual particles of the quantum vacuum still do not allow us to reveal our (constant velocity) motion between two inertial frames. However, this is not the case for an accelerated observer or for observers in gravitational fields).

Since the contents of the cells or their locations are not physically observable to us, they cannot be used to help us setup a universal absolute reference frame for motion. However, there does exist a universal reference frame in the CA, and this frame is completely hidden from experimentation, which we call the 'CA absolute reference frame'. We associate with this frame an **absolute space** and **absolute time**. Everyday objects like this desk, which is a very large collection of elementary particles, occupies a specific volume of cells in CA space. These cell patterns are (most likely) shifting through our cell space at some specific rate. Therefore, there does exist a kind of Newtonian absolute space and absolute time scale, but these are hidden from the viewpoint of an observer living in the CA. We will find that in EQMG theory, the idea of absolute CA space and CA time becomes very important in considerations of inertial and gravitation frames. Even more important to observers, is the state of the virtual particles of the quantum vacuum. These virtual particles can act to produce forces for observers in a state of acceleration. We will see that Mach's principle and Einstein's weak principle of equivalence depend on the existence of virtual particles. We will return to this subject later, when we formulate EQMG theory.

In order to provide a full explanation for the postulates of special relativity, a detailed model for matter and space is required for the cellular automata theory of our universe. Since this model has not been found yet, we can use some simple assumptions about the nature of matter in a CA. In cellular automata, the clock rate specifies the time interval in which all the cells are updated, and acts as the synchronizing agent for the cells. Matter is known to consist of atoms and molecules, which themselves consist of elementary particles bound together by forces. An elementary particle in motion is represented in CA theory by a shifting numeric information pattern, that is free to 'roam' from cell to cell. Recall that space consists of cells or storage locations for numbers in the cellular automata, and particles (number patterns) freely 'move' in this cell space. From these simple ideas, it can be seen that there must be a maximum rate that number patterns are able to achieve. This is due to the following two reasons. First, there is fixed, constant rate in which cells can change state. Secondly in CA theory, information can only be transferred sequentially, from one cell to adjacent cell, and only one cell at a time per clock cycle. This is simply a limitation of the structure of the cellular automata computer model. The CA structure provides the most massively parallel computer model known. It is the CA's high degree of parallelism that is responsible for this limitation, because a particular cell state can only be affected by its immediate neighbors. Information can only evolve after each 'clock' period, and information can only shift from cell to adjacent cell. These facts result in a *definite* **maximum speed limit** for information transfers on the CA.

This maximum speed limit might represent light velocity, which is the fastest speed any particle can go (figure 15). *(NOTE: Later, we shall see that this maximum speed actually represents the raw or 'low-level' light velocity, defined as the velocity of light in between encounters with virtual particles. We will see that the scattering of photons with the virtual particles of the quantum vacuum reduces the speed of the photons to the familiar observable light velocity).* This maximum speed limit can be calculated if the precise quantization scale of space and time on the cellular automata level is known. Let us assume for now that the quantization of space and time corresponds *exactly* to the plank distance and the time scales. This means that the shifting of one cell represents a change of one fundamental plank distance $L_P$: $1.6 \times 10^{-35}$ meters, and that the time required for the shift of one cell is one fundamental plank time $T_P$: of $5.4 \times 10^{-44}$ seconds. Let us further assume that a photon represents the fastest of all the information patterns that shifts around in the CA. In fact, we propose that the photon information pattern is *only* capable of shifting one cell per clock period, and not at any other rate, and therefore exits at one speed with respect to the cells. The value for the speed of light can then be derived simply as the ratio of (our) distance over (our) time for the information pattern transfer rate. The maximum information transfer velocity is thus (figure 15):

$$V_P = L_P / T_P = 3 \times 10^8 \text{ meter/sec} = c \tag{9.1}$$

Therefore, $V_P = c$, the speed of light. The velocity of light can also be expressed as one plank velocity, which is defined in units of plank length divided by plank time. (There are plank units for mass, temperature, energy, etc as detailed in ref. 10).

Thus, the fastest rate that the photon can move (shift) is an increment of one cell's distance, for every clock cycle. If two or more clock cycles are required to shift information over one cell, then the velocity of the particle is lower than the speed of light.

To summarize, in cellular automata theory the maximum speed simply represents the *fastest* speed in which the cellular automata can transfer information from place to place. Matter is information in the cellular automata, which occupies the cells. The cells themselves provide a means where information can be stored or transferred, and this concept corresponds to what we call the 'low level' discrete space. 'Low level' time corresponds to the time evolution of the state of the cellular automata, which is governed by the 'clock period'. To put it another way, the rate of transfer of information in any cellular automata is limited, and infinite speeds are simply not possible. Of course, this rules out action at a distance, which is why CA theory is manifestly compatible with special relativity.

In passing, it is interesting to note that in the famous 2D Geometric CA, called Conway's game of life, there exists a stable, coherent 'L' shaped pattern commonly known as a 'glider' pattern. This pattern is always contained in a 3 x 3 cell array, and the glider completes a kind of an internal 'oscillation' in four clock cycles. Thus, in four clock cycles it returns to it's initial 'L' shaped starting pattern. This glider travels in 2D cell space, at *one fixed speed*! It is also the fastest moving pattern known in Conway's game of life. The glider particle in some sense resembles the photon particle in our universe! It has an internal oscillation, and it only moves at one fixed velocity. However, the similarity ends here, because in the game of life, the glider only moves in four fixed directions.

Special Relativity predicts through the work of Minkowski, that space and time must be united into a special four-dimensional space-time structure. This unification can be thought of as being the only way to restore the mathematical concept of space in special relativity. In this way, observers in any state of uniform motion can agree on the measure of a mathematical 'distance' between two events. According to special relativity, if a distance "d" is measured in frame A, then this distance is generally different than that measured in another frame B which is in the state of constant velocity motion 'v' with respect to frame A. As a result, the ordinary 3D formula for the distance between two points $(x_1,y_1,z_1)$ and $(x_2,y_2,z_2)$ at time t, which is given by $d^2 = (x_2-x_1)^2 + (y_2-y_1)^2 + (z_2-z_1)^2$ will generally not be agreed on by all observers. This is because of the special relativistic distance contraction and time dilation that occurs in co-moving frames. But Minkowski showed that if a four dimensional coordinate system is properly chosen with the following line element: $ds^2 = dx^2 + dy^2 + dz^2 - c^2dt^2$ , (or which is represented by a 4D coordinates given by (x,y,z,ict) where i=√-1) then distance and time measurements become invariant under any inertial frame of reference. In other words, all inertial observers agree on measurements of space and time in this coordinate system regardless of the relative velocity. This is what we refer to as the high level 4D space-time continuum of special relativity. This is the space-time that is accessible to measurement, and generally depends on the state of relative motion of the observer. Most importantly, this space-time system is *relative*, and depends on the state of motion of an observer.

It can be shown that the constancy of light velocity, and the principle of relativity (Einstein's first and second postulate) leads directly to the famous Lorentz Transformations, a set of equations that allows us to relate space and time measurements between two different inertial reference frames. The Lorentz transformations between an observer 'B' in (x*,y*,z*,t*) moving at velocity 'v' with respect to the reference observer 'A' in (x,y,z,t) are given by (ref. 35):

$x^* = (x - vt) * (1 - v^2/c^2)^{-1/2}$ (9.2)
$y^* = y$
$z^* = z$
$t^* = (t - (v/c^2) x) * (1 - v^2/c^2)^{-1/2}$

These equations are derived by examining the motion of light that is spreading in a spherical wave front, starting at time t=0 (in the frame of observer 'A'), from the perspective of both inertial observers. Based on the constancy of light velocity in all inertial frames, the equations for the spherical wave front at time t for both observers (recall that the equation of a light sphere is: $x^2 + y^2 + z^2 = r^2$ , where r=ct and r=ct* and the velocity of light is the same for both observers) is given by:

$x^2 + y^2 + z^2 = c^2t^2$ … light sphere seen by observer 'A' after time t (9.3)
$x^{*2} + y^{*2} + z^{*2} = c^2t^{*2}$ … light sphere seen by observer 'B' after time t* (9.4)

It is a matter of pure algebra (which we will not repeat here, see ref. 35 for the detailed calculations) to derive the Lorentz transformation from these equations. The Lorentz transformation is at the heart of special relativity, from which we can derive the relativistic velocity addition formula, Lorentz length contraction, Lorentz time dilation, and many other results. Notice that the light velocity 'c' in the above equations is the measured light velocity, measured with identically constructed rulers and clocks by both observers.

How can we translate the behavior of photons existing in CA absolute space and separate absolute time, into a statement concerning the *measurable* light velocity? In other words, into a statement based on an inertial observer's actual measurement data for his light velocity measurement. Furthermore, how do we compare these readings with respect to other inertial observers, who also use *actual measuring instruments*? A definition of a space and time measurement must be defined, along with a method of comparing these measurements among different observers in different inertial reference frames. This definition is required, because light velocity is defined as the measured distance that light moves, divided by the measured time that is required to cover this distance.

First we must define an inertial reference frame in far space, away from gravitational fields. Imagine a three dimensional grid of identically constructed clocks, placed at regular intervals measured with a ruler (figure 17 illustrates what an inertial reference frame is), in the three dimensional space (ref. 36). Local observers are stationed at each of the clocks. Thus, the definition of an inertial frame is a whole set of observers uniformly distributed in space as we have described. All observers in a given reference frame agree on the position and the time of some event. Only one observer would actually be close enough to record the event (an event is defined as something that occurs at a single point in space, at a single instant in time). The data collected by all the observers are communicated to the others at a later time (by any means). Notice how light naturally enters in the definition of an inertial reference frame. Light is required by observers to literally 'see' the clock readings.

Now we are in a position to evaluate the Lorentz transformations from our low-level CA definition of space, time, and constancy of light velocity. Light is an absolute constant in absolute CA space and time units. No matter what the state of motion of the source, whether it is an inertial source of even if it is accelerated, the light moves as an absolute constant that is unaffected by the source. We must now translate this statement about *measured* light velocity, into the actual reality of the CA with imaginary observers with highly specialized measuring instruments capable of measuring plank distance and time units (which is not possible in our reality). Let us introduce an absolute, discrete (3D space) integer array: [$x(k),y(k),z(k)$], where information changes state at every $t(k)$ (figure 18). These units represent our absolute space and time measurements (but in practice, we cannot actually make these measurements). The origin is an arbitrarily chosen cell (which can be looked at as being at absolute rest on the CA). A shift of data from one cell in any space direction to next, for example from $x(5)$ to $x(6)$, represents one plank distance unit (pdu), and if this take one clock unit, it happens in one plank time unit (ptu). The velocity of light represents one plank velocity unit (pvu) in our absolute units. We intend to show that when two different inertial observers measure light velocity using *absolute* space and time units, both observers *measure* light velocity as being one plank velocity. However, space and time *measurements* between our two inertial observers, do *not* compare in our absolute units. We will show that this is the same situation we find in special relativity, for two observers with *real* measuring instruments in space-time.

Imagine two inertial observers with a relative velocity '$v_r$' in the CA absolute units. Both observers are in a state of constant velocity motion with respect to our absolute cell coordinate system. Observer 'A' contains a green light source and moves with absolute velocity $v_a$ with respect to the cell rest frame. Observer 'B' moves with absolute velocity $v_b$, and is moving away from our observer 'A' (so that $v_b > v_a$), and $v_r = v_b - v_a$. Both observers carry measuring instruments capable of measuring space and time in absolute units (figure 16). Of course, this is not actually possible with real observers.

Observer 'A' measures the velocity of light of his green light source, with his measuring instruments. He uses a ruler of length 'd' in absolute units, and measures the number of CA clock cycles it takes for the wave front of the green light to move the length of the ruler. Because observer 'A' is moving with velocity $v_a$, with respect to the absolute frame, his measurement of length and time are distorted. Recall that light simply shifts from cell to cell, in every clock unit immediately after leaving the source (figure 16). His measurement of length using the wave front moving across his ruler, appears longer, because of his motion $v_a$. Thus, observer 'A' distance measurement appears longer by: 'd + $v_a$d' pdu, where $v_a$ is less than one. (for example, if $v_a$ = ½ pvu, and d=1,000,000 pdu, then the distance measured is 1,000,000 + ½ 1,000,000 pdu). In comparison, an observer at absolute rest would measure a distance of 'd' pdu. Similarly, the clock measures a longer time, because it takes longer for the wave front to reach the end of the receding ruler. Therefore, the time required to transverse the ruler is: 'd + $v_a$d' ptu (in our example, the time taken for light to traverse the ruler is 1,000,000 + ½ 1,000,000 ptu). Thus, the *measured* light velocity in absolute CA units is: $(d + v_a d) / (d + v_a d) = 1$ pvu, the velocity of light. Similarly, for observer 'B' moving at velocity $v_b$, the measured velocity of the green light he receives in his reference frame is: $(d + v_b d) / (d + v_b d) = 1$ pvu, again equal to light velocity in absolute units. Thus, both observers conclude that light is a universal constant, equal to one pvu, no matter what the state of motion of the light source in an inertial frame! This is similar to the same situation in ordinary space-time.

What happens if observer 'A' sends his measurements to observer 'B' (by any means, carrier pigeon for example)? First, will observer 'B' conclude that the color of light received from 'A' is green? Secondly, will the measured distances and times be equal? It is obvious from the above analysis, that the measurements are not equal, unless $v_a = v_b$! Furthermore, observer 'B' concludes that the received light is shifted towards the red. Why? Observer 'B' examines the light received from 'A'. A 'wave marker' passes by him, and he then finds that the next 'wave marker' appears to take a longer time to arrive, compared to when both observers are both at absolute rest. Thus, the light appears to have a longer wavelength that is shifted towards the red, when compared to observer 'A'. The actual spacing between 'wave markers' is constant, and was determined by the energy of observer 'A's light emitting equipment. Note that observer 'A's measurement of his light wavelength at velocity $v_a$ is actually different from the wavelength measurement when he is at absolute rest, when measured in absolute units!

Let us now examine the results of the same experiment with measurements made in *ordinary* space-time, with ordinary measuring equipment like clocks and rulers. A common reference is required to make comparison measurements of length and time, since the absolute coordinate system is *not* available. Based on our definition of reference frames (as a grid of observers), light becomes the natural choice for comparative space-time measurements. Observer 'A' decides to define length in terms of the green light from his light source, where one basic length unit (bdu) ≡ 1000 wavelengths of green light, from which he has constructed a standard ruler of this length. Similarly, observer 'A' chooses to define the time of one basic time unit (btu) as the elapsed time required to receive 1000 cycles (or 1000 audible clicks from each wave crest, for example) of the green reference light, from which he constructs a calibrated standard clock. Observer 'B' has the identically constructed ruler and clock. Now, as before, observer 'B' has a relative velocity of $v_r$, with respect to 'A'. What happens when observer 'B' makes measurements on the incoming green light, sent by observer 'A'?

Now we do *not* have the luxury of absolute units to arbitrate between the two observers. Furthermore, no observer can be regarded as being at absolute rest! Both observers have an equal right to formulate the laws of physics of motion in his own frame. Observer 'A' measures the light velocity as follows: The green light travels distance '$D_a$' in time $T_a$, and therefore the measured light velocity is: $c = D_a/T_a$. Observer 'B' uses his identically constructed standard clock and standard ruler to measure the incoming green light. Does his measuring instruments measure the velocity of light the same as observer 'A'. The answer is yes. Recall that observer 'A''s velocity does not affect the light velocity at all. It is an absolute constant, and cannot be affected by the source motion. Recall that observer 'A' specifies the wavelength of light, through his source apparatus. Once set, the wavelength of light propagates as a constant, not affected by the source (as described in our CA model of light above). Therefore, observer 'B' measures the light velocity as follows:

$(D_a + kD_a) / (T_a + kT_a) = [(1 + k)/(1+k)] (D_a/T_a) = D_a/T_a = c$ as for observer 'A'.  (9.5)

The motion of observer 'B' ruler adds a length of $kD_a$ cycles of light to his measurement distance, and adds the same $kT_a$ time delay, leaving the measured light velocity the same as 'A'. Observers 'A' and 'B' decide to compare their space and time measurements, with their identically constructed ruler and clock. Do these measurements agree? It is very clear that they do not!

Observer 'B' performs similar measurements, with identically constructed equipment on the incoming green light. Observer 'B' notices that the wavelength of the green light is shifted to the red, as we just discussed. Thus, his standard ruler of a length of one bdu contains *less* than 1000 wavelengths of the incoming light, because each wavelength is longer than 'A's (recall that 1 bdu is the length containing 1000 wavelengths of light). Similarly, he notices that when he listens for 1000 audible clicks (which should correspond to one btu of time), more than one btu of time elapses on his identically constructed clock (because each click takes a longer time to arrive). When the results of these measurements are compared by any means (by carrier pigeon, for example) observer 'B' concludes that his time has been dilated, and his distances have contracted compared to observer 'A's measurement. Incidentally, if observer 'B' has the green light source and shines it towards 'A', observer 'A' would conclude the same thing. How do we mathematically compute the values of these space-time comparisons? One may be tempted to apply a one dimensional Doppler-type analysis to deduce the quantity of space-time distortion. This, however, would *not* yield the correct answer. The above analysis is applicable for all the 3 dimensions that light can travel in space. Therefore, one must correct for light moving in all directions. This is precisely how Einstein derived the Lorentz transformations! In other words, the velocity of light measured in all directions of an expanding spherical wave front is what we take to be a constant. Thus, by showing that the velocity of light propagates as a constant in all directions in CA absolute space and time, we find that all inertial observers measure light velocity as a constant. However, they do not agree on the actual values of the space and time measurements. In this way, the principle of relativity leads us directly to the Lorentz transformation.

In summary, by postulating that on the lowest level of the CA, photons are information patterns 'moving' by a simple shifting from cell to adjacent cell at every clock 'cycle' in any given direction. As a consequence of this we found that:

(1) Light propagates in a kind of absolute, quantized 3D space, and separate 1D time (plank units) of the cellular automata, whose velocity is totally *unaffected by the source motion*. The light source determines the energy, and therefore the wavelength of the light. Once the light leaves the source, the wavelength and velocity is an absolute constant, specified in absolute CA units.
(2) In absolute CA space and CA time units, observers have an absolute velocity. The actual cell addresses of the information on the CA form the absolute 'rest' frame (which is not directly accessible by experiment). Hypothetical measurements in these absolute units yield light velocity and wavelength to be a constant, no matter what the state of motion of the source.
(3) When two (or more) inertial observers, with real measuring instruments are employed, and the measurements are made in the familiar 4D space-time defined by relativity theory, we have shown that all observers *measure* the velocity of light as a constant. However, when two (or more) inertial observers compare their space and time measurement (which is required to measure velocity, the measurements can be communicated by any means), they find that the measurements do not agree.
(4) We showed that the *measured* light velocity is constant in *all* space directions, which still remains only a postulate of special relativity. The Lorentz transformation directly follows from this through simple algebra. The Lorentz transformations form the core of special relativity, and yields the familiar results of relativity such as: time dilation, Lorentz contraction, velocity addition, and so on.

In regards to inertial frames, one might be tempted to consider that the virtual particles of the quantum vacuum might act as some sort of an abstract universal reference frame. One might think that the virtual particles in the neighborhood of a point might be used to gauge your constant velocity motion, a frame that would have been unknown to Einstein when he formulated special relativity. However, the virtual particles have completely random velocities, move in completely random directions, and most importantly are short lived and unobservable. Furthermore, one cannot 'tag' the virtual particles with labels, and follow the progress of all the virtual particles in order to judge your own motion with respect to the average motion of the virtual particles! Therefore, it is impossible to tell your state of constant velocity motion with respect to the vacuum, unless a force or some other vacuum phenomena makes it's presence felt. It is a well-known experimental fact that virtual particles introduce no new forces for inertial observers. However, this is definitely not the case for an accelerated frame, where we are concerned with the state of the acceleration vectors of the virtual particles with respect to a Newtonian accelerated mass (F=MA). Here forces are present, which originate from the electromagnetic interaction of the quantum vacuum with the matter. We will see that this becomes the basis for the formulation of EMQG for accelerated reference frames, and also for gravitational reference frames!

Acceleration is a special motion, because an accelerated observer can detect his state of acceleration (inside a closed box, for example) by simply measuring the force exerted on him with an accelerometer. He does not need to compare his motion against some other reference frame to find out if he is accelerating. Newton was well aware of this fact, which led him to postulate the existence of 'absolute' space. Therefore, it appears that an accelerated test mass does *not* require another reference frame to gauge motion, and therefore acceleration has a special status in physics. However, it will be shown through the new quantum principle of inertia (discussed later), acceleration also has a special hidden reference frame that was unknown to both Einstein and Newton when they formulated their famous theories of motion. The reference frame in question here is the state of accelerated motion of the test mass with respect to the virtual particles of the quantum vacuum. However, it is not the velocity of the particles that sets up this abstract reference frame, it is the net statistical average **acceleration** of the virtual particles of the quantum vacuum near the test mass that forms the absolute reference frame. These concepts affect the meaning of inertial mass. Therefore, we elaborate on the meaning of inertial mass in the next section.

## 10. SPECIAL RELATIVITY - INERTIAL MASS AND INERTIAL FORCE

**"In contrast to the Newtonian conception, it is easy to show that in relativity the quantity force, is not codirectional with the acceleration it produces ... It is also easy to show that these force components have no simple transformation properties ...."**

- M. Hammer

Quantum Inertia (QI) provides (section 13) a new understanding of Newtonian momentum. We will show that it is only *inertial force* (and forces in general) that is truly a *fundamental* concept of nature, not momentum or conservation of momentum. The Newtonian momentum, which is defined by 'mv', is simply a bookkeeping value used to keep track of the inertial mass 'm' (defined as F/A) in the state of constant velocity motion 'v' *with respect to another mass* that it might collide with at some future time. In this way, momentum is a relative quantity. Momentum simply represents information (with respect to some other mass) about what will happen in later (possible) force reactions. This fits in with the fact that *inertial* mass cannot be measured for constant velocity mass in motion (in outer space for example, away from all other masses) without introducing some **test** acceleration. If a mass is moving at a constant velocity, there are **no** forces present from the vacuum. Furthermore, since momentum involves velocity, it requires some other inertial reference frame in order to gauge the velocity 'v'. The higher the velocity that a mass 'm' achieves, the greater will be the subsequent deceleration (and therefore the greater the subsequent inertial force present) during a later collision (when it meets with some another object). If the velocity doubles with respect to a wall ahead, for example, then the deceleration doubles in a later impact. Before doubling the velocity, the acceleration $a_0 = (v_0 - 0)/t$; and after doubling, $a = (2v_0 - 0)/t = 2a_0$. Therefore we find that $f = 2f_0$, the force required from the wall (assuming the time of collision is the same). Similarly, if the mass is doubled, the force required from the wall doubles, or $f=2f_0$. Recall that inertial force comes from the **opposition of the quantum vacuum to the acceleration of mass** (or deceleration as in this case). Similarly, the kinetic energy '$1/2mv^2$' of a mass moving at a constant relative velocity 'v', it is also a bookkeeping parameter (defined as the product of <u>force</u> and the <u>time</u> that a force is applied). This quantity keeps track of the subsequent energy reactions that a mass will have when later accelerations (or decelerations) occur with respect to some other mass. It is important to remember that it is the **quantum vacuum force** that acts against an inertial mass to oppose any change in its velocity that is truly fundamental.

We therefore conclude that according to principles of QI theory, the inertial force is <u>absolute</u>. We have also seen that acceleration **can** be considered <u>absolute.</u> By this we mean that it is only the acceleration 'a' of a mass 'm' with respect to the net statistical average acceleration of the virtual particles of the quantum vacuum that accounts for inertial force. Therefore, we conclude that inertial mass can also be considered to be <u>absolute,</u> and follows the simple Newtonian relationship 'M=F/A'. Since inertial force, acceleration, and mass can all be considered to be absolute in this framework, we must closely reexamine the principles of special relativity in regards to the variation of inertial mass with the relative velocity of another inertial frame. Relativity is based on the premise that all constant velocity motion is relative, and also on the postulate of the constancy of light velocity. According to special relativity (which restricts itself to frames of constant velocity, called inertial frames), the inertial mass 'm' is relative, and varies with the relative velocity 'v' with respect to a constant velocity observer, in accordance with the following formula: $m = m_0 / (1-v^2/c^2)^{1/2}$. Here m is defined as the inertial mass measured in the other frame with velocity v, and $m_0$ is defined as the rest mass (inertial mass measured in the same frame as the mass) and 'c' is the velocity of light. It appears on the surface that QI and special relativity are not compatible in regards to the meaning of inertial mass. From the point of view of quantum inertia, Einstein's definition of inertial mass cannot be *fundamentally* correct, because it is not related to the quantum vacuum process described above for inertia. This is because we cannot associate the relative velocity 'v' directly to any quantum vacuum process. Recall that it is only the acceleration 'a' of a mass 'm', with respect to the net statistical average acceleration vectors of the virtual particles of the quantum vacuum that is the source of inertial mass.

Most special relativistic textbook accounts of inertial force and mass are based on the so-called 'conservation of momentum approach' (ref. 20). The conservation of momentum is assumed to be a fundamental aspect of nature. In order for momentum to be conserved with respect to all constant velocity reference frames, the mass must vary. To see this, recall that momentum is defined as mass times velocity, or 'mv', and that the momentum is important in a collision only because it provides bookkeeping of the mass and relative velocity. The **relative** velocity between the two colliding masses will determine the amount of deceleration in the impact as follows: $a=(v_f - v_i)/\Delta t$, were $v_f$ is the final velocity, and $v_i$ the initial velocity. Also, the mass is important because the subsequent force (and therefore energy $E = F \Delta t$) is determined by $F= m a$ through the quantum vacuum process described above. The more mass particles contained in a mass, the greater the resistance to the acceleration of the mass. Therefore, the product of mass and velocity is an indicator of the amount of **future** energy to be expected in a collision (or interaction) of the two masses. The total incoming momentum is defined as the momentum of the in-going masses $(m_1v_1 + m_2v_2)$, the total out-going momentum is $(m_1v_1' + m_2v_2')$. Here the two masses $m_1$ and $m_2$ are moving at velocities $v_1$ and $v_2$ before the collision, with respect to an observer, and velocity $v_1'$ and $v_2'$ after the collision. In Newtonian mechanics, the total momentum is conserved for any observer in a constant velocity reference frame. Therefore, $(m_1v_1 + m_2v_2) = (m_1v_1' + m_2v_2')$, even though different observers in general will disagree with each of the relative velocities of a pair of masses that are colliding. This is what we mean by conservation of momentum. In special relativity, if we do not modify the definition of inertial mass, we would find that different observers in different constant velocity frames ***disagree*** on the conservation of momentum for colliding masses. However, it can be shown (ref. 20) that if the mass of an object 'm' (from the point of view of an observer in constant velocity motion 'v' with respect to a mass $m_0$, measured by an observer at relative rest) is redefined as follows:

$m = m_0 / (1-v^2/c^2)^{1/2}$, then the ***total*** momentum of the collision remains conserved as in Newtonian mechanics.

How does special relativity treat the definition of inertial mass and inertial force? Since Einstein was aware that acceleration is not invariant in different inertial frames, he knew that Newton's law had to be modified.

To quote A.P. French (ref. 21):

*"… the discovery and specification of laws of force is a central concern of physics. It is certainly important, therefore, to know how to transform forces and equations of motion so as to give a description of them from the point of view of different inertial frames. Since in special relativity the acceleration is not invariant, we know that we cannot enjoy the simplicity of Newtonain mechanics, but we can certainly arrive at some useful and meaningful statements.*

*The starting point, which we indeed made use of in the initial stages of our approach to relativity is Newton's law in the form*

*$F = dp/dt = d(mv)/dt$     where $m = m_0 (1-v^2/c^2)^{-1/2}$*

*We take this as a definition of F. It is a natural extension (and simplest extension) of the non-relativistic result. It is not a statement that can be independently proved."*

French goes on to analyze the consequences of this relativistic force definition for components parallel and transverse with respect to the direction of acceleration:

*"... for the case where $F_{0x}$ is applied parallel to v, causing an acceleration $a_{0x}$ ... $F_x = F_{0x}$ This is a striking result. Despite the change of the measures of mass and acceleration in the two frames, the measure of the x component of force remains the same.*

*When we make a similar calculation for the transverse force, we find that this invariance does not hold. ... i.e.     $F_y = 1/(1-v^2/c^2)^{1/2} \; F_{0y}$*

*In the above results one can discern the feature that in general force and acceleration are not parallel vectors. ... Only in the instantaneous rest frame of a body can one guarantee that F, as defined by the time derivative of momentum, is in the same direction as the acceleration."*

Einstein had to modify Newton's inertial law during his program to revise all physics in order to be relativistic, and was not aware of the existence of the quantum vacuum at that time. When Einstein considered this law, he found that in addition to incorporating his new relative mass definition formula above, he had to contend with relative accelerated motion. Contrary to popular belief, special relativity *does* address the problem of accelerated motion, which can be measured by any observer in an inertial reference frame. Therefore, in order to allow different observers in different states of constant velocity motion to measure inertial forces, Newton's law of motion must be changed. Since space and time are involved in measuring acceleration relative to an observer, and therefore acceleration must also be relative.

As we have seen in our QI analysis, **inertial mass is absolute**. Furthermore, there exists absolute acceleration of a mass, which is defined as the state of acceleration of the matter particles making up that mass, with respect to the (net statistical) average acceleration of the virtual particles of the quantum vacuum. (Note: Since virtual particles interact with each other, not all the individual accelerations of the virtual particles with respect to the mass will be the same, hence the statistical nature of this statement). What about the applied force? As the force is applied, an acceleration results which causes the velocity of the mass to increase. What if the velocity of the mass approaches light speed with respect to the applied force? Is the force still as effective in further increasing the velocity of the mass?

## 10.1 RELATIVISTIC MASS VARIATION FROM PARTICLE EXCHANGES

Classical physics is based on the assumption that forces between two bodies can act upon each other instantaneously through direct contact. Furthermore, the resulting action is independent of the relative velocity of the two bodies from which the force acts. When Einstein proposed special relativity, he abolished all action-at-a-distance including forces acting instantaneously. However, forces were still treated by Einstein within the framework of classical Newtonian physics. In his program to make classical physics relativistic, he accepted Newton's law of inertial force without modification (in fact, the law F=MA was postulated as still being correct). Modern physics now treats **all** forces as a quantum particle exchange process. Note: The gravitational force is a special case where two exchange particles are involved (section 15). We will find that two different force exchange particles are involved simultaneously: the photon and the graviton particle. As an example, consider the electric force exchange process, which involves the photon particle as described by Quantum Electrodynamics (QED). Here the charged particles (electrons, positrons) act upon each other through the exchange of force particles, which are photons. In the language of computer science, electromagnetic forces can be viewed as being 'digital'. What appears to be a smooth force variation is really the result of countless numbers of photon exchanges, each one contributing a 'quanta' of electromagnetic force.

To see how the exchange process works for electromagnetic forces, we will examine the classical Coulomb force law in the rest frame of two stationary charges. The electric force from the two charged particles decrease with the inverse square of their separation distance (the inverse square law: $F = kq_1q_2/r^2$, where k is a constant, $q_1$ and $q_2$ are the charges, and r is the distance of separation). QED accounts for the inverse square law by the existence of an exchange of photons between the two electrically charged particles. The number of photons emitted by a given charge (per unit of CA time) is fixed and is called the charge of the particle. Thus, if the charge doubles, the force doubles because twice as many photons are exchanged during the force interaction. This force interaction process causes the affected particles to accelerate either towards or away from each other depending on whether the charge is positive or negative (different charges transmit photons with a slightly different wave functions). It is interesting to note that certain cellular automata patterns exhibit behaviors like charge. For example, in the famous 2D geometric CA called Conway's game of life there exists a class of CA patterns called 'guns', which constantly emit a steady stream of 'glider' patterns indefinitely. This CA emission process is constant without any degradation of the original gun pattern. This resembles the charge property possessed by electrons, where photons are constantly emitted without any change of state of the electron.

The strength of the electromagnetic force depends on the quantity of the electric charge, and also depends on the distance of separation between the charges in the following way: each charge sends and receives photons from every direction. But, the number of photons per unit area, emitted or received, decreases by the factor $1/4\pi r^2$ (the surface area of a sphere) at a distance 'r', because the photon emission process take place in all directions. Thus, if the distance doubles, the number of photons exchanged decreases by a factor of four. This process can be easily visualized on a 3D geometric CA. Imagine that an electron is at the center of a sphere and sends out virtual photons in all directions. Imagine that a second electron somewhere on the surface of a sphere at a distance 'r' from the emitter, absorbing some of the exchange photons. The absorption of the exchange photons causes an outward acceleration, and thus a repulsive force. If the charge is doubled on the electron, there is twice as many photons appearing at the surface of the sphere, and twice the force acting on the electron. Thus, this accounts for the linear product of charge terms in the inverse square law. In QED, photons do not interact with photons (by a force exchange interaction). As a result, in-going and out-going photons do not affect each other during the exchange process.

We will go into more detail on the consequences of the particle exchange process for gravity, and the connection with cellular automata later (section 15.2). For now we are interested in the consequences of the force exchange process for special relativity, where the exchange particle has a <u>finite, and fixed velocity of propagation</u> (the speed of light, with the exception of the weak nuclear force where some bosons carry mass). To our knowledge, no one has examined the consequences of particle exchanges from the point of view of forces acting on each other in different inertial frames, where exchange particle propagates at the speed of light. At the time that Einstein developed special relativity, the force exchange process was unknown. The basic idea we want to develop here is that the quantity of force transmitted between two objects very much depends on the received <u>flux rate</u> of the exchange particles. In other words, the number of particles exchanged per unit of <u>time</u> represents the magnitude of the force transmitted between the particles. For example, imagine that there are two charged particles at relative rest in an inertial reference frame. There are a fixed number of particles exchanged per second at a separation distance 'd'. Now imagine that particle B is moving away at a slow constant relative velocity 'v' with respect to particle A. If the relative velocity v<<c the exchange process appears almost the same as when the two particles are at rest. This is because the velocity of light is very high when compared to 'v', and the flux rate is unaffected. Now imagine that the relative recession velocity v -> c, which is comparable to the velocity of the exchange particle. Does the received flux rate of particle B get altered from the perspective of particle B's frame? The answer is yes, and this follows from another result of special relativity: the Lorentz Time Dilation!

It is clear from Lorentz time dilation that the timing of the exchange particle will be altered when there is a very high relative velocity away from the source. Recall the Lorentz time dilation formula of special relativity: $t = t_0 / (1 - v^2/c^2)^{1/2}$, which states that the timing of events varies with relative velocity 'v'. If the timing of the exchange particles is altered, then the flux rate is altered as well, since flux has units of numbers of particles per unit time.

Now assume that particle A emits a flux of $\Phi_a$ particles per second, as seen by an observer in particle A's rest frame. When the force exchange particles are transmitted to particle B, particle B sees the flux rate decrease because of time dilation. Therefore, we find that particle B receives a smaller quantity of exchange particles per second $\Phi_b$ then when the particles are at relative rest. Thus, particle A acts like it transmits a smaller flux rate $\Phi_b$, such that $\Phi_b = \Phi_a \ (1 - v^2/c^2)^{1/2}$. Since the force due to the particle exchange is directly proportional to the flux of particles exchanged, we can therefore write:

$$F = F_0 \ (1 - v^2/c^2)^{1/2} \tag{10.1}$$

where is $F_0$ is the magnitude of the force when particle A and B are at relative rest, and F is the resulting smaller force acting between particle A and B when the receding relative velocity is 'v'. Thus, we can conclude that when a force acts to cause an object to recede away from the source of the force, the force <u>reduces</u> in strength. With a similar line of reasoning, we find that the force increases in strength when a force acts to cause an object to move towards the source of the force.

We are now in a position to see the apparent relationship between the inertial mass and velocity. Since all forces are due to particle exchanges, we can use the method developed above to study the forces acting between to inertial frames. First, at relative rest where v=0, we have F= $F_0$. The rest mass '$m_0$' is defined by Newton's law: $F_0 = m_0 \ a$, where 'a' is a test acceleration that is introduced to measure the inertial rest mass. Now, assume that there is a relative velocity 'v' between the applied force and the mass 'm', which causes the mass to recede. Therefore, we can write:

$$F = F_0 \ (1 - v^2/c^2)^{1/2} = m \ a \tag{10.2}$$

where the force is reduced in magnitude for the reasons discussed above, and the mass 'm' is considered absolute (or m= $m_0$, as in Newtonian Mechanics). In EMQG, we believe that equation 10.2 represents the actual physics of the force interaction. However, if one takes the position that the force does not vary with velocity, but that the mass is what actually varies, then the above equation can be interpreted as:

$F = F_0 = m \ a = m_0 \ (1 - v^2/c^2)^{-1/2} \ a$ , and $m = m_0 \ (1 - v^2/c^2)^{-1/2}$ as given by Einstein.

So we see that we are in a situation where it is experimentally impossible to distinguish between the following two approaches: inertial mass variation with high velocity (Einstein) versus the force variation with high velocity (EMQG). What velocities can a mass achieve through the application of an accelerating force? According to our analysis above, the answer is that the limiting speed is the speed of the exchange particles, or light velocity. At this limit, the accelerating force effectively becomes zero!

It is, however, convenient to associate the variation of force with an increase in relativistic mass as Einstein proposed, for two important reasons. First this restores the conservation of total momentum in collisions for all inertial observers (in fact, this is how Einstein derived his famous mass-velocity relationship). Secondly, if a mass is accelerated to the relativistic velocity 'v' with respect to observer 'A' by some given force, and this force is removed, there will be no way to determine the subsequent energy release when a collision occurs later. In other words, when this mass collides with another object, a rapid deceleration occurs with a large release of energy (which is force multiplied by time). This energy release is greater then what can be expected from Newton's laws. In fact, the large energy release is due to the effective increase in the force during the collision due to increased numbers of force exchange particles acting to reduce the speed of the colliding mass.

The force $F = F_0 (1 - v^2/c^2)^{1/2}$ tends to zero as the velocity v -> c. This means that any force becomes totally ineffective as the mass is accelerated to light velocity with respect to the source. As we have seen, this is attributed to the force resulting from exchange particles, which become totally ineffective in propagating from the source to the receiver, as the velocity of the receiver with respect to the source approach the velocity of the exchange particle. In order to clarify these ideas, we will analyze an actual experiment that was performed to confirm relativistic mass increase effect.

10.2   ANALYSIS OF 'ULTIMATE SPEED' EXPERIMENT FOR ELECTRONS

The textbook titled 'Special Relativity' by A.P. French (ref. 21) describes an actual experimental test of relativistic mass and energy, which was performed for a film called 'The Ultimate Speed'. This experiment is designed to measure special relativistic effects such as mass increase, momentum, and energy of an electron accelerated to relativistic velocities. The experiment consists of an electron gun, a linear accelerator, an oscilloscope to measure the electron time of flight, and an aluminum disk to stop the electrons and signal the arrival. A direct calorimetric measurement of the electrons is made in order to compare the energy transferred by the electrons during collision with the target. We will give our interpretation of this experiment in terms of the quantum vacuum interactions involved for mass, and the variation of accelerating force with relative velocity discussed above.

The electrons are emitted from the electron gun and accelerated through an electrical potential in the laboratory frame. We assume that the electrical force acts in a direction that is parallel to the direction of motion. An electrical force is required to accelerate the electrons in order to overcome the resistance offered by the charged virtual particles of the quantum vacuum, which is simply the electron's inertia. (Note: This is the definition of inertial mass in EQMG, where each charged masseon particle inside the electron contributes elementary quanta of inertia). This resistance to acceleration is the <u>absolute inertial mass</u> (or rest mass in relativity), where absolute mass is the *actual* resistance to acceleration originating from the vacuum. As the electrons are accelerated, photon particle exchanges occur between the charged electrons and the accelerating plates maintained at a high electrical potential. When the velocity of the electrons with respect to the accelerating plates in the laboratory frame is much smaller then the speed of light, the force exchange process is unaffected. In other words the accelerating electrical force is just as effective in accelerating the electrons as if the electrons are at rest, with the force being qV (q is the electron charge, and V is the accelerating potential. As the velocity of the electron approaches light velocity however, the timing of arrival of the exchange particles is delayed in the electron frame as described previously. The force on the electron is reduced to:

$F = F_0 (1 - v^2/c^2)^{1/2}$ , where $F_0$ is the magnitude of the force at low velocities.

Table 1-1 of ref. 21 tabulates the time of flight, the kinetic energy, and the velocity of the electron. It was observed that the velocity of the electron increased more slowly as the electron achieved relativistic speeds, in spite of the large forces applied. To quote A. P. French:

*"… the kinetic energy is raised by a factor of 30, so one might have looked for a factor of 5.5 in the speed (according to Newtonian Mechanics). Instead, there is an increase of only 15% (in the speed). The increase of v between 1.5 and 4.5 MeV is barely detectable within the accuracy of the experiment. One might therefore question whether the electrons are in fact being given the energy calculated from the value of qV …"*

In order to verify that the electrons actually have the energies calculated, the experimenter makes a *direct* calorimetric measurement of the collision energy of the electrons with the aluminum target plate. They conclude that the energy given to the electron (force multiplied by time) is definitely carried by the electrons during this collision, through their heat dissipation measurement.

What actually happens in the experiment however, is that the force becomes less and less effective in increasing the velocity of the electrons during the relativistic phase of the electron flight, in the reference frame of the accelerating electrons. In other words, the electrons achieve little increase in speed (a = f / m, and since the force 'f' decreases in EMQG, the acceleration 'a' decreases, m remaining constant) for the force applied. Where does the extra energy come from during the collision (as compared to Newtonian mechanics)? The collision of the electrons with the aluminum target is really an extremely rapid deceleration, and therefore involves the quantum vacuum. The electrons are rapidly decelerated by the aluminum target, which creates a relative acceleration between the charged electrons and the background, charged virtual particles of the quantum vacuum. However, in the reference frame of the relativistic electrons, the number of exchange particles per second increases (as compared to the slow velocity phase of the electrons), creating a larger force operating against the target. One can view the exchange particles as being 'red shifted' in terms of frequency (particles per second) during the acceleration phase of the electron trajectory giving a decrease in force, and 'blue-shifted' during the deceleration phase in collision giving an increase in force (and thus energy).

To summarize, **inertial mass is absolute.** The applied force that is used to accelerate a mass is reduced at relativistic speeds, and depends on the relative velocity between the applied force and the accelerated mass. The absolute acceleration of a mass is defined as the state of acceleration of the matter particles making up that mass with respect to the (net statistical) average acceleration of the surrounding virtual particles of the quantum vacuum.

## 10.3   EQUIVALENCE OF MASS AND ENERGY: $E = Mc^2$

One of the most important results of special relativity is the equivalence between mass and energy. This is represented in perhaps the most famous formula in all of physics: $E=Mc^2$. This formula implies that photons carry mass, since they carry energy. Photons are capable of transferring energy from one location to another, as by solar photons for example. Do photons really have mass?

You might think that if a particle has energy, it automatically has mass; and if a particle has mass, then it must emit or absorb gravitons (and thus composed of masseons). This reasoning is based of course, on $E=mc^2$. Einstein derived this formula from his famous light-box thought experiment (ref. 21). In his thought experiment, a photon is emitted from a box, causing the box to recoil and thus to change momentum. In quantum field theory this momentum change is traceable to a fundamental QED vertex, where a electron (in an atom in the box) emits a photon, and recoils with a momentum equivalent to the photon's momentum '$m_p c$". Therefore, we can conclude that the photon behaves as if it has an effective inertial mass '$m_p$' given by: $m_p = E/c^2$ in Einstein's light box. For simplicity, lets consider a photon that is absorbed by a charged particle like an electron at rest. The photon carries energy and is thus able to do work. When the photon is absorbed by the electron with mass '$m_e$', the electron recoils, because there is a definite momentum transfer to the electron given by $m_e v$, where v is the recoil velocity. The electron momentum gained is equivalent to the effective photon momentum lost by the photon $m_p c$. In other words, the electron momentum '$m_e v_e$' received from the photon when the photon is absorbed is equivalent to the momentum of the photon '$m_p c$', where $m_p$ is the effective photon mass. If this electron later collides with another particle, the same momentum is transferred. The rest mass of the photon is defined as zero. Thus, the effective photon mass is a measurable inertial mass.

**Note**: the recoil of the light box is a backward acceleration of the box, which works against the virtual particles of the quantum vacuum. Thus, when one claims that a photon has a real mass, we are really referring to the photon's ability to impart momentum. This momentum can later do work in a quantum vacuum inertial process. We will see in section 13.1 that although the photon carries inertial mass (and gravitational mass), it does not posses any low-level mass charge and does not contain any masseon particles.

Einstein's derivation of $E=mc^2$ was unnecessarily complex (ref. 21) because of his reluctance to utilize results from quantum theory. Although he was one of the founders of the (old) quantum theory, he remained skeptical about the validity of the theory throughout his whole career. In EMQG, we treat the energy-mass equivalence as a *purely* quantum process, and not as a result of special relativity. Although Einstein derived this law when he developed special relativity, it can be derived purely from quantum theory.

As we hinted, the ability of a photon to transfer momentum (and thus carry energy) can be traced to a QED vertex, where a packet of momentum is transferred from the photon to an electron. Let us assume that the effective mass of the photon is $m_c$. Furthermore, the photon has a velocity c, momentum p, energy E, a wavelength $\lambda$, and a frequency $\nu$. Therefore, by using the properties of the photon below (where h is plank's constant):

P=mc   (from classical physics)                          (CLASSICAL)     (10.31)
c=$\nu\lambda$   (definition of frequency and wavelength)        (CLASSICAL)     (10.32)
E=h$\nu$   (from Plank's energy-frequency law)             (QUANTUM)       (10.33)
$\lambda$=h/p (from DeBroglie wavelength-momentum law)       (QUANTUM)       (10.34)

Therefore, c/$\nu$ = h/p = h/(mc) (using 10.32, 10.34, and 10.31).

Thus, c/(E/h) = h/(mc) (using 10.33), or $E=mc^2$. Thus, a very simple derivation of the energy-mass relationship is possible from quantum mechanics.

## 11.   GENERAL RELATIVITY, ACCELERATION, GRAVITY AND CA

**"The general laws of physics (and gravitation) are to be expressed by equations which hold good for all systems of coordinates."**

                                                                                  **- Albert Einstein**

We will see that from the perspective of EMQG, Einstein's gravitational field equations are a set of observer dependent equations for observers that are subjected to gravity and/or to acceleration. These equations are based on *measurable* 4D space-time. The core of Einstein's theory is the principle of equivalence and the principle of general covariance, which allow an observer in any state of motion (and coordinate system) to describe gravity and acceleration. However, CA theory places little significance to an observer unless the observer interferes with the interaction being measured. In a CA, physical processes continue without regards to the presence of an observer, where events unfold in absolute space and time. We will reconcile these two different views of gravity later in section 15. First, we review the general theory of relativity.

**POSTULATES OF GENERAL RELATIVITY**

General relativity is a classical field theory founded on all the postulates and results of special relativity, as well as on the following new postulates:

**(1) PRINCIPLE OF EQUIVALENCE (STRONG) - The results of any given physical experiment will be precisely identical for an accelerated observer in free space as it is for a non-accelerated observer in a perfectly uniform gravitational field. A weaker form of this postulate states that: objects of the different mass fall at the same rate of acceleration in a uniform gravity field.**

**(2) PRINCIPLE OF COVARIANCE - The general laws of physics can be expressed in a form that is independent of the choice of space-time coordinates and the state of motion of an observer.**

As a consequence of postulate 1, the inertial mass of an object is equivalent to it's gravitational mass. Einstein uses this principle to encompass gravity and inertia into his single framework of general relativity in the form of a metric theory of acceleration and gravity, based on quasi-Riemann geometry.

These postulates, and the additional assumption that when gravitational fields are present nearby, space-time takes the form of a quasi-Riemannian manifold endowed with a metric curvature of the form $ds^2 = g_{ik} \, dx^i \, dx^j$, led Einstein to discover his famous gravitational field equations given below:

$$R_{ik} - (1/2) \, g_{ik} \, R = (8\pi G / c^2) \, T_{ik} \quad \text{Einstein's Gravitational Field Equations} \quad (11.1)$$

where, $g_{ik}$ is the metric tensor, $R_{ik}$ is the covariant Riemann curvature tensor. The left-hand side of the above equation is called the Einstein tensor or $G_{ik}$, which is the mathematical statement of space-time curvature that is reference frame independent and generally covariant. The right hand side $T_{\alpha\beta}$ is the stress-energy tensor which is the mathematical statement of the special relativistic treatment of mass-energy density, G is Newton's gravitational constant, and c the velocity of light.

For comparison purposes, we will now present the EMQG equations (which we will derive later) for the classical gravitational field where the gravitational field is not *too strong*, or *too weak*:

$$\nabla^2 \phi - (1/c^2) \, \partial^2 \phi / \partial t^2 = 4\pi G \, \rho(x,y,z,t) \quad (11.2)$$

where $\phi$ represents the classical Newtonian potential in absolute CA space and time units and $\rho(x,y,z,t)$ represents the **absolute** mass density distribution (that can be time varying) as measured from an observer at relative rest from the center of mass. This is a modified Poisson's equation, where the first term corresponds to the Poisson term, and the second term corresponds to the delay in the propagation of the graviton particles originating from the mass distribution. In EMQG, all distance units are expressed in absolute cellular automata space units in a 3D rectangular cell grid, and time as a count of the elapsed clock cycles. In other words, space is measured by counting the number of cells between two points (cells). Time is measured by counting the number of clock cycles that has elapsed between two events. The acceleration vector **a** for an average virtual particle at point (x,y,z) in CA space from the center of mass can be obtained from the gravitational potential $\phi$ at this point by the derivative of the potential as follows:

$$\mathbf{a} = \nabla \phi \tag{11.3}$$

More will be said about this later when EMQG is developed fully.

Einstein's law of gravitation (eq 11.1) cannot be arrived at by any 'rigorous' proof. The famous physicist S. Chandrasekhar writes (ref 37):

*"... It seems to this writer that in fact no such derivation exists and that, at the present time, no such can be given. ... It is the object of this paper to show how a mixture of physical reasonableness, mathematical simplicity, and aesthetic sensibility lead one, more or less uniquely, to Einstein's field equations."*

The principle of equivalence (in its strong form) is incorporated in the above framework by the assertion that all accelerations that are caused by either gravitational or inertial forces are **metrical** in nature. More precisely, the presence of acceleration caused by either an inertial force or a gravitational field modifies the geometry of space-time such that it is a quasi-Riemannian manifold endowed with a metric.

Furthermore, point particles move in gravitational fields along geodesic paths governed by the equation:

$$d^2x^i / ds^2 + \Gamma_{jk}^i (dx^j / ds)(dx^k / ds) = 0 \quad \text{... Equation for the geodesics} \tag{11.4}$$

The most striking consequence of general relativity is the existence of curved 4D space-time specified by the metric tensor $g_{ik}$. We will find that in EMQG theory, the meaning of the geodesic is very simple; it is the path taken by light or matter through the falling virtual particles undergoing acceleration, in the absence of any other external forces. We will see that curvature can be completely understood at the particle level. Furthermore, we will see that the principle of equivalence is a pure particle interaction process, and not a fundamental rule of nature. Before we can show this, we must carefully review the principle of equivalence from the context of general relativity theory.

## 12. THE PRINCIPLE OF EQUIVALENCE AND GENERAL RELATIVITY

*"I have never been able to understand this principle (principle of equivalence) ... I suggest that it be now buried with appropriate honors."*

                                                                        - Synge: Relativity- The General Theory

It should be noted that Einstein did not explain the origin of inertia in general relativity. Instead he relied on the existing Newtonian theory of inertia. Inertia was described by Newton in his famous law: F=MA; which states that an object resists being accelerated. A force (F) is required to accelerate an object of mass (M) to an acceleration (A). Since acceleration is a form of motion, it would seem that a reference frame is required in order to gauge this motion. But this is not the case in Newtonian physics. All observers agree as to which frame is actually accelerating by finding out which frames has a force associated with it. Only non-accelerated frames are relative. Einstein did not elaborate on this anomaly, or provide a reason why the inertial and gravitation masses are equal. This still remains as a postulate in his theory. The principle of equivalence has been tested to great accuracy. The equivalence of inertia and gravitational mass has been verified to an accuracy of one part in about $10^{-15}$ (ref 24).

Einstein's general theory of relativity is considered a "classical" theory, because matter, space, and time are treated as continuous classical variables. It is known however, that matter is made of discrete particles, and that forces are caused by particle exchanges as described by quantum field theory. A more complete theory of gravity should encompass a detailed quantum process for gravity involving particle interactions only. We will return to both special and general relativity in later sections, where a completely new interpretation and formulation of general relativity is given in the context of EMQG theory.

Inertia ought to be explained at the particle level as well, and should somehow be tied in to quantum gravity in a deep way according to the principle of equivalence. But, until recently there has been no adequate explanation for the origin of inertia. The next section summarizes some recent work on this problem, which has become the basis of EMQG.

### 13. THE QUANTUM THEORY OF INERTIA

**"Under the hypothesis that ordinary matter is ultimately made of subelementary constitutive primary charged entities or 'partons' bound in the manner of traditional elementary Plank oscillators, it is shown that a heretofore uninvestigated Lorentz force (specifically, the magnetic component of the Lorentz force) arises in any accelerated reference frame from the interaction of the partons with the vacuum electromagnetic zero-point-field (ZPF). ... The Lorentz force, though originating at the subelementary parton level, appears to produce an opposition to the acceleration of material objects at a macroscopic level having the correct characteristics to account for the property of inertia."**

**- B. Haisch, A. Rueda, H. E. Puthoff**

According to CA theory, there must be a localized explanation for all global phenomena such as acceleration and gravity. Inertia and gravity should originate from the small-scale particle interactions such that a global law emerges from the activity. Recall that CA theory is based on the local rules for the local cellular neighborhood, and these rules are repeated on a vast scale for all the cells in the universe. Many of our existing physical theories are general, global principles or general observations of nature. Both gravity and inertia have only been described successfully by "classical theories", applicable on global scales. In EQMG, both inertia and gravity have a detailed, particle level explanation based on the local "conditions" at the neighborhood of a given matter particle, and is thus manifestly compatible with the philosophy of a cellular automata theory and the principle of locality in special relativity.

In a recent theory (ref. 5) proposed by Haisch, Rueda, and Puthoff (known here as the HRP Theory of Inertia), it was shown that inertia comes from the buzz of activity of the virtual particles that fills even a perfect vacuum. It is this ever-present sea of energy that resists the acceleration of mass, and so creates inertia. Thus, they have found the low-level quantum description of inertia that is manifestly compatible with CA theory. Inertia is now described as being purely the result of quantum particle interactions. Haisch, Rueda, and Puthoff have come up with a new version of Newton's second law: F=MA. As in Newton's theory, their expression has 'F' for force on the left-hand side and 'A' for acceleration on the right. But in the place of 'M', there is a complex mathematical expression tying inertia to the properties of the vacuum. They found that the fluctuations in the vacuum interacting with the charge particles of matter in an accelerating mass give rise to a magnetic field, and this in turn, creates an opposing force to the motion. Thus, electromagnetic forces (or photon exchanges) is ultimately responsible for the force of inertia! The more massive an object, the more 'partons' it contains; and the more partons a mass contains means more individual (small) electromagnetic forces from the vacuum are present and the stronger the reluctance to undergo acceleration. But, when a mass is moving at a **constant** velocity, inertia disappears, and there is no resistance to motion in any direction as required by special relativity.

In their theory, inertia is caused by the magnetic component of the Lorentz force which arises between what the author's call 'parton' particles in an accelerated reference frame interacting with the background vacuum electromagnetic zero-point-field (ZPF). The author's use the old fashion term originated by Feynman called the 'parton', which referred to the elementary constituents of the nuclear particles such as protons and neutrons. It is now known that Feynman's partons are quarks, and that the proton and neutron each contain three quarks of two types: called the 'up' and 'down' quarks.

We have found it necessary to make a small modification of HRP Inertia theory in our investigation of the principle of equivalence. In EMQG, the modified version of inertia is known here as the "Quantum Inertia", or QI. In EMQG, a new elementary particle is required to fully understand inertia, gravitation, and the principle of equivalence. **All** matter, including electrons and quarks, must be made of nature's most fundamental mass unit or particle which we call the 'masseon' particle. These particles contain one fixed, fundamental 'quanta' of both inertial and gravitational mass. The masseons also carry one basic, smallest unit or quanta of electrical charge as well, of which they can be either positive or negative. Masseons exist in particle or anti-particle form (called anti-masseon), that can appear at random in the vacuum as masseon/anti-masseon particle pairs of opposite electric charge. The earth consists of ordinary masseons (no anti-masseons), of which there are equal numbers of positive and negative electric charge varieties. The masseon particle model will be elaborated later. Instead of the 'parton' particles (that make up an inertial mass in an accelerated reference frame) interacting with the background vacuum electromagnetic zero-point-field (ZPF), we postulate that the real masseons (that make up an accelerated mass) interacts with the surrounding, virtual masseons of the quantum vacuum, electromagnetically. However, the detailed nature of this interaction is not known at this time. For example, why is it that for constant velocity motion the forces add to zero, but when acceleration is introduced the forces add up to Newton's inertial force? Since the answers to these questions are not known, we treat the Quantum theory of Inertia as a postulate of EMQG.

We will see that quantum inertia is deeply connected with the subject of quantum gravity. EMQG explains why the inertial mass and gravitational mass are identical in accordance with the weak equivalence principle. The weak equivalence principle translates to the simple fact that the mass (m) that measures the ability of an object to produce (or react to) a gravitational field ($F=GMm/r^2$) is the same as the inertial mass value that appears in Newton's F=ma. In EMQG, this is not a chance coincidence, or a given fact of nature, which is assumed to exist, *a prior*. Instead, equivalence follows from a deeper process occurring inside a gravitational mass due to interactions with the quantum vacuum, which are *very similar* in nature to the interactions involved in inertial mass undergoing acceleration.

Since this new quantum theory of the inertia has still not been fully developed or confirmed yet, we raise QI to the level of a postulate. This is assigned as the third postulate of EMQG theory. The virtual particles of the quantum vacuum can be considered to be a kind of absolute reference frame for accelerated motion only. This frame is simply represented as the resultant acceleration vector given by the sum of all the acceleration vectors of the virtual particles of the quantum vacuum in the immediate neighborhood of a given charged masseon particle in the accelerated mass. This quantum vacuum reference frame gauges absolute acceleration. We do not need to measure our motion with respect to this frame in order to confirm that a mass is accelerated, we simply need to measure if an inertial force is present. We will see that this new, local quantum vacuum reference frame resolves Mach's paradox in regards to what reference frame nature uses to gauge accelerated mass.

13.1 THE BASIC MASS DEFINITIONS IN EMQG

Based on quantum inertia (section 13) and the quantum principle of equivalence (section 15.8) EMQG proposes three *different* mass definitions for an object, listed below:

(1) **INERTIAL MASS** is the measurable mass defined in Newton's force law F=MA. This is considered as the absolute mass in EMQG, because it results from force produced by the relative (statistical average) acceleration of the charged virtual particles (masseons) of the quantum vacuum with respect to the charged particles that make up the inertial mass (masseons). To some extent, the virtual particles of the quantum vacuum forms Newton's absolute reference frame. In special relativity this mass is equivalent to the rest mass.

(2) **GRAVITATIONAL MASS** is the measurable mass involved in the gravitational force as defined in Newton's law $F=GM_1M_2/R^2$. This is what is measured on a weighing scale. This is also considered as absolute mass, and is (almost) exactly the same as inertial mass. The same quantum process in inertia is also occurring in gravitational mass (section 15.8).

(3) LOW LEVEL **GRAVITATIONAL 'MASS CHARGE'** which is the origin of the pure gravitational force, is defined as the force that results through the exchange of graviton particles between two (or more) quantum particles (masseons). This type of mass is analogous to 'electrical charge', where photon particles are exchanged between electrically charged particles. Note: this force is very hard to measure because it is masked by the background quantum vacuum electromagnetic force interactions, which dominates over the graviton force processes.

These three forms of mass are not necessarily equal! It turns out (section 15.8) that the inertial mass is almost exactly the same as gravitational mass, but not perfectly equal. All quantum mass particles (fermions which are made from masseons) have all three mass types defined above. But bosons (only photons and gravitons are considered here) have only the first two mass types. This means that photons and gravitons transfer momentum, and do react to the presence of inertial frames and to gravitational fields, but they do not emit or absorb gravitons. Gravitational fields affect photons, and this is linked to the concept of space-time curvature, described in detail later (Section 16). It is important to realize that gravitational fields deflect photons (and gravitons), but not by force particle exchanges directly. Instead, it is due to a scattering process (described later).

You might think that if a particle has energy, it automatically has mass; and if a particle has mass, then it must emit or absorb gravitons. This reasoning is based on Einstein's famous equation $E=mc^2$, which is derived purely from considerations of inertial mass (and Einstein's principle of equivalence extended to gravitational fields). In his famous thought experiment, a photon is emitted from a box, causing a recoil to the box in the form of a momentum change, and from this he derives his famous $E=mc^2$. In quantum field theory this momentum change is traceable to a fundamental QED vertex, where a electron (in an atom in the box) emits a photon, and recoils with a momentum equivalent to the photon's momentum '$m_p c$'. We have analyzed Einstein's thought experiment from the perspective of EMQG and concluded that the photon behaves as if it has an effective inertial mass '$m_p$' given by: $m_p = E/c^2$ in Einstein's light box. For simplicity, lets consider a photon that is absorbed by a charged particle like an electron at rest. The photon carries energy and is thus able to do work. When the photon is absorbed by the electron with mass '$m_e$', the electron recoils, because there is a definite momentum transfer to the electron given by $m_e v$, where v is the recoil velocity. The electron momentum gained is equivalent to the effective photon momentum lost by the photon $m_p c$. In other words, the electron momentum '$m_e v_e$' received from the photon when the photon is absorbed is equivalent to the momentum of the photon '$m_p c$', where $m_p$ is the effective photon mass. If this electron later collides with another particle, the same momentum is transferred. The rest mass of the photon is defined as zero. Thus, the effective photon mass is a measurable inertial mass. Note: the recoil of the light box is a backward acceleration of the box, which works against the virtual particles of the quantum vacuum. Thus, when one claims that a photon has a real mass, we are really referring to the photon's ability to impart momentum. This momentum can later do work in a quantum vacuum inertial process.

Does the photon have an effective gravitational mass? By this we mean; does it behave as if it carries a measurable gravitational mass in a gravitational field like the earth (as given by $E/c^2$)? The answer is yes! For example, when the photon moves parallel to the earth's surface, it follows a parabolic curve and deflects downwards. You might guess that this deflection is caused by the graviton exchanges originating from the earth acting on this photon, and that this deflection is the same as that inside an equivalent rocket accelerating at 1g. The amount of deflection is equivalent, but according to Einstein this is a direct result of the space-time curvature near the earth and in the rocket. Our work on the equivalence principle has shown however, that this is not true. The photon deflection is caused by a different reason, but ends up giving the same result. In the rocket, the deflection is simply caused by the accelerated motion of the rocket floor, which carries the observer with it. This causes the observer to perceive a curved path (described as curved space-time). In a gravitational field, however, the deflection of light is real, and caused by the scattering of photons with the downward accelerating virtual particles. The photon scatters with the charged virtual particles of the quantum vacuum, which are accelerating downwards (statistically). The photon moving parallel to the surface of the earth undergoes continual absorption and re-emission by the falling virtual (electrically charged) particles of the quantum vacuum. The vacuum particles induce a kind of 'Fizeau-like' scattering of the photons (Note: this scattering is present in the rocket, but does not lead to photon deflection because only the rocket and observer are accelerated). The photons are scattered because of the electromagnetic interaction of the photons with the falling charged virtual particles of the vacuum. Since the downward acceleration of the quantum vacuum particles is the same as the up-wards acceleration of the floor of the rocket, the amount of photon deflection is equivalent. Under the influence of a gravitational field, photons take on the same downward (component) as the accelerating (charged) virtual particles of the vacuum. This, of course, violates the constancy of the speed of light; which we will explore further in section 16. For now, one should note that downward acceleration component is picked up by the photons only during the time the photons are absorbed by the quantum vacuum particles (and thus exist as charged virtual particles). In between virtual particle scattering, the photon motion is still strictly Lorentz invariant, and light velocity is still an absolute constant.

A similar line of reasoning as above applies to the motion of the graviton particle. The graviton has inertial mass because like the photon, it can transmit a momentum to another particle when absorbed in the graviton exchange process during a gravitational force interaction (although considerably weaker then photon exchanges). Like the photon, the graviton deflects when moving parallel to the floor of the rocket (from the perspective of an observer on the floor) and therefore has inertial mass. The graviton also has a gravitational mass (like the photon) when it moves parallel to the earth's surface, where it deflects under the influence of a gravitational field. Again, the graviton deflection is caused by the scattering of the graviton particle with the downward acceleration of the virtual 'mass-charged' particles of the quantum vacuum through an identical 'Fizeau-like' scattering process described above. Unlike the photon however, the scattering is caused by the 'mass charge' interaction (or pure graviton exchanges) of the quantum vacuum virtual particles, and not the electric charge as before. The end result is that the graviton has an effective gravitational mass like the photon. Again a graviton does not exchange gravitons with another nearby graviton, just as a photon does not exchange photons with other photons.

To summarize, both the photon and the graviton do ***not*** carry low level 'mass charge', even though they both carry inertial and gravitational mass. The graviton exchange particle, although responsible for a major part of the gravitational mass process, does not itself carry the property of 'mass charge'. Contrast this to conventional physics, where the photon and the graviton both carry a non-zero mass given by $M=E/C^2$. According to this reasoning, the photon and the graviton both carry mass (since they carry energy), and therefore both must have 'mass charge' and exchange gravitons. In other words, the graviton particle not only participates in the exchange process, it also undergoes further exchanges while it is being exchanged! This is the source of great difficulty for canonical quantum gravity theories, and causes all sorts of mathematical renormalization problems in the corresponding quantum field theory. Furthermore, in gravitational force interactions with photons, the strength of the force (which depends on the number of gravitons exchanged with photon) varies with the energy that the photon carries! In modern physics, we do not distinguish between inertial, gravitational, or low level 'mass charge'. They are assumed to be equivalent, and given a generic name 'mass'. In EMQG, the photon and graviton carry measurable inertial and gravitational mass, but neither particle carries the 'low level mass charge', and therefore do not participate in graviton exchanges.

We must emphasize that gravitons do not interact with each other through force exchanges in EMQG, just as photons do not interact with each other with force exchanges in QED. Imagine if gravitons did interact with other gravitons. One might ask how it is possible for a graviton particle (that always moves at the speed of light) to emit graviton particles that are also moving at the speed of light. For one thing, this violates the principles of special relativity theory. Imagine two gravitons moving in the same direction at the speed of light that are separated by a distance d, with the leading graviton called 'A' and the lagging graviton called 'B'. How can graviton 'B' emit another graviton (also moving at the speed of light) that gets absorbed by graviton 'A' moving at the speed of light? As we have seen, these difficulties are resolved by realizing that there are actually three different types of mass. There is measurable inertial mass and measurable gravitational mass, and low level 'mass charge' that cannot be directly measured. Inertial and gravitational mass have already been discussed and arise from different physical circumstances, but in most cases give identical results. However, the 'low level mass charge' of a particle is defined simply as the force existing between two identical particles due to the exchange of graviton particles only, which are the vector bosons of the gravitational force. Low level mass charge is not directly measurable, because of the complications due to the electromagnetic forces simultaneously present from the quantum vacuum virtual particles.

It would be interesting to speculate what the universe might be like if there were no quantum vacuum virtual particles present. Bearing in mind that the graviton exchange process is almost identical to the photon exchange process, and bearing in mind the complete absence of the electromagnetic component in gravitational interactions, the universe would be a very strange place. We would find that large masses would fall faster than smaller masses, just as a large positive electric charge would 'fall' faster then a small positive charge towards a very large negative charge. There would be no inertia as we know it, and basically no force would be required to accelerate or stop a large mass.

## 14. APPLICATIONS OF QUANTUM INERTIA: MACH'S PRINCIPLE

*"... it does not matter if we think of the earth as turning round on its axis, or at rest while the fixed stars revolve around it ... The law of inertia must be so conceived that exactly the same thing results from the second supposition as from the first."*

*E. Mach*

Ernst Mach (ca 1883) proposed that the inertial mass of a body does not have any meaning in the absence of the rest of the matter in the universe. In other words, acceleration requires some other reference frame in order to determine accelerated motion. Thus, it seemed to Mach that the only reference frame possible was that of the average motion of all the other masses in the universe. This implied to Mach that the acceleration of an object must somehow be dependent on the sum total of all the matter in the universe. To Mach, if all the matter in the universe were removed, the acceleration, and thus the force of inertia would completely disappear since no reference frame is available to determine the actual acceleration.

A spinning elastic sphere bulges at the equator due to the centrifugal force. The question that Mach asked was how does the sphere 'know' that it is spinning, and therefore must bulge. If all the matter in the universe was removed, how can we be sure that it really rotates? Therefore, how would the sphere 'know' that it must bulge or not? Newton's answer would have been that the sphere somehow felt the action of Newtonian absolute space. Mach believed that the sphere somehow 'feels' the action of all the cosmic masses rotating around it. To Mach, centrifugal forces are somehow gravitational in the sense that it is the action of mass on mass. To Newton, the centrifugal force is due to the rotation of the sphere with respect to absolute space. To what extent that Einstein's general theory of relativity incorporates Mach's ideas is still a matter of debate (ref. 38). EMQG (through the quantum inertia principle) takes a similar view as Newton, where Newton's absolute space is replaced by the virtual particles of the vacuum. Mach was never unable to develop a full theory of inertia based on his idea of mass affecting mass.

Mach's ideas on inertia are summarized as follows:

(1) A particle's inertia is due to some (unknown) interaction of that particle with all the other masses in the universe.
(2) The local standards of non-acceleration are determined by some average value of the motion of all the masses in the universe.
(3) All that matters in mechanics is the relative motion of all the masses.

Quantum inertia theory fully resolves Mach's paradox by introducing a new universal reference frame for gauging acceleration: the net statistical average acceleration vector of the virtual particles of the quantum vacuum with respect to the accelerated mass. In other words, the cause of inertia is the interaction of each and every particle with the quantum vacuum. Inertial force actually *originates* in this way. It turns out that the distant stars do affect the local state of acceleration of our vacuum here through the long-range gravitation force. Thus, our local inertial frame is slightly affected by all the masses in the distant universe. However, in our solar system the local gravitational bodies swamp out this effect. (This long-range gravitational force is transmitted to us by the graviton particles that originate in all the matter in the universe, which will distort our local net statistical average acceleration vector of the quantum virtual particles in our vacuum with respect to the average mass distribution). Thus, it seems that Mach was correct in saying that acceleration here depends somehow on the distribution of the distant stars (masses) in the universe, but the effect he predicted is minute.

14.1    MORE APPLICATIONS OF QI: NEWTON'S LAWS OF MOTION

We are now in a position to understand the quantum nature of Newton's classical laws of motion. According to the standard textbooks of physics (ref. 19) Newton's three laws of laws of motion are:

**1. An object at rest will remain at rest and an object in motion will continue in motion with a constant velocity unless it experiences a net external force.**

**2. The acceleration of an object is directly proportional to the resultant force acting on it and inversely proportional to its mass. Mathematically: ΣF = ma, where 'F' and 'a' are the vectors of each of the forces and accelerations.**

**3. If two bodies interact, the force exerted on body 1 by body 2 is equal to and opposite the force exerted on body 2 by body 1. Mathematically: $F_{12} = -F_{21}$.**

Newton's first law explains what happens to a mass when the resultant of all external forces on it is zero. Newton's second law explains what happens to a mass when there is a nonzero resultant force acting on it. Newton's third law tells us that forces always come in pairs. In other words, a single isolated force cannot exist. The force that body 1 exerts on body 2 is called the action force, and the force of body 2 on body 1 is called the reaction force.

In the framework of EMQG theory, Newton's first two laws are the direct consequence of the (electromagnetic) force interaction of the (charged) elementary particles of the mass interacting with the (charged) virtual particles of the quantum vacuum. Newton's third law of motion is the direct consequence of the fact that all forces are the end result of a boson particle exchange process.

Newton's First Law of Motion:

In EMQG, the first law is a trivial result, which follows directly from the quantum principle of inertia (postulate #3). First a mass is at relative rest with respect to an observer in deep space. If no external forces act on the mass, the (charged) elementary particles that make up the mass maintain a *net acceleration* of zero with respect to the (charged) virtual particles of the quantum vacuum through the electromagnetic force exchange process. This means that no change in velocity is possible (zero acceleration) and the mass remains at rest. Secondly, a mass has some given constant velocity with respect to an observer in deep space. If no external forces act on the mass, the (charged) elementary particles that make up the mass also maintain a *net acceleration* of zero with respect to the (charged) virtual particles of the quantum vacuum through the electromagnetic force exchange process. Again, no change in velocity is possible (zero acceleration) and the mass remains at the same constant velocity.

Newton's Second Law of Motion:

In EMQG, the second law is the quantum theory of inertia discussed above. Basically the state of *relative* acceleration of the charged virtual particles of the quantum vacuum with respect to the charged particles of the mass is what is responsible for the inertial force. By this we mean that it is the tiny (electromagnetic) force contributed by each mass particle undergoing an acceleration 'A', with respect to the net statistical average of the virtual particles of the quantum vacuum, that results in the property of inertia possessed by all masses. The sum of all these tiny (electromagnetic) forces contributed from each charged particle of the mass (from the vacuum) is the source of the total inertial resistance force opposing accelerated motion in Newton's F=MA. Therefore, inertial mass 'M' of a mass simply represents the total resistance to acceleration of all the mass particles.

Newton's Third Law of Motion:

According to the boson force particle exchange paradigm (originated from QED) all forces (including gravity, as we shall see) result from particle exchanges. Therefore, the force that body 1 exerts on body 2 (called the action force), is the result of the emission of force exchange particles from (the charged particles that make up) body 1, which are readily absorbed by (the charged particles that make up) body 2, resulting in a force acting on body 2. Similarly, the force of body 2 on body 1 (called the reaction force), is the result of the absorption of force exchange particles that are originating from (the charged particles that make up) body 2, and received by (the charged particles that make up) body 1, resulting in a force acting on body 1. An important property of charge is the ability to readily emit and absorb boson force exchange particles. Therefore, body 1 is both an emitter and also an absorber of the force exchange particles. Similarly, body 2 is also both an emitter and an absorber of the force exchange particles. This is the reason that there is both an action and reaction force. For example, the contact forces (the mechanical forces that Newton was thinking of when he formulated this law) that results from a person pushing on a mass (and the reaction force from the mass pushing on the person) is really the exchange of photon particles from the charged electrons bound to the atoms of the person's hand and the charged electrons bound to the atoms of the mass on the quantum level. Therefore, on the quantum level there is really is no contact here. The hand gets very close to the mass, but does not actually touch. The electrons exchange photons among each other. The force exchange process works both directions in equal numbers, because all the electrons in the hand and in the mass are electrically charged and therefore the exchange process gives forces that are equal and opposite in both directions.

# 15. ELECTRO-MAGNETIC QUANTUM GRAVITY

*"All (the universe) is numbers"*

*- Pythagoras*

*"Subtle is the lord…"*

*- Einstein*

## 15.1 INTRODUCTION

EMQG theory is now developed fully. We have found that both Special and General Relativity must be modified to be compatible with EMQG theory. Gravity is one of the four basic forces of nature. The highly successful standard model of particle physics does not account for gravity. The standard model addresses the electromagnetic, weak and strong nuclear forces within the framework of quantum field theory. In quantum field theory, forces are thought to originate as the exchange of force particles (vector bosons) which are represented by the quanta of the associated classical field. In EMQG, it is found that two fundamental particle exchange processes are responsible for gravity; one particle exchange being very familiar, while the second particle exchange type has been postulated but not yet been successfully detected. The particles involved are the photon and graviton exchange particles. We will see that the in EMQG, the photon and the graviton are almost identical in their physical properties, except for their relative strengths. The particle exchange process (in general) fits very well into the general framework of CA theory without much modification. The boson acts like the go between particle, shifting from cell to cell until it is absorbed by a destination particle. This transfers an acceleration (or force) without action at a distance. We will now examine the nature of forces as particle exchanges on the CA in more detail.

## 15.2 FORCES, PARTICLE EXCHANGES, AND CA THEORY

The theory that best describes the quantization of the electromagnetic force field is called Quantum Electrodynamics (QED). Here the charged particles (electrons, positrons) act upon each other through the exchange of force particles, which are called photons. The photons represent the quantization of the classical electromagnetic field. In classical electromagnetic theory, the force due to two charged particles decreases with the inverse square of their separation distance (Coulomb's inverse square law: $F = kq_1q_2/r^2$, where k is a constant, $q_1$ and $q_2$ are the charges, and r is the distance of separation). QED accounts for this inverse square law by postulating the exchange of photons between the charged particles. The number of photons emitted and absorbed by a given charge (per unit of CA time) is fixed and is called the charge of the particle. Thus, if the charge doubles, the force doubles because twice as many photons are exchanged during the force interaction. This force interaction process causes the affected particles to accelerate either towards or away from each other depending on if the charge is positive or negative (because different charges transmit photons with slightly different wave functions). Certain known cellular automata roughly exhibit behaviors like this. For example, in the famous 2D geometric CA called Conway's game of life there exist a large variety of CA patterns types generally called 'guns'. They are constantly emitting a steady stream of 'gliders' as they travel through CA space. This emission process is constant without any degradation to the original gun pattern. This resembles the property possessed by and electron called electric charge, where photons are constantly emitted without any degradation or change to the electron.

The strength of the electromagnetic force varies as the inverse square of the distance of separation between the charges in the following way: each charge sends and receives photons from every direction. But, the number of photons per unit area, emitted or received, decreases by the factor $1/4\pi r^2$ (the surface area of a sphere for a 3D geometric CA) at a distance 'r' due to the photon emission pattern spreading in all directions. Thus, if the distance doubles, the number of photons exchanged decreases by a factor of four. This process can easily be visualized on a 3D geometric CA. Imagine that an electron particle is at the center of a sphere sending out virtual photons in all directions. Imagine that another electron is on the surface of a sphere at a distance 'r' from the emitter, which absorbs some of these photons. The absorption of these photons causes an outward acceleration, and thus a repulsive force. If the charge is doubled on the central electron, there is twice as many photons appearing at the surface of the sphere, and twice the force acting on the other electron. This accounts for the linear product of charge terms in the numerator of the inverse square law. In QED, photons do not interact with each other (through force exchanges). As a result, in-going and out-going photons do *not* affect each other during the exchange process, by the exchange of force particles.

Someone that is not fully versed in modern quantum field theory may question why two oppositely charged particles can be attracted to each other, while each is absorbing an exchange particle. On face value, classical thinking would imply that the momentum transfer would cause the particles to always move apart! The typical QED textbooks 'explain' this fact by the mathematics of momentum transfer at the vertices of the associated Feynman fundamental process. Certainly, classical models cannot explain this process, nor can classical models explain why photons are constantly emitted without degradation to the original electron, simply because all that is involved is a purely numerical CA process. We speculate that in the context of CA theory, the constant emission of photons (which maintains the charge of a particle) happens without degradation of the original electron pattern. This is possible because the original electron is a 'numeric' pattern which can remain stable indefinitely during this emission process (we have seen CA counterparts to this process in the familiar 'Game of Life' CA). Similarly, we speculate that the absorption of vector boson information patterns alters the internal state of the numeric pattern in such a way as to change the state of motion of that pattern (or causes acceleration). We believe that there must be a lot of *hidden activity* in the Feynman fundamental vertices of QED, and that the details are hidden from the physicist because of the purely numeric aspect of this process. In fact, in 'Conway's game of life', we discovered a CA pattern that is called a 'loop' which evolves into something resembling a two particle exchange process. Two larger internal oscillating CA patterns are seen to move apart while 'glider' particles (which are small, high-speed oscillating patterns reminiscent of photons) are exchanged. This pattern is something like a CA prototype pattern of a particle exchange process leading to a force. However, we found that this is not a perfect model, because the gliders are not emitted in every direction. Also, the particle exchange gives a constant velocity outward motion (and not accelerated motion as required). To date, no one has found a perfect particle exchange process that looks identical to real physical particles in any Cellular Automata simulations. However, we believe that something like this is happening on the plank scale in real particle exchanges in our universe.

15.3    GRAVITATION ORIGINATES FROM GRAVITON EXCHANGES

For gravitational forces, it is experimentally observed that the force originating from two particles possessing mass decreases with the inverse square of their separation distance, and is given by Newton's inverse square law: $F = Gm_1m_2 / r^2$, where G is the gravitational constant, $m_1$ and $m_2$ are the masses, and r is the distance of separation. For electromagnetic forces, it is also experimentally observed that the force originating from two particles possessing charge also decreases with the inverse square of their separation distance, and is given by Coulomb's inverse square law: $F = kq_1q_2 / r^2$, where k is Coulomb's constant, $q_1$ and $q_2$ are the masses, and r is the distance of separation. It can be seen that the two force laws are very similar in form. QED theory accounts for Coulomb's law by the photon exchange process. Following the lead from the highly successful QED, EMQG postulates that we replace the concept of electrical 'charges' exchanging 'photons' with the idea that 'mass charges' exchange gravitons. Hence, gravitational mass at a fundamental level is simply the ability to emit or absorb gravitons, and pure low-level gravitational mass is interpreted as 'mass charge'.

For gravity there are gravitons instead of photons, which are the force exchange particles of gravity. Like charge, it is the property called mass-charge that determines the number of exchange gravitons. The larger the mass, the greater the number of gravitons exchanged. Like electromagnetism, the strength of the gravity force decreases with the inverse square of the distance. This conceptual framework for quantum gravity has been around for some time now, but how are we to merge these simple ideas to be compatible with the framework of general relativity? We must be able to explain the Einstein's Principle of Equivalence and the physical connection between inertia, gravity, and curved space-time all within the general framework of graviton particle exchange. General Relativity is based on the idea that the forces experienced in a gravitational field and the forces due to acceleration are equivalent, and both are due to the space-time curvature.

In classical electromagnetism, if a charged particle is accelerated towards an opposite charged particle, the rate of acceleration depends on the electrical charge value. If the charge is doubled, the force doubles, and the rate of acceleration is doubled. If quantum gravity were to work in the exact same way, we would expect that the rate of acceleration of a mass near the earth would double if the mass doubles. The reason for this expectation is that the exchange process for gravitons should be very similar to electromagnetism. In other words, if the 'mass-charge' is doubled, the gravitational force is doubled. The only difference between the two forces is that gravity is a lot weaker by a factor of about $10^{-40}$. The weakness of the gravitational forces might be attributed either to the very small interaction cross-section of the graviton particle as compared to the photon particle, or to a very weak coupling constant (the absorption of a single graviton causing a minute amount of acceleration), or both.

Unfortunately, if the graviton exchange process worked exactly like QED, it would not reproduce the known nature of gravity. First, there is the problem of variation of mass with velocity as described by special relativity as $m = m_0 (1-v^2/c^2)^{-1/2}$. At face value, this would mean that the number of gravitons exchanged depends on **velocity** of the gravitational mass, which does not easily fit into the framework of a QED approach to quantum gravity. Secondly, if two masses are sitting on a table with mass 'M' and mass '2M', the forces against the tabletop varies with the mass, just as you would expect in a QED-like exchange of graviton particles. If the mass doubles, the force on the table doubles. Yet, the rate of acceleration is the same for these two masses in free fall. Why? Since twice the number of gravitons is exchanged under mass '2M', you would expect twice the force, and therefore mass '2M' would arrive early. Matter has inertia, and this complicates everything. In almost all quantum gravity theories inertia appears as a separate process that is 'tacked' on in an ad hoc manner. The principle of equivalence merely raises this relationship between inertia and gravitation to the status of a postulate as in Einstein's theory of general relativity.

All test masses accelerate at the same rate ($g=9.8$ m/sec$^2$ on the earth) no matter what the value of the test mass is. This is a direct consequence of the principle of equivalence. Mathematically, this follows from Newton's two **different** force laws: inertia and gravity as follows:

$F_i = ma$ ..... (Inertial force) (15.31)

$F_g = GmM / r^2$ ..... (Gravitational force) (15.32)

In free fall, an object (mass m) in the presence of the earth's pull (mass M) is force free, i.e. $F_i = F_g$. Note that the same mass value 'm' appears in the two mass definition formulas (for some *mysterious* reason) for equations 15.31 and 15.32.

Therefore, $ma = GmM/r^2$ or $a = GM/r^2$ ..... Equivalence Principle (15.33)

From equation 15.33, we see that the rate of acceleration does not depend on the test mass m. All test masses accelerate at the same rate. Thus, inertia and gravity are intimately connected in a deep way because the measure of mass m is the same for acceleration as for gravity. What is mass? In EMQG, gravitational mass originates from a low-level graviton exchange process originating from 'mass charge', where there the emission rate is constant. In fact, mass is quantized in exactly the same way as electric charge in QED. (There exists a fundamental unit of mass charge that is carried by the masseon particle, the lowest quanta of mass).

Recall the new quantum theory for inertia given in the previous section. We have found an explanation for inertia based on low-level quantum processes. The quantum source of the force of inertia is the resistance to acceleration offered by the virtual electrically charged particles of the quantum vacuum. What is unique about EMQG theory, is that the same virtual electrically charged particle processes are also present near a large gravitational mass. It is the interactions of these electrically charged virtual particles with the real electrically charged particles in the mass that accounts for the bulk of the gravitational force, and for the principle of equivalence. Yet we have retained the same simple QED type model for the fundamental low-level gravitational interactions through the graviton exchange process.

This new quantum theory of gravity is called EMQG or Electro-Magnetic Quantum Gravity, primarily because gravity involves a strong electromagnetic component. It is based on the four postulates of section 15.6 (these four postulates should be derivable from the low-level quantum descriptions of electromagnetic photon exchanges and graviton exchanges in the future, but at this time this is not fully developed). We shall accept these postulates, and examine the consequences. Before we do this, we must first characterize the masseon particle, which is required to fully understand the principle of equivalence.

## 15.4  THE PHYSICAL PROPERTIES OF THE MASSEON PARTICLE

In order to understand the principle of equivalence on the quantum level, we must postulate the existence of a new elementary particle. This particle is the most elementary form of matter or anti-matter, and carries the lowest possible quanta of low level gravitational 'mass charge'. This elementary particle is called the masseon particle (and also comes as anti-masseons, the corresponding anti-particle). The masseon is postulated to be the most elementary mass particle and readily combines with other masseons through a new, unknown hypothetical force coupling which we call the 'primal force'. Presumably, the primal force comes in positive and negative 'primal charge' types. The proposed mediator of this force is called the 'primon' particle. Since the masseon has not yet been detected, we can safely assume that the primal force is very strong. It is not necessary to understand the exact nature of the primal force to achieve the important results of EMQG. Suffice it to say that the primal force binds together masseon particles to make all the known fermion particles of the standard model. The masseon carries the lowest possible quanta of positive gravitational 'mass charge'. Low level gravitational 'mass charge' is defined as the (probability) fixed rate of emission of graviton particles in close analogy to electric charge in QED. Recall that the graviton is the vector boson of the pure gravitational force. Gravitational 'mass charge' is a fixed constant in EMQG, and is analogous to the fixed electrical charge concept. Gravitational 'mass charge' is **not** governed by the ordinary physical laws of *observable* mass, which appear as 'm' in the various physical theories. This includes Einstein's special relativity theory:

$$E=mc^2 \text{ or } m = m_0 (1 - v^2/c^2)^{-1/2} \qquad (15.41)$$

This is why we call it gravitational 'mass charge' or sometimes called the *low-level* mass of a particle, and this should not be confused with the ordinary observable inertial or observable gravitational mass. It will be assumed that when the low level mass is used in this paper, we are talking about the low level gravitational 'mass charge' property of a particle, and the associated graviton exchange process.

Masseons simultaneously carry a positive gravitational 'mass charge', and either a positive or negative electrical charge (defined exactly as in QED). Therefore, masseons also exchange photons with other masseon particles. It is important to note that the graviton exchange process is responsible for the low-level gravitational interaction only, which is not directly accessible to our measurements (as we will see later), and is also masked by the presence of the electromagnetic force component in all gravitational measurements. Masseons are fermions with half integer spin, which behave according to the rules of quantum field theory. Gravitons have a spin of one (not spin two, as is commonly thought), and travel at the speed of light. This paper addresses the gravitational and electromagnetic force interactions only, and the strong and weak nuclear forces are ignored here. Presumably, masseons also carry the strong and weak 'nuclear charge' as well.

Anti-masseons carry the lowest quanta of negative gravitational 'mass charge'. Anti-masseons also carry either positive or negative electrical charge, with electrical charge being defined according to QED. An anti-masseon is always created with ordinary masseon in a particle pair as required by quantum field theory (specifically, the Dirac equation). In EMQG, the anti-masseon is the negative energy solution of the Dirac equation for a fermion, where now the **mass** is taken to be 'negative' as well. Ordinarily, the standard model requires that the mass of any anti-particle is always positive, in order to comply with the principle of equivalence, or $M_i=M_g$. In EMQG, the principle of equivalence is not taken to be an absolute law of nature, and is definitely grossly violated for anti-particles (for reasons that will become clear in section 15.4). The anti-particles have positive inertia mass and *negative* gravitational mass, or $M_i=-M_g$.

Thus, a beautiful symmetry exists between EMQG and QED for gravitational and electromagnetic forces. The masseon-graviton interaction becomes almost identical to the electron-photon interaction. There are only two differences between these forces. First, the ratio of the strength of the electromagnetic over the gravitational forces is on the order of $10^{40}$. Secondly, there exists a difference in the nature of attraction and repulsion between positive and negative gravitational 'mass-charges' (as detailed in the table #1 and 2).

In QED, the quantum vacuum as a whole is electrically neutral because the virtual electron and positron (negative electron) particles are always created in particle pairs with equal numbers of positive and negative electrical charge. In EMQG, the quantum vacuum is also gravitationally neutral for the **same** reason. At any given instant of time, there is a 50-50 mixture of virtual gravitational 'mass charges', which are carried by the virtual masseon and anti-masseon pairs. These masseon pairs are created with equal and opposite gravitational 'mass charge'. This is the reason why the cosmological constant is zero (or very close to zero). Half the graviton exchanges between quantum vacuum particles result in attraction, while the other half result in repulsion. To see how this works, we will closely examine how masseons and anti-masseons interact.

The following tables summarize the fundamental electron and masseon force interactions:

**TABLE #1   EMQG MASSEON - ANTI-MASSEON GRAVITON EXCHANGE**

|  | (DESTINATION) | |
|---|---|---|
| (SOURCE) | MASSEON | ANTI-MASSEON |
| **MASSEON** | attract | attract |
| **ANTI-MASSEON** | repel | repel |

**TABLE #2   QED ELECTRON - ANTI-ELECTRON PHOTON EXCHANGE**

|  | (DESTINATION) | |
|---|---|---|
| (SOURCE) | ELECTRON | ANTI-ELECTRON |
| **ELECTRON** | repel | attract |
| **ANTI-ELECTRON** | attract | repel |

In QED, if the source particle is an electron, it emits photons whose wave function induces repulsion when absorbed by a destination electron, and induces attraction when absorbed by a destination anti-electron. Similarly, if the source is an anti-electron, it emits photons whose wave function induces attraction when absorbed by a destination electron, and induces repulsion when absorbed by a destination anti-electron.

In EMQG, if the source particle is a masseon, it emits gravitons whose wave function induces attraction when absorbed by a destination masseon, and induces attraction when absorbed by a destination anti-masseon. If the source is an anti-masseon, it emits gravitons whose wave function induces repulsion when absorbed by a destination masseon, and induces repulsion when absorbed by a destination anti-masseon. This subtle difference in the nature of graviton exchange process is responsible for some major differences in the way that low-level gravitational 'mass charge' and the electrical charges operate.

It is convenient to think of the photon as occurring in photon and anti-photon varieties (the photon is its own anti-particle). Similarly, the graviton comes in graviton and anti-graviton varieties. Thus, we can say that the masseons emit gravitons, and anti-masseons emit anti-gravitons. The absorption of a graviton by either a masseon or anti-masseon induces attraction. The absorption of an anti-graviton by either a masseon or anti-masseon induces repulsion. Similarly, we can say the electrons emit photons and anti-electrons emit anti-photons. The absorption of a photon by an electron induces repulsion, and the absorption of a photon by an anti-electron induces attraction. The absorption of an anti-photon by an electron induces attraction, and the absorption of an anti-photon by an anti-electron induces repulsion.

## 15.5   THE QUANTUM VACUUM AND VIRTUAL MASSEON PARTICLES

What virtual particles are present in the quantum vacuum? In QED, it is virtual electrons and anti-electrons (and virtual muons and tauons), along with the associated virtual photons. In the standard model of particle physics the quantum vacuum consists of all varieties of virtual fermion and virtual boson particles representing the known virtual matter and virtual force particles, respectively. This includes virtual electrons, virtual quarks, virtual neutrinos for fermions, and virtual photons, virtual gluons, and virtual W and Z bosons for the bosons. In EMQG, we restrict ourselves to the study of gravity and electromagnetism. Therefore, the EMQG quantum vacuum consists of the virtual masseons and virtual anti-masseons, and the associated virtual photons and virtual graviton particles (sometimes, virtual masseon combine to form virtual electrons, etc). Recall that ordinary matter consists only of real masseons bound together in certain combinations to form the familiar elementary particles. We now ask how the virtual electrons/positrons of the QED vacuum behave in the vicinity of a real electrical charge. We want to compare this with virtual masseon and virtual anti-masseon near a large real mass-charge like the earth in our EMQG formulation.

First, we review how the QED quantum vacuum is affected by the introduction of a real negative electrical charge. According to QED, the nearby virtual particle pairs become **polarized** around the central charge. This means that the virtual electrons of the quantum vacuum are repelled away from the central negative charge, while the virtual positrons are attracted towards the central negative charge. Thus for real electrons the vacuum polarization produces charge screening, which reduces the charge of a real electron, when measured over relatively long distances. According to QED, each electron is surrounded by a cloud of virtual particles that winks in and out of existence in pairs lasting a tiny fraction of a second, and this cloud is always present and acts like an electrical shield against the real charge of the electron. Recently, a team of physicists led by D. Koltick of Purdue University in Indiana reported that charge screening of an electron has been observed (ref. 33) experimentally at the KEK collider. They fired high-energy particles at electrons and found that the effect of this cloud of virtual particles was reduced the closer a high-energy particle penetrated towards the electron. They report that the effect of the higher charge for an electron that has been penetrated by particles accelerated to an energy 58 giga-electron volts, was equivalent experimentally to a fine structure constant of 1/129.6. This agreed well with their theoretical prediction of 1/128.5.

Next we study how the EMQG quantum vacuum is affected by the introduction of a large mass. According to EMQG, the quantum vacuum virtual masseon particle pairs are **not** polarized near a large mass, as we found for electrons (as can be seen from table #1 above). The virtual masseon and anti-masseon pairs are *both* attracted towards the mass. This *lack* of polarization which results is the main difference between electromagnetism and gravity. A large gravitational mass (like the earth) does **not** produce vacuum polarization of virtual particles. In gravitational fields, all the virtual masseon and virtual anti-masseon particles of the vacuum have a net average statistical acceleration directed towards a large mass, and produces a net inward (acceleration vectors only) flux of quantum vacuum virtual masseon/anti-masseon particles that can, and do affect other masses placed nearby. In contrast to this, an electrically charged object *does* produce vacuum polarization in QED; where the positive and negative electric charges accelerates towards and away, respectively from the charged object. Hence, there is no energy contribution to other electrical test charges placed nearby (from the vacuum particles only), because the charged vacuum particles contributes equal amounts of force contributions from both directions.

We will see that in gravitational fields like the earth, the *lack* of vacuum polarization is responsible for the weak equivalence principle. This is because the electrically charged quantum vacuum masseons/anti-masseon particles can act in unison against a test mass dropped on the earth. Had there been vacuum polarization for masseons, the vacuum particles would act in the two opposite directions, and hence no net vacuum action would result against a test mass. Now we are in a position to state the basic postulates of EMQG theory.

15.6   THE BASIC POSTULATES OF EMQG

We expect that when EMQG theory becomes more fully developed at a future date, some or all these postulates might be proven from the basic principles of quantum field theory (perhaps with the exception of postulates 1 and perhaps 2). Notice that we did not include the principle of equivalence as one of our postulates. This is because equivalence is not a fundamental principle. Instead equivalence is simply a consequence of quantum particle interactions, from which we will derive in section 15.8. The basic postulates of EMQG are:

## POSTULATE #1: CELLULAR AUTOMATA

The universe is a vast cellular automata computation, which has an inherently quantized absolute 3D space consisting of 'cells', and absolute time (section 3). The numeric information in a cell changes state through the action of the numeric content of the immediate neighboring cells (26 neighbors) and the local mathematical rules, which are repeated for each and every cell. The action of absolute time (through 'clock cycles') synchronizes the state transition of all the cells. The number of 'clock cycles' elapsed between two different numeric states signifies the amount of absolute time elapsed. The cells are interconnected (mathematically) to form a simple 3D geometric CA (section 3). Matter, forces, and motion are the end result of information changing in the cells as absolute time progresses. Gravity, motion, and any other physical process do **not** affect low-level absolute 3D space and absolute time in any way. Photons propagate in the simplest possible manner on the CA, they shift from cell to adjacent cell on each and every 'clock cycle', in a given direction. This rate represents the maximum speed that information can be moved on the CA cycle' (section 9). The quantization scale is not known yet, but must be much finer then the Plank Scale of distance and time (section 4.2).

## POSTULATE #2: GRAVITON-MASSEON PARTICLES

The masseon is the most elementary form of matter (or anti-matter), and carries the lowest possible quanta of low level, gravitational 'mass charge'. The masseon carries the lowest possible quanta of positive gravitational 'mass charge', where the low level gravitational 'mass charge' is defined as the (probability) fixed rate of emission of graviton particles in close analogy to electric charge in QED. Gravitational 'mass charge' is a fixed constant and analogous to the fixed electrical charge concept. Gravitational 'mass charge' is **not** governed by the ordinary physical laws of *observable* mass, which appear as 'm' in the various physical theories, including Einstein's special relativity mass-velocity relationship: $E=mc^2$ or $m = m_0 (1 - v^2/c^2)^{-1/2}$. Masseons simultaneously carry a positive gravitational 'mass charge', and either a positive or negative electrical charge (defined exactly as in QED). Therefore, masseons also exchange photons with other masseon particles. Masseons are fermions with half integer spin, which behave according to the rules of quantum field theory. Gravitons (which are closely analogous to photons) have a spin of one (not spin two, as is commonly thought), and travel at the speed of light. Anti-masseons carry the lowest quanta of negative gravitational 'mass charge'. Anti-masseons also carry either positive or negative electrical charge, with electrical charge being defined according to QED. An anti-masseon is always created with an ordinary masseon in a particle pair as required by quantum field theory (specifically, the Dirac equation). The anti-masseon is the negative energy solution of the Dirac equation for a fermion, where now the **mass** is taken to be 'negative' as well, in clear violation of the principle of equivalence. Another important property exhibited by the graviton particle is the **principle of superposition**. This property works the same way as for photons. The action of the gravitons originating from all sources acts to yield a net vector sum for the receiving particle. EMQG treats graviton exchanges by the same successful methods developed for the behavior of photons in QED. The dimensionless coupling constant that governs the graviton exchange process is what we call 'β' in close analogy with the dimensionless coupling constant 'α' in QED, where $β ≈ 10^{-40} α$.

## **POSTULATE #3:     QUANTUM THEORY OF INERTIA**

The property which Newton called the inertial mass of an object, is caused by the resistance to acceleration of all the individual, electrically charged masseon particles that make up the mass. This resistance force is caused by the electromagnetic force interaction (where the details of this process are unknown at this time) occurring between the electrically charged virtual masseon/anti-masseon particle pairs created in the surrounding quantum vacuum, and all the real masseons particles making up the accelerated mass. Therefore inertia originates in the photon exchanges with the electrically charged virtual masseon particles of the quantum vacuum. The total inertial force $F_i$ of a mass is simply the sum of all the little forces $f_p$ contributed by each of the individual masseons, where the sum is: $F_i = (Σ f_p) = MA$ (Newton's law of inertia).

## **POSTULATE #4:     PHOTON FIZEAU-LIKE SCATTERING IN THE VACUUM**

Photons have an absolute, fixed velocity resulting from its special motion on the CA, where photons simply shift from cell to adjacent cell on every CA 'clock cycle' (section 9). This 'low level' photon velocity (measured in CA absolute space and time units) is *much higher* (by an unknown amount) than the observed light velocity (300,000 km/sec). This is because photons travelling in the vacuum (in an inertial frame) takes on a path through the quantum vacuum, that is the end result of a vast number of electromagnetic scattering processes with the surrounding electrically charged virtual particles. Each scattering process introduces a small random delay in the subsequent remission of the photon, and results in a cumulative reduction in the velocity of photon propagation. Real photons that travel near a large mass like the earth, takes on a path through the quantum vacuum that is the end result of a large number of electromagnetic scattering processes with the falling (statistical average) electrically charged virtual particles of the quantum vacuum. The resulting path is one where the photons maintain a net statistical average acceleration of zero with respect to the electrically charged virtual particles of the quantum vacuum, through a process that is very similar to the Fizeau scattering of light through moving water. Through very frequent absorption and re-emission (which introduces a small delay) by the accelerated charged virtual particles of the quantum vacuum, the apparent light velocity assumes an accelerated value with respect to the center of mass *in absolute* CA space and time units. (The light velocity is still an absolute constant when moving in between virtual particles, and is always created at this fixed constant velocity). The accelerated virtual particles of the quantum vacuum (that appears in gravitational and accelerated reference frames) can be viewed as a special Fizeau-like fluid, which affects the motion of matter and light in the direction of the fluid, which is ultimately responsible for 4D space-time curvature (section 16).

**Explanatory Notes:**

**POSTULATE #1**

This Cellular Automata postulate has led us to formulate EMQG through the belief that all physical phenomena originate from quantum interactions that are strictly 'local'. CA theory has been discussed in detail in section 3. The most important results of this study that are required for EMQG theory are briefly summarized below:

- On the lowest distance scales, physical space and time must be quantized where space is represented by the 'cells' of the CA. Time is represented by the 'clock' periods which act to evolve the numeric state of the CA. Matter particles are complex information patterns that roam around in 3D CA space. All forces are the result of the exchange of other complex information patterns called vector boson particles. Particles can only exist at finite densities, before the information patterns loss their identity.

- There is a maximum speed limit for the transfer of information from cell to cell, which causes our universe to have a universal speed limit (section 9). This speed limit is the raw light velocity, or light velocity between virtual particle scattering, which is an absolute constant and always created at this one fixed velocity. This speed also represents the speed limit for which any boson information patterns (for example, gravitons and photons) can move. In the vacuum, away from all gravitational and accelerated frames, the commonly measured light velocity of 300,000 km/sec represents the scattered light velocity, which is much smaller than the raw light velocity existing on the CA level, between vacuum particle scattering.

- All the laws of physics must be local, and derivable from the local rules at the cellular level of the CA. The rules for the numeric evolution of a cell is contained in each and every cell, and is the same for all other cells existing everywhere in the vast 3D CA cell space.

- True classical motion does not exist. Motion really is the transfer of information from cell to cell. In some sense, motion is 'digital'. Because the numeric content of the cells themselves and their locations are unobservable, motion is also relative. In other words, another nearby reference information pattern is required to gauge relative 'motion'.

- The motion of a light source will not affect the velocity of light, since light consists of numeric information simply being transferred from cell to an adjacent cell at every 'clock' cycle. The source motion cannot affect this shifting process. Nor can any other physical process.

**POSTULATE #2**

- This postulate is the quantum origin of the low-level gravitational force, and automatically implies that the virtual particles of the quantum vacuum are falling because of the presence of a nearby massive object, by the action of the large graviton flux. The masseon is required as the constituent of all elementary mass particles (fermions), in order to completely explain the weak equivalence principle. The low-level 'mass-charge' of a particle is defined as the fixed flux rate of gravitons emitted or absorbed by a masseon particle per unit time, and is similar to the concept of electrical charge for photon exchanges. In QED, charge is defined slightly differently as the probability of emission or absorption of photons. For elementary particles, this flux rate is a fixed constant corresponding to the combined mass of a certain number of masseon particles, which carry the basic unit of mass. Because, there is a difference of about $10^{-40}$ in the strength of the graviton and photon coupling for a single masseon particle, the photon exchange process dominates when considering mass interactions near a gravitational field. The domination of the total gravitational force by the electromagnetic force component is the biggest difference between EMQG and rival theories of quantum gravity.

- An elementary charged particle that is not interacting with other particles, will move through the background quantum vacuum virtual particles in such a way as to maintain a net statistical average acceleration of zero with respect to the quantum vacuum particles, due to very frequent electromagnetic photon exchanges.

**POSTULATE #3**

This postulate has been described in detail in section 13, and is a slightly modified version of the HRP theory of inertia of reference 5. Instead of inertia being caused by a Lorentz force from the ZPF (virtual photons), inertia results from the (unspecified) electromagnetic force that is present when a electrically charged masseon is accelerated with respect to the electrically charged virtual masseon particles that fills the background quantum vacuum.

**POSTULATE #4**

- In section 16.6, we show that the action on the real photons by the charged virtual particles of the quantum vacuum falling towards the earth leads to the same quantity of 4D space-time curvature predicted by the Einstein's general relativity. The vacuum acts like a special Fizeau-like fluid unknown to Einstein, which acts on the propagation of light (and matter). The actual *measured* light velocity for an observer under the influence of the gravitational field still remains unaffected, while space and time measurements compared with a distant observer do not agree.

- Real photons move along what Einstein called the "Geodesic" path of light, which represent the "straight" minimal length paths in curved 4D space-time for light near a large mass. Einstein based this conclusion on the postulate of constancy of light velocity. Thus, from the perspective of general relativity we can say that light moves perfectly '**straight**', but that the space-time that it moves in is '**curved**'. Now according to postulate #4, the geodesic path for photons can be reinterpreted as the path that maintains a net average acceleration of zero of the real photons with respect to the virtual particles of the quantum vacuum.

• On face value, the EMQG postulate 4 seems to violate the first postulate of special relativity, where the speed of light is constant with respect to any inertial reference frame. One must bare in mind that when comparing light velocity in far empty space with light velocity near the earth (based on photon scattering theory), that the state of the background quantum vacuum is different in these two cases. Near the earth, it is falling or accelerated. It flows towards the center of earth with an accelerated motion, whereas in far space it does not. The velocity of light that is affected by this accelerated flow is the photon velocity based on absolute CA space and time units.

• An observer accelerated in a rocket at 1g (far from gravitational fields) also observes light deflecting parallel to the floor. This light deflection is caused for a different reason. Taking the net average acceleration of the vacuum particles as the reference frame for accelerated motion, the path of the real photons is a straight path (again the result of photon scattering). However, the rocket now has a relative acceleration of 1 g with respect to the vacuum. The rocket carries the observer upward, leading to the perception of a curved path for the photon propagation, and therefore a curved 4D space-time. (As a consequence, we will see that equivalence is not a basic law of nature, but simply a matter of circumstance).

In order to understand the postulates of EMQG and how to apply them, we will consider the simplest example of a gravitational field, the field around a large spherically symmetrical non-rotating, non-charged mass (like the earth). We will derive the principle of equivalence and curved 4D space-time from these postulates.

## 15.7  VIRTUAL PARTICLE FIELD NEAR A SPHERICAL MASS IN EMQG

Our first application of EMQG theory is to determine the quantum nature of the gravitational field for a spherically symmetrical large mass. In general relativity, Einstein's field equation has been solved for this special case (ref. 39), and the solution is called the Schwarzchild metric and given by:

$$ds^2 = dr^2 / (1 - 2GM/(rc^2)) - c^2 dt^2 (1 - 2GM/(rc^2)) + r^2 d\Omega^2 \qquad (15.71)$$

where  $d\Omega^2 = d\theta^2 + \sin^2\theta \, d\phi^2$

This is a complete mathematical description of the space-time curvature (the metric in polar coordinates)) near the large spherical mass in spherical coordinates. This equation describes the path that light or matter takes through curved 4 D Minkowski space-time.

We will find that this solution is a very good *approximation* to the gravitational field. There are, however, hidden quantum processes involving the virtual particles in EMQG that are responsible for this curvature, and for the <u>very tiny</u> inaccuracy of this metric due to a slight violation of the principle of equivalence.

A large spherical mass turns out to be an excellent example for EMQG, because it has a very simple motion associated with the virtual particles that make up the surrounding quantum vacuum (in absolute CA units). The normal background motion of virtual particle creation and annihilation process near a massive spherical distribution of matter is distorted when compared to the vacuum in empty space far removed from any matter. Surrounding a large spherically symmetrical mass like the earth, the virtual particles created in the quantum vacuum have a net average acceleration vector that is directed downward towards the earth's center (figure 13) along radius vectors. (Note: We are ignoring the mutual interactions of the vacuum particles, which are why this statement is statistical in nature.) The cause of this downward acceleration of the vacuum is graviton exchanges between the earth and the virtual masseon particles of the quantum vacuum (postulate #2), which propagate at the speed of light (in absolute units). At any one instant, the vacuum particles have random velocity vectors (figure 12) which point in all directions, even including the *up* direction. However, the acceleration vectors are generally coordinated in the downward direction as in figure 13. The closer the virtual particles are to the earth, the greater the acceleration, as you would expect from the inverse square law of graviton exchanges. The average net statistical acceleration of this stream of virtual particles is directed downward, and varies with the height 'r'. This accelerated vacuum 'stream' plays the most important role in the dynamics of gravity in EMQG, and also naturally ties in with the problem of inertia and the equivalence principle. In fact, we will see that the average net acceleration vector of the vacuum particles at each point in space surrounding the earth, at its interaction with test masses and light is equivalent to the Schwarzschild 4D space-time metric given above. This is because the average net acceleration vector of the charged virtual particles at each point in space surrounding the mass guides the motion of the electrically charged free masseon particles or photons through the electromagnetic force. We will show this mathematically in section 16.6.

The magnitude and direction of the net average statistical acceleration of the virtual particles at point r above the earth (the direction is along the radius vectors) can be easily found from Newton's inverse square law of gravitation ($a = GM/r^2$). It is also possible to calculate this from the basic EMQG equations for a general, slow mass distribution (equation 17.22 and 17.23), which is derived later (section 17). This complex calculation of the state of motion of the vacuum particles is not required in the case for large spherically symmetrical masses like the earth, because of the simplicity of the virtual particle motion.

When a small test mass moves through the space surrounding the earth, the electromagnetic interactions between the real charged masseon particles in the mass with respect to the virtual charged particles quantum vacuum dominates over the pure graviton exchange process between the mass and the earth. This electromagnetic component plays the **major** role in the dynamics of motion of a nearby test mass. From postulate #2, the real masseon particles consisting of the earth exchanges gravitons with the virtual masseons of the quantum vacuum, causing a downward acceleration of the quantum vacuum of 1g. If we now introduce a test mass near the earth, according to postulate #2 all the real masseons making up the test mass will fall at the same average rate as that of the net statistical average of virtual particles of the vacuum. This is due to the relatively strong electromagnetic force acting between the electrically charged virtual masseons of the vacuum and the real masseons of the test mass.

Therefore, based on the Newtonian principles of how ordinary matter falls, the net average acceleration 'a' of a virtual particle in the vacuum with respect to height of the test mass, along the radius vector '**r**' towards the center of the earth is given by:

The net statistical average acceleration vector:   $\mathbf{a} = GM / \mathbf{r}^2$ (15.72)

where **r** is the distance vector along the radius from the center of the earth to a typical virtual particle, G is the Newtonian Universal Gravitation constant, and M is the mass of the earth.

**Note:** We have not proved that equation (15.72) is correct. Instead, it is based on the observation of the motion of a test mass near the earth. However, this equation can be derived from the semi-classical EMQG equations of motion.

To fully account for the gravitational field around a spherically symmetrical massive object, the motion of light near the object must also be accounted for. We will find that the altered behavior of light near a massive object drastically modifies the nature of equation (15.72). We will see that this equation is based on absolute cellular automata 3D space and time units. We will find that the relativistic curved 4D space-time is an emergent phenomena from this process, because of the way that light and matter behaves in this 'accelerated stream of virtual particles' near the earth. This alters the nature of equation 15.72, which now has to be specified in absolute CA units. This is because the acceleration (a=dv/dt, and velocity v=dx/dt) involves distance and time measurements. We will return to this important issue of the meaning of curved space-time after deriving the principle of equivalence from the fundamental postulates of EMQG.

## 15.8   EMQG AND THE PRINCIPLE OF EQUIVALENCE: INTRODUCTION

*"The principle of equivalence performed the essential office of midwife at the birth of general relativity, but, as Einstein remarked, the infant would never have got beyond its long clothes had it not been for Minkowski's concept [of space-time geometry]. I suggest that the midwife be now buried with appropriate honors and the facts of absolute space-time faced."*                                                              *- Synge*

The principle of equivalence means different things to different people, and to some it means nothing at all as can be seen in the quotation above. The equality of inertial and gravitational mass is only known to be true strictly through observation and experience. Is this equivalence exact, though? Since the principle of equivalence cannot be currently traced to deeper physics, we can never say that these two mass types are *exactly* equal. Currently, we can only specify the accuracy to which the two mass types have been shown *experimentally* to be equal.

How is the principle of equivalence defined? Well, there are two main formulations of the principle of equivalence. The strong equivalence principle states that the results of any given physical experiment will be precisely identical for an accelerated observer in free space as it is for a non-accelerated observer in a perfectly uniform gravitational field. A weaker form of this postulate restricts itself to the laws of motion of masses only. In other words, the laws of motion of identical masses on the earth are identical to the same situation inside an accelerated rocket (at 1g). Technically, this holds only at a point near the earth. It can be stated that objects of the different mass fall at the same rate of acceleration in a uniform gravity field. In regards to the strong equivalence principle, Synge writes:

*"… I never been able to understand this Principle … Does it mean that the effects of a gravitational field are indistinguishable from the effects of an observer's acceleration? If so, it is false. In Einstein's theory, either there is a gravitational field or there's none, according as the Riemann tensor does not or does vanish. This is an absolute property. It has nothing to do with any observer's world line … The principle of equivalence performed the essential office of midwife at the birth of general relativity, but, as Einstein remarked, the infant would never have got beyond its long clothes had it not been for Minkowski's concept [of space-time geometry]. I suggest that the midwife be now buried with appropriate honors and the facts of absolute space-time faced."*

Few physicists would doubt the validity of his statement. Synge has hit on an important point in regards to the nature of the equivalence principle and space-time. He is right to say that "*either there is a gravitational field or there's none, according as the Riemann tensor does not or does vanish. This is an absolute property* (of space near masses)". What he means is that the Riemann tensor describing curvature is there, or is not there, depending on whether or not there is a large mass present to distort space-time. (in other words, whether there exists a global space-time curvature or not). The existence of a ***global*** space-time curvature reveals whether you are in a gravitational field. In an accelerated frame, the space-time curvature is local to your motion only, and is not global property of space-time.

According to EMQG, if a large mass is present, the mass emits huge numbers of graviton particles, and distorts the surrounding virtual particles of the quantum vacuum. In an accelerated frame, there are very few gravitons, and the quantum vacuum is not affected. However, an observer in the accelerated frame 'sees' the quantum vacuum accelerating with respect to his frame, and hence the space-time distortion. However, the quantum vacuum *still remains undisturbed*. Thus in EMQG, the equivalence principle is regarded as being a coincidence due to quantum vacuum appearing the same for accelerated observers and for observers in gravitational fields.

Recently, some theoretical evidence has appeared to suggest that the strong equivalence principle does *not* hold in general. First, if gravitons could be detected experimentally with a new and sensitive graviton detector (which is not likely to be possible in the near future), we would be able to distinguish between an inertial frame and a gravitational frame with this detector. This is possible because inertial frames would have virtually no graviton particles present, whereas the gravitational fields like the earth have enormous numbers of graviton particles. Thus, we have performed a physics experiment that can detect whether you are in a gravitational field or an accelerated frame. Secondly, recent theoretical considerations of the emission of electromagnetic waves from a uniformly accelerated charge, and the lack of radiation from the same charge subjected to a static gravitational field leads us to the conclusion that the strong equivalence principle does not hold for radiating charged particles. Stephen Parrott (ref 23) has done an extensive analysis of the electromagnetic energy released from an accelerated charge in Minkowski space and a stationary charge in Schwarzchild space. He writes in his paper on "Radiation from a Uniformly Accelerated Charge and the Equivalence Principle":

*"It is generally accepted that any accelerated charge in Minkowski space radiates energy. It is also accepted that a stationary charge in a static gravitational field does not radiate energy. It would seem that these two facts imply that some forms of Einstein's Equivalence Principle do not apply to charged particles.*

*To put the matter in an easily visualized physical framework, imagine that the acceleration of a charged particle in Minkowski space is produced by a tiny rocket engine attached to the particle. Since the particle is radiating energy, that can be detected and used, conservation of energy suggests that the radiated energy must be furnished by the rocket. We must burn more fuel to produce a given accelerated world line than we would to produce the same world line for a neutral particle of the same mass. Now consider a stationary charge in Schwarzchild space-time, and suppose a rocket holds it stationary relative to the coordinate frame (accelerating with respect to local inertial frames). In this case, since no radiation is produced, the rocket should use the same amount of fuel as would be required to hold stationary a similar neutral particle. This gives an experimental test by which we can determine locally whether we are accelerating in Minkowski space or stationary in a gravitational field - simply observe the rocket's fuel consumption."*

He does a detailed analysis of the energy in Minkowski and Schwarzchild space-time, and shows that strong principle of equivalence does not hold for charged particles.

As for the weak equivalence principle, we can now only specify the accuracy as to which the two different mass types have been shown *experimentally* to be equal in an inertial and gravitational field. In EMQG, we show that the equivalence principle follows from lower level physical processes, and the basic postulates of EMQG. We will see that mass equivalence arises from the equivalence of the force generated between the net statistical average acceleration vectors of the matter particles inside a mass interacting with the surrounding quantum vacuum virtual particles inside an accelerating rocket. The *same* force occurs between the matter particles and virtual particles for a mass near the earth. We will find that equivalence is *not* perfect, and breaks down when the accuracy of the measurement approaches $10^{-40}$!

Basically, the equivalence principle arises from the *reversal* of the net statistical average acceleration vectors between the charged matter particles and virtual charged particles in the famous Einstein rocket, with the same matter particles and virtual particles near the earth. To fully understand the hidden quantum processes in the principle of equivalence on the earth, we will detail the behavior of test masses and the propagation of light near the earth. Equivalence is shown to hold for both stationary test masses and for free-falling test masses.

First we derive the principle of equivalence for the motion of ordinary masses. Next, we show that the quantum principle of equivalence holds for elementary particles. Next, we will demonstrate that equivalence also holds for large spherical masses with considerable self-gravity (and self-energy) such as the earth with a hot molten core, and the moon with a considerably colder core, with respect to a third mass like the sun. We will see that if both the earth and the moon fall towards the sun, they would arrive at the same time to a high degree of precision in the framework of EMQG. Finally, we examine the principle of equivalence and curved Minkowski 4D space-time curvature.

15.9    MASSES INSIDE AN ACCELERATED ROCKET AT 1g

In figure #1, there are two different masses at rest on the floor of a rocket which is accelerated upwards at 1 g far from any gravitational sources. The floor of the rocket experiences a force under the mass '2M' that is twice as great as for the mass 'M'. In Newtonian physics, the inertial mass is defined in precisely this way, the force 'F' that occurs when a mass 'M' is accelerated at rate 'g' as given by F=Mg. The quantum inertia explanation for this is that the two masses are accelerated with respect to the net average statistical motion of the virtual particles of the vacuum by the rocket. Since mass '2M' has twice the masseon particle count as mass 'M', the sum of all the tiny electromagnetic forces between the virtual vacuum and the masseon particles of mass '2M' is twice as great as compared to mass 'M', i.e. for mass 'M', $F_1$=Mg and for mass '2M', $F_2$=2Mg=2$F_1$. Because the particles that make up the masses do not maintain a net zero acceleration with respect to the virtual particles, a force is always present from the rocket floor (figure 1).

In figure #2, the two different masses (M and 2M) have just been released and are in free fall inside the rocket. According to Newtonian physics, no forces are present on the two masses since the acceleration of both masses is zero (the masses are no longer attached to the rocket frame). The two masses hit the rocket floor at the same time. The quantum inertia explanation for this is trivial. The net acceleration between all the real masseons that make up both masses and virtual masseon particles of the vacuum is a net (statistical average) value of zero. The rocket floor reaches the two masses at the same time, and thus unequal masses fall at the same rate inside an accelerated rocket.

15.9   MASSES INSIDE A GRAVITATIONAL FIELD (THE EARTH)

In figure #3, there are the same two masses (2M and M) which are at rest on the surface of the earth. The surface of the earth experiences a force under mass '2M' that is twice as great as for that under mass 'M'. The reason for this is that the two stationary masses do not maintain a net acceleration of zero with respect to the net statistical average acceleration of the virtual masseons in the neighborhood. This is because the virtual particles are all accelerating towards the center of the earth ($\mathbf{a}=GM/\mathbf{r}^2$) due to the graviton exchanges between the real masseons consisting of the earth and the virtual masseons of the vacuum. Since mass '2M' has twice the masseon particles as mass 'M', the sum of all the tiny electromagnetic forces between the virtual masseon particles of the vacuum and the real masseon particles of mass '2M' is twice as great as that for mass 'M'. Thus, a force is required from the surface of the earth to maintain these masses at rest, mass '2M' having twice the force of mass 'M'. The physics of this force is the same as for figure #1 in the rocket, but with the acceleration frames of the virtual charged masseons and the real charged masseon particles of the mass being reversed (with the exception of the direct graviton induced forces on the masses, which is negligible). Equivalence between the inertial mass 'M' on a rocket moving with acceleration 'A', and gravitational mass 'M' under the influence of a gravitational field with acceleration 'A' can be seen to follow from Newton's laws as follows:

$F_i = M(A)$      ...inertial force opposes the acceleration A of the mass 'M' in rocket.
$F_g = M(GM_e/r^2)$ ...gravitational force where **$GM_e/r^2$ is now virtual particle acceleration**.

Under gravity, the magnitude of the gravitational field acceleration is $A=GM_e/r^2$, which is the same as the magnitude of the acceleration of the rocket. From the reference frame of an average accelerated virtual particle on earth, a virtual masseon particle 'sees' the real masseon particles of the stationary mass M accelerating in exactly the same way as an average stationary virtual masseon in the rocket 'sees' the accelerated mass particles in the rocket. In other words, the vacuum state appears the same from both of these reference frames. We have illustrated equivalence in a special case; between an accelerated mass M and a stationary gravitational mass 'M'. Equivalence holds because $GM_e/r^2$ represents the net statistical average downward acceleration vector of the virtual masseons with respect to the earth's center, and is **equal** to the acceleration of the rocket. Newton's law of gravity was rearranged here to emphasize the form F=MA for *gravitational mass* so that we can see that the **same** electromagnetic force summation process for real masseons of the mass occurs under gravity as it does for accelerated mass. Thus the same processes at work in inertia are also present in gravitation.

This example shows why both the masses of figure 1 are equivalent to the masses in figure 3. The force magnitude is the same because the calculation of the force involves the same sum of all the tiny electromagnetic forces between the virtual charged masseon particles and the real masseon particles of the mass. The only difference in the physics of the masses in figure 1 is that the relative motions of all the tiny electromagnetic force vectors are reversed. The other difference is that large numbers of graviton particles (that originate from the earth's mass) slightly unbalances perfect equivalence between the masses falling on the earth. The larger mass has the largest graviton flux.

**Note:** There is a *very small* discrepancy in the equivalence principle for unequal masses in free fall near the earth which is caused by the excess graviton exchange force for the heavier mass. This discrepancy in the free fall rate of test masses near the earth is extremely minute in magnitude because there is a ratio of about $10^{40}$ in the field strength existing between the electromagnetic and gravitational forces. In principle it could be measured by extremely sensitive experiments, if two test masses are chosen with a very large mass difference.

In figure #4, two different masses are in free fall near the surface of the earth, and no external forces are present on the two masses. The two masses hit the earth at the same time. The net statistical average acceleration of the real masseon particles that make up the masses and virtual charged masseon particles of the vacuum is still zero, because this process is dominated by the electromagnetic force (the direct graviton exchanges are negligible). The electromagnetic forces between the virtual particles and the matter particles of the test mass dominates the interactions, because the electromagnetic force is $10^{40}$ times stronger than the graviton component. Although mass '2M' has twice the gravitational force due to twice the number of graviton exchanges, this is totally swamped out by the electromagnetic interaction, and the accelerated virtual particles and the test masses are in a state of electromagnetic equilibrium as far as acceleration vectors are concerned. Both masses fall at the same rate (neglecting the slight imbalance of the note above).

15.10  MICROSCOPIC EQUIVALENCE PRINCIPLE OF PARTICLES

Does the weak equivalence principle hold for an elementary particle? For example, does a neutron and an electron simultaneously dropped on the surface of the earth fall at the same rate (ignoring stray electrical charge effects)? Is this equivalent to the same experiment performed inside a rocket that is accelerating at 1 g? The answer to all these questions is yes. In fact, the equivalence principle has actually been experimentally verified for the case of a neutron in a gravitational field (ref. 40).

An astute observer may have questioned why **all** the virtual particles (virtual neutrons, virtual electrons, virtual quarks, etc, all consisting of different masses) are accelerating downwards towards the earth with the same acceleration in our EMQG model. Certainly inside an accelerated rocket an observer stationed on the floor will view *all* the virtual particles of the quantum accelerating with respect to him at the same rate, no matter what the masses of the virtual particles are. This is simply because the floor of the rocket moves upward at 1g, giving the *illusion* (to an observer on the floor of the rocket) that virtual particles of different mass are accelerating at the same rate. Since the masses of the different types of virtual particles are **all different** according to the standard model of particle physics, why are they all falling at the same rate on the earth? Here, the cause of the acceleration is graviton exchanges with the earth. Since we are trying to derive the equivalence principle from fundamental concepts, we cannot invoke this principle to state that the virtual particles must be accelerating at the same rate.

We can trace why all quantum vacuum virtual particles are accelerating at the same rate on the earth to the existence of the virtual masseon particle. All particles with mass (virtual or not) are composed of combinations of the fundamental "masseon" particle, which carries just one fixed quanta of mass (postulate #2). Since all virtual masseon particles exchange the same fixed flux of gravitons with the earth, the virtual masseons are all accelerated at the same rate. However, masseons can bind together to form the familiar particles of the standard model such as virtual electrons, virtual positrons, virtual quarks, etc. or even unknown species of virtual particles. According to postulate #2, masseons carry both gravitational 'mass charge' and ordinary electrical charge. However, the electromagnetic interactions (photon exchanges) will work to equalize the fall rate (from the point of view of acceleration vectors) of virtual masseons that momentarily combine to make virtual particles like virtual electrons and virtual quarks. If a virtual quark consists of say 100 bound masseons (the actual number is not known), the graviton exchanges would normally be cumulative, and 100 times more acceleration will be imparted to the virtual quark than a single virtual masseon. However, virtual masseons dominate the quantum vacuum since they are the fundamental mass particle, and do not have to bond with other masseons to exist. Therefore the lone, unbound virtual masseon is by far the most common virtual mass particle in the quantum vacuum (this is illustrated in figures 5 and 6).

No matter how many virtual masseons combine to give other virtual particles, the local electromagnetic interaction between the far more numerous virtual masseons and the virtual quark (or any other virtual particle) will equalize the fall rate. This process works like a *microscopic version* of the EMQG weak principle of equivalence, for falling virtual particles, with the same action occurring on the particle level as what happens for large falling masses discussed in section 15.8. Figures 5 and 6 shows the microscopic equivalence principle at work for a free falling mass in an accelerated frame and in a gravitational field. To summarize, the electromagnetic forces from the free virtual masseons of the quantum vacuum (all falling at the same rate), dominates over the more familiar virtual particles that consist of combinations of masseons (like the virtual neutrons, electrons, quarks, and all other virtual particles). The virtual quark would normally fall faster than the virtual electron and the virtual electron faster than an individual virtual masseon. This is because many virtual masseons bound together exchange many more gravitons with the earth. However, the electromagnetic interaction between the far numerous virtual masseons of the vacuum, and the virtual masseons combined inside the virtual neutrons, quarks, and virtual electrons acts to equalize the fall rate, causing all virtual particles to fall at the same rate. Since the quantum vacuum background appears the same from the perspective of a mass on the surface of the earth, as for the same mass inside an accelerated rocket equivalence still holds.

## 15.11 THE INERTION: AN ELEMENTARY QUANTA OF INERTIA

Recall that the real neutron has inertial mass because it is composed of real electrically charged masseon particles, which interact electromagnetically with the charged virtual masseon particles of the quantum vacuum (the sum of positive and negative electrically charged masseons inside a neutron is equal). Each real masseon inside the neutron contributes a fundamental unit or 'quanta' of inertia to the neutron, because of the electromagnetic force interaction with the immediate surrounding virtual masseons of the vacuum. Therefore, the sum of the real masseon force contribution to the inertial mass of the neutron defines the neutron's inertial mass. We propose that the electromagnetic force existing between *one* real masseon particle accelerating at a rate of 1g with respect to the surrounding quantum vacuum be given the status of a new universal constant. We call this new constant the '**inertion**' constant, or "i". The inertion thus represents the lowest possible quanta of inertia force. In order to determine the numerical value of the inertion, one must determine the number of masseons inside a neutron for example, and divide this into the observable (inertial) mass of the neutron.

How does the background quantum vacuum look like from the frame of reference of the neutron particle sitting on the surface of the earth, and from the frame of a neutron on the floor of a rocket accelerating at 1g? The answer is that it is the <u>same</u>. As we have found, the virtual masseon and anti-masseon particles of the quantum vacuum are accelerated by the graviton exchanges with the earth. In the accelerating rocket the motion of the virtual masseons and anti-masseons are now an illusion caused by the motion of the rocket floor. The motion is the same, but the cause is definitely different. Therefore, we can conclude that the 'inertion' also represents the lowest possible quanta of gravitational mass on the earth.

## 15.12 EQUIVALENCE PRINCIPLE FOR THE SUN-EARTH-MOON SYSTEM

Will the weak equivalence principle hold for the following imaginary scenario, where the earth and the moon are simultaneously in free fall towards the sun with an acceleration of gravity '$G_{sun}$' of the sun? In other words, would the earth and moon arrive at the same time on the surface of the sun? Would they also arrive on the floor at the same time when free falling inside a huge rocket undergoing acceleration $G_{sun}$ in space (far from any other large masses)? This question is at the heart of the so-called metric theories of gravity. In metric theories such as general relativity, all objects with any kind of internal composition follow the natural curvature of space-time. This includes objects with considerable internal energy sources and self-gravity. Any deviation from perfect equivalence would constitute what is called the Nordtvedt effect (ref. 41), after the discoverer in 1968. Because the earth's core is molten and very hot, the earth contains a significant internal energy. The earth is also a large source of gravitational energy as well, which significantly distorts the nearby virtual particles of the quantum vacuum. Contrast this with the moon, which is relatively cool and less energetic, with considerably less gravitational energy.

To see if the weak equivalence applies here, we start with the situation where the earth and moon are 'dropped' simultaneously from a height 'h' inside a huge rocket (with negligible mass) accelerating with the same $G_{sun}$ as exists on the surface of the sun (figure 7). The result of this experiment is obvious. They both arrive on the floor of the rocket at the same time. Actually, it is the floor of the rocket that accelerates upwards and meets both bodies at the same time! However, now the virtual particles of the quantum vacuum are disturbed near these large bodies (by graviton exchanges), particularly the acceleration vectors of the virtual particles in close proximity with the earth and the moon (now tending to point towards the centers of the two bodies in these regions). This fact, however, does not affect the results of this experiment. The results are no different than if two small masses are dropped inside the rocket; again because the floor moves up to meet them at the same time. However, one must note the virtual particle pattern of figure 8.

The situation near the surface of the sun, where the earth and moon are 'dropped' from the same height 'h', is far more complex than inside the rocket. Now all three bodies disturb the virtual particles of the quantum vacuum! The sun sets up a strong $GM_s/r^2$ acceleration field consisting of virtual particles, which applies over long distances and points towards the center of the sun. The earth and moon also produce their own fields in their vicinity, although much weaker (figure 7). The sun's acceleration dominates over the surrounding space, except near the moon and near the earth; where some of the virtual particles actually are moving away from the sun (as happens near the surface of the night side of the earth). How can equivalence possibly hold in this scenario? Recall that we stated that it is the electromagnetic action of the virtual particles of the vacuum on the real particles inside the bodies that determines the motion of the earth and the moon undergoing gravitational acceleration. However, in different regions of the earth, the virtual particles are accelerating in different directions! Part of the answer to this problem is an important property exhibited by the graviton particle (postulate #2): the **principle of superposition**. This is a property also shared by the photon particle. The action of the gravitons originating from all three sources on a given virtual particle of the quantum vacuum yields a net acceleration that is the net vector sum of the action of all the gravitons received by the virtual particle.

To explain equivalence, we must first recall that equivalence only holds in a sufficiently small region of space (technically, at a given point above the sun) when compared to the equivalent accelerated reference frame. This is because the acceleration of the sun varies with the distance 'r' from the sun's center, whereas inside an accelerated rocket it does not vary with height. Secondly, we must recall that the motion of the virtual particles in the rocket is also disturbed near the vicinity of the earth and the moon. In fact, inside the rocket the virtual particles are also directed along the radius vectors of both the earth and the moon in their vicinity (figure 7). Yet, the sun and the moon still reach the floor of the rocket at the same time. Therefore, the quantum vacuum can be disturbed in the case of the free fall of the earth and the moon towards the sun, provided the total virtual particle pattern can be shown to match the case of the accelerated rocket.

A close study of figures 7 and 8 reveals that the quantum vacuum pattern *is the same* when viewed in the *correct* reference frame. In figure 7, the observer is stationed on the surface of the sun, so that we can see the reason why both bodies are attracted to the sun. Recall that the electromagnetic interaction between the acceleration vectors of the falling virtual masseons and the real masseons in the earth and moon is the primary reason for the attraction. The direct graviton action is negligible in comparison. Now in order to compare the two experiments of figure 7 and 8, the reference frame for the sun experiment should be equivalent to the rocket. In the rocket experiment, the frame chosen for our observer is outside the rocket (the quantum vacuum has a relative acceleration of zero) in order to understand the results. Therefore for the sun, the observer's frame should be in free fall, thus restoring the relative acceleration of the quantum vacuum to zero just as for the observer outside the rocket. When this is done, we have to correct the acceleration vectors of the virtual particles near the earth and moon in figure 7. It is easy to show that the result of this operation gives an identical result as figure 8. Therefore, equivalence holds in both experiments.

15.13  LIGHT MOTION IN A ROCKET: SPACE-TIME EFFECTS

We will examine three scenarios for the motion of light in a rocket which is accelerated upwards at 1 g (far from any gravitational sources). First, we study light moving from the floor of the rocket to the ceiling where it is detected by an observer. Next we look at light moving from the ceiling of the rocket to the floor where it is detected by an observer. Finally, we examine light moving parallel with the floor of the rocket, where it follows a curved path (figure 11).

(A) LIGHT MOVING FROM THE FLOOR TO THE CEILING OF THE ROCKET

Here the light is positioned on the floor of the rocket which is being accelerating upwards at 1 g, and propagates in a straight line up to the observer on the ceiling. Meanwhile, the rocket has accelerated upwards while the light is in flight. What happens to the light? According to general relativity, an observer outside the rocket examines the light moving upward at the speed of light in a straight path. Meanwhile, according to general relativity, an observer inside the rocket stationed on the ceiling also observes the light moving upwards in a straight line towards him. He also observes that the light is red-shifted. He makes a measurement of the light velocity of the incoming red-shifted light with his measuring instruments (which were calibrated within his reference frame). He observes that the velocity of the red-light light is the same on the ceiling as he found when he previously checked his internal light sources with his calibrated instruments. In other words, the speed of light does not vary under all these circumstances. Closer examination reveals that the clocks in the ceiling differ from the clocks stationed on the floor. In particular, the clock on the floor of the rocket runs slower than one on that on the ceiling. Distances measurements are also affected. General relativity explains all these observations with the 4D space-time curvature existing inside accelerated frames. We will return to this example with our EMQG interpretation of these measurements.

(B) LIGHT MOVING FROM THE CEILING TO THE FLOOR OF THE ROCKET

Here the light is positioned on the ceiling of the rocket which is accelerating upwards at 1 g, and propagates in a straight line down to the observer on the floor. Meanwhile, the rocket has accelerated upwards while the light is in flight. What happens to the light? According to general relativity, an observer outside the rocket observes the light moving downwards at the speed of light in a straight path. Meanwhile, according to general relativity, an observer inside the rocket stationed on the floor also observes the light moving downwards in a straight line towards him. He also observes that the light is blue-shifted. He makes a measurement of the light velocity of the incoming blue-shifted light with his measuring instruments (which were calibrated within his reference frame). He observes that the velocity of the blue-light light is the same on the floor as he found when he previously checked his internal light sources and with his calibrated instruments. In other words, the speed of light does not vary under all these circumstances. Closer examination reveals that the clocks in the floor differ from the clocks stationed on the ceiling. In particular, the clock on the ceiling of the rocket runs faster than one on that on the floor. Distances measurements are also affected. Again, general relativity explains all these observations with the 4D space-time curvature existing inside accelerated frames. We will return to this example with our EMQG interpretation of these measurements.

(C) LIGHT MOVING PARALLEL TO THE FLOOR OF THE ROCKET

Here the light leaves the light source on the left wall of the rocket which is accelerating upwards at 1 g, and propagates in a straight line towards the observer on the right wall (figure 11). Meanwhile, the rocket has accelerated upwards while the light is in flight. Therefore an observer in the rocket observes a curved light path. An observer outside the rocket sees a straight light path. According to general relativity, the space-time inside the rocket is curved (in the direction of motion), and light moves along the natural geodesics of curved 4D space-time. Meanwhile, the observer outside the rocket lives in flat-space time, and therefore observes light moving in a perfect straight line, which is the geodesic path in flat 4D space-time. We will return to this example with our EMQG interpretation of these measurements. Next we will examine all three scenarios on the surface of the earth.

## 15.14   LIGHT MOTION NEAR EARTH'S SURFACE - SPACE-TIME EFFECTS

We will examine the same three scenarios for the motion of light on the surface of the earth (1g), which is the same as for the rocket (1g) according to the principle of equivalence. We will ignore the variation of acceleration with height found on the earth, as well as the slight change in the direction of acceleration caused by acceleration vectors being directed along radius vectors. First, light moves from the floor of the room on the surface of the earth to the ceiling, where it is detected. Next, light is moving from the ceiling of the room to the surface of the earth where it is detected by an observer. Finally, light is moving parallel with the earth's surface from the left side of the room to the right, and follows a curved path (figure 10).

(A) LIGHT MOVING FROM THE FLOOR TO THE CEILING ON EARTH

Light is positioned on the floor of a room on the surface of the earth, and propagates in a straight line up to the observer on the ceiling. What happens to the light? According to general relativity, an observer outside the room in free fall observes the light moving upward at the speed of light in a straight path. Meanwhile, according to general relativity, an observer inside the room stationed on the ceiling also observes the light moving upwards in a straight line. He also observes that the light is red-shifted. He makes a measurement of the light velocity with his measuring instruments (which were calibrated within his reference frame) and observes that the velocity of light is the same on the floor as he found when he measured received light speed in his internal reference frame with the same instruments. In other words, the speed of light does not vary in all cases. Closer examination reveals that clocks measured in his reference frame differ from the clocks on the floor. In particular, the clock on the floor of the room runs slower than the one on the ceiling. Distances are also affected. In general relativity, all these conclusions follow directly from 4D space-time curvature.

(B) LIGHT MOVING FROM THE CEILING TO THE FLOOR ON EARTH

Here the light is positioned on the ceiling of the room on the surface of the earth, and propagates straight down to the observer on the floor. What happens to the light? According to general relativity, an observer outside the room in free fall observes the light moving downward at the speed of light in a straight path. Meanwhile, according to general relativity, an observer inside the room stationed on the floor also observes the light moving downwards in a straight line. He also observes that the light is blue-shifted. He makes a measurement of the light velocity with his measuring instruments (which were calibrated within his reference frame) and observes that the velocity of light is the same on the floor as he found when he measured received light speed in his internal reference frame with the same instruments. In other words, the speed of light does not vary in all cases. Closer examination reveals that clocks measured in his reference frame differ from the clocks on the ceiling. In particular, the clock on the ceiling of the room runs faster than the one on the floor. Distances are also affected. In general relativity, all these conclusions follow directly from 4D space-time curvature.

(C) LIGHT MOVING PARALLEL TO THE SURFACE OF THE EARTH

Here the light leaves the light source on the left wall of the room on the earth and propagates in a curved path towards the observer on the right wall (figure 10). Meanwhile, an observer in free fall towards the earth's surface sees light moving in a straight path. According to general relativity, and light moves along the natural geodesics of curved 4D space-time in the room. Meanwhile, an observer in free fall lives in flat 4D space-time, and hence an observer sees straight-line paths for light. In general relativity, all these conclusions are identical as for the observer accelerated in the rocket at 1g in accordance with the principle of equivalence. Now we look at 4D space-time curvature from the perspective of EMQG.

## 16. EMQG AND THE PROBLEM OF SPACE-TIME CURVATURE

*"The relativistic treatment of gravitation creates serious difficulties. I consider it probable that the principle of the constancy of the velocity of light in its customary version holds only for spaces with constant gravitational potential."*

- **Albert Einstein (in a letter to his friend Laub, August 10, 1911)**

In this section, we contrast the two different approaches to the problem of space-time curvature, and the propagation of light in a gravitational field: Einstein's General Relativity and EMQG theory. First we will derive the gravitational time dilation equation using the solution to Einstein's gravitational field equations for a spherical mass called the Schwarzschild metric. Next, we fully develop the EQMG theory of space-time curvature. From this, we calculate the quantity of space-time curvature using EMQG theory, and show that the results are the same.

### 16.1 GENERAL RELATIVISTIC 4D SPACE-TIME CURVATURE

General relativity accounts for the motion of light under all scenarios (section 16.1) for a large spherical mass. General Relativity postulates space-time curvature in order to preserve the constancy of the light velocity in an accelerated frame or in a gravitational field. The solution of Einstein's gravitational field equation for the case of spherical mass distribution is given by the Schwarzchild metric (ref. 39):

$$ds^2 = dr^2 / (1 - 2GM/(rc^2)) - c^2 dt^2 (1 - 2GM/(rc^2)) + r^2 d\Omega^2 \qquad (16.11)$$

where $d\Omega^2 = d\theta^2 + \sin^2 \theta \, d\phi^2$

This is a complete mathematical description of the space-time curvature near the large spherical mass in spherical coordinates in differential form called the 4D space-time metric. From this, it is easy to show (ref. 39) that the comparison of time measurements between a clock outside a gravitational field (called proper time $t(\infty)$) to a clock at distance r from the center of a spherical mass distribution (called the coordinate time $t(r)$) is given by:

$$t(r) = \left(1 - 2GM/(rc^2)\right)^{-1/2} t(\infty) \tag{16.12}$$

which follows from Schwarzchild metric directly.

Using the relationship $(1 - x)^{-1/2} \approx 1 - x/2$ when $x \ll 1$, and realizing the quantity $2GM/rc^2$ is very small (for the earth this is $\approx 10^{-9}$) we can write this as:

$$t(r) \approx \left(1 - GM/(rc^2)\right) t(\infty) \tag{16.13}$$

This gives the amount of time dilation between a clock on the earth "$t(r)$" compared to a clock positioned at infinity "$t(\infty)$". From this, we see that clocks on the earth run slower then at infinity.

Similarly, from the metric, we find that the distance at point $s(r)$ follows as:

$$s(\infty) = \left(1 - 2GM/(rc^2)\right)^{-1/2} s(r), \tag{16.14}$$

or we can also write this as:

$$s(r) \approx \left(1 - GM/(rc^2)\right)^{-1} s(\infty) \tag{16.15}$$

This gives the amount of space distortion for rulers on the earth "$s(r)$" compared to rulers positioned at infinity "$s(\infty)$".

## 16.2 EMQG AND 4D SPACE-TIME CURVATURE

In order to understand space-time curvature and the principle of equivalence in regards to the equivalence of all light motion in an accelerated rocket compared with that on the surface of the earth, we must examine the effects of the background virtual particles on the propagation of light. The big question to consider here is this:

**Does the general downward acceleration of the virtual particles of the quantum vacuum near a large mass affect the motion of nearby photons? Or is the deflection of photons truly the result of an actual space-time geometric curvature (which holds down to the tiniest of distance scales), as required by the constancy of the light velocity in Einstein's special relativistic postulate**?

The answer to this important question hinges on whether our universe is truly a curved, geometric Minkowski 4D space-time on the smallest of distance scales, or whether curved 4D space-time results merely from the activities of quantum particles interacting with other quantum particles. EMQG takes the second view! According to postulate 4 of EMQG theory, light takes on the same general acceleration as the net statistical average value of quantum vacuum virtual particles, through a 'Fizeau-like' scattering process. By this we mean that the photons are frequently absorbed and re-emitted by the electrically charged virtual particles, which are (on the average) accelerating towards the center of the large mass. When a virtual particle absorbs the real photon, a new photon is re-emitted after a small time delay in the same general direction as the original photon. This process is called photon scattering (figure 10). We will see that photon scattering is central to the understanding of space-time curvature.

The velocity of light in an ordinary moving medium is already known to differ from its value in an ordinary stationary medium. Fizeau (1851) demonstrated this experimentally with light propagating through a current of water flowing with a constant velocity. Later (1915), Lorentz identified the physics of this phenomena as being due to his microscopic electromagnetic theory of photon propagation. Einstein attributed this to the special relativistic velocity addition rule. In EMQG, we propose that in gravitational fields (and in accelerated motion) the moving water of Fizeau's experiment is now replaced by the accelerated virtual particles of the quantum vacuum. Like in the Fizeau experiment, photons scatter by the accelerated motion of the virtual particles of the quantum vacuum.

Imagine what would happen if Fizeau placed a clock inside his stream of moving water. Would the clock keep time properly, when compared to an observer with an identically constructed clock placed outside the moving water? Of course not! The very idea of this seems almost ridiculous. Yet we are expected to believe that the flow of virtual particles does not affect clocks and rulers under the influence of a gravitational field, as compared to the identical circumstance in far space. If Einstein knew the nature of the quantum vacuum at the time he proposed general relativity theory, he might have been aware of this connection between gravity, space-time curvature, and accelerated virtual particles.

In our review of special relativity, we have seen (section 9) the importance of the propagation of light in understanding the nature of space and time measurements. Recall that the definition of an inertial frame in space is a vast 3D grid of identically constructed clocks placed at regular intervals with a ruler. Therefore, we will closely examine the behavior of light near the earth.

In order to understand the connection between light propagation and space-time curvature near *a large gravitational field* with EMQG theory, we find it useful to review the behavior of a high-speed, non-relativistic test particle moving at 1/100 light velocity c near the earth. For the case where the test particle moves from the floor (distance r from the center of the earth) to the ceiling (height r+h, where h is small) on the *earth* (where the acceleration is 1g), we find that with an non-relativistic initial velocity $v_0 = 1/100$ c, the velocity of the particle at the detector is approximately ($v_0$ - gt). Here we can ignore all relativistic effects. The particle has a downward acceleration of 1g, which is *independent* of it's mass. Since t ≈ h/c, we find that the final velocity of the particle with respect to the detector is given approximately by: **$v_0(1 - gh/c^2)$**, according to Newtonian physics. The reason for the change in the velocity of the high-speed particle (with respect to the detector) is that as the particle moves up, there is a change in the Newtonian gravitational potential on earth. This decelerates at –1g the particle as it moves towards the detector.

For the case where the same particle moves with velocity $v_0 \approx 1/100$ c from the floor of **an accelerated rocket** (same 1g) to the ceiling at the same small height h, the final velocity of the particle with respect to the detector is again ($v_0$ -gt), but for a *different* reason. As before, since t ≈ h/c, and $v_0 = 1/100$ c, we find that the final velocity of the particle with respect to the detector is given approximately by: **$v_0(1 - gh/c^2)$** as in a gravitational field. The reason for the change in the velocity of the high speed particle (as viewed by the detector) is that as the particle moves up, it is now the detector itself which attains the velocity (-gt) with respect to the particle during the time t, due to the acceleration of the rocket. The end result is the same, but different physical processes are occurring.

When viewed from the principles of EMQG theory, there is really only **one** reason for the equality of the final rocket and earth velocities of our high-speed particle. In both cases, the elementary particles that make up the high-speed particle maintained a net statistical average *acceleration of zero* with respect to the virtual quantum vacuum particles. Thus, for the final velocity of the particle $v_0(1 - gh/c^2)$ on the rocket, the acceleration 1g represents the relative acceleration of the detector with respect to the high-speed particle which is in equilibrium with the non-accelerating quantum vacuum. And for the final velocity of the particle $v_0(1 - gh/c^2)$ on the earth, the acceleration 1g represents the relative acceleration of the detector with respect to the high speed particle which is also in equilibrium with the surrounding, but now falling, quantum vacuum virtual particles. Thus, on the earth, both the vacuum particles and the high-speed mass particle are *falling* at 1 g. Meanwhile inside the rocket it is only the detector that has the 1g acceleration with respect to the vacuum particles. We can see from this analysis that the equivalence of mass applies for the rocket and for the earth. We must stress that although equivalence exists, the physical process is actually different.

We will now take a bold step and assume that for the case on the surface of the earth the equation: $c(1 - gh/c^2)$ holds for the propagation of photons moving upwards, but only for *very short distances*. Technically this is true only at a point, which means that this equation must be written in differential form. We ignore the special relativistic postulate of the constancy of light velocity for now, and address this problem later. This means that photons continuously vary their velocity (the velocity of light is still an absolute constant between vacuum scattering events) by scattering with the falling virtual particles, as they propagate up or down. The scattering process will be described in detail later. If this picture is true, why is it that we do not observe this variation in light velocity in actual experiments on the earth?

First we must carefully understand what is meant by light velocity. Velocity is *defined* as distance divided by time, or c=d/t. Light has very few observable characteristics in this regard: we can measure velocity c (the ratio of d/t); frequency $\nu$; wavelength $\lambda$; and we can also measure velocity by the relationship $c=\nu\lambda$. It is important to note that all these observables are related. We know that $\nu = 1/t$ (t is the period of one light cycle) and $\lambda=d$ (the length of one light cycle). Thus, $c=d/t$ and $c=\nu\lambda$ are equivalent expressions. If we transmit green light to an observer on the ceiling of a room on the earth, and he claims that the light is red shifted, it is impossible for him to tell if the red shift was caused by the light velocity changing, or by space and time distortions which causes the timing and length of each of the light cycles to change. For example, if the frequency is halved, or $\nu_f = (1/2)\nu_i$ and the wavelength doubles $\lambda_f = 2\lambda_i$ (and you were not aware of both changes), then the velocity of light remains unchanged ($c=\nu\lambda$). However, if the velocity of light is halved, and you were not aware of it, then you could conclude that the frequency is halved, $\nu_f = (1/2)\nu_i$ and the wavelength doubles $\lambda_f = 2\lambda_i$. To illustrate this point, we will now examine what happens if an observer on the floor feeds a ladder (which represents the wave character of light) with equally spaced rungs to an observer on the ceiling, where each observer cannot see what the other observer does with the ladder.

Imagine a perfect ladder with equally space rungs of known length being passed up to you at a known velocity, such that it is impossible to tell the motion of the ladder other than by observing the rungs moving past you. If the rung spacing are made larger, you would conclude that either the ladder is slowing down, or that the spacing of the ladder rungs was increased. But it would be impossible to tell which is which. Let us assume that you make a measurement on the moving rungs, and observe a spacing of 1 meter between any two rungs. Then you observe that two rungs move past you every second. You therefore conclude the velocity of the ladder is 2 m/sec. Now, suppose that the ladder is fed to you at half speed or at 1 m/sec, and that you are not aware of this change in velocity. You could conclude that the velocity halved from your measurements, because you now observe that one rung appears in view for every second that elapses instead of two rungs, and that the velocity was thus reduced to 1 m/sec. However, you could just as well conclude that your space and time was altered, and that the velocity of the ladder is constant or unaffected. Since you observe only one rung in view per second instead of the usual two rungs, you could claim that the rung spacing on the ladder is enlarged (red-shifted) or doubled by someone, and that the velocity still remains unaltered. From this, you conclude that the frequency is halved, and that time measurements that will be based on this ladder are now dilated by a factor of two.

Which of these two approaches is truly correct? It is impossible to say by measurement, unless you know before hand what trait of the ladder was truly altered. For photons, the same problem exists. No known measurement of photons in an accelerated rocket or on the surface of the earth can reveal whether space and time is affected, or whether the velocity of light has changed. In EMQG theory, the variable light velocity approach is chosen for several reasons. First, the *equivalence of light motion in accelerated and gravitational frames now becomes **fully understood*** as a dynamic process having to do with motion (for gravity, hidden virtual particle motion), just as we found for ordinary matter in motion. Secondly, the *physical basis of the curvature* of Minkowski 4D space-time near a large mass now becomes clear. It arises from the interaction of light and matter with the background accelerated virtual particle processes. This process can be visualized as a fluid flow (for acceleration only) affecting the motion of light and matter. Finally, *the physical **action** that occurs between the earth and the surrounding space-time curvature now becomes clearly understood.* The earth acts on the virtual particles of the quantum vacuum through graviton exchanges, causing them to accelerate towards the earth. The accelerated virtual particles act on light and matter to produce curved 4D space-time effects. The physical process involved is photon scattering.

Since photon scattering is essential to our 4D space-time curvature approach we will examine scattering in some detail. First we review the conventional physics of light scattering in real moving, and real non-moving transparent matter such as water or glass. After this review, we will examine photon scattering due to the virtual particles of the quantum vacuum.

### 16.3  SCATTERING OF PHOTONS IN REAL, TRANSPARENT MATTER

It is a well known result of classical optics that light moves slower in glass than in air. Furthermore, the velocity of light in air is slower than that of its vacuum velocity. It also has been known for over a century that the velocity of light in a moving medium differs from its value in the same, stationary medium. Fizeau demonstrated this experimentally in 1851 (ref. 41). For example, with a current of water (with refractive index of the medium of n=4/3) flowing with a velocity V of about 5 m/sec, the relative variation in the light velocity is $10^{-8}$ (which he measured by use of interferometry). Fresnel first derived the formula (ref. 41) in 1810 with his ether dragging theory. The resulting formula relates the longitudinal light velocity '$v_c$' moving in the same direction as a transparent medium of an index of refraction 'n' defined such that 'c/n' is the light velocity in the stationary medium, which is moving with velocity 'V' (with respect to the laboratory frame), where c is the velocity of light in the vacuum:

Fresnel Formula: $v_c = c/n + (1 – 1/n^2) V$                               (16.31)

Why does the velocity of light vary in a moving (and non-moving) transparent medium? According to the principles of special relativity, the velocity of light is a constant in the vacuum with respect to all inertial observers. When Einstein proposed this postulate, he was not aware of the fact that the vacuum is not empty. However, he was aware of Fresnel's formula and derived it by the special relativistic velocity addition formula for parallel velocities (to first order). According to special relativity, the velocity of light relative to the proper frame of the transparent medium depends only on the medium. The velocity of light in the stationary medium is defined as 'c/n'. Recall that velocities u and v add according to the formula:  $(u + v) / (1 + uv/c^2)$
Therefore:

$v_c = [ c/n + V ] / [ 1 + (c/n)(V)/c^2 ] = (c/n + V) / ( 1 + V/(nc) ) \approx c/n + (1 – 1/n^2) V$
                                                                                                                                       (16.32)

The special relativistic approach to deriving the Fresnel formula does not say much about the actual quantum processes going on at the atomic level. At this scale, there are several explanations for the detailed scattering process in conventional physics. Because light scattering is central to EMQG theory, we will investigate these different approaches in more detail below:

## 16.31   CLASSICAL PHOTON SCATTERING THEORY IN MATTER

The Feynman Lectures on Physics gives one of the best accounts of the classical theory of the origin of the refractive index and the slowing of light through a transparent material like glass (ref. 42, chap. 31 contains the mathematical details). We will summarize the important points of the argument below:

(1) The incoming source electromagnetic wave (light) consists of an oscillating electric and magnetic field. The glass consists of electrons bound elastically to the atoms, such that if a force is applied to an electron the displacement from its normal position will be proportional to the force.
(2) The oscillating electric field of the light causes the electron to be driven in an oscillating motion, thus acting like a new radiator generating a new electromagnetic wave in the same direction as the source wave. This new wave is always delayed, or retarded in phase. This delay results from the time delay required for the bound electron to oscillate to full amplitude. Recall that the electron carries mass, and therefore inertia, and therefore time is required to move the electron.
(3) The total resulting electromagnetic wave is the sum of the source electromagnetic wave plus the new phase-delayed electromagnetic wave, where the total resulting wave is phase-shifted.
(4) The resulting phase delay of the electromagnetic wave is the cause of the reduced velocity of light in a medium such as glass.

## 16.32   LORENTZ SEMI-CLASSICAL THEORY OF PHOTON SCATTERING

The microscopic theory of the light propagation in matter was developed as a consequence of Lorentz's non-relativistic, semi-classical electromagnetic theory. We will review and summarize this approach to photon scattering, which will not only prove useful for our analysis of the Fizeau effect, but will provide insight into the 'Fizeau-like' scattering of photons near large gravitational fields in EMQG theory.

To understand what happens in photon scattering inside a moving medium, imagine a simplified one-dimensional quantum model of the propagation of light in a refractive medium. The medium consisting of an idealized moving crystal of velocity 'V', composed of evenly spaced point-like atoms of spacing 'l'. When a photon traveling between atoms at a speed 'c' (vacuum light speed) encounters an atom, that atom absorbs it and another photon of the same wavelength is emitted after a time lag '$\tau$'. In the classical wave interpretation, the scattered photon is out of phase with the incident photon. We can thus consider the propagation of the photon through the crystal is a composite signal. As the photon propagates, part of the time it exists in the atom (technically, existing as an electron bound elastically to some atom), and part of the time as a photon propagating with the undisturbed light velocity 'c'. When it exists as a bound electron, the velocity is 'V'. From this, it can be shown (ref. 41, an exercise in algebra and geometry) that the velocity of the composite signal '$v_c$' (ignoring atom recoil, which is shown to be negligible) is:

$$v_c = c \ [1 + (V\tau/l) (1 - V/c)] / [1 + (c\tau/l) (1 - V/c)] \tag{16.321}$$

If we set V=0, then $v_c = c / (1 + c\tau/l) = c/n$. Therefore, $\tau/l = (n - 1)/c$. Inserting this in the above equations give:

$$v_c = [(c/n) + (1 - 1/n) V (1 - V/c)] /[1 - (1 - 1/n)(V/c)] \approx c/n + (1 - 1/n^2) \ V$$
(to first order in V/c). $\tag{16.322}$

Again, this is Fresnel's formula. Thus the simplified non-relativistic atomic model of the propagation of light through matter explains the Fresnel formula to the first order in V/c through the simple introduction of a scattering delay between photon absorption and subsequent re-emission. This analysis is based on a semi-classical approach. What does quantum theory say about this scattering process? The best theory we have to answer this question is QED.

## 16.33   QUANTUM FIELD THEORY OF PHOTON SCATTERING IN MATTER

The propagation of light through a transparent medium is a very difficult subject in quantum field theory (or QED). It is impossible to compute the interaction of a collection of atoms with light exactly. In fact, it is impossible to treat even one atom's interaction with light exactly. However, the interaction of a real atom with photons can be approximated by a simpler quantum system. Since in many cases only two atomic energy levels play a significant role in the interaction of the electromagnetic field with atoms, the atom can be represented by a quantum system with only two energy eigenstates. In the book "Optical Coherence and Quantum Optics" a thorough treatment of the absorption and emission of photons in two-level atoms is given (ref. 43, Chap. 15, pg. 762). When a photon is absorbed, and later a new photon of the same frequency is re-emitted by an electron bound to an atom, there exists a time delay before the photon re-emission. The probabilities for emission and absorption of a photon is given as a function of time $\Delta t$ for an atom frequency of $\omega_0$ and photon frequency of $\omega_l$ :

Probability of Photon Absorption is:   $K \ [ \sin (0.5(\omega_l - \omega_0) \Delta t) / ( 0.5(\omega_l - \omega_0)) ]^2$
Probability of Photon Emission  is:   $M \ [ \sin (0.5(\omega_l - \omega_0) \Delta t) / ( 0.5(\omega_l - \omega_0)) ]^2$

$$\text{(16.331)}$$

(where K and M are complex expressions defined in ref. 43)

The important point we want to make is that the probability of absorption or emission depends on the length of time $\Delta t$ (where the probability of the emission is zero, if the time $\Delta t = 0$). In other words, according to QED a *finite time* is required before re-emission of the photon. There are other factors that affect the probability, of course. For example, the closer the frequency of the photon matches the atomic frequency, the higher the probability of re-emission in some given time period. The question we want to address next is the effect of the virtual particles of the quantum vacuum on the propagation velocity of real (non-virtual) photons.

### 16.4 SCATTERING OF PHOTONS IN THE QUANTUM VACUUM

The above analysis can now be used to help us understand how photons travel through the virtual particles of the quantum vacuum. First we investigate the propagation of photons in the vacuum in far space, away from all gravitational fields. The virtual particles all have random velocities and move in random directions, and have random energies $\Delta E$ and life times $\Delta t$, which satisfies the uncertainty principle: $\Delta E\, \Delta t > h/(2\pi)$. Imagine a real photon propagating in a straight path through the virtual particles in a given direction. The real photon will encounter an equal number of virtual particles moving in a certain direction, as it does from the exact opposite direction. The end result is that the quantum vacuum particles do not contribute anything different than if all the virtual particles were at relative rest. Thus, we can consider the vacuum as some sort of stationary matter medium, with a very high density.

Is the progress of the real photon delayed as it travels through the quantum vacuum, where it encounters many electrically charged virtual particles? The answer to this question depends on whether there is a time delay between the absorption, and subsequent re-emission of the photon by a given virtual particle. Based on our arguments above, we postulate that the photon is delayed as it travels through the quantum vacuum (EMQG Postulate #4). The uncertainty principle definitely places a lower limit on this time delay. In other words, according to the uncertainty principle the time delay cannot be exactly equal to zero! Our examination of the physics literature has not revealed any previous work on the time delay analysis of photon propagation through the quantum vacuum, or any evidence to contradict our hypothesis of photon vacuum delay (presumably because of the precedent set by Einstein's postulate of light speed constancy).

We will take the position that the delays due to photon scattering through the quantum vacuum reduces the 'raw light velocity $c_r$' (defined as the photon velocity between vacuum particle scattering) to the average light velocity 'c' in the vacuum of 300,000 km/sec that we observe in actual experiments. Furthermore, we propose that the quantum vacuum introduces a vacuum index of refraction 'n' such that $c = c_r / n$. What is the raw light velocity? It is unknown at this time, but it must be significantly larger than 300,000 km/sec. The vacuum index of refraction 'n' must be very large because of the high density of virtual particles in the vacuum. What happens if the entire quantum vacuum is accelerated? How does the motion of a photon get affected? These questions turn out to have a deep connection to space-time curvature.

## 16.5   PHOTON SCATTERING IN THE ACCELERATED QUANTUM VACUUM

Anyone who believes in the existence of the virtual particles of the quantum vacuum (which carry mass), will acknowledge the existence of an accelerated state of virtual particles of the quantum vacuum near any large gravitational field. The graviton-masseon postulate states that gravitons from the real masseons on the earth exchange gravitons with the virtual masseons (both the virtual masseons and anti-masseons), causing a downward acceleration. The virtual particles of the quantum vacuum (now accelerated by a large mass) acts on light (and matter) in a similar manner as a stream of moving water acts on light in the Fizeau effect. How does this work mathematically? Again, it is impossible to compute the interaction of an accelerated collection of virtual particles of the quantum vacuum with light exactly. However, a simplified model can yield useful results. We will proceed using the semi-classical model proposed by Lorentz, above. We have defined the raw light velocity '$c_r$' (section 9) as the photon velocity in between virtual particle scattering. Recall that raw light velocity is the shifting of the photon information pattern by one cell at every clock cycle on the CA, so that in fundamental units it is an absolute constant (section 9). Again, we assume that the photon delay between absorption and subsequent re-emission by a virtual particle is '$\tau$', and the average distance between virtual particle scattering is 'l'. The scattered light velocity $v_c(t)$ is now a function of time, because we assume that it is constantly varying as it move downwards towards the surface in the same direction of the virtual particles. The virtual particles move according to: a =gt, where g = $GM/R^2$.

Therefore we can write the velocity of light after scattering (equation 16.321) with the accelerated quantum vacuum:

$$v_c(t) = c_r \ [1 + (gt\tau/l) \ (1 - gt/c_r)] / [1 + (c_r\tau/l) \ (1 - gt/c_r)] \qquad (16.51)$$

If we set the acceleration to zero, or gt = 0, then $v_c(t) = c_r \ / (1 + c_r\tau/l) = c_r/n$. Therefore, $\tau/l = (n – 1)/c_r$. Inserting this in the above equation gives:

$$v_c(t) = [(c_r/n) + (1 – 1/n) \ gt \ (1 - gt/c_r)] / [1 - (1 – 1/n)(gt/c_r)] \approx c_r/n + (1 – 1/n^2) \ gt$$
to first order in $gt/c_r$. $\qquad (16.52)$

Since the average distance between virtual charged particles is very small, the photons (which are always created at velocity $c_r$) spend most of the time existing as some virtual charged particle undergoing downward acceleration. Because the electrically charged virtual particles of the quantum vacuum are falling in their brief existence, the photon *effectively* takes on the same downward acceleration as the virtual vacuum particles (postualte #4). In other words, because the index of refraction of the quantum vacuum 'n' is so large, and $c = c_r/n$ we can write in equation 16.52:

$$v_c(t) = c_r/n + (1 - 1/n^2) gt = c + gt = c (1 + gt/c) \text{ if } n \gg 1. \tag{16.53}$$

Similarly, for photons going against the flow (upwards): $v_c(t) = c (1 - gt/c)$ (16.54)

We will see that this formula for the variation of light velocity near a large gravitational field leads to the correct amount of general relativistic space-time curvature (section 16.6).

Einstein, himself briefly considered the hypothesis of variable light velocity near gravitational fields shortly after releasing his paper on the deflection of light in gravitational fields (ref. 33), as can be seen in the quotation at the beginning of this section. It would be interesting to contemplate what Einstein might have concluded if he new about the existence of virtual particles undergoing downward acceleration near a massive object (or in accelerated frames). Since Einstein was aware of the work by Fizeau on the effect of light velocity by a moving media, he might have been able to explain the origin of space-time curvature at the quantum level.

Now let us imagine that two clocks that are identically constructed, and each calibrated with a highly stable monochromatic light source in the same reference frame. These clocks keep time by using a high-speed electronic divider circuit that divides the light output frequency by "n" such that an output pulse is produced every second. For example, the light frequency used in the clock is precisely calibrated to $10^{15}$ Hz; this light frequency is converted in to an electronic pulse train of the same frequency, where it is divided by $10^{15}$ to give an electronic pulse every second. Another counter in this clock increments every time a pulse is sent, thus displaying the total time elapsed in seconds on the clock display. Now, let us place these two clocks in a gravitational field on earth with one of them on the surface, and the other at a height "h" above the surface. The clocks are compared every second to see if they are still running in unison by exchanging light signals. As time progresses, the clocks loose synchronism, and the lower clock appears to run slower. According to general relativity, light always maintains a constant speed, and space-time curvature is responsible for the difference in the timing of the two clocks. Recalling the accelerated Fizeau-like quantum vacuum fluid, we can derive the same time dilation effect by assuming that the light velocity has exactly the same downward acceleration component of the background falling quantum vacuum virtual particles.

## 16.6 SPACE-TIME CURVATURE FROM SCATTERING THEORY

We are now in a position to formulate the EMQG equations for the time dilation near a large gravitational mass based on the Fizeau-like quantum vacuum fluid. We assume that light is moving upward from the surface of the earth. As the photon moves upward from point r to point r+Δr it decelerates at -1g according to equation 16.54:

$$c(r+\Delta r) = c(r)\ (1 - g\Delta t / c) \tag{16.61}$$

Since $\Delta t = \Delta r /c$ for small distances, we can now write:

$$c(r+\Delta r) = c(r)\ (1 - g\Delta r / c^2) \tag{16.62}$$

Since, $g = GM/r^2$ at point r above the center of the earth, we can write this as:

$$c(r+\Delta r) = c(r)\ (1 - GM\ \Delta r / r^2 c^2) \tag{16.63}$$

Since, the only observable property of light that we can be *sure* about is the red shift, as discussed in section 16.2, and $c = \nu \lambda$, it follows:

$$\nu(r+\Delta r) = \nu(r)\ (1 - GM\ \Delta r / r^2 c^2) \tag{16.64}$$

from which the wavelength appears longer by the same factor, or

$$\lambda(r+\Delta r) = \lambda(r)\ (1 + GM\ \Delta r / r^2 c^2) \tag{16.65}$$

To find the total change in frequency from point r to infinity, we integrate $GM\ \Delta r / r^2 c^2$:

$$\int_r^\infty GM / (r^2 c^2)\ dr = GM / (r c^2)\text{ and therefore,} \tag{16.66}$$

$$\nu(\infty) = \nu(r)\ (1 - GM / r c^2) \tag{16.67}$$

But, since $\nu = 1/t$ by definition, therefore time is affected as follows:

$$1 / t(\infty) = (1 / t(r))\ (1 - GM / (r c^2)) \tag{16.68}$$

Finally, we have:

$$t(r) = \left(1 - GM / (r c^2)\right) t(\infty) \tag{16.69}$$

which is the exactly the same expression for time dilation from the Schwarzchild metric.

Similarly, wavelength received at infinity is increased by the following expression:

$$\lambda(\infty) = \lambda(r)\ (1 + GM / r c^2) \tag{16.691}$$

Now, an observer at infinity can use the light signal from the surface of the earth to make measurements of distance in his reference frame at infinity. For example, suppose that in his own reference frame, a reference laser light source is used to measure a given reference length, and say that this corresponds to 1,000,000 wavelengths or $10^6 \lambda_r$, where $\lambda_r$ is the reference wavelength. Subsequently, he uses the light received from the surface of the earth from an identically constructed reference laser light source ($\lambda_r$) to measure the same length, and finds that when he counts the standard 1,000,000 wavelengths the reference length has shortened (because of the wavelength increase). In general he concludes that the distances at infinity $s(\infty)$ are contracted by the amount:

$$s(\infty) = s(r)\ (1 - GM / r c^2) \tag{16.692}$$

compared to distances $s(r)$ on the surface of the earth. Finally, we can write:

$$s(r) = \left(1 - GM / (r c^2)\right)^{-1} s(\infty) \tag{16.693}$$

which is the exactly the same expression for length that we found from the Schwarzchild metric. This equation specifies the amount of distortion for rulers on the earth "$s(r)$" compared to rulers positioned at infinity "$s(\infty)$". Thus by postulating that it is the light velocity that is actually varying (and not space-time curvature), we are led to the same amount of red shift, and the same amount of space-time curvature.

We can now see that in order to formulate a theory of gravity involving observers with measuring instruments (such as clocks and rulers) we must take into account how these measurements are affected by the local conditions of the quantum vacuum. Our analysis above shows that quantum vacuum can be viewed as a Fizeau-like fluid undergoing downward acceleration near a massive object, which affects the velocity of light. Indeed, not only is the velocity of light affected, it is *all* the particle exchange processes including graviton exchanges. Therefore, we find that the accelerated Fizeau-like 'quantum vacuum fluid' effects all forces. This has consequences for the behavior of clocks, which are constructed with matter and forces. After all, nobody questions the fact that a mechanical clock submerged in moving water cannot keep proper time with respect to an external clock. Similarly, a clock near a gravitational field (with a Fizeau-like, quantum vacuum fluid flow inside the clock) also cannot be expected to keep proper time with respect to an observer outside the gravitational field. The accelerated Fizeau-like 'quantum vacuum fluid' moves along radius vectors directed towards the center of the earth, and thus has a specific direction of action. Therefore, the associated space-time effects should also work along the radius vectors (and not parallel to the earth). This is precisely the nature of curved 4D space-time near the earth, as we will see (section 15.7).

For the case of light moving parallel to the earth's surface, the light path is the result of a tremendous number of photon to virtual particle scattering interactions (figure 10). Again in between virtual particle scattering, the light velocity is constant and 'straight'. The total path is curved as shown in figure 10. The path the light takes is called a geodesic in general relativity. In EMQG, this path simply represents the natural path that light takes through the accelerated vacuum. For the case of light moving parallel to the floor of the accelerated rocket (figure 11), the path for light is also the result of virtual particle scattering, but now the quantum vacuum is not in a state of relative acceleration.

Therefore, the path is straight for the observer outside the rocket. The observer inside the rocket sees a curved path simply because he is accelerating upwards.

We now see why Einstein's gravitational theory takes the form that it does. Because of the continuously varying frequency and wavelength of the light with height, Einstein interpreted this as a variation of space and time with height. We postulated that the scattering of light with the falling vacuum changes the light velocity in absolute CA units, which cause the *measurements of space and time* to be affected. As we have already seen, these two alternative explanations **cannot** be distinguished by direct experimentation. This is why the principle of the constancy of light velocity is still a postulate in general relativity (through the acceptance of special relativity).

We are now in a position to understand the concept of the geodesic proposed by Einstein. ***The downward acceleration of the virtual electrically charged masseons of the quantum vacuum serves as an effective 'electromagnetic guide' for the motion of light (and for test masses) through space and time***. This 'electromagnetic guide' concept replaces the 4D space-time geodesics that guide matter in motion in relativity. For light, this guiding action is through the electromagnetic scattering process of section 16.4. For matter, the electrically charged virtual particles guide the particles of a mass by the electromagnetic force interaction that results from the relative acceleration. Because the quantum vacuum virtual particle density is quite high, but not infinite (at least about $10^{90}$ particles/$m^3$), the quantum vacuum acts as a very effective reservoir of energy to guide the motion of light or matter.

The *relative nature* of 4D space-time can now be easily seen. **Whenever the background virtual particles of the quantum vacuum are in a state of relative acceleration with respect to an observer, the observer lives in curved 4D space-time.** Why should the reader accept this new approach, when both approaches give the same result? The reason for accepting EMQG is that the action between a large mass and 4D space-time curvature becomes quite clear. The reason that 4D space-time is curved in an accelerated reference is also clear, and very much related to the gravitational case.

The relative nature of curved 4D space-time also becomes very obvious. An observer inside a gravitational field would normally live in a curved 4D space-time. If he decides to free-fall, he cancels his relative acceleration with respect to the quantum vacuum, and 4D space-time is restored to flat 4D space-time for the observer. The principle of general covariance no longer becomes a principle, but merely results for the deep connection between the quantum vacuum state for accelerated frames and gravitational frames. Last, but not least, the principle of equivalence is completely understood as a reversal of the (net statistical) relative acceleration vectors of the charged virtual masseons of the quantum vacuum, and real masseons that make up a test mass. We have seen EMQG at work for spherically symmetrical and non-rotating masses. What about the nature of the virtual particle acceleration field around an arbitrary mass distribution in any state of motion?

## 17. THE EMQG GRAVITATIONAL FIELD EQUATIONS

By treating the quantum vacuum as a continuous fluid surrounding an absolute mass-density ρ, we can formulate classical EMQG equations of motion for an arbitrary mass distribution. Here we derive the EMQG gravitational equations of the motion of the virtual masseon particles of the quantum vacuum near an arbitrarily shaped large mass with absolute mass density ρ. We further assume that there is a large enough mass involved which will significantly disturb the nearby quantum vacuum particles, and that the graviton flux is not so high as to disrupt the principle of equivalence. If the graviton flux is extremely high, it can compete with the normal electromagnetic forces in the vacuum and disturb equivalence. Furthermore, since the density of the virtual particles in the quantum vacuum is so high (at least $10^{90}$ particles per cubic meter), the variation of the virtual particle acceleration from point to point in space can be considered as a classical continuous field. Therefore, the methods of vector calculus can be used with the assumption that CA space and separate CA time form a perfect continuum, and that the acceleration of the virtual particles from place to place is a mathematical vector field. The EMQG field equations are formulated in absolute CA space and time units, and thus not directly observable (space-time effects mask these results).

We will start by reviewing the classical equations of gravitation as given by Newton and Poisson and see how these relate to EMQG theory. Central to development of the EMQG equations of virtual particle motion is the ***concept of forces as particle exchanges***.

### 17.1 THE CLASSICAL NEWTONIAN GRAVITATIONAL FIELD

*'... and the Newtonian scheme was based on a set of assumptions, so few and so simple, developed through so clear and so enticing a line of mathematics that conservatives could scarcely find the heart and courage to fight it.'*
*- Isaac Asimov*

Here is a brief review of the classical laws of gravitation based on the classical concept of a force field. The field concept can be traced to Newton's instantaneous law of gravity for two point masses repeated for the large collection of particles in the mass.

Newton's law of gravitation:  $F = G M_1 M / r^2$ (17.1)

which is the mathematical form of Newton's gravitational law for the force F between two point masses $M_1$ at (x,y,z) and M at (x', y',z') directed along the line between the two points, r is the distance between the two particles; $r = [(x-x')^2 + (y-y')^2 + (z-z')^2]^{1/2}$, and G is Newton's gravitational constant $G=6.673\times10^{-11}$ m$^3$ kg$^{-1}$ s$^{-2}$. This can be stated in a concise form as follows:

**"*Every particle in the universe attracts every other particle with a force which is directly proportional to the product of the two masses and inversely proportional to the*** *square of the distance between them; the direction of the force being in the line joining the two mass particles."* **This force acts instantaneously**.

According to the particle exchange paradigm, this law works the same way as Feynman's photon exchange process in Coulomb's law of electrical attraction. Newton's inverse square law is a result of the geometry of the graviton exchange process. The exchange particle flux density spreads out on the surface of a sphere (area = $4\pi r^2$), and the flux is directly proportional to the product of the magnitudes of the masses. Stated in terms of masseons, the product of the number of masseon particles contained in each of the point masses determines the number of gravitons exchanged. Furthermore, since gravitons move at the speed of light, there is a delay in transmitting the force of gravity that was overlooked by Newton, which will be closely examined later. The graviton particles do not interact with each other, just like the photons do not interact with other photons (through some force exchange process).

If we now let $M_1$ be a test particle of unit mass, then dividing the force of gravity by $M_1$ provides the *gravitational attraction* **g** produced by a mass $M_1$ at the location of the test particle P(x,y,z) along the unit vector **r**:

$$\mathbf{g}(x,y,z) = -G M \mathbf{r} / r^2 \tag{17.2}$$

where **r** is a unit vector directed from the mass M to the test mass at point P(x,y,z). Because **g** has units of force divided by mass (or acceleration), it is sometimes called the *gravitational acceleration*. Any particle of any mass value at point P(x,y,z) will have the same acceleration due to the principle of equivalence. Therefore, **g** represents the average acceleration vector of the virtual masseon particles in the quantum vacuum at that point. It is known that the gravitational attraction g is an irrotational classic field because $\nabla \times \mathbf{g} = 0$. From the Helmholtz theorem (ref. 44), gravitational attraction is a conservative field and can be represented as the gradient of the Newtonian scalar potential field $\phi(\mathbf{x},t)$:

$$\mathbf{g} = \nabla \phi, \text{ where } \phi = G M / r \tag{17.3}$$

Newton's law treats gravitational forces between particles as vectors. When this law is generalized to a large number of particles interacting, the concept of a mass distribution as a collection of a large number of particles emerges. But for this concept to work properly, the gravitational potential must obey the ***principle of superposition***:

**The gravitational potential of a collection of masses is the sum of the gravitational attractions of the individual masses. The net force on a test particle is the vector sum of the forces due to all the mass particles in space.**

The principle of superposition gives us one of the most important properties of the graviton particles (postulate #2): *graviton particles do not exhibit force interactions with other graviton particles*. Thus the total graviton interaction is the vector sum of the individual graviton interactions. This works the same way as the superposition principle works for photons in QED. The principle of superposition can be applied to find the resultant gravitational attraction as the limit is taken towards a continuous distribution of matter. A continuous distribution of matter with mass m is defined as a collection of a great many very small masses dm = ρ(x,y,z) dv, where ρ(x,y,z) is defined as the mass-density of the distribution, and dv is the change in volume.

We follow the work of Bernard F. Schutz (ref. 45) for the Newtonian gravitational potential. The force on a unit test mass at the coordinate (x,y,z) is the vector sum of an infinite and continuous distribution of particles in the mass. The concept of a vector force field can be thought of as the force that will be applied to a unit mass at point (x,y,z). This force is usually written in the form of the gravitational potential $\phi(\mathbf{x},t)$ and this can calculated by solving Poisson's equation for the mass distribution:

**Poisson's Equation:** $\nabla^2 \phi = 4\pi G\, \rho(x,y,z,t)$ (17.4)

where $\phi$ is the Newtonian gravitational potential field, ρ is the mass density function of the source mass which is a function of position (x,y,z) and of time t (the mass distribution can be in motion). In Newtonian physics, the gravitational potential $\phi$ follows the variation of the mass distribution ρ instantaneously. For example, when the earth orbits the sun, the gravitational potential of the earth follows the earth exactly, no matter what the speed of the orbit.

Poisson's equation has the solution for the Newtonian potential $\phi_N(\mathbf{x},t)$ given by:

**Newtonian Potential:** $\phi_N(\mathbf{x},t) = -G \int \rho(\mathbf{y},t)\, r^{-1}\, d^3y$ , where $r \equiv |\mathbf{x} - \mathbf{y}|$ (17.5)

This can be thought of as a superposition of the 1/r potential fields of each of the mass elements given by mass $\rho\, d^3y$ at position $(\mathbf{y},t)$. The vector **r** represents the distance between the unit test mass at (x,y,z) and the mass element $\rho(\mathbf{y},t)\, d^3y$. These equations are formulated with Newton's version of absolute space and time, which we will discuss later.

17.2  EMQG GRAVITATIONAL FIELD EQUATION FOR A SLOW MASS

Based on the above arguments, we want to formulate the equations of motion of the virtual particles of the quantum vacuum near a slow mass. The particle exchange process responsible exists in absolute 3D space and time on the cellular automata level. We can approximate the discrete nature of CA 3D space by a 3D-space continuum based on the simple 3D Cartesian coordinate system. Thus, Newton's absolute space and time is replaced by cellular automata units of cells and 'clock' cycles. The origin (0,0,0) is arbitrary and usually taken at the center of mass (which may be moving with respect to the absolute CA cell space).

In EMQG, the gravitational potential is of secondary importance. What we really need to know to calculate anything is the net statistical average acceleration of a virtual particle at point (x,y,z), and the direction of acceleration at that point in space (which is determined by the absolute mass distribution). This acceleration is independent of the mass of a virtual particle, by virtue of the microscopic principle of equivalence in section 15.10. Hence, we can formulate a classical vector acceleration field '**a**', which would represent the acceleration that any (small) mass placed at any point would achieve. Here we assume that the velocity of the mass distribution is much less than the speed of the graviton particles (v << c), so that variations of mass density $\rho(\mathbf{y},t)$ with time causes the Newtonian potential $\phi_N(\mathbf{x},t)$ to follow it instantaneously. We will derive the acceleration vector '**a**' from the gravitational potential $\phi_N(\mathbf{x},t)$ at a point (x,y,z).

Fortunately, it is easy to convert from Newtonian gravitational potential $\phi_N(\mathbf{x},t)$ at a point (x,y,z) to the acceleration '**a**' of a typical virtual particle at that same point. First, the vector force **F** components are calculated from the potential vector at point (x,y,z) as follows:

$\mathbf{F} = \nabla\phi$, from which: $F_x = \partial\phi / \partial x$,  $F_y = \partial\phi / \partial y$, $F_z = \partial\phi / \partial z$ (17.21)

The force vector **F** has the above (x,y,z) vector force components for the gravitational attraction which is originating from the mass distribution $\rho(\mathbf{y},t)$. This force is what acts on a unit of test mass (m=1) at the point (x,y,z). Thus, the acceleration vector of a typical virtual particle in the quantum vacuum at point (x,y,z) is also given by the same '**a**' by virtue of the principle of equivalence, i.e. all masses fall at the same rate at a given point. Therefore, the (x,y,z) components of the acceleration vector of a typical virtual particle located at point (x,y,z) due to the graviton exchange with the mass distribution ρ is given by:

**THE EMQG EQUATIONS FOR THE VIRTUAL PARTICLE ACCELERATION FIELD AT A POINT (X,Y,Z) FOR A SLOW MOVING MASS DISTRIBUTION:**

We have: $\mathbf{a} = \nabla\phi$, or $A_x = \partial\phi_N / \partial x$,  $A_y = \partial\phi_N / \partial y$,  $A_z = \partial\phi_N / \partial z$ (17.22)

where, $\phi_N(\mathbf{x},t) = - G \int \rho(\mathbf{y},t) \, r^{-1} \, d^3y$ , $r \equiv |\mathbf{x} - \mathbf{y}|$ (17.23)

We are still using Newton's version of absolute distance and time, which are equivalent to cellular automata space and time units. We are also using Newton's absolute mass or mass-density. The origin (0,0,0) is taken at the center of mass, which may be moving with respect to the cells in CA space via the simple Galilean coordinate transformation. This formulation acknowledges the absolute nature of CA space and time, which is observer independent. Because of the 3D geometric cellular automata connectivity, the cell structure is best represented in Cartesian coordinate system as is done here.

17.3    THE EMQG FIELD EQUATIONS FOR A HIGH SPEED MASS

Here we will determine the equations for the acceleration of the virtual particles and their direction at each point in space, as determined by a high speed (v approaches c) mass distribution. Here we assume that the velocity of the mass distribution can be comparable to the speed of the graviton particles, so that there will be a delay or retardation between the variations of the mass density $\rho(\mathbf{y},t)$ with time, and thus the corresponding Newtonian potential $\phi_N(\mathbf{x},t)$. The delay is due to the velocity of the graviton particles, which moves at the speed of light. The gravitons propagate from the mass distribution to a unit test mass at point (x,y,z) which are also occupied by a dense collection of virtual particles. The virtual particles in turn, are responsible for the subsequent force of gravity on that unit test mass by means of the electromagnetic force on a test mass, as discussed previously. This retardation can be easily introduced in the Newtonian potential function $\phi_N(\mathbf{x},t)$ as follows:

$$\phi_R(\mathbf{x},t) = - G \int \rho(\mathbf{y},t - r/c) \, r^{-1} \, d^3y \tag{17.31}$$

Here, a change in $\rho$ at $\mathbf{y}$ ought to be felt at $\mathbf{x}$ only after a time $|\mathbf{x} - \mathbf{y}| / c$, the propagation delay of the graviton particles. This leads to a modified high speed potential field $\phi_R(\mathbf{x},t)$. Again, this can be thought of as a superposition of the 1/r potential fields of each mass element given by mass $\rho \, d^3y$ at position ($\mathbf{y}$,t). This superposition is justified because the graviton flux satisfies linear superposition. As before, the vector $\mathbf{r}$ represents the distance between the unit test mass at (x,y,z) and the mass element $\rho(\mathbf{y},t) \, d^3y$.

It can be shown that $\phi_R$ satisfies the following equation:

$$\nabla^2 \phi_R - (1/c^2) \, \partial^2 \phi_R / \partial t^2 = 4\pi G \, \rho(x,y,z,t) \tag{17.32}$$

The solution to equation 17.32 is the gravitational potential $\phi_R(\mathbf{x},t)$ at a point (x,y,z). Again, the virtual particle acceleration vector $\mathbf{a}$ at that point is the derivative of the potential. Therefore the (x,y,z) components of the acceleration vector of a typical virtual particle located at point (x,y,z) due to the graviton exchange with the mass distribution $\rho$ is given by:

**THE EMQG EQUATIONS FOR THE VIRTUAL PARTICLE ACCELERATION FIELD AT A POINT (X,Y,Z) FOR A HIGH SPEED MASS DISTRIBUTION:**

We have, $\mathbf{a} = \nabla \phi_R$, or $A_x = \partial \phi_R / \partial x$, $A_y = \partial \phi_R / \partial y$, $A_z = \partial \phi_R / \partial z$ $\tag{17.33}$

where $\phi_R(\mathbf{x},t) = - G \int \rho(\mathbf{y},t - r/c) \, r^{-1} \, d^3y$ $\tag{17.34}$

Again, we must emphasize that these equations are <u>not</u> formulated in the Minkowski four-dimensional curved space-time formalism of general relativity, which is derived for an arbitrary observer with his space-time measuring instruments and chosen coordinate system. Instead, these equations are based on *absolute* cellular automata space and time units. The x, y, and z distances are measured as a count of the number of cells occupying a given length in space, and the time t as a count of the number of 'clock' cycles that has elapsed between two events on the CA. The origin (0,0,0) is usually the center of the mass distribution. These equations approximately represent the inner workings of the gravity on the cellular automata as seen by the cells themselves, and are independent of a physical observer and his instruments. Because of this, these equations cannot be verified directly by experiment.

Therefore, these equations are not generally covariant because they are not formulated for an arbitrary observer in any reference frame with measuring instruments made from matter. They are formulated in the specific coordinate system of the CA cell space, and are all written in vector form (not as tensors). The mass-density $\rho$ is also treated as the absolute mass-density, which is independent of the observer. The source of the Newtonian gravitational field is the mass-density which is given by $\rho$ = mass/volume. According to relativity, mass and volume are observer dependent, i.e. they vary with the state of motion of the observer. For example, if an observer is moving at relativistic speeds with respect to the mass distribution, the mass varies according to $m = m_0 \, (1-v^2/c^2)^{-1/2}$, where v is the relative velocity. Similarly, the volume is Lorentz contracted, which is also an observer dependent entity. Thus, the mass-density varies from observer to observer. In EMQG, there exists absolute space, absolute time, and an absolute mass distribution (or collection of masseon particles) that occupies a definite number of cells. An event takes a definite number of 'clock' cycles. EMQG is not formulated for an arbitrary observer. How would one formulate these same laws of gravity from the perspective of an arbitrary observer in an arbitrary state of motion, who chooses an arbitrary coordinate system for his measurements? It turns out that Einstein has already accomplished this task beautifully in his gravitational field equations.

## 17.4  EINSTEIN'S GENERAL COVARIANT GRAVITATIONAL EQUATION

Einstein's goal was to propose a theory of gravity while retaining the basic postulates of special relativity in regards to the speed of light being a universal constant. Simple considerations of the motion of light inside an accelerated rocket and the principle of equivalence led him to *postulate* the curvature of four-dimensional Minkowski space-time. This postulate allows him to retain light as an absolute constant (light still moves perfectly 'straight'), while proposing that the space-time that light moves in is curved. This 4D space-time curvature guides light to move along the curved geodesic paths of the background space-time, which are the 'straight lines' of Riemann geometry. This idea, along with the principle of general covariance, led Einstein to formulate his theory of gravitation in the form of quasi-Riemannian geometry. Einstein started with Poisson's equation, just as we have done in EMQG. Through the use of tensors and the principle of equivalence, Einstein was able to formulate 4D space-time curvature in a form that was generally covariant. This allowed observers to *switch* between accelerated and gravitational frames at will, or to switch between different coordinate systems, and yet have the same general form for the gravitational equations.

The postulates of general relativity, along with the special theory of relativity and the mathematics of Riemann Geometry lead to the famous Einstein gravitational field equation:

$$G_{\alpha\beta} = (8\pi G / c^2) \, T_{\alpha\beta} \quad \text{.... Einstein's Tensor Gravitational Field Equations} \quad (17.41)$$

$G_{\alpha\beta}$ is the Einstein tensor, which is the mathematical statement of space-time curvature, that is reference frame independent, and generally covariant. $T_{\alpha\beta}$ is the stress-energy tensor, which is the mathematical statement of the special relativistic mass-energy density (and observer dependent), and G is Newton's gravitational constant, and c the velocity of light. The constant G here still reflects the Newtonian aspects of general relativity. The constant $8\pi G/c^2$ is chosen to adjust the strength of coupling between matter and 4D space-time, so that it corresponds to the correct amount of Newtonian gravitational force.

Thus, Einstein has accomplished an *observer* dependent formulation of gravity. In contrast, EMQG formulates the law of gravity that is **not** observer dependent. Instead, it is based on the only truly important units of measurement: the cellular automata absolute units of space and time, which is not directly accessible to measurement. Einstein concluded that space-time is four-dimensional, and curved in order to be compatible with a constant light velocity in **all** frames. Einstein observed that when light moves parallel to the surface of the earth, light will curve (also for an accelerated observer). Either he abandons the postulate of constancy of light velocity, and allows light move in curved paths, or he must have space and time curved somehow, by the unknown action of the nearby mass distribution, in which case light velocity can be constant and follow a geodesic path. Einstein chose the later approach, but could never find the physical action that causes matter to curve 4D space-time.

## 18. EINSTEIN'S SPECIAL RELATIVITY REVISITED

After presenting EMQG theory in full, we need to briefly revisit special relativity in order to clarify a few points. First, we review some of the important results of EMQG theory that affect the results of special relativity theory. The light velocity (absolute units) near a large gravitational field is modified by 'Fizeau-like' scattering process. This process changes the light velocity (in absolute units) over classical distance scales near a gravitational field, when compared to the measured light velocity over the same distance in deep space (in absolute units). However, we maintain that the 'low-level' or absolute light velocity, defined as the velocity between the quantum vacuum scattering events in absolute units, remains unchanged. In fact, low-level light velocity is an absolute constant (and is equal to 1 pvu, which is 1 pdu divided by 1 ptu, section 9), and is a characteristic of the simplest 'motion' on the CA. We also proposed that the *measured* light velocity, by an observer in the same reference frame as the gravitational field with ordinary measuring instruments remains unaffected compared to a distant observer. In other words, the measured light velocity is constant. This is because of the associated space-time effects that have to do with the accelerated 'Fizeau-fluid' and it's effect on measurements with 'variable' light velocity (section 16.4). We also concluded that when acceleration (or gravity) is involved, mass can be defined as being *absolute mass*. In other words, inertial mass and gravitational mass can be treated as being absolute, contrary to conventional special relativity theory. In these cases, the absolute reference frame for the *accelerated* motion of the mass is with respect to the net (statistical) average acceleration vector of the virtual particles of the quantum vacuum near the mass (not the CA absolute space and time). In other words, we are able to measure our absolute acceleration **without** reference to *any* other observer. However, we still maintained that there are **no** *measurable* absolute reference frames to gauge constant velocity motion (including the virtual particles of the quantum vacuum and including absolute plank units), which is in accordance with the principles of special relativity.

Let us address the problem of the constancy of the speed of light in detail. It is *quite clear* experimentally that the measured light velocity is constant in every circumstance, no matter what the state of motion. The light constancy principle is applicable to constant velocity observers, and according to general relativity, for accelerated and gravitational observers as well. The light constancy principle has a high price to bear, however. One must give up the idea of absolute time and space and accept relativistic, curved, 4D space-time in gravitational frames and in accelerated frames. EMQG theory is based on a different approach: *the 'Fizeau-like' scattering of photons with the virtual particles of the quantum vacuum, which alters the low-level light velocity under the influence of a gravitational field*. However, this scattering of photons still occurs in special relativity, where light propagates through the virtual particles of the quantum vacuum (away from all gravitational and accelerated frames). Does this scattering change the light velocity, even when light propagates in a vacuum? The answer is yes, but with a qualifier. The light velocity between scattering operations still remains an absolute constant.

Photons, which travel at one velocity in absolute CA space, travel at an absolute constant rate (in absolute units). However, as the photons travel they interact with virtual particles of the quantum vacuum along the way. This process is known as scattering, and is an integral part of classical optics and QED theory. Scattering introduces delays, which is responsible for the index of refraction in classical optics and the subsequent change in light velocity when light propagates from air to glass. The virtual particles also scatter light, since they are charged (virtual masseons). Therefore, the virtual particles contribute to the reduction of light velocity over large distance scales, with the same inter-particle delay process found in classical optics for glass affecting light.

Therefore, we conclude that the speed of light measured in a universe where the quantum vacuum is *absent* of all virtual particles would have to be **greater** than the measured light velocity in our own universe. This is because in our universe, the real photons scatter off the virtual particles, where each virtual particle introduces a **small random time delay** before another real photon is re-emitted (Note: in addition to this, photons participate in many second-order, and higher scattering processes, in accordance with QED theory).

Thus, the velocity of light that we observe in our universe is kind of a net average (statistical) value derived from the raw 'low-level' photon velocity **minus** a velocity penalty due to the total average delay in the vacuum scattering process. Over classical distance scales, the average light velocity is constant and is highly repeatable, due to the remarkable regularity of the quantum vacuum from place to place. It is also important to note that in a quantum vacuum scattering process, the photon that is remitted at a QED particle vertex (for example a photon emitted from an electron) always leaves at fixed constant velocity (in absolute CA space and time units). This is true no matter what the state of the motion of the vertex particle, whether it has constant velocity or accelerated motion. However, for accelerated motion, all virtual particles are statistically accelerating in the same direction and with the same average value. So for accelerated motion, the tiny random delays involved in the absorption/re-emission of the photon from a virtual particle become cumulative and add up as a change in the low-level light velocity. In between virtual particle scattering, we emphasize that **the velocity of light is an absolute constant**!

On quantum distance scales Minkowski 4D space-time gives way to the secondary (quantized) absolute 3D space and separate absolute (quantized) time required by CA theory. The curved, 4D space-time paradigm is now replaced by the pure particle interaction process. Here particles occupy definite locations on the CA cell, and the numeric state evolves by the universal CA 'clock'. All interactions are *absolute,* because they depend on absolute CA processes. However, we cannot probe this scale because we are unable to access the absolute cell locations, and numeric contents of the cells. In this realm, the photon particle (as well as the graviton) is an information pattern that shifts from cell to adjacent cell with an absolute constant 'velocity'. In fact, this is the *simplest* type of 'motion' possible on the universal CA.

Another thing has to be clarified is the maximum velocity that matter can obtain in the vacuum under the influence of gravitational fields (or accelerated frames). We have seen that in inertial frames, the force $F = F_0 (1 - v^2/c^2)^{1/2}$ tends to zero as the velocity v -> c. This means that any force becomes totally ineffective as the mass is accelerated to light velocity with respect to the source of the force. We have seen that force is due to exchanged particles or bosons, which become totally ineffective in propagating from the source to the receiver, as the velocity of the receiver with respect to the source approaches the velocity of the force exchange particle. However, inside gravitational fields the measured light velocity 'c' is equivalent to the scattered light velocity for the particular gravitational field (or accelerated frame). For example, on the earth the scattered light velocity (in absolute units) is higher when light propagates downward and lower when moving upward. In both cases, the maximum velocity that matter can achieve is the same as the speed of the force exchange particles. The force exchange particle scatters in the same fashion as does a real photon. Therefore, the limiting speed for matter becomes the same as the limiting speed for the propagation of the photon and graviton force exchange particles.

## 19. APPLICATIONS OF EMQG THEORY

In this section we will look at some preliminary applications of EMQG theory in understanding various gravitational phenomena. In the applications we will be looking at, we will see simple and intuitive visual models that illustrate how EMQG theory is applied to real gravitational physics problems. These applications were chosen because we found that simple models were lacking in conventional general relativistic treatment of these subjects. At this point in the development of EMQG, many of the mathematical details are still lacking in these applications.

### 19.1 GRAVITATIONAL WAVES

Although there are a lot of similarities between QED and EMQG theory, we found that Gravitational Wave (GW) physics is significantly different than Electromagnetic Wave (EM) physics. For a periodically accelerating large mass (or masses), the fluctuating graviton flux is responsible for the initial periodic disturbance in the net statistical average acceleration of the virtual particles of the quantum vacuum in the immediate vicinity (with respect to the original mass center). For example, a close pair of relativistic neutron stars undergoing orbital acceleration would periodically disturb the state of the virtual particles of the quantum vacuum, through a periodically fluctuating graviton flux rate. The time-varying graviton flux is caused by the time-varying distances between the neutron stars due to their mutual acceleration around the center of mass.

Thus, the outward propagating GW is really a time-varying periodic increase and decreases in the net acceleration of the virtual particle acceleration vectors (at a given point) with respect to the neutron stars, that was started by the time varying graviton flux. Once the GW is initiated, however, it is self-sustaining as is propagates throughout space. Based on the principles of EMQG, we can easily see that it is the strong electromagnetic component contained in the gravitational waves that primarily acts upon other test masses (composed of real electrically charged masseon particles) placed in the path of the GW. As we found in the equivalence principle, the direct graviton exchanges between the source and the test mass are negligible in the total GW interaction. The GW periodic excitation propagates primarily through electromagnetic means, because the electrically charged virtual masseon particles in the quantum vacuum are constantly redistributing their net acceleration vector imbalance amongst themselves. This happens by electromagnetic force interactions because some virtual masseon particle regions have higher acceleration than their surroundings, creating an electromagnetic imbalance which tends to spread to stabilize the vacuum through photon exchanges. This wave-like electromagnetic disturbance spreads outwards in all directions, and is called a gravitational wave.

According to EMQG theory, whenever we have a disturbance of the state of relative acceleration of the virtual particles of the quantum vacuum, we have curved 4D space-time (section 16.6). In other words, the GW wave affects the nature of 4D space-time along its path of propagation. Thus, the common visualization of a GW as 'rippling space-time' is now interpreted as a periodic disturbance of the net average acceleration vectors of the virtual particles of the quantum vacuum that spreads out like a wave. The periodic acceleration vector disturbances acts on both the motion of light (and on matter) to distort 4D space-time. The periodically fluctuating, acceleration vectors of the virtual particles of the quantum vacuum cause a *'periodically fluctuating absolute light velocity'* in the path of the GW, through the Fizeau-like scattering process discussed in section 16.6. This is interpreted as a periodically fluctuating, 4D space-time curvature.

In summary, gravitational waves cause periodically varying 4D space-time curvature by the following three-step process:

- First, the rapidly accelerating periodic mass movement causes the graviton flux from the source mass to periodically fluctuate at a given point.
- Secondly, the resulting periodically fluctuating, acceleration vectors of the virtual masseon particles radiates outwards in all directions, and this is what we call a GW.
- Finally, the periodically fluctuating acceleration vectors of the virtual masseon particles of the quantum vacuum acts on any light or on matter, and thus induces a corresponding periodically fluctuating 4D space-time curvature spreading outward. The periodically fluctuating 4D space-time curvature can be detected over great distances.

It is known in the context of general relativity theory that the GW has a very large stiffness, in analogy with Hooke's law (ref. 13). Thus, the GW carries a large amount of energy density. This fact follows directly from Einstein's Gravitational Field equations. This simple fact is easily explained in EMQG as arising from the very large energy density of the quantum vacuum particle disturbance itself. The virtual particles of the quantum vacuum have a large particle density that is much greater than $10^{90}$ particles per cubic meter. Because of this, the fluctuating quantum vacuum disturbance (GW) carries a large energy density in the form of the huge numbers of virtual particles involved, and this is quite capable of explaining the stiffness of the GW. In fact, this periodic fluctuating vacuum disturbance (GW) is quite capable of vibrating a large solid aluminum cylinder after traveling for hundreds of light years! The amplitude is much reduced, however, because of the spreading of the GW over enormous distances. It is easy to see that the GW carries energy, because the GW is capable of doing work on distant test masses through the ability to exert inertial force.

Using the high-speed EMQG field equations, we can get a deeper understanding of GW physics. Recall the expression for the Newtonian potential $\phi_R(\mathbf{x},t)$ for fast moving mass in absolute CA units is given by:

$$\nabla^2 \phi_R - (1/c^2) \partial^2 \phi_R / \partial t^2 = 4\pi G \rho(x,y,z,t) \qquad (19.1)$$

The solution to equation 19.10 is the gravitational potential $\phi_R(\mathbf{x},t)$ at a point (x,y,z). The virtual particle acceleration vector **a** at that point is the derivative of the potential. Hence the (x,y,z) components of the acceleration vector of a typical virtual particle located at point (x,y,z) due to the graviton exchange with the mass distribution ρ is given by:

$$\mathbf{a} = \nabla \phi_R, \text{ or } A_x = \partial \phi_R / \partial x, \quad A_y = \partial \phi_R / \partial y, \quad A_z = \partial \phi_R / \partial z \tag{19.2}$$

where $\phi_R(\mathbf{x},t) = - G \int \rho(\mathbf{y}, t - r/c) \, r^{-1} \, d^3y$ (19.3)

Again we emphasize that the acceleration components of the virtual particles based on the absolute CA space and time units, and the solution to this is given by:

The gravitons propagate from the mass distribution to a unit test mass located at point **x** at (x,y,z), which are also occupied by a dense collection of virtual particles. The virtual particles in turn, are responsible for the subsequent force (of measurable gravity) on a unit test mass, by means of the electromagnetic force as discussed previously.

The partial differential equation for the potential energy resembles the familiar classical scalar wave equation. We have already seen (section. 17) that the second term of the potential is the retardation effect due to the finite propagation delay of the graviton exchange particles (light velocity). The gravitons originate from the mass concentration at **y**, which has an **absolute** mass (technically, 'mass charge') given by $\rho \, d^3y$, where ρ is the absolute mass density. The resulting acceleration of the virtual particles located at point **x** is given by $\mathbf{a} = \nabla \phi_R$. The gravitons originating from the accelerated mass at **y** and absorbed by a typical virtual particle located at **x**, is responsible for the initial gravity wave disturbance.

To see how this happens, let us take the spatial gradient of $\phi_R$ (following Schultz, ref. 45):

$$\nabla \phi_R(\mathbf{x},t) = - G \int [\rho/r - (1/c)(\partial \phi / \partial t)] ((\mathbf{x}-\mathbf{y})/r^2) \, d^3y \tag{19.4}$$

The region of integration is limited to the bounded region where $\rho \neq 0$. If we consider $\nabla \phi_R(\mathbf{x},t)$ far from the source, where $r \approx |\mathbf{x}|$, and r is so large that the first term of the above equation is negligible compared with the second. Then we have:

$\mathbf{n} \nabla \phi_R \approx - (1/c) \, \partial \phi_R / \partial t$ where $\mathbf{n} = \mathbf{x} / |\mathbf{x}|$, the unit radial vector from the origin, (19.5)

inside the source, to the virtual particle field point **x**. According to this equation, we are far from the source, and the typical length scale λ on which $\phi_R$ changes is c times the typical time scale p on which $\phi_R$ (and the source ) changes as $(\partial \phi_R / \partial t) \approx \phi_R / p$. This is characteristic of a wave traveling at velocity c. The potential $\phi_R$ has a spatial dependence which 'forgets' the distance |x| to the source, and is sensitive only to $\partial \phi_R / \partial t$. The net average acceleration of the virtual particles of the quantum vacuum at a point far from the source, is given by: $\mathbf{a} = \nabla \phi_R$. Therefore the acceleration of the virtual particles satisfies:

$\mathbf{n} \, \mathbf{a} \approx - (1/c) \, \partial \phi_R / \partial t$, where $\mathbf{n} = \mathbf{x} / |\mathbf{x}|$, the unit radial vector from the origin (19.6)

This is what is called a gravitational wave. From this equation, we can see that the net acceleration of the virtual particles of the quantum vacuum fluctuates periodically with the rate of change of the Newtonian potential. This disturbance spreads outwards, and is capable of doing work as it passes through a test mass in the way.

## 19.2 EMQG AND THE GRAVITOMAGNETIC FIELD

In EMQG, the graviton particle and the photon particle are very closely related (postulate #2 gives the properties of the graviton). Since the graviton is ultimately responsible for the gravitational field, should we not expect to find a new force field in gravitational physics that is analogous to the magnetic field of electromagnetism for moving masses? The answer turns out to be yes. In electromagnetism, the photon exchange process generates a magnetic field perpendicular to the direction of the moving electrical charge. Similarly, a moving mass can produce an analogous 'gravitomagnetic' force field, also directed perpendicular to the motion of the moving mass. In fact, there exists a whole gambit of equivalent electromagnetic phenomena related to magnetism that occurs with gravitational interactions such as: gravito-Magnetic fields in general (ref. 46), gravito-Induction (ref. 48), gravito-Meissner effect (ref. 47), and gravitoLorentz force (ref. 49). The gravitomagnetic fields are also a consequence of Einstein's field equations, because of the close parallels between his gravitational field equation and Poisson's equation (section 17.4). The classical textbook on gravitation by Misner, Thorne, and Wheeler (ref. 50) and reference 27 gives a good treatment on the subject.

According to general relativity (ref. 27) the weak field, slow motion limit gravitational field equations can be written in the form which emphasizes the gravitomagnetic field. When this is done, the equations for the gravitoelectric and gravitomagnetic fields become almost identical to Maxwell's equations for the electric and magnetic fields:

$$\nabla \cdot \mathbf{E_g} \approx -4\pi G\rho \; , \; \nabla \times \mathbf{E_g} \approx 0 \; , \; \nabla \cdot \mathbf{H_g} \approx 0 \; , \; \nabla \times \mathbf{H_g} \approx 4[-4\pi G\rho \mathbf{v}/c + (1/c)\partial \mathbf{E_g}/\partial t] \qquad (19.21)$$

where $d\mathbf{v}/dt = \mathbf{E_g} + (\mathbf{v}/c) \times \mathbf{H_g}$.

$\mathbf{E_g}$ is the gravitoelectric field and is the same as the ordinary Newtonian gravitational acceleration, $\mathbf{H_g}$ is the gravitomagnetic field which in the solar system is about $10^{12}$ times weaker than $\mathbf{E_g}$, and $\rho$ is the matter density and $\mathbf{v}$ is the velocity of the mass, and $\mathbf{E_g} = -\nabla \phi$, and $\mathbf{H_g} = \nabla \times \gamma$ . The potentials $\phi$ and $\gamma$ are constructed in a similar manner as in electromagnetism (where the electromagnetic four-vector potential $A_\alpha$ is decomposed into an electrical scalar potential and $\psi = -A_0$ and a magnetic vector potential $A=A_j$), and are related to the space-time metric $g_{uv}$ by: $\phi \approx -0.5(g_{00} + 1)c^2$ , $\gamma_i \approx g_{0i}$. The gravitoelectric scalar potential $\phi$ is essentially the same as the time part $g_{00}$ of the space-time metric, and the gravitomagnetic vector potential $\gamma$ **is** essentially the time-space part of the space-time metric.

Contrast these equations with the classical Maxwell's equations of electromagnetism, and the following differences are found:

(1) The minus sign appears in the source terms.
(2) Replacement of the charge density with mass-density multiplied by G.
(3) A factor of four in the strength of $\mathbf{H_g}$.
(4) The replacement of the charge current by $G\rho \mathbf{v}$

For a rotating massive body like the earth, the gravitomagnetic field looks like the magnetic dipole field of a rotating electrically charged sphere. If one places a spinning object like a gyroscope in this field, there will be precession of the spin axis (just as a current loop in a magnetic field). This phenomena is sometimes called inertial frame dragging (we will look at an alternative way of viewing this effect in the next section).

If the existence of the gravitomagnetic fields becomes fully confirmed, we would take this as further evidence for the similarity between the photon and graviton particle exchanges. This would support the view that nature's two long-range forces (electromagnetism and gravity) operate in virtually the same way. The gravitomagnetic force component is far weaker than the observable gravitoelectric force (which corresponds to the ordinary Newtonian force of gravity), because this force is produced by pure graviton exchanges. However, the pure low-level gravitoelectric force is mixed up with the ordinary strong electrical charged component of gravity existing in all gravitational processes from the electrically charged virtual particles of the quantum vacuum. Since the electrical force from the virtual particles far exceeds the pure graviton processes, the pure gravitoelectric force component (caused by pure graviton exchanges only) is completely disguised and hard to observe. The ordinary measurable force of gravity we perceive everyday is dominated by the electromagnetic forces from the electrically charged virtual particles of the quantum vacuum, which also equalizes the fall rate of all masses.

## 19.3    EMQG AND FRAME DRAGGING: LENSE-THIRRING EFFECT

We now apply the principles of EMQG to calculate the amount of inertial frame dragging (Lense-Thirring effect) on the earth. The Lense-Thirring effect is a tiny perturbation of the orbit of a particle caused by the spin of the attracting body, first calculated by the physicists J. Lense and H. Thirring in 1918 using general relativity (ref. 25). Einstein's general relativity predicts the perturbation in the vicinity of the spinning body, but the effect has not been accurately verified experimentally. However, recent work in ref. 26 using the LAGEOS and LAGEOS II earth orbiting satellites has rendered an unconfirmed experimental value that agrees with theory to an accuracy of about 20 %. The Lense-Thirring effect has also been interpreted as being due to gravitomagnetic fields (section 19.2), and also tied in with Mach's principle (ref. 51). It is hoped that with the launch of the Gravity Probe B (co-developed by Stanford University) by NASA the Lense-Thirring effect will be measured to an unprecedented accuracy of 1% or better. The Gravity Probe B (there was a different Gravity Probe A launched earlier by NASA) is a drag free satellite carrying four ultra-precise gyroscopes that will be put in a polar orbit around the earth at a height of about 400 miles (ref. 28).

An important consequence of the Lense-Thirring effect is that the orbital period of a test mass around the earth depends on the direction of the orbit! A test mass that has an orbit which revolves around the earth in the same direction of the spin rotation would be longer then the orbital period of the same test mass revolving opposite to the direction of the spin of the earth. The difference in the orbital period of the two test masses becomes smaller with increasing height until it disappears when the orbits are at infinity.

The Lense-Thirring effect can be thought of as a kind of 'a dragging of inertial frames', first named by Einstein himself. What this means is that the free fall (gravity-free) motion of a test mass is modified in the presence of a large massive spinning sphere, as compared to the identical case of the non-rotating sphere. If a test mass is in free-fall, the contribution of gravity to the motion of the mass is canceled out (at a point), and the resulting motion of the test mass is purely inertial. However, the inertial motion of the mass is different by the presence of the Lense-Thirring effect, as compared to a non-rotating earth. As a result of the Lense-Thirring effect, the earth 'drags the local inertial reference frame'. Therefore the small orbiting gyroscopes in Gravity Probe B with spin axes oriented towards a distant star in a polar orbit around the earth will have is its spin axis changed by the Lense-Thirring effect as it orbits.

To investigate inertial frame dragging with EMQG theory we will start with the conceptually simpler case where we have a large rectangular mass that is very thick and heavy. This thought experiment avoids the complications of a rotating mass. Imagine that you are observer A on the surface of this large rectangular mass, and you are interested in whether there is inertial frame dragging. In our thought experiment we will give this large rectangular mass a relativistic and constant linear velocity (say 1/10 light velocity) in the forward direction perpendicular to the surface (with respect to an external observer B). In EMQG theory, our first step is to determine the motion of the virtual particles of the quantum vacuum. To determine this, we need to know the graviton flux and direction. Since the rectangular mass is very heavy it emits a huge graviton flux, which travels outwards. The gravitons are absorbed by the virtual masseons of the quantum vacuum above the mass, which causes them to accelerate towards the source in straight lines (postulate #2). The graviton flux falls off with distance above the surface and therefore the acceleration of the virtual particles varies with height. However the massive rectangle has a relativistic velocity perpendicular to the graviton flux. Thus the resulting velocity vector of the gravitons (with respect to observer A) is the vector sum of two velocity components: the graviton velocity and the mass velocity. The graviton flux moves along straight lines, which are deflected by an angle in the direction of motion that depends on the velocity of light (graviton speed) and the velocity of the massive rectangle. The virtual particles absorb some of the graviton flux, which results in a downward acceleration in the same direction as the gravitons. The magnitude of the virtual particle acceleration depends on the received graviton flux rate.

We have seen that the magnitude of the virtual particle acceleration at a point is given by the graviton flux rate, and the direction is the direction of the graviton flux at that point. Since the graviton flux rate decreases with increasing distance from the surface (at some given rate) due to spreading, the virtual particles accelerate downwards in slanted lines with the acceleration vectors decreasing with height. If observer A drops a mass, the mass takes on the same net downward acceleration as the virtual particles of the vacuum, and the path is on the slanted line. This is inertial frame dragging. With EMQG theory we have a very simple physical picture for the root cause of inertial frame dragging. The path of the accelerated virtual particles of the quantum vacuum dictates the direction and magnitude of the 4D space-time curvature (section 16.6). Normally it is perpendicular to the surface, but with relativistic velocity the direction is altered.

Now we are in a position to study the Lense-Thirring effect for a rotating mass like the earth. The effect is more pronounced as the angular velocity of the rotating mass increases. As we have seen above, the basic reason for inertial frame dragging is the finite speed of propagation of the graviton particle (the speed of light). This allows time for a large rapidly spinning mass to rotate a small amount while the graviton is still in flight as it propagates outwards. ***The finite velocity of the graviton particle along with the downward $GM/R^2$ acceleration component of the charged virtual particles of the quantum vacuum is entirely responsible for inertial frame dragging***.

We now examine inertial frame dragging for a weak gravitational field such as the earth (the strong field inertial frame dragging is a very formidable problem in EMQG). In order to simplify our analysis, we will assume that most of the mass of the earth is concentrated in a small spherical region at the earth's center, the rest of the earth's volume containing a negligible quantity of mass. The graviton moves outward at the speed of light from the spinning core (neglecting the virtual particle scattering effects, which reduces its absolute velocity as it propagates upwards near the earth). Gravitons are emitted in huge numbers from the earth. Since we know that gravitons are physically very similar to photons (postulate #2), we can predict the characteristics of this graviton flux. The graviton flux can be visualized as being the same as a rotating flashlight emitting light outwards as it rotates. Since the velocity of the source does not effect the velocity of light, we conclude that the velocity of the spinning core does not affect the motion of the gravitons. The inset in the upper right hand corner of figure 9 shows the view of the earth looking down at the north pole, where the earth is rapidly rotating at an angular velocity close to light velocity. The non-rotating observer 'B' is stationed above the equator, and 'sees' the graviton flux leaving the equator along the radius vectors. It is important to note that when the graviton leaves the rotating source, the motion of the graviton is totally de-coupled from the source. The gravitons move at a constant velocity, and are not affected by the motion of the source in any way. However, gravitons are affected by the state of accelerated motion of the virtual particles. The note at the end of this section gives more detail on this important point, and the reason why inertial frame dragging is highly non-linear for very massive objects rotating at relativistic speeds. For now, we can neglect this effect for the earth.

As the gravitons propagate outward, they encounter virtual masseon particles in the quantum vacuum. Since masseons posses 'mass-charge', the vacuum is accelerated down (postulate #2). The virtual masseons are therefore accelerated in the same direction of the motion of the graviton particle flux. The magnitude of the acceleration depends on the graviton flux at a point 'r' from the center, and is given by $GM/r^2$. If a test mass (composed of real electrically charged masseon particles) is dropped onto the surface of the earth, the electromagnetic interactions (postulate #3) between the real masseons of the mass and the virtual masseons causes the test mass to accelerate downwards at 1g in the direction of the graviton flux. Thus, from the perspective of observer 'B', the mass falls along the radius vectors. What does observer 'A' see?

The main drawing in figure 9 shows the view of the earth looking down at the north pole, where the earth is rapidly rotating at an angular speed close to light velocity. Observer 'A' is situated on the surface of the equator and 'sees' the graviton flux leaving the equator in curved paths. Why is the graviton flux curved? Recall that the graviton flux is moving in 'straight' line paths along the radius vectors. However, observer 'A' is carried along with the rotation of the earth. Therefore from his frame of reference the gravitons appear to curve. The average acceleration vector of the virtual particles follows this curved path, with the acceleration vectors increasing in magnitude the closer to the center. The magnitude of the acceleration of the virtual particles is approximately $GM/r^2$. The test mass moves along the curved path in figure 9, guided by the motion of the virtual particles by photon exchanges (postulate #3). These curved paths also represent the path that light will take if it propagates straight up (in other words, geodesic paths). In absolute CA units, the light velocity varies upwards along these curved paths from observer's A reference frame, due to the Fizeau-like scattering with the quantum vacuum (section 16.6). Thus, these paths represent the direction of the 4D space-time curvature. These paths are deflected when compared to the non-rotating earth.

The equation of the outward propagating graviton curve (in the equatorial plain) for a clockwise rotating earth turns out to be Archimedes' Spiral, and takes the form $r = k\theta$ in polar coordinates, where k is a constant (r is the distance that the graviton travels). The constant k depends only on the velocity of the graviton 'c' and the velocity $v = 2\pi R/T_p$ of the earth's rotation, where $T_p$ is the rotation period of the earth and R is the earth's radius, and the time t of transit. If k is small then the spiral has a high curvature, and if k is large the curvature is small. The ratio 'c/v' determines the value of the spiral constant k. If c were to be very large, then $k \gg 1$ which causes the gravitons to 'unwind' slowly ; and if v were to be very large, then $k \ll 1$ which causes the gravitons to 'unwind' rapidly.

Based on these considerations, it is easy to show that the equation for the spiral is:

$r = c^2 \theta / (2\pi v)$   where $v = 2\pi R/T_p$ , and $\theta = 2\pi t v/c$ (19.31)

The Archimedes spiral is normalized such that if v=c, $\theta = 2\pi$, and t=1 second then the value of r is 300,000 km after one rotation. To verify that eq. 19.31 is correct, let us see how r varies with $\theta$. When the velocity of rotation is v=c, then according to eq. 19.31 the equation for the spiral is $r = [c/(2\pi)]\theta$. When the graviton travels for one rotation of the observer A, the spiral path appears to twist once, with $\theta = 2\pi$ radians, and r = 300,000 km. Similarly, when the graviton travels for two rotations of the observer A, the spiral path appears to twist once, with $\theta = 4\pi$ radians, and r = 600,000 km. If v=c/2, then if t = 2 seconds, $\theta = 2\pi$, and r = 600,000 km.

Let us calculate the shape of the spiral for the observer A on the earth. The earth has a radius $R = 6.37 \times 10^6$ meters, and a rotation period $T_p$ = 24 hours. Therefore, the v = 463 m/sec. The ratio c/v = 647,948. Therefore, the equation of the spiral is: $r = 103,176\ c\theta$. We wish to solve for the angle $\theta$ at the earth's surface, where $r = 6.37 \times 10^6$ meters. Therefore: $\theta = 6.37 \times 10^6/103176c = 2.05 \times 10^{-7}$ radians. Recalling that 1 radian = $180/\pi$ °. Therefore, $\theta = 1.18 \times 10^{-5}$ degrees = 42.5 milli-arcseconds. This **agrees well** with the prediction based on general relativity (ref. 25). However, the methods of general relativity are significantly more complex. The result we obtained is derived from the simple deflection of the downward accelerating virtual particle path.

We have seen that the angle $\theta$ represents the deflection of the downward accelerating virtual particles of the quantum vacuum with respect to the non-rotating earth. Furthermore, the deflection angle varies with height (along a spiral path). Recall that the direction of 4D space-time curvature is the same as the direction of the virtual particle acceleration. Therefore, this angle represents the shifting of the direction of 4D space-time curvature. Now we are in a position to understand inertial frame dragging near the earth. The motion of any free falling (gravity-free) test mass is modified in the presence of the curved virtual particle path. If a test mass is in free-fall, the contribution of gravity to the motion of the mass is canceled out (at a point), and the resulting motion of the test mass is purely inertial. Therefore, the free fall is deflected by an angle of 42.5 milli-arcseconds. Therefore, the earth appears to drag the local inertial reference frame, with the drag effect varying with height.

**NOTE**: The gravitons *do* scatter with the virtual masseon particles of the quantum vacuum, because masseons possess 'mass-charge' and interacts with the quantum vacuum in a similar manner as light (postulate #2). This scattering is small for a mass the size of the earth. However, for objects with a very strong gravitational field that is rotating at relativistic speeds, the calculation of the Lense-Thirring effect (based on Einstein's gravitational field equation) is an extremely difficult problem. In general relativity, this is due to the notorious non-linearity of Einstein's gravitational field equation. In EMQG, the non-linearity manifests itself in the way that gravitons and the quantum vacuum interact in producing strong gravitational fields. The graviton flux determines the motion of the surrounding virtual particles of the quantum vacuum. However, the virtual particles of the quantum vacuum affect the motion of the gravitons. This makes it very difficult to calculate frame dragging under the conditions of extreme curvature and rapid rotations, such as might exist in a black hole. In general relativity, the Lense-Thirring calculation is based on the weak field approximation of Einstein's equation (ref. 25). In EMQG, the Lense-Thirring calculation is also done for weak fields and slow rotation, which simplifies the motion of graviton particles.

## 20. GENERAL RELATIVITY LIMITATIONS AND EMQG RESOLUTIONS

Most physicists assume that Einstein's theory of general relativity has no inconsistencies or flaws, and can be applied over any distance scale to any size mass distribution. Part of this belief comes from the mathematical elegance of the theory, and partly from the spectacular experimental agreement with theory, especially over the past 25 years (ref. 6).

In fact, general relativity is known to have a number of mathematical problems, and fails to answer some very important physical questions regarding gravitation. Furthermore, it appears that quantum theory and general relativity are incompatible. In spite of over 70 years of intensive research, no full theory of quantum gravity has been formulated successfully. We will briefly summarize some of these problem areas in general relativity, and show how EMQG is able to resolve these issues.

### (1) GENERAL RELATIVITY FAILS TO PROVIDE A PHYSICAL CAUSE-EFFECT OR ACTION THAT COUPLES MATTER-ENERGY TO THE LOCAL 4D SPACE-TIME CURVATURE.

**GENERAL RELATIVITY**: Einstein's gravitational field equations dictate how much curvature of 4D space-time there is, based on how much matter-energy there is. It does not tell you *why* there is curvature. What is the cause of 4D space-time curvature? What is it about matter that causes the local space-time to curve? In general relativity there is no physical reason or action to relate matter to space-time. Instead, curvature just happens. Einstein appealed to Newton and Poisson's theory of gravity to adjust the amount of coupling between matter-energy and 4D space-time, without providing any further insights into this question.

**EMQG:** There is definitely a physical action that couples matter to the amount of space-time curvature. It is called the graviton particle! The total amount of absolute matter determines the total number of masseon particles in the mass. Double the absolute mass, and you double the number of masseons. Double the number of masseons, and you double the flux rate of the gravitons exchanged with the virtual masseon particles of the quantum vacuum. The absorption of gravitons accelerates the virtual particles of the quantum vacuum downwards. The graviton flux spreads out from the earth and the number of graviton particles per unit area falls as $1/4\pi r^2$, where 'r' is the distance from the center. The decrease in the graviton flux with height reduces the downward acceleration of the virtual masseon particles of the quantum vacuum (with height). Since the virtual masseon particles accelerating downwards also possess electrical charge, there is an electromagnetic interaction with light and matter moving in the vicinity of the earth. We call this interaction the 'Fizeau-like' scattering from the virtual particles, which acts like an accelerated flowing fluid medium in analogy with the Fizeau experiment for light propagation with moving water (section 16.5). Therefore, the variation of the acceleration vectors of the virtual particles with height also represents the variation of 4D space-time curvature with the height. The action of the graviton particle is not instantaneous, however, since the gravitons move at light velocity. The finite speed of the graviton particle is responsible for the Lense-Thirring effect, and also partially responsible for gravitational waves.

### (2) GENERAL RELATIVITY ONLY POSTULATES THE EXISTENCE OF THE PRINCIPLE OF EQUIVALENCE.

**GENERAL RELATIVITY**: Why should there be equivalence between inertial and gravitational mass? After all, both these mass types are defined differently in Newtonian physics. Inertial mass involves accelerated motion, but there appears to be no motion involved in the definition of gravitational mass. In general relativity, equivalence has still remained a postulate as first proposed by Einstein, in spite of 70 years of investigation.

**EMQG:** The principle of equivalence turns out to be derivable from lower level quantum particle processes, and has been investigated in great detail in section 15.8. Equivalence and quantum inertia forms the central core of EMQG theory. We have found that equivalence arises from the *same electromagnetic interaction* of the electrically charged virtual masseon particles of the quantum vacuum with the real electrically charged masseon particles of the test mass, which also occurs in inertia (section 13). We have also found that equivalence is *not* perfect! It breaks down when both an extremely large mass and a tiny mass are dropped on the earth, with the larger mass arriving first.

### (3) GENERAL RELATIVITY FAILS TO ACCOUNT FOR THE ORIGIN OF INERTIA AND CANNOT FULLY ACCOUNT FOR MACH'S PRINCIPLE.

**GENERAL RELATIVITY**: General Relativity incorporates inertia in a purely Newtonian way, without modification. Since inertial mass is quantized in the form of elementary particles, there must be some explanation as to why the individual particles resist any change in motion. Also, Mach's principle is still not fully incorporated or understood in general relativity. This principle strikes at the very heart of the meaning of motion. For an in depth discussion of Mach's principle and it's relationship to general relativity refer to section 14.

**EMQG:** Inertia (called quantum inertia) is postulated to be the electromagnetic interaction (of which the details are still unknown) between the electrically charged virtual masseon particles of the quantum vacuum and the accelerated real electrically charged masseon that make up a mass (section 13). Although quantum inertia is still a postulate (postulate #3) of EMQG, we are reasonably sure that the details of this process will be found soon. Mach's principle follows directly from the principles of quantum inertia. The virtual particles of the quantum vacuum **are** Mach's invisible reference frame to gauge accelerated motion (from the point of view of acceleration only). In fact, the state of relative acceleration of a mass particle with respect to the background vacuum particles is the actual **origin** of inertial force, and this process is also present in gravitational interactions! However, the quantum vacuum cannot be used as an absolute reference frame for constant velocity motion, and therefore velocity remains relative in EMQG, just as it is in relativity.

(4) **GENERAL RELATIVITY FAILS TO BE QUANTIZED.**

**GENERAL RELATIVITY**: In spite of brilliant attempts at unification of general relativity and quantum theory, there still exists no fully accepted quantum gravity theory. It was known even when Einstein proposed general relativity that matter is quantized, and comes in the form of atoms and molecules. These in turn are quantized in the form of elementary particles. Particles behave according to the rules of quantum field theory. Yet, the energy-momentum-stress tensor of a mass is treated as a classical continuum. Forces are also known to originate from particle exchange processes. Is the gravitational force an exception to this general rule? Furthermore, there is growing evidence that space-time is also quantized. Yet, no one has found a consistent theory of quantum gravity that reduces to general relativity in the classical limit, leading one to believe that the foundations of general relativity need to be modified.

**EMQG:** EMQG theory *is* the quantization of general relativity. EMQG is totally based on the quantum particle nature of matter and forces, which behave according to the rules of quantum field theory. The 4D space-time curvature of general relativity is a manifestation of the changed state of the quantum vacuum, both under acceleration and under the influence of a nearby gravitational field. EMQG restores the concept of absolute space and separate absolute time in the form of the inherently quantized CA space and CA time, which is not directly observable with our measuring instruments. This CA space and time is inherently quantized because of the very nature of the cellular automata model, which is the fastest possible parallel computer model (section 3). The relativistic 4D Minkowski space-time of general relativity is retained in EMQG as the end result of a purely quantum particle interaction process (section 16.6). Thus, in EMQG the curved 4D space-time geometry paradigm is replaced by a total, quantum particle interaction paradigm, where the quantum particles live in absolute CA space and CA time.

(5) **THE AD HOC NATURE OF SPACE AND TIME IN GENERAL RELATIVITY**

**GENERAL RELATIVITY**: 4D relative Minkowski space-time appears to behave in an 'ad hoc' manner in general relativity theory. According to general relativity, the universe is a geometric Riemann 4D curved space-time on <u>all</u> distance scales. Yet, the 4D space-time curvature is not absolute for accelerated or gravitational observers in any state of motion. Instead, 4D space-time depends very much on the particular physical circumstance or motion involved. In other words, 4D space-time is relative! Yet the presence of a large mass causes an absolute amount of 4D space-time curvature given by the metric. In the case of an accelerated frame, the amount of curvature is not absolute, but depends only on the amount of acceleration possessed by the observer!

For example, an observer 'A' stationed on the surface of the earth finds himself embedded in 4D, curved space-time. Yet, another observer 'B' in free fall near the earth finds himself in perfectly flat 4D space-time. Let us suppose that observer 'B's free fall path takes him directly past observer 'A', whom he can even momentarily touch as he falls past him. At the moment of contact the space-time is *still different* for *both* observers! This seems to be a very strange way to construct the fundamental geometry of our universe. Even though both observers can actually physically touch each other for a brief moment, each observer still lives in different 4D space-times! What is even stranger about space-time is that observer 'A' has a space-time curvature directed along the radius vector only (as can be seen from the Schwarzchild metric). In other words, the curvature of the space and time components of the metric is directional! No curvature results in directions that are parallel to the surface of the earth (we ignore the earth's surface shape). To add to the confusion, if we imagine that observer 'A' decides to take on a new acceleration (2 g's) parallel to the earth's surface, he now lives in a curved space in both the parallel and perpendicular directions. However, now the curvature is unequal in value in both the parallel and perpendicular directions!

**EMQG:** At the most fundamental distance scales, there exists only absolute CA 3D space, and separate absolute CA time, which are both inherently quantized by the structure of the CA model. This absolute space and time is ***not*** curved in the presence of a gravitational field, nor is space and time unified as it is in relativity! Furthermore, absolute space and time is *not* relative, and the state of motion of an observer does **not** affect this coordinate system at all! The measurable relativistic 4D space-time is one layer above the absolute space consisting of cells, and absolute time consisting of 'clock' cycles. We have seen that the 4D relativistic, curved space-time, which is present in accelerated frames and gravitational frames, is an abstraction. It results from the behavior of light and matter in the presence of the accelerated virtual particles of the quantum vacuum, which can be looked at as a kind of a special 'Fizeau-like' fluid (section 16.4). Generally for observers in gravitational and accelerated frames, the virtual particles are in a state of accelerated motion. Observer 'B' can cancel this curved 4D space-time simply by taking the same acceleration as the virtual particles of the quantum vacuum. This restores the virtual particles of the quantum vacuum (from the point of view of his reference frame) to the equivalent state that the observer would 'see' in flat, special relativistic, 4D space-time. Similarly, observer 'A' senses directional space-time curvature simply because the direction of virtual particle acceleration is directed downward, and not side to side. However, if he decides to accelerate in a direction parallel to the surface of the earth, he will observe the virtual particles of the quantum vacuum accelerating towards him from two different directions, downwards and towards him. In other words, there are two acceleration vectors from the quantum vacuum, one vector directed downwards with an acceleration of 1g, and the other equal in magnitude to the amount of forward acceleration of 2g. This results in unequal space-time curvature in both directions, because the accelerated Fizeau-like quantum vacuum fluid has two different acceleration components in the x and y direction. This introduces two different and independent 4D space-time curvatures.

### (6) **GENERAL RELATIVITY PREDICTS SPACE-TIME SINGULARITIES.**

**GENERAL RELATIVITY**: Space-time singularities and black holes are a direct consequence of the mathematics of general relativity theory. However, it is <u>very</u> doubtful that these mathematical monsters really exist. In fact, a singularity usually means that a theory has somehow broken down.

**EMQG:** In EMQG, we cannot have a space-time singularity. EMQG is based on the idea of 3D absolute CA space and absolute CA time, that are inherently quantized in the form of cells on an a cellular automata. This model prohibits an infinite density of particles to accumulate in one place due to gravitational collapse (and thus predicts that Quantum Field Theory will break done at very high particle densities). The density of particles that can be supported cannot exceed the number of cells per cubic meter. However, it is not clear what will happen when particle densities become this great.

### (7) **GENERAL RELATIVITY FAILS TO ACCOUNT FOR THE VALUE OF THE COSMOLOGICAL CONSTANT**

**GENERAL RELATIVITY**: Einstein originally introduced the cosmological constant in 1918, which was a term that he added to his gravitational field equations to make the universe remain in the steady state (or non-expanding, as was believed in his time). This new term corresponded to a nonzero energy momentum tensor of the vacuum. Einstein later abandoned this constant when Hubble discovered in the 1920's that the universe was actually expanding. In fact, Einstein considered the addition of this term to his field equations as the biggest blunder of his life. The cosmological constant represents the measure of the mass-energy density contained in empty space alone. It has been said that the cosmological constant is the *most* striking problem in contemporary fundamental physics (ref. 52). In fact, some theoretical predictions of quantum field theory differs by observations by at least $10^{45}$ and possibly, by 120 orders of magnitude! Experimentally, it is safe to say that its actual value is very close to zero. In fact, S. Hawking (ref. 52) once stated that:

*"The cosmological constant is probably the quantity in physics that is most accurately measured to be zero: observations of departures from the Hubble law for distant galaxies place an upper limit of the order of $10^{-120}$."*

On the other hand, quantum field theory predicts that there ought to be plenty of contributions from the virtual particles of the quantum vacuum.

**EMQG:** It is true that the quantum vacuum at any instant of time contains an enormous number of virtual particles, which have both mass and electrical charge. Why is it that their presence is not felt electrically or gravitationally? The cosmological constant is essentially zero because of the symmetrical production of virtual masseon and anti- masseon particle pairs in the quantum vacuum. According to the general principles of quantum field theory, these virtual masseon particle pairs are always created in opposite electrically charged particle pairs, which explains the electrical neutrality of the vacuum. We propose in section 15.4 that the particle pair creation principle in the quantum vacuum is also applicable to 'gravitational mass charge' as well (postulate #2). Virtual masseon particle pairs are also created in *opposite* gravitational 'mass-charge' pairs. At any instant of time, the quantum vacuum has almost exactly equal numbers of positive and negative electrically charged and 'gravitationally charged' masseon particles, which leads to ***neutrality of both electrical and gravitational forces in the quantum vacuum and a cosmological constant of zero.***

(8) **GENERAL RELATIVITY DOES NOT CONSERVE ENERGY-MOMENTUM.**

**GENERAL RELATIVITY**: It is known that general relativity does not assign an energy-momentum-stress tensor to the gravitational field. There are also serious problems with local energy-momentum conservation (ref. 53). The absence of the energy-momentum-stress tensor is due to the geometric aspect of gravity in general relativity. We know that gravitational waves transfer energy by the rippling of 4D space-time to far away destinations. For example, a pair of orbiting neutron stars emits gravitational waves, which cause them to slowly spiral inwards. This orbital energy loss is precisely the energy carried away by the gravity waves. Thus, gravity waves can do work on a distant mass. However, how can rippling 4D space-time carry energy?

**EMQG:** In general relativity, curved space-time actually does represent the geometry of the universe. This **is** the source of the difficulty in energy conservation in general relativity. In EMQG, curved space-time represents the distorted distribution of the net statistical average acceleration vectors of the virtual electrically charged masseon particles of the quantum vacuum from place to place. These charged vacuum masseon particles interact through electromagnetic forces by photon exchanges with a nearby test mass (which consists of electrically charged masseons). It is this electromagnetic interaction between the quantum vacuum and a test mass that represents the energy interactions in gravitation, and the energy content of curved 4D space-time. The quantum vacuum has a very large energy density, but it is not infinite. Energy and momentum conservation is always obeyed on the quantum particle level in EMQG.

21.    **EXPERIMENTAL TESTS OF EMQG THEORY**

EMQG proposes several new experimental tests that give results that are different from conventional general relativistic physics. Although most of these experiments are very difficult to perform, they do provide a solid basis for testing the principles of EMQG theory.

*(1)* EMQG opens up a new field of investigation, which we call anti-matter gravitational physics. We propose that if two sufficiently large pieces of anti-matter are manufactured to allow measurement of the mutual gravitational interaction, then the gravitational force will be found to be repulsive! The force will be equal in magnitude to $-GM^2/r^2$ where M is the mass of each of the equal anti-matter masses, r is their mutual separation, and G is Newton's gravitational constant). This is in clear violation of the principle of equivalence, since in this case $M_i = - M_g$, instead of masses $M_i = M_g$. Antimatter that is accelerated in far space has the same inertial mass '$M_i$' as ordinary matter, but when interacting gravitationally with another antimatter mass it is repelled ($M_g$). (**Note:** The earth will *attract* bulk anti-matter because of the large abundance of gravitons originating from the earth of the type that induce attraction). This means that no violation of equivalence is expected for anti-matter dropped on the earth, where anti-matter falls normally (recall that virtual masseons and anti-masseons are both attracted to the earth (postulate #2). However, an antimatter earth will repel an antimatter mass dropped on the earth. Recent attempts at measuring earth's gravitational force on anti-matter (e.g. anti-protons will not reveal any deviation from the principle of equivalence).

*(2)* For an extremely large test mass and a very small test mass dropped simultaneously on the earth (in a vacuum free of air resistance), there will be an extremely small difference in the arrival time of the masses, in slight violation of the principle of equivalence. This effect is on the order of $\approx \Delta N \times \delta$, where $\Delta N$ is the difference in the number of masseon particles in the two masses, and $\delta$ is the ratio of the gravitational to electric forces for one masseon. This experiment is very difficult to perform on the earth, because $\delta$ is extremely small ($\approx 10^{-40}$), and $\Delta N$ cannot be made sufficiently large. To achieve a difference of $\Delta N = 10^{30}$ particles between the small and large mass requires dropping a molecular-sized atomic cluster and a large military tank simultaneously in the vacuum in order to give a measurable deviation. Note: For ordinary objects that might seem to have a large enough difference in mass (like dropping a feather and a tank), the difference in arrival time may be obscured by background interference, or by quantum effects like the Heisenberg uncertainty principle which restrict the accuracy of time measurements.

*(3)* If gravitons can be detected by the invention of a graviton detector/counter, then there will be experimental proof for the violation of the strong principle of equivalence. The strong equivalence principle states that all the laws of physics are the same for an observer situated on the surface of the earth as it is for an accelerated observer on a rocket (1 g). The graviton detector will find a tremendous difference in the graviton count in these two cases, because gravitons are vastly more numerous here on the earth.

*(4)* Since mass has a strong electromagnetic force component, a sensitive mass measurement near the earth might be disrupted by experimentally manipulating the electrically charged virtual particles of the nearby quantum vacuum through electromagnetic means. If a rapidly fluctuating magnetic field (or rotating magnetic field) is produced under a mass it might effect the instantaneous virtual charged particle spectrum, and disrupt the tiny electromagnetic forces contributed by each electrically charged masseon of the mass. This may reduce the measured gravitational mass of an object in the vicinity (this would also affect the inertial mass). In a sense, this device would act like a primitive weak "anti-gravity" device. The virtual particles are constantly being "turned-over" in the vacuum at different rates depending on the energy, with the high frequency particles (and therefore, high-energy particles) being replaced the quickest. If a magnetic field is made to fluctuate fast enough so that it does not allow the new virtual particle pairs to replace the old and smooth out the disruption, the spectrum of the vacuum will be altered. According to conventional physics, the energy density of virtual particles is infinite, which means that all frequencies of virtual particles are present. In EMQG there is a definite upper cut-off to the frequency, and therefore the highest energy according to the Plank's law: $E=h\upsilon$, where $\upsilon$ is the frequency that a virtual particle can have. This frequency cutoff is very roughly on the order of the plank distance scale. We can therefore state that the smallest wavelength that a virtual particle can have is on the order of about $10^{-35}$ meters, e.g. the plank wavelength (or a corresponding maximum Plank frequency of about $10^{43}$ hertz for very high velocity ($\approx c$) virtual particles). Unfortunately for our "anti-gravity" device, it is technologically impossible to disrupt the highest frequencies. According to the uncertainty principle, the relationship between energy and time is: $\Delta E \times \Delta t < h$. This means that the high frequency end of the spectrum consists of virtual particles that "turns-over" the fastest. To give measurable mass change the higher frequencies of the vacuum must be disrupted, which requires magnetic fluctuations on the order of at least $10^{20}$ cycles per seconds. Therefore, only lower frequencies virtual particles of the vacuum can be practically affected, and only small changes in the measured mass can be expected with today's technology. As a result of this, a relationship should exist between the amount of gravitational (or inertial) mass loss and the frequency of electromagnetic fluctuation or disruption. The higher the frequency the greater the mass loss. Work on the Quantum Hall Effect (ref. 29) by Laughlin has suggested that the electron density in a two-dimensional sheet under the influence of a strong magnetic field causes the electrons to move in concert, with very high speed swirling vortices created in the resulting 2D electron gas. In ordinary magnetic fields, electrons are merely 'pushed' around, while a strong magnetic field causes the electrons to swirl in high-speed 'whirlpools'. There is also a possibility that this 'whirlpool' phenomena holds for the virtual particles of the quantum vacuum under the influence of a strongly fluctuating magnetic field. These high-speed whirlpools might disrupt the high frequency end of the spectral distribution of electrically charged virtual particles in small pockets. Therefore, there might be a greater mass loss under these circumstances. Recent experiments on mass reduction with rapidly rotating magnetic fields are inconclusive at this time. Reference 30 gives an excellent and detailed review of the various experiments on reducing the gravitational force with superconducting magnets.

## 22. DIRECTIONS FOR FUTURE WORK IN EMQG

EMQG theory presented here is an outline for (hopefully) a full theory of quantum gravity, and as such is not complete. In this section we present some areas that require further research. We have listed what we know so far, in order to assist future investigators that might take up this search. Be cautioned that these ideas are very speculative.

### 22.1 INVESTIGATION OF THE UNIVERSAL CELLULAR AUTOMATA

The Holy Grail of modern physics would be the discovery of a set of equations that would describe all of physics, and this is often called the theory of everything. In EMQG theory it would be the discovery of the exact structure and mathematical laws that govern the Cellular Automata that is our universe. Much work needs to be done before this can be accomplished. The connection between quantum field theory and CA theory, and the exact nature of quantum particles on the CA needs to be developed. Currently, we have proposed a preliminary model for the CA as a simple, geometric 3D CA. This model may turn out to be wrong when a full quantum theory is developed, that is compatible with the CA model.

### 22.2 QUANTUM THEORY AND CELLULAR AUTOMATA

Currently, we have done some very preliminary work on two possible models of the quantum wave function of a non-relativistic quantum particle with a DeBroglie wavelength given by $\gamma = h/mv$. We have started to investigate two tentative models for the quantum wave function on the CA.

In our first approach, the wave-particle duality some how emerges from a CA numeric information pattern directly. The wavelength of the wave function corresponds directly to the wavelength of a periodic, fluctuating numeric information wave pattern existing on the CA cells. In other words, the wave function directly represents pure oscillating numeric information that represents the state of motion of a non-relativistic particle in absolute CA space and time. Oscillating patterns are very common in the game of life, which is a simple geometric 2D CA. Patterns have been found with many different periods and sizes. This model readily explains why the wave function has not been detected directly by experiment (diffraction and interference experiments are an indirect observation). One cannot detect the numeric state of the cells directly through experiment. This model also explains the probabilistic nature of the wave function. Since we cannot know the numeric state of an information pattern at a given time, we cannot predict the exact future evolution of a particle pattern.

However, there are some difficulties with this approach as well. In simple 2D CA models like the game of life, it is very rare to find large coherent patterns on the CA. For example, a radio photon (somewhere in the AM radio band) has a wavelength the size of a house! How can this huge number of cells oscillate coherently on the CA? How does a wave function this size collapse when an observation occurs? In EMQG the measurable speed of light does not correspond to the low-level speed of light in absolute CA units of measure. Therefore, the speed that numerical information patterns evolve on the CA is much faster than the speed of light by an unknown amount. Recall that the low-level light velocity represents the simple shifting of information from cell to adjacent cell in every clock cycle, which is the maximum speed possible for information changes (section 9). Therefore, if there is an actual wave function collapse (this is still disputed), then the collapse can happen at speeds much greater than light velocity. This is because numeric information can be processed much faster than the measured light velocity (recall that this speed depends on the exact nature of photon scattering, and therefore depends on the index of refraction of the quantum vacuum).

Our second approach utilizes the idea that since inertial mass is involved in the DeBroglie wavelength, there exists the possibility that the DeBroglie wavelength may be the result of the interactions with the quantum vacuum. Recall that inertial mass is defined as the result of electromagnetic interactions of the quantum particles that make up a mass with the virtual masseons of the quantum vacuum. Therefore, the wave function might result from a 'point-like' information pattern in motion with respect to the virtual particles of the quantum vacuum. The motion of the 'point-like' particle might somehow induce a periodic oscillation of the immediate virtual particles of the quantum vacuum. In this way, the quantum wave function represents the periodic fluctuation of the virtual particles of the quantum vacuum near a 'point-like' particle. This model readily explains the probabilistic nature of the wave function as the inability to compute the exact state of the huge number of virtual particles of the quantum vacuum during an interaction. This model is somewhat reminiscent of the DeBroglie/Bohm pilot wave theory, where a particle is guided by an unobservable wave, which in turn affects the particle it is guiding. However, this approach to quantum theory also has its difficulties.

## 22.3 QUANTUM FIELD THEORY OF MASSEONS AND GRAVITONS

A full quantum field theory of the graviton exchange process must be developed in order to complete EMQG. This theory is expected to closely resemble QED theory for the electron and photon, because of the close similarities discussed in section 15.4. Therefore, a fully renormalizable quantum field theory of the masseon-graviton particles is possible. The Feynman diagrams and rules must also be developed. The properties of the masseon given in section 15.4 must be fully developed. In particular, how does the masseon particle fit in with the other forces of nature (the strong and weak nuclear forces) in the standard model? What is the exact nature of the force that binds masseons together to form the particles of the standard model? How does this fit in with the various unification schemes like super-symmetry, super-gravity, etc? Why are there large gaps in the allowed mass of the particles of the standard model? We call this the mass hierarchy problem. In other words, there is a large jump in the mass of particles as you go from the lepton family (an electron, for example) to the baryon (an up quark, for example) family of particles, with no other particle types in between. One solution to the hierarchy problem is to postulate that leptons and quarks are made up of masseons in some kind of orbital arrangements, where only certain orbital arrangements are allowed. Thus, the quark has tightly bound orbits with highly relativistic masseon orbital speeds, and the electron with lower speeds. The mass would look higher from our frame of reference, when the particle orbit has higher speeds.

## 22.4 BLACK HOLES AND EMQG THEORY

Do black holes exist in EMQG theory? Is there a singularity at the center of a black hole? Is there an event horizon around black holes? If there is an event horizon, how do graviton particles escape the event horizon, even though photons cannot escape? Is Minkowski 4D space-time frozen at the event horizon? It is clear in EMQG theory that a singularity cannot exist, because there is an upper limit to the particle density possible on a CA. It does seem that an event horizon is possible in EMQG. If enough matter is contained in the black hole it is conceivable that the escape velocity can approach the speed of light. In other words, there ought to be situations where the graviton flux is so large, that the virtual particles of the quantum vacuum achieve enormous accelerations (perhaps $10^{20}$ g's of gravity). In this situation, virtual particles created in the quantum vacuum will quickly approach light velocity in a tiny fraction of a second, and become relativistic. Based on the light scattering postulate, photons will not be able to escape. What becomes of the matter trapped inside the event horizon? The fate of quantum particles that achieve these enormous densities is not known in EMQG. Much work is needed in this area.

## 22.5 COSMOLOGY AND EMQG THEORY

Is the universe expanding in the general relativistic sense, from an initial singularity at the time t=0 of the Big Bang? How does the general relativistic Robertson-Walker Metric fit in with EMQG theory? It is quite clear in EMQG that at time t=0, there was no singularity. It is also clear that all the particles that exist today, were not present at t=0. Therefore, there must have been a 'matter creation' era, which halted at some early time. In EMQG, it is apparent that the cosmological 4D space-time curvature is strictly a quantum vacuum process, where all the matter/energy contained in the universe is involved. The visualization of expanding 4D space-time from an initial singularity (as demanded by the 'expanding balloon' model), where galaxies represented as spots on the balloon is obviously not compatible with the principles of EMQG. The curvature of 4D space-time is the result of the activities of the virtual particles of the quantum vacuum, which act on light and matter through a Fizeau-like scattering process (section 16.5).

Einstein's curved 4D space-time does not represent the actual geometry of the universe. The actual geometry of the universe is simply a collection of a vast number of cells in the CA. However, the outward motion of matter can represent the expansion of the universe from the initial creation area. Currently, we believe that Milne cosmology (ref. 54) is the best model for cosmology. Here the observed characteristics of the red shift can be explained by the dynamics of the expansion motion. Here, the expansion does not follow the perfect cosmological principle, and there exists reference frames that are privileged. The cosmological redshift, and isotropy are explained by this model without recourse to expanding 4D space-time. Again, this is an area that requires much work.

## 22.6     WORMHOLES AND TIME MACHINES

The wormhole is a short cut through space and time. It is synonymous with a spacewarp, or a warping, bending, or folding of 4D space-time. These objects are usually associated with very intense gravitational fields, and therefore highly curved 4D space-time. Wormholes come in at least two types: inter-universe wormholes, which connect 'our' universe with 'another' universe, and intra-universe wormholes, which connect two distant regions of our universe with each other. A time machine is an object or system that permits one to travel into the past. Paradoxes arise because once back in the past one should be able to influence one's own future (which is one's own past) by some physical means. One example of this is the famous 'grandfather paradox'.

In EMQG, wormholes and time machines are simply not possible! Recall that the curvature of 4D space-time is the result of the activities of the virtual particles of the quantum vacuum, which act on light and matter through a Fizeau-like scattering process (section 16.5). Curved, 4D space-time does not represent the actual geometry of the universe. The actual geometry of the universe cannot be warped, since it is simply a collection of a vast number of cells in the CA. In particular, if such enormous concentrations of matter are brought together that can cause the required curvatures, the laws of general relativity might breakdown. For example, the principle of equivalence may break down (and therefore, general relativity). Recall that the principle of equivalence is not perfect, and the direct graviton component of the interaction slightly imbalances the fall of a very large and a small mass (section 15.8). If the mass concentration is sufficiently large, the graviton component can compete with the normally strong electromagnetic component of the vacuum, and cause a significant imbalance. Similarly, time machines are not possible. A time machine would have to return an observer to a previous state of the CA, which cannot be done. To do so would require the ability to manipulate the CA on a numerical level.

## 23.     CONCLUSIONS

We have presented two new paradigms for physical reality, which restores a great unity to all physics. First, we have argued that our universe is a vast Cellular Automata (CA) simulation, the most massively parallel computer model known. This CA structure is a simple 3D geometric CA. All physical phenomena, including space, time, matter, and forces are the result of the interactions of *information* patterns, governed by the mathematical laws and the connectivity of the CA. Because of the way the CA functions, all the known global laws of the physics must result from the local mathematical law that governs each cell, and each cell contains the same mathematical law. We have seen that the CA structure automatically presents our universe with a speed limit for 'motion'. Quantum field theory requires that all forces are the result of the boson particle exchange process. This particle exchange paradigm fits naturally within CA theory, where the boson exchange represents the transfer of boson information patterns between (fermion) matter particles. All forces (gravity is no exception) originate from exchange processes dictated by quantum field theory.

Secondly, we have argued that quantum gravity (called EMQG) involves *two* pure force exchange processes. *Both* the photon and graviton exchanges occurring simultaneously inside a large gravitational field. Both particle exchange processes follow the particle exchange paradigm that was introduced in QED. We modified a new theory of inertia first introduced in ref. 5, which we call HRP inertia. In HRP inertia, inertia is the resulting electromagnetic force interaction of the charged 'parton' particles making up a mass with the background virtual photon field, which they called the zero point fluctuations (or ZPF). We modified HRP inertia, which we now call Quantum Inertia (or QI). The modification involves the introduction of new particle of nature called the masseon, which composes all (fermion) mass particles. The masseon is electrically charged (as well as possessing mass-charge). Quantum Inertia is based on the idea that inertial force is due to the tiny electromagnetic force interactions (not fully defined at this time) originating from each charged masseon particle of real matter undergoing relative acceleration with respect to the virtual, electrically charged masseon particles of the quantum vacuum. These tiny forces is the source of the total resistance force to accelerated motion in Newton's law 'F = MA', where the sum of each of the tiny masseon forces equals the total inertial force. The exact detail of the tiny electromagnetic forces is not known, and hence this remains a postulate of EMQG. This new approach to classical inertia automatically resolves the problems and paradoxes of accelerated motion introduced in Mach's principle, by suggesting that the virtual particles of the quantum vacuum serve as Newton's universal reference frame (which he called absolute space) for accelerated motion only. Thus, Newton was correct that it is the relative accelerated motion with respect to absolute space that somehow determines the inertia of a mass, but absolute space is totally useless in determining the *absolute* velocity of a mass.

The relative accelerated motion of the virtual masseons of the quantum vacuum with respect to the average motion of the real masseon particles that are bound in a mass can be used as the reference frame to define absolute mass (which is equivalent to special relativistic rest mass). This is contrary to special relativity, and to the known mass-velocity formula of special relativity. However, we have shown that the quantity of force transmitted between two objects in different inertial reference frames depends on the flux rate of the exchange particles. In other words, the number of particles exchanged per unit of time represents the magnitude of the force transmitted between the particles. If the relative velocity v<<c the exchange process appears almost the same as when the two particles are at rest, which is because the velocity of light is very high when compared to 'v', and the flux rate remains unaffected. Now as the relative recession velocity v -> c, it is comparable to the velocity of the exchange particle, and this effects the received flux rate. From the Lorentz time dilation, the timing of the exchange particle is altered, as given by $t = t_0 / (1 - v^2/c^2)^{1/2}$. The timing of the exchange particles is altered, and therefore the flux rate is altered as well, since flux has units of numbers of particles per unit time. Therefore, the magnitude of the force is reduced by: $F = F_0 (1 - v^2/c^2)^{1/2}$; where is $F_0$ is the magnitude of the force when at relative rest. Thus, we have concluded that the force is actually reduced in strength, not the mass that increases. The reduced force is less effective in accelerating a mass, which can be interpreted as a mass increase.

We found that gravity also involves the same 'inertial' electromagnetic force component that exists in an accelerated mass, which reveals the deep connection between inertia and gravity. Inside large gravitational fields there exists a similar quantum vacuum process that occurs for inertia, where the roles of the real charged masseon particles of the mass and the virtual electrically charged masseons of the quantum vacuum are reversed. Now it is the charged virtual particles of the quantum vacuum that are accelerating, and the mass particles are at relative rest. Furthermore, the general relativistic Weak Equivalence Principle (WEP) results from this common physical process existing at the quantum level in both gravitational mass and inertial mass. Gravity involves *both* the electromagnetic force (photon exchanges) and the pure gravitational force (graviton exchanges) that are occurring simultaneously. However, for a gravitational test mass, the graviton exchange process (only found in minute amounts in inertial reference frames) occurring between a large mass, the test mass, and the surrounding vacuum particles upsets perfect equivalence of inertial and gravitational mass, with the gravitational mass being slightly larger. One of the consequences of this is that if a very large, and a tiny mass are dropped simultaneously on the earth, the larger mass would arrive slightly sooner. Since this is in violation of the WEP, the strong equivalence principle is no longer applicable.

We proposed that all (fermions) matter particles get their quantum mass numbers from combinations of just one fundamental matter (and anti-matter) particle called the 'masseon' particle. The masseon has one fixed, (lowest) quanta of mass, which we called low level 'mass charge'. The masseon also carries either a positive or negative (lowest) quanta of electric charge. Furthermore, we proposed a new universal constant "i", defined as the inertial force existing between the quantum vacuum and a single charged masseon particle, which is accelerating at 1 g. This force represents the lowest possible quanta of inertial force (at 1g) and gravitational force on the earth. The masseon particle generates a fixed flux of gravitons (in analogy to electrical charge), with the flux rate being unaffected by relativistic motion. In EMQG, graviton exchanges are physically similar to photon exchanges in QED, with the same concept of positive and negative gravitational 'mass charge' carried by masseons and anti-masseons, with the ratio of the graviton to photon exchange force being $10^{-40}$. We found that graviton exchanges occur between a large mass and the surrounding virtual particles of the quantum vacuum, and they also directly occur between the large mass and a test mass. The electromagnetic force (<u>photon exchanges)</u> between the virtual particles and the test mass (occurring in inertial frames and in gravitational frames) is responsible for the equivalence of inertial and gravitational mass. The pure gravitational force (<u>graviton exchanges)</u> is responsible for the distortion of the (net statistical average) acceleration vectors of the virtual particles of the quantum vacuum near the earth (with respect to the earth). We also found that because there are equal numbers of virtual masseon and anti-masseon particles existing in the quantum vacuum everywhere (at any given instant of time) the cosmological constant must be very close to zero.

Since the state of the electrically charged virtual masseons of the quantum vacuum are very important in considerations of inertia and gravitation (and is responsible for the equivalence principle), we introduced a new paradigm for the origin of 4D, curved Minkowski space-time near a large mass. We found that 4D space-time is simply a consequence of the behavior of matter (fermions) and energy (photons) under the influence of the (net statistical average) downward accelerated 'flow' of the charged virtual particles of the quantum vacuum. This accelerated flow can be thought of as a special 'Fizeau-like fluid' (unknown to Einstein when he was developing relativity). Like in the Fizeau experiment (performed with constant velocity water) the behavior of light, clocks, and rulers are now affected by the <u>accelerated</u> 'flow' of the virtual particles of the quantum vacuum with respect to a large mass (and in accelerated frames). This accelerated flow can now **act** on motion of matter and light, to distort space and time. Furthermore, we have shown mathematically that the amount of space and time curvature based on EMQG corresponds to the same amount predicted by the Schwartzchild metric for the earth. This conclusion was based on the concept that photons scatter off the virtual particles of the quantum vacuum, thus maintaining the *same acceleration* as the downward 'flow' of virtual particles (in absolute CA units). Photons, however, still move at an absolute constant speed between the virtual particle scattering.

We have concluded that the speed of light in a universe where the quantum vacuum is absent of all virtual particles is greater than the light velocity in our actual universe. This is because in our universe the real photons scatter off the virtual particles, which introduce a **small random delay** before another real photon is re-emitted. In addition, the photon undergoes second-order (and higher) scattering processes (according to QED theory), which also contribute to extra delays. Thus the velocity of light that we observe in our universe is a net average statistical value based on the 'raw low-level' photon velocity **minus** an unknown but very large velocity penalty due to the total scattering process. Over classical distance scales, the average light velocity is constant in all directions because of the immensely large number of interactions that occur, and the remarkable regularity of the quantum vacuum.

On quantum distance scales Minkowski 4D space-time gives way to the secondary (quantized) absolute 3D space and separate absolute (quantized) time required by CA theory. Curved 4D space-time is replaced by a new paradigm where curvature is a result of pure particle interaction processes. Particles occupy definite locations on the CA cells, and particle states are evolved by a universal 'clock'. All interactions are *absolute,* because they depend on absolute space and time units on the CA. However, we cannot probe this scale, because we are unable to access the absolute cell locations, and numeric contents of the cells. In this realm, the photon particle (as well as the graviton) is an information pattern, that moves (shifts) with an absolute constant 'velocity', since it merely shifts from cell to neighboring cell in every 'clock' cycle of the CA.

Gravitational waves (GW) do *not* work the same way as the Electro-Magnetic Waves (EMW) in QED. The GW' is not directly quantized by gravitons. A periodic accelerating mass causes a corresponding periodic variation in the graviton flux at any point surrounding the mass. This is responsible for the initial periodic disturbance in the (net average statistical) acceleration vectors of the virtual particles of the quantum vacuum at that point. This periodic disturbance of the average acceleration vectors of the nearby virtual particles with respect to the original mass *is* the actual GW, which can carry energy to a very distant detector. We found that the GW carries a large energy density due to the huge numbers of virtual particles involved, which is quite capable of explaining the large stiffness of the GW predicted by general relativity. The 'stiffness' of the GW can induce a vibration in a large solid aluminum cylinder after traveling for hundreds of light years! Once the periodic virtual particle disturbance is started it is self –propagating, primarily by the mutual quantum vacuum virtual particle electromagnetic force interactions. The undisturbed acceleration vectors of the virtual particles in the path of the GW, now becomes disturbed (electromagnetically) when the GW arrives.

We have studied the Lense-Thirring effect (or inertial frame dragging) for a rotating mass like the earth using EMQG. The basic reason for inertial frame dragging is the finite speed of propagation of the graviton particle (the speed of light). This allows time for a spinning mass like the earth to rotate an observer a small amount while the graviton is still in flight as it propagates outwards. The finite velocity of the graviton particle along with the downward $GM/R^2$ acceleration component of the charged virtual particles of the quantum vacuum is entirely responsible for inertial frame dragging. The angle of deflection of the downward accelerating virtual particles of the quantum vacuum with respect to the non-rotating earth represents the amount of deflection of inertial frames, with the deflection angle varies with height. Recall that the direction of 4D space-time curvature is the same as the direction of the virtual particle acceleration. Therefore, this angle represents the shifting of the direction of 4D space-time curvature. The motion of any free falling (gravity-free) test mass is modified in the presence of the curved trajectory of virtual particles. If a test mass is in free-fall, the contribution of gravity to the motion of the mass is canceled out (at a point), and the resulting motion of the test mass is purely inertial. Therefore, the free fall frame is deflected by an angle that varies with height. We have calculated this angle at the earth's surface to be 42.5 milli-arc seconds which agrees well with general relativity, with a much simpler approach.

Thus, through a small reformulation of special relativity and the development of a new theory of quantum gravity that eliminates the classical ideas of space-time and matter in general relativity, we conclude that our universe is a vast, 3D Cellular Automata computer. In the process, we have discovered the hidden quantum interactions behind inertia and the principle of equivalence.

## 24. REFERENCES

(1) **FRACTALS, CHAOS, AND POWER LAWS** by Manfred Schroeder, 1991, Chap. 17. This contains an excellent introduction to Cellular Automata theory. Published by W.H. Freeman and Company.
(2) **CELLULAR AUTOMATA: THEORY AND EXPERIMENT** Edited by H. Gutowitz, 1991. Contains many reprints from Physica D. On page 254 there is an excellent atricle by Edward Fredkin, (the originator of Cellular Automata theory applied to fundamental physics) titled 'DIGITAL MECHANICS'.
(3) **LINEAR CELLULAR AUTOMATA** by H. V. McIntosh, May 20, 1987, LANL Archives.
(4) **WHAT HAS AND WHAT HASN'T BEEN DONE WITH CELLULAR AUTOMATA** by by H. V. McIntosh, Nov 10, 1990, LANL Archives.
(5) **INERTIA AS A ZERO-POINT-FIELD LORENTZ FORCE** by B. Haisch, A. Rueda, and H.E. Puthoff; Physical Review A, Feb. 1994. This landmark paper provides the first known proposal that inertia can be understood as the interactions of matter with the surrounding virtual particles, and proposes a solution to Mach's paradox.
(6) **THEORY AND EXPERIMENT IN GRAVITATIONAL PHYSICS** by C. M. Will, Chap. 2, pg. 24, Cambridge University Press, 1985.
(7) **GENERAL RELATIVITY AND EXPERIMENT** by T. Damour, Dec. 8, 1994, LANL Archives.
(8) **ESSENTIAL RELATIVITY: SPECIAL, GENERAL, AND COSMOLOGICAL**, $2^{nd}$ Edition, Springer-Verlag, 1977. Chap. 1. One of the best reviews of the status of inertia, the principle of equivalence, and Mach's principle.
(9) **THE RECURSIVE UNIVERSE: CHAOS, COMPLEXITY, AND THE LIMITS OF SCIENTIFIC KNOWLEDGE** by W. Poundstone, 1988, Oxford Univ. Press. Chap. 2 contains a very good survey of the Game of Life.
(10) **CAN WE SCALE THE PLANK SCALE?** by D. J. Gross, Physics Today, June 1989, pg.9.
(11) **THE ENERGETIC VACUUM** by H.E. Puthoff, Speculations in Science and Technology, vol. 13, No. 4, pg. 247-257, 1990.
(12) **Physical Review Letters**, Vol. 78, pg5.
(13) **THE DETECTION OF GRAVITATIONAL WAVES** edited by D.G. Blair, Chap. 1.
(14) **RELATIVITY OF MOTION IN VACUUM** by M. Jaekel, .., LANL archives, quant-ph/9801071, Jan.30 1998.
(15) **QUANTUM FIELDS IN CURVED SPACE** by N.D. Birrell & P.C.W. Davies, Cambridge Monographs, chap. 3.3, pg. 48
(16) **SOV. PHYS. – DOKL. 12, 1040** by A.D. Sakharov, 1968 and **THEOR. MATH. PHYS. 23, 435 (1975)** by A.D. Sakharov.
(17) **PARTICLE CREATION BY BLACK HOLES** by S. W. Hawking, Commun. Math. Phys. 43, 199-220 (1975).
(18) **QUANTUM FIELDS IN CURVED SPACE** by N.D. Birrell & P.C.W. Davies, Cambridge Monographs, chap. 8.2, pg. 264.
(19) **PHYSICS FOR SCIENTISTS AND ENGINEERS** by R. Serway, Chap. 5.
(20) **ESSENTIAL RELATIVITY** by W. Rindler, Chap. 5.
(21) **SPECIAL RELATIVITY** by A.P. French, Chap. 7, pg.214.
(22) **GENERAL RELATIVITY** by I.R. Kenyon, Chap. 2.
(23) **RADIATION FROM A UNIFORMLY ACCELERATED CHARGE AND THE EQUIVALENCE PRINCIPLE** by S. Parrott, LANL Archives, GR-QC/9303025V4, Jan. 23, 1996.

(24) **GENERAL RELATIVITY AND EXPERIMENT** by T. Damour, LANL Archives, GR-QC / 9412024, Dec. 8, 1994.
(25) **PHYS. Z. 19**, 156-63 Lense J and Thirring H (English Translation 1984 Gen. Rel. Grav. 16 711-50).
(26) **TEST OF LENSE-THIRRING ORBITAL SHIFT DUE TO SPIN** by I. Ciufolini, F. Chieppa, D. Lucchesi, and F. Vespe, Class. Quantum Grav. (1997) 2701-2726.
(27) **NEAR ZERO** Edited by J.D. Fairbank, Chap. VI.2 – Gravitomagnetism by Kip S. Thorne.
(28) **NEAR ZERO** Edited by J.D. Fairbank, Chap. VI.3(A) – The Stanford Relativity Gyroscope Experiment, C. W. F. Everitt.
(29) **SPLITTING THE ELECTRON** by B.Daviss, New Scientist, Jan. 31, 1998.
(30) **SUPERCONDUCTORS AND GRAVITY SHIELDING: REFERENCES AND EXPERIMENTS,** Internet Web address: **Error! Reference source not found.**.
(31) **RELATIVITY: THE GENERAL THEORY** by J.L. Synge, 1971, North-Holland, Amsterdam, p. IX.
(32) **VERH. DEUTSCH. PHYS. GES. 18, 83**, 1916, W. Nernst.
(33) **KOLTICK**, Measurement of electron charge screening.
(34) **DIGITAL MECHANICS: An Informational Process based on Reversible Cellular Automata** by Edward Fredkin, Physica D 45 (1990) 254-270.
(35) **INTRODUCTION TO THE THEORY OF RELATIVITY** by P.G. Bregmann, Chap. IV, pg.33.
(36) **UNIVERSITY PHYSICS** by H. Benson, Wiley, Chap. 39, pg. 797
(37) **ON THE "DERIVATION" OF EINSTEIN'S FIELD EQUATIONS** by S. Chandrasekhar, AJP Volume 40, pg. 224 (1972).
(38) **ESSENTIAL RELATIVITY** by W. Rindler, Chap. 1, pg. 10 (Rise and fall of Absolute Space).
(39) **GRAVITATION AND COSMOLOGY** by S. Weinberg, Chap. 8, pg. 179.
(40) **VERIFICATION OF THE EQUIVALENCE OF GRAVITATIONAL AND INERTIAL MASS FOR THE NEUTRON** by L. Koester, Physical Rev. D, Vol. 14, Num. 4, pg.907 (1976).
(41) **DOES THE FIZEAU EXPERIMENT REALLY TEST SPECIAL RELATIVITY** by G. Clement, Am. J. Phys. 48(12), Dec. 1980.
(42) **THE FEYNMAN LECTURES ON PHYSICS** by Feynman, Leighton, and Sands, Vol. 1, Chap. 31 The Origins of the Refractive Index.
(43) **OPTICAL COHERENCE AND QUANTUM OPTICS** by L. Mandel and E. Wolf., Cambridge
(44) **POTENTIAL THEORY IN GRAVITY AND MAGNETIC APPLICATIONS** by R.J. Blakely, Cambridge.
(45) **GRAVITATIONAL WAVES ON THE BACK OF AN ENVELOPE** by B. Schutz, AM. J. Phys. 52(5), pg. 412 (1984).
(46) **BLACK HOLES: THE MEMBRANE PARADIGM** edited by K. Thorne, R. Price, D. Macdonald.
(47) **GRAVITATIONAL MEISSNER EFFECT** by R.P. Lano, HEP-TH / 9603077, Mar. 12, 1996.
(48) **GRAVITOMAGNETIC INDUCTION DURING THE COALESCENCE OF COMPACT BINARIES** by S. L. Shapiro, Phys. Review Letters, Vol. 77, Num. 22, pg. 4487 (1996).
(49) **GRAVITATIONAL LORENTZ FORCE AND THE DESCRIPTION OF THE GRAVITATIONAL INTERACTION** by V.C. Andrale and J.G. Pereira, gr-qc / 9703059 (1997).
(50) **GRAVITATION** by C.W. Misner, K.S. Thorne, J.A. Wheeler, Freeman, 1973.
(51) **THE LENSE-THIRRING EFFECT AND MACH'S PRINCIPLE** by H. Bondi and J. Samuel, gr-qc / 9607009 (Jul. 4, 1996).
(52) **THE COSMOLOGICAL CONSTANT IS PROBABLY ZERO** by S.W. Hawking, Physics Letters, Vol. 134B, Num. 6 (1984), pg. 403.

(53) **FAREWELL TO GENERAL RELATIVITY** by K. Dalton, physics / 9710001 (Oct. 1, 1997) concerns itself with the problem energy conservation and the failure to energy, momentum, and stress of the gravitational field.

(54) **PRINCIPLES OF PHYSICAL COSMOLOGY** by P.J.E. Peebles, Princeton Series, 1993, pg. 199.

## 24. FIGURE CAPTIONS

The captions for the figures are shown below:

Figures 1 to 4: Principle of equivalence for Stationary Mass on the Earth and in a Rocket
Figure 5: Virtual Particles Bonded under free fall in a Rocket (1g)
Figure 6: Virtual Particles Bonded under free fall near the Earth
Figure 7: Virtual Particle Pattern for the Earth and Moon in Free Fall near the Sun
Figure 8: Virtual Particle Pattern for the Earth and Moon in Free Fall in a Rocket
Figure 9: Inertial Frame Dragging or Lense-Thirring Effect
Figure 10: Motion of Real Photons in the Presence of Virtual Particles Near Earth
Figure 11: Motion of Real Photons in Rocket Accelerating at 1g
Figure 12: Figure too large, Available from Authors
Figure 13: Figure too large, Available from Authors
Figure 14: Block Diagram of Relationship of CA and EMQG with Physics
Figure 15: Simplified Motion of a Photon Information Pattern
Figure 16: Light Velocity Measurement from two Different Observers
Figure 17: Definition of an Inertial Reference Frame
Figure 18: Figure too large, Available from Authors

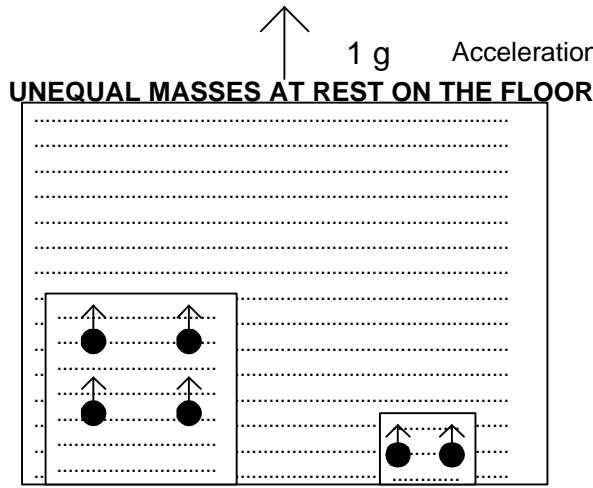
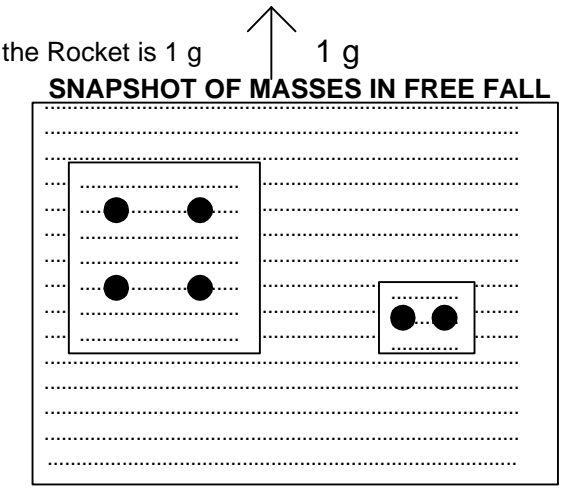

**Figure #1** - Masses '2M' and 'M' at rest on the floor of the rocket

**Figure #2** - Masses '2M' and 'M' in free fall inside of a rocket

**LEGEND**:
- **.** = A virtual particle of the quantum vacuum (taken as the rest frame)
- ↑• = A real mass particle undergoing relative upward acceleration of 1g
- • = A real matter particle at relative rest with respect to the vacuum

Equivalence

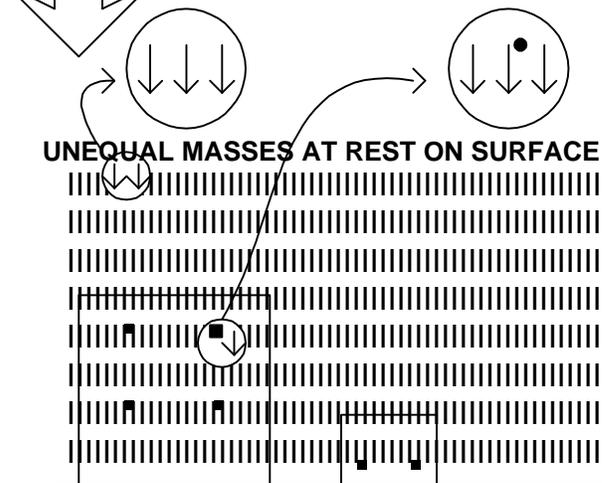
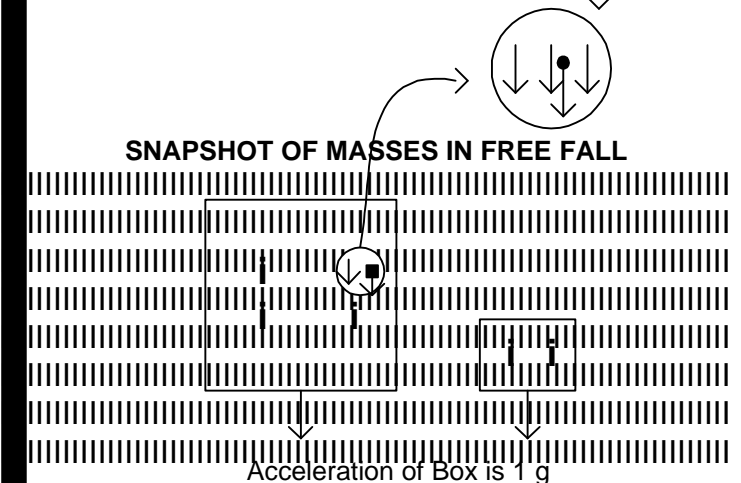

Surface of the Earth where gravity produces a 1 g acceleration

**Figure #3** - Masses '2M' and 'M' at rest on Earth's surface

**Figure #4** - Masses '2M' and 'M' in free fall above the Earth

**LEGEND:**
- **l** = Relative downward acceleration (1g) of a virtual particle
- **i** = Relative downward acceleration (1g) of a real matter particle
- **.** = A real stationary matter particle (with respect to the earth's center)

## **FIGURES 1 TO 4 - THE PRINCIPLE OF EQUIVALENCE FOR A STATIONARY MASS ON THE EARTH AND INSIDE A ROCKET**

FIGURE #5 - VIRTUAL PARTICLES BONDED UNDER FREE FALL IN A ROCKET (1G)

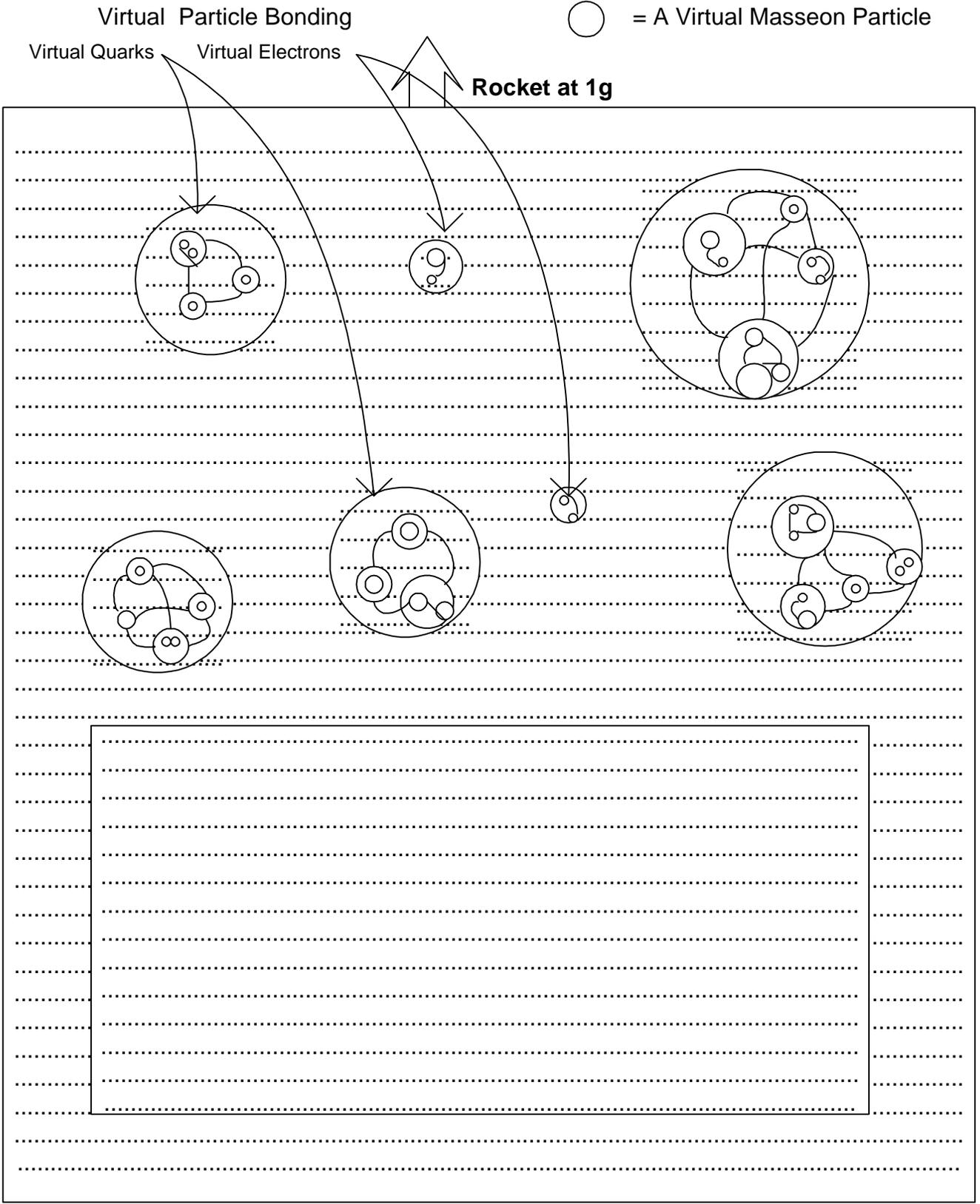

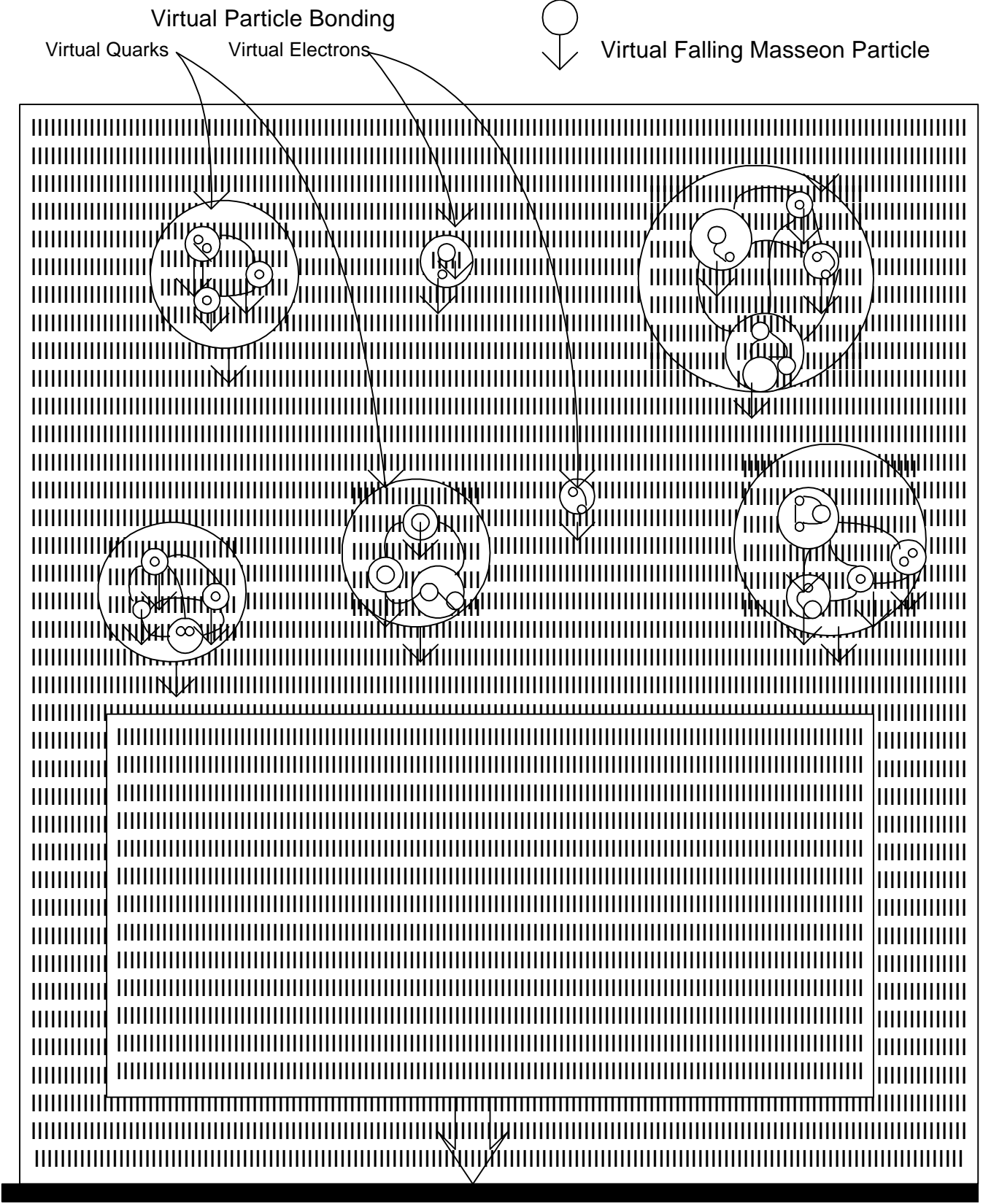

FIGURE #6 - VIRTUAL PARTICLES BONDED UNDER FREE FALL NEAR THE EARTH

### **Figure #7 - VIRTUAL PARTICLE PATTERN FOR THE EARTH AND MOON IN FREE FALL NEAR THE SUN**

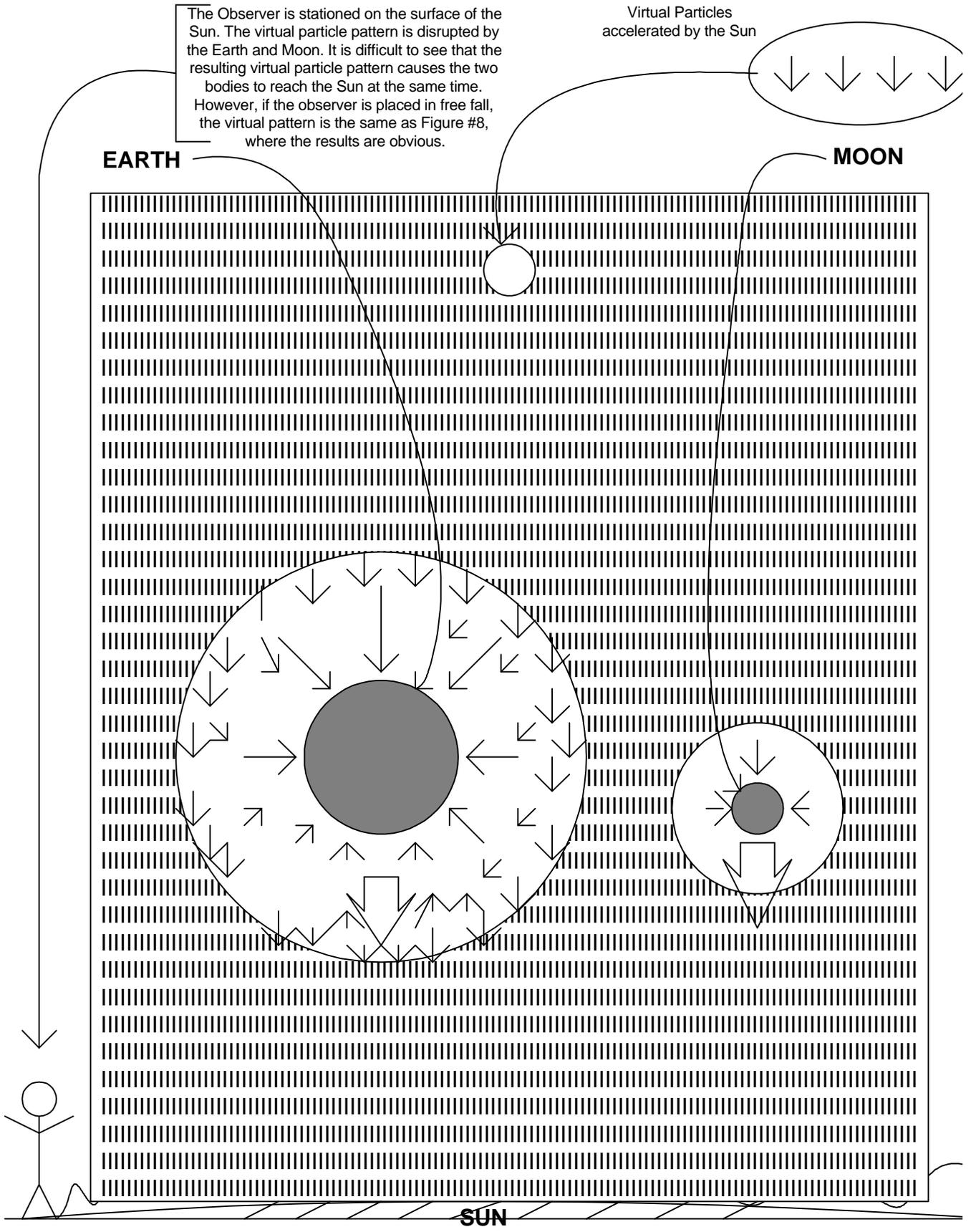

The Observer is stationed on the surface of the Sun. The virtual particle pattern is disrupted by the Earth and Moon. It is difficult to see that the resulting virtual particle pattern causes the two bodies to reach the Sun at the same time. However, if the observer is placed in free fall, the virtual pattern is the same as Figure #8, where the results are obvious.

Virtual Particles accelerated by the Sun

EARTH

MOON

SUN

**Figure #8 - VIRTUAL PARTICLE PATTERN FOR THE EARTH AND MOON IN FREE FALL IN A ROCKET**

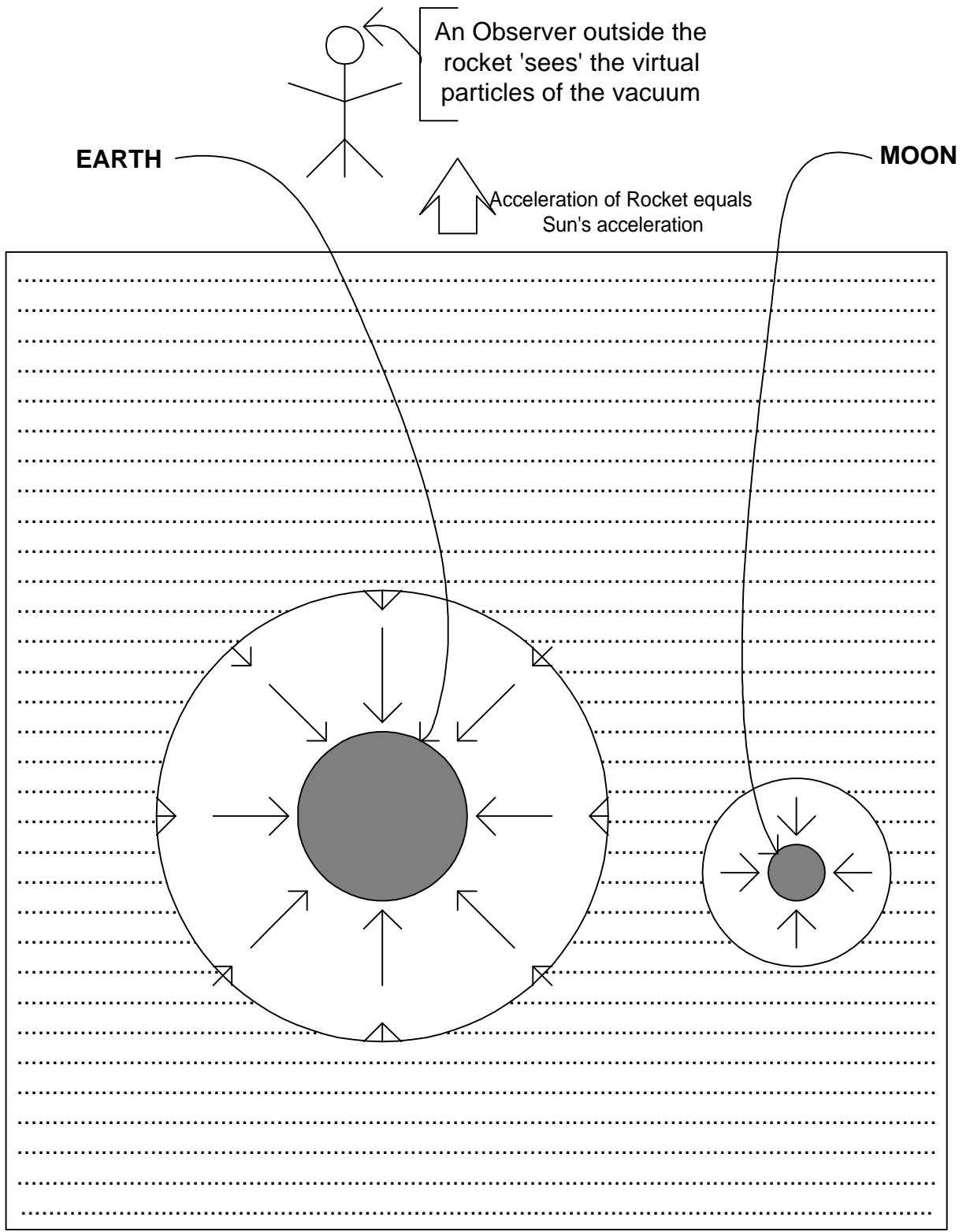

The Earth and Moon arrive on the floor of the huge rocket at the same time (the floor simply moves up to meet these bodies). But now, the virtual masseon particles near these two bodies are distorted and interacting with the real masseons in these two bodies.

# Figure #9 - INERTIAL FRAME DRAGGING OR LENSE-THIRRING EFFECT

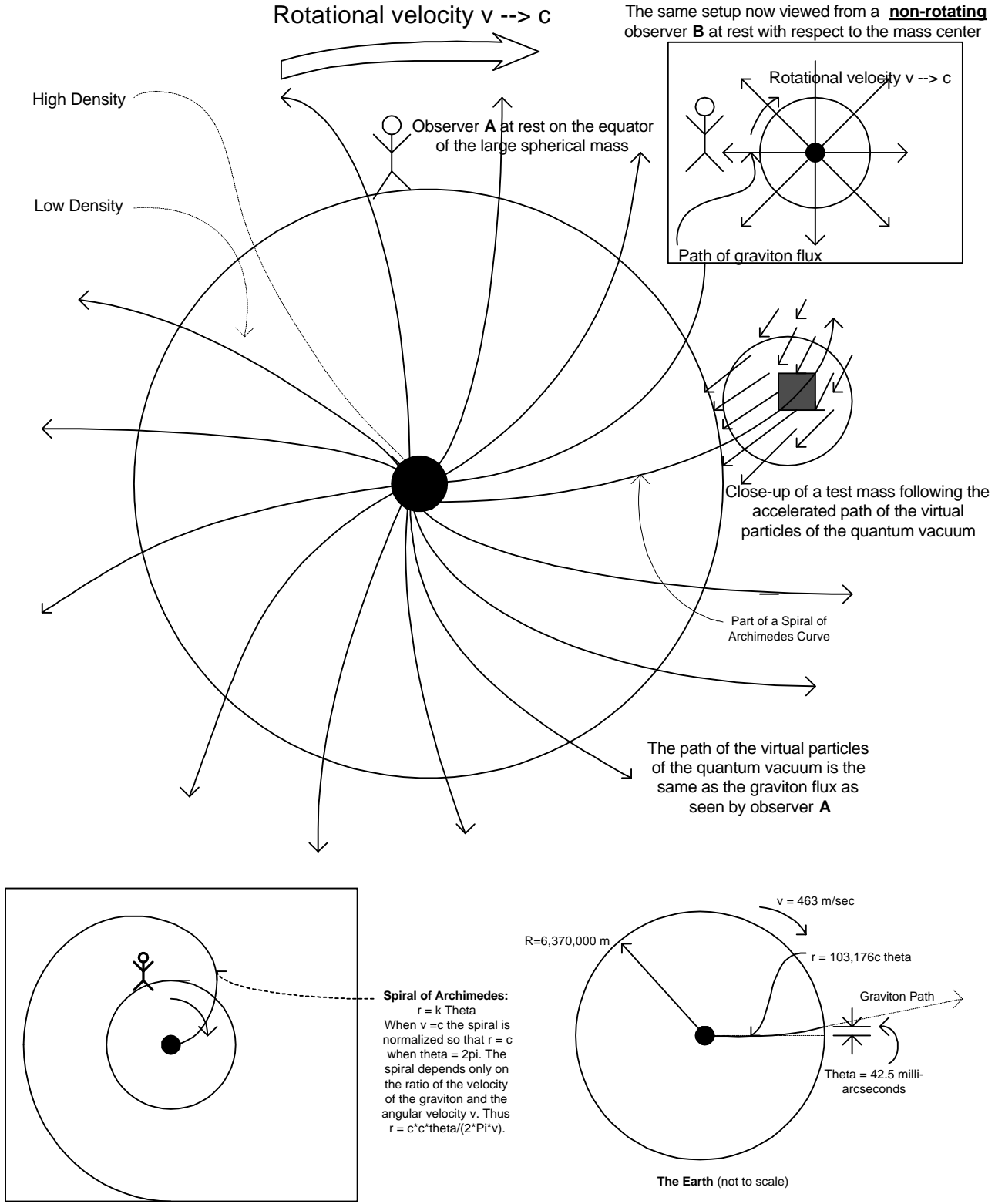

### Figure #10 - MOTION OF REAL PHOTONS IN THE PRESENCE OF VIRTUAL PARTICLE NEAR EARTH

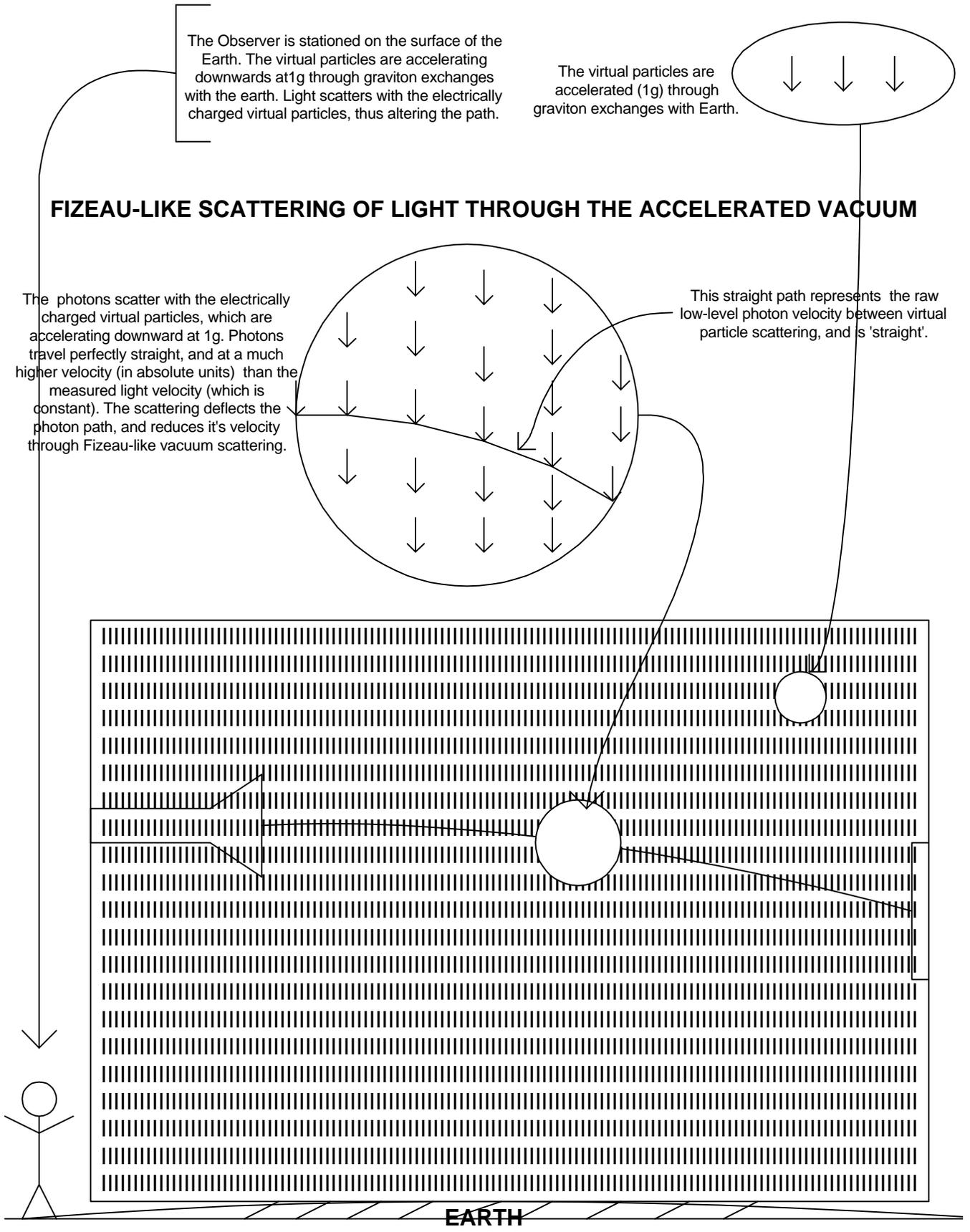

The Observer is stationed on the surface of the Earth. The virtual particles are accelerating downwards at 1g through graviton exchanges with the earth. Light scatters with the electrically charged virtual particles, thus altering the path.

The virtual particles are accelerated (1g) through graviton exchanges with Earth.

## FIZEAU-LIKE SCATTERING OF LIGHT THROUGH THE ACCELERATED VACUUM

The photons scatter with the electrically charged virtual particles, which are accelerating downward at 1g. Photons travel perfectly straight, and at a much higher velocity (in absolute units) than the measured light velocity (which is constant). The scattering deflects the photon path, and reduces it's velocity through Fizeau-like vacuum scattering.

This straight path represents the raw low-level photon velocity between virtual particle scattering, and is 'straight'.

**EARTH**

### **Figure #11 - MOTION OF REAL PHOTONS IN A ROCKET ACCELERATING AT 1g**

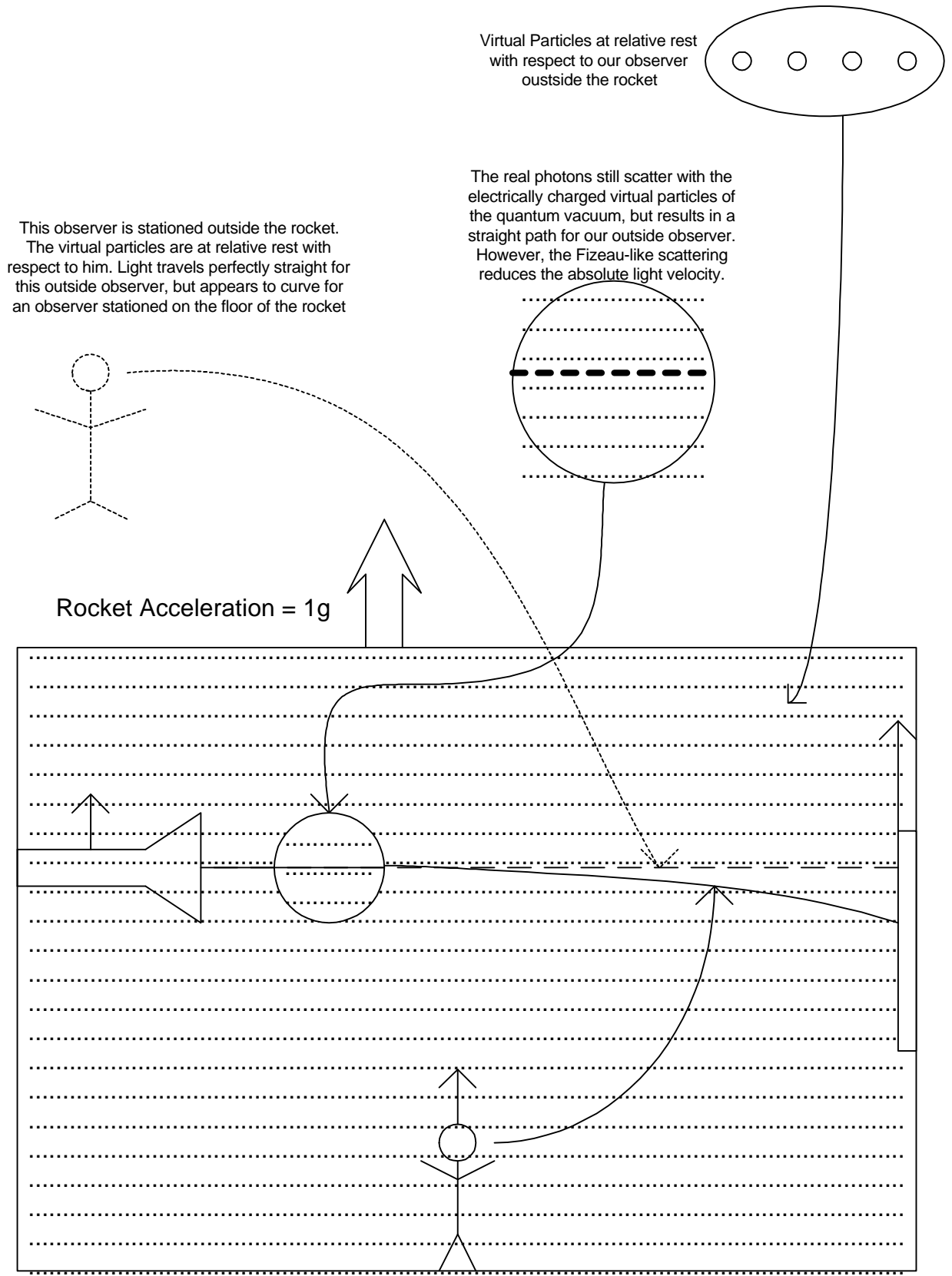

**Figure #14 - BLOCK DIAGRAM OF RELATIONSHIP OF CA AND EMQG WITH PHYSICS**

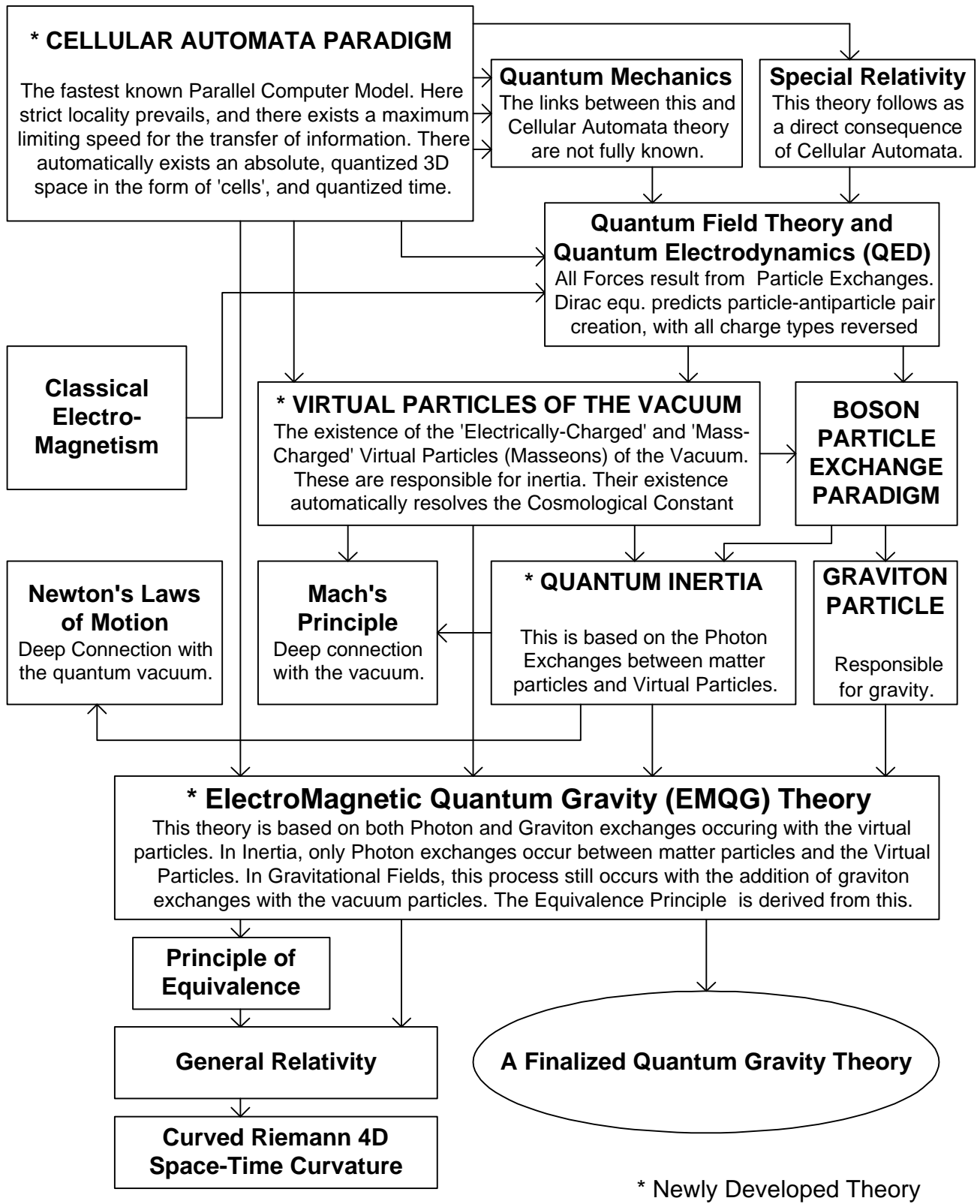

* Newly Developed Theory

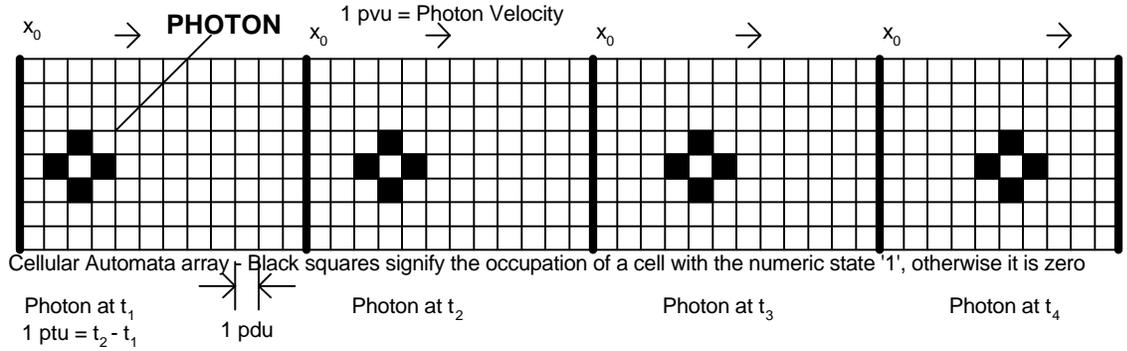

**Figure #15** - Simplified model of the motion of the photon information pattern on the CA.
The photon information pattern moves 1 plank unit to the right at every plank 'clock cycle'
(Note: The photon is actually an oscillating wavepattern (the wavefunction not shown in this simplified diagram)

**Absolute CA units:** 1 pdu is the shifting of information by 1 cell; 1 ptu is the time to shift 1 cell; 1 pvu = photon velocity

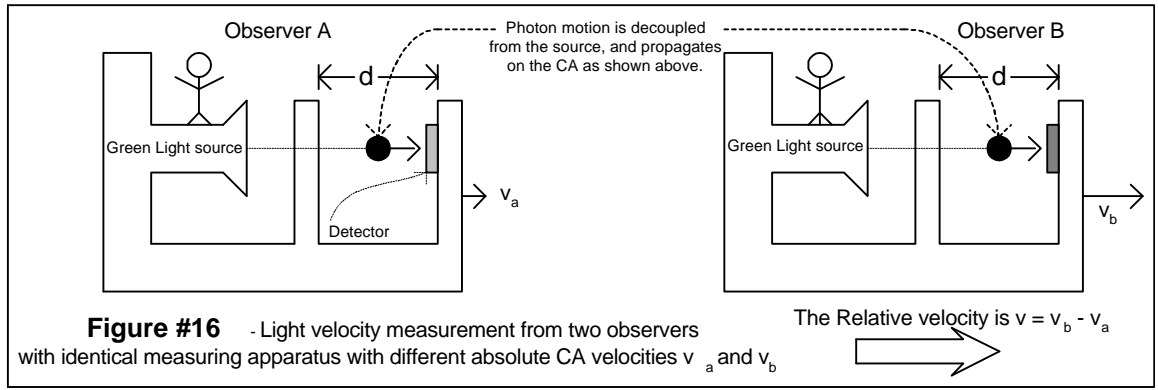

**Figure #16** - Light velocity measurement from two observers with identical measuring apparatus with different absolute CA velocities $v_a$ and $v_b$

The Relative velocity is $v = v_b - v_a$

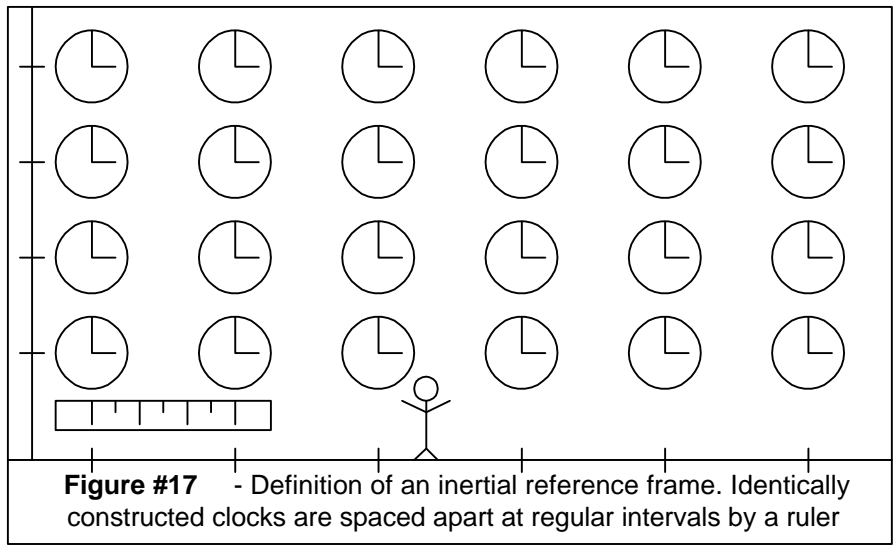

**Figure #17** - Definition of an inertial reference frame. Identically constructed clocks are spaced apart at regular intervals by a ruler